\documentclass[longauth]{aa}  

\usepackage{graphicx}
\usepackage{bm}
\usepackage{txfonts,xcolor}
\usepackage{multirow}
\usepackage{booktabs}
\usepackage{arydshln}

\begin{document} 

   \title{The GRAVITY young stellar object survey}
     \subtitle{XIII. Tracing the time-variable asymmetric disk structure in the inner AU of the Herbig star \object{HD\,98922}}

\author{GRAVITY Collaboration\thanks{GRAVITY is developed in a collaboration by the Max Planck Institute for Extraterrestrial Physics, LESIA of the Paris Observatory, and IPAG of the Université Grenoble Alpes/CNRS, the Max Planck Institute for Astronomy, the University of Cologne, the Centro de Astrofísica e Gravitação, and the European Southern Observatory.}:
          V. Ganci\inst{1,2}
          \and L. Labadie\inst{1}
          \and K. Perraut\inst{3}
          \and A. Wojtczak\inst{1}
          \and J. Kaufhold\inst{1}
          \and M. Benisty\inst{3}
          \and E. Alecian\inst{3}
          \and G. Bourdarot\inst{8}
          \and W. Brandner\inst{4}
          \and A. Caratti o Garatti\inst{4,5,6}
          \and C. Dougados\inst{3}
          \and R. Garcia Lopez\inst{4,5}
          \and J. Sanchez-Bermudez\inst{4,9}
          \and A. Soulain\inst{3}
          \and A. Amorim\inst{12,14} 
          \and J.-P. Berger\inst{3}  
          \and P. Caselli\inst{7}     
          \and Y. Cl\'enet\inst{10} 
          \and A. Drescher\inst{7}
          \and A. Eckart\inst{1,2}
          \and F. Eisenhauer\inst{7}
          \and M. Fabricius\inst{7}
          \and H. Feuchtgruber\inst{7}
          \and P. Garcia\inst{12,13}
          \and E. Gendron\inst{10} 
          \and R. Genzel\inst{7}
          \and S. Gillessen\inst{7}
          \and S. Grant\inst{7}
          \and G. Hei\ss el\inst{10}  
          \and T. Henning\inst{4}  
          \and M. Horrobin\inst{1}              
          \and L. Jocou\inst{3}
          \and P. Kervella\inst{10} 
          \and S. Lacour\inst{10} 
          \and V. Lapeyr\`ere\inst{10} 
          \and J.-B. Le Bouquin\inst{3}
          \and P. L\'ena\inst{10}
          \and D. Lutz\inst{7}
          \and F. Mang\inst{7}
          \and N. Moruj\~ao\inst{12,13}
          \and T. Ott\inst{7}
          \and T. Paumard\inst{10} 
          \and G. Perrin\inst{10} 
          \and D. Ribeiro\inst{7}
          \and M. Sadun Bordoni\inst{7}
          \and S. Scheithauer\inst{4}
          \and J. Shangguan\inst{7}
          \and T. Shimizu\inst{7}
          \and C. Straubmeier\inst{1}
          \and E. Sturm\inst{7}
          \and L. Tacconi\inst{7} 
          \and E. van Dishoeck\inst{7,8}
          \and F. Vincent\inst{10}
          \and J. Woillez\inst{11}
}

    \institute{I. Physikalisches Institut, Universit\"at zu K\"oln, Z\"ulpicher Str. 77, 50937, K\"oln, Germany\\
    \email{ganci@ph1.uni-koeln.de}
    \and
    Max-Planck-Institute for Radio Astronomy, Auf dem H\"ugel 69, 53121 Bonn, Germany
    \and
    Univ. Grenoble Alpes, CNRS, IPAG, 38000 Grenoble, France
    \and
    Max Planck Institute for Astronomy, K\"onigstuhl 17, 69117 Heidelberg, Germany
    \and 
    School of Physics, University College Dublin, Belfield, Dublin 4, Ireland
    \and
    INAF -- Osservatorio Astronomico di Capodimonte, via Moiariello 16, 80131 Napoli, Italy
    \and
    Max Planck Institute for Extraterrestrial Physics, Giessenbachstrasse, 85741 Garching bei M\"unchen, Germany
    \and
    Sterrewacht Leiden, Leiden University, Postbus 9513, 2300 RA Leiden, The Netherlands
    \and
    Instituto de Astronom\'ia, Universidad Nacional Aut\'onoma de M\'exico, Apdo. Postal 70264, Ciudad de M\'exico 04510, Mexico
    \and
    LESIA, Observatoire de Paris, PSL Research University, CNRS, Sorbonne Universit\'es, UPMC Univ. Paris 06, Univ. Paris Diderot,Sorbonne Paris Cit\'e, France
    \and
    European Southern Observatory, Karl-Schwarzschild-Str. 2, 85748, Garching bei M\"unchen, Germany
    \and
    CENTRA – Centro de Astrof\'{\i}sica e Gravita\c{c}\~{a}o, IST, Universidade de Lisboa, 1049-001 Lisboa, Portugal 
    \and
    Faculdade de Engenharia da Universidade do Porto, Rua Dr. Roberto Frias, s/n, 4200-465 Porto, Portugal
    \and
    Universidade de Lisboa – Faculdade de Ci\^{e}ncias, Campo Grande, 1749-016 Lisboa, Portugal
}

\date{Received xx, xxxx; accepted xx, xxxx}

  \abstract
   {Temporal variability in the photometric and spectroscopic properties of protoplanetary disks is common in young stellar objects. However, evidence pointing toward changes in their morphology over short
timescales has only been found for a few sources, mainly due to a lack of high-cadence observations at high angular resolution. Understanding this type of variation could be important for our understanding of phenomena related to disk evolution.
   }
   {We study the morphological variability of the innermost 
   circumstellar environment of \object{HD\,98922}, focusing on its dust and gas content.
   }
   {Multi-epoch observations of \object{HD\,98922} at milliarcsecond resolution with VLTI/GRAVITY in the K-band at low (R=20) and high (R=4000) spectral resolution are combined with VLTI/PIONIER archival data covering a total time span of 11 years. We interpret the interferometric visibilities and spectral energy distribution with geometrical models and through radiative transfer techniques using the code MCMax. We investigated high-spectral-resolution quantities (visibilities and differential phases)  to obtain information on the properties of the HI Brackett-$\gamma$ (Br$\gamma$)-line-emitting region.
   }
   {
   Comparing observations taken with similar \textit{(u,$\varv$)} plane coverage, we find that the squared visibilities do not vary significantly, whereas we find strong variability in the closure phases, suggesting temporal variations in the asymmetric brightness distribution associated to the disk.
   Our observations are best fitted by a model of a crescent-like asymmetric dust feature located at $\sim$1\,au and accounting for $\sim$70\,\% of the near-infrared (NIR) emission. The feature has an almost constant magnitude and orbits the central star with a possible sub-Keplerian period of $\sim$12\,months, although a 9\,month period is another, albeit less probable, solution. The radiative transfer models show that the emission originates from a small amount of carbon-rich ($25\%$) silicates, or quantum-heated particles located in a low-density region. Among different possible scenarios, we favor hydrodynamical instabilities in the inner disk that can create a large vortex. 
   The high spectral resolution differential phases in the Br$\gamma$ line show that the hot-gas compact component is offset from the star and in some cases is located between the star and the crescent feature. The scale of the emission does not favor magnetospheric accretion as a driving mechanism. The scenario of an asymmetric disk wind or a massive accreting substellar or planetary companion is discussed. 
   }
   {With this unique observational data set for \object{HD\,98922}, we reveal morphological variability in the innermost 2\,au of its disk region. This property is possibly common to many other protoplanetary disks, but is not commonly observed due to a lack of high-cadence observation. It is therefore important to pursue this approach with other sources for which an extended dataset with PIONIER, GRAVITY, and possibly MATISSE is available.
   }


   \maketitle


\section{Introduction}
\label{sec:Int}

In the last decade, our knowledge of protoplanetary disks around young stars has grown considerably thanks to the drastic improvement of observing facilities. It is nowadays established that such disks show different substructures  across the optical to submillimeter
wavelength range, such as rings, gaps, spiral arms, vortices, warps, and shadows on scales of tens of au  (\cite{Huang2018}, \cite{Garufi2018} and references therein). 
Thanks to long-baseline infrared interferometry, the morphology of inner disks at scales of less than 1 au has been revealed (e.g., \citealt{Monnier2005,Eisner2014,Lazareff2017,Perraut2019,Kluska2020}). 
Temporal photometric variability is a common property of YSOs (e.g., \citealt{Kospal2012, Rice2015, Wolk2018, Guarcello2019, Robinson2019}) and may originate, for instance, from the presence of cool or hot spots on the stellar surface, variable accretion, or changes in the inner disk structure leading to partial occultation of the central star and variable dimming of the system as in the case of "dippers" or UX-Ori type objects. 
The question of time-variable morphology of the inner disk has been tackled for a handful of sources (e.g., \citealt{Kluska2016, Kobus2020, Sanchez2021}). The scarcity of such studies is mainly due to the fact that the long-period high-cadence observations of the same object required for such analyses are seldom available.
In this context, observations with a large temporal baseline using the PIONIER \citep{Lebouquin2011} and GRAVITY instruments \citep{Abuter2017} at the VLTI provide a unique opportunity to probe the origin of the variability in the brightness distribution of the innermost regions of YSOs.\\
\\
In the present paper, we study \object{HD\,98922}, a B9Ve/A2III Herbig star \citep{Hales2014,Garatti2015} 
 characterized by a spectral energy distribution (SED) with a high near-infrared (NIR) excess and a low far-infrared (FIR) excess \citep{Garufi2022}. Classifications of the star and estimates of its parameters have shown discrepancies over the years. Indeed, the distance estimates were drastically divergent before the Gaia era, with values going from $\sim$\,450\,pc \citep{Garatti2015} to $\sim$\,1150\,pc \citep{Leeuwen2007}, which affected the derivation of parameters such as luminosity, mass, and age. 
For instance, \cite{Lee2016} suggested classification as a post-main sequence giant, but accurate Gaia parallax information supports the classification of \object{HD\,98922} as a Herbig Be star \citep{Vioque2018,Arun2019}. 
Using VLT/SPHERE, \cite{Garufi2022} set a lower limit of at least 200\,au in radius for the physical extent of the dust disk. However, comparison with an ALMA image suggests an even larger radius of $\lesssim$\,500\,au \citep{Garufi2022}. At scales of smaller than $10\,$au, \cite{Menu2015} constrained the N-band-emitting dust disk radius to $\sim$\,7.2\,au using VLTI/MIDI interferometric observations. 
\begin{table}[t]
\centering
\caption{\centering \object{HD\,98922} stellar parameters}
\begin{tabular}{llcc} 
\hline
\hline
Parameter & Unit & Value & Reference \\
\hline
Distance                        & pc                     & $650.9 \pm 8.8$   & 1 \\
Age                     & Myr                    & $[0.2, 0.7]$      & 2 \\
M$_{\star}$             & M$_{\odot}$            & $[5.0, 7.0]$      & 2,3 \\
R$_{\star}$             & R$_{\odot}$            & $11.45\pm0.36$    & 3 \\
log\,L$_{\star}$        & L$_{\odot}$            & $3.16\pm0.02$     & 3 \\
T$_{\mathrm{eff}}$              & K                      & $10500 \pm 125$   & 3 \\
log\,$g$                        & cm\,s$^{-2}$           & $3.5 \pm 0.2\,$   & 4 \\
$[$Fe/H$]$                      &                        & $-0.5 \pm 0.2$    & 4 \\
$\varv\,$sin$i$                 & km\,s$^{-1}$                   & $39.0 \pm 5.3$    & 5 \\
P$_{rot}$                       & d                   & 4--8    & 5 \\

log\,$\dot{\rm M}_{acc}$& M$_{\odot}$\,yr$^{-1}$ & $[-7.0, -5.0]$    & 6,3 \\
\hline
\end{tabular}\label{tab:StellarParameters}
\tablefoot{ 
(1) \cite{Gaia2021}; (2) \cite{Garufi2022}; (3) \cite{Guzman2021}; (4) \cite{Garatti2015}; (5) \cite{Aarnio2017}; (6) \cite{Fairlamb2015}.}
\end{table}\\
Furthermore, the system shows a CO-rich circumstellar disk, which  \cite{Hales2014} suggest is geometrically flat with an inner radius of $\sim$\,1\,au, extending to $\sim$\,200\,au. 
Other authors instead suggested a system with a flared CO disk with an inner radius of $\sim$\,5\,au and a flattened dust disk \citep{Vanderplas2015}. This transitional disk scenario was suggested because of the relatively strong polycyclic aromatic hydrocarbon (PAH) emission of \object{HD\,98922} and its similarities with two other objects, \object{HD\,101412} and \object{HD\,95881}, for which such a disk structure was already proposed \citep{Fedele2008, Verhoeff2010}. At smaller spatial scales, a rotating gaseous disk inside the dust sublimation radius was proposed to explain the [OI] emission-line profiles \citep{Acke2005}, while a wind or outflow was suggested to explain the P\,Cygni profiles in the H$\alpha$, Si\,II, and  He\,I lines \citep{Grady1996, Oudmaijer2011}. Finally, the Br$\gamma$ emission line was found to arise from a compact region of $\sim$\,0.65\,au in radius, possibly tracing magnetospheric accretion \citep{Kraus2008} or a disk wind \citep{Garatti2015}.\\
\\
Previous interferometric works (\citealt{Lazareff2017,Perraut2019,Kluska2020}) revealed the noncentrosymmetric brightness distribution of the inner disk of \object{HD\,98922}. 
Here, we exploit a VLTI/GRAVITY and VLTI/PIONIER multi-epoch data set to study the temporal morphological variability of the innermost circumstellar environment of this source, looking at the NIR continuum emission and the hot hydrogen gas. 
In Sect.~\ref{sec:Observations}, we present the observations and describe the interferometric data. In Sect.~\ref{sec:continuum_analysis}, we present the methodology used to analyze the continuum data and our results. In Sect.~\ref{sec:brg_analysis}, we present the methodology used to analyze the Br$\gamma$-emitting gas data and the ensuing results. In Sect.~\ref{sec:RT_modeling}, we present a new radiative transfer model for the source. In Sect.~\ref{sec:Discussion}, we discuss some possible interpretations of our results   in detail, and finally, in Sect.~\ref{sec:Summary}, we summarize our main findings.

\section{Observations and data reduction}\label{sec:Observations}
\subsection{The source}\label{ssec:source}

\object{HD\,98922} is a $\sim$6\,M$_\odot$ pre-main sequence star of less than 1\,Myr of age that shows a high accretion rate. We adopt the distance of 650.9\,$\pm$\,8.8\,pc from Gaia DR3 \citep{Gaia2022}, which we assume throughout the paper. The most up-to-date stellar parameters are listed in Table~\ref{tab:StellarParameters}. 
HD\,98922 is a group II source \citep{Juhasz2010} in the Meeus classification \citep{Meeus2001}, which is typically associated with a flat disk morphology. The K\,band emission was measured to be inside $\sim$1.5\,au \citep{Perraut2019} and the H\,band emission inside $\sim$1.2\,au \citep{Lazareff2017}.
\subsection{Observations}\label{ssec:small-obs}
\object{HD\,98922} was observed with VLTI/PIONIER \citep{Lebouquin2011} using the four 1.8 m Auxiliary Telescopes (ATs) in 21 different epochs between 2011 and 2016. Data were obtained using small- to large-baseline configurations for different epochs. The data consist of low-spectral-resolution (R$\, \approx \, $40) interferometric observables in the H\,band. 
The observations span a spatial frequency range between about 5 M$\lambda$ and 90 M$\lambda$ with a maximal angular resolution of $\lambda/2$B\,$\sim$\,1.25 mas for the longest baseline of 138.7 m, which corresponds to 0.81\,au at 650.9\,pc.
In total, 45 files were acquired and four files were discarded due to bad weather conditions. 
The description of the data  per epoch can be found in Table~\ref{tab:ObsLogPIONIER} along with the observation logs.\\
 \object{HD\,98922} was observed at 13 different epochs between 2017 and 2022 using GRAVITY, with the so-called astrometric, large-, medium-, and small-baseline configurations. 
The data consist of high-spectral-resolution (R$\, \approx \, $4000) observables recorded by the science channel (SC) detector across the K-band with individual integration times of 30\,s, as well as low-spectral-resolution (R$\, \approx \, $20) observables recorded using the fringe tracker (FT) detector. 
The spatial frequency ranges between about 5 M$\lambda$ and 65 M$\lambda$ with a maximal angular resolution of $\lambda/2$B\,$\sim$\,1.72 mas for the longest baseline of 129 m, which corresponds to 1.12\,au at the distance of HD\,98922. Each observation block corresponds to 5 minutes of observing time on the object. In total, 97 files were acquired. Detailed information on the GRAVITY observations is given Table~\ref{tab:ObsLogGRAVITY}. In summary, the PIONIER and GRAVITY dataset spans an 11 year period from 2011 to 2022.
\subsection{Data reduction}\label{sec:datared}
The PIONIER data are archival reduced and calibrated data retrieved from the JMMC Optical interferometry DataBase$\footnote{available at \texttt{http://oidb.jmmc.fr/index.html}}$. 
The errors bars were calculated using the reduction and calibration pipeline. 
These range from 0.25$^{\circ}$ to 10.1$^{\circ}$ for the closure phases and from 0.001 to 0.18 for the squared visibilities, depending on the atmospheric conditions.\\
The GRAVITY data were reduced and calibrated using the GRAVITY data reduction software \citep{Lapeyrere2014}. 
For the low-resolution FT data, we discarded the first spectral channel, which is typically affected by the metrology laser operating at 1.908\,$\mu$m. 
Following \cite{Perraut2019}, we applied a floor value on the error bars of 2\% for the squared visibilities and 1$^{\circ}$ on the closure phases as the error bars computed by the pipeline might be underestimated. \\
For the high-resolution science (SC) data, the pipeline produces single files containing 
the spectrum, the calibrated visibilities, and the calibrated differential phases. 
The observables for one epoch result from the averaging of N single files as listed in the fourth column of Table~\ref{tab:ObsLogGRAVITY}. Each averaged file is wavelength-calibrated using the position of the telluric absorption lines bracketing the Br$\gamma$ emission line and then corrected for the radial velocity of the 
star and the Earth's motion with respect to the LSR. 
The wavelength calibration is discussed in more detail in Appendix~\ref{apx:spctr-corr}.
The error bars are calculated by the reduction pipeline for the single files and are propagated through the averaging process. 
These range from 0.3\%\ to 1\% for the spectrum, from 0.001 to 0.01 for the squared visibilities, and from 0.5$^{\circ}$ to 2$^{\circ}$ for the differential phases, depending on the observation epoch.

\begin{figure*}
    \centering
    \includegraphics[width=\textwidth]{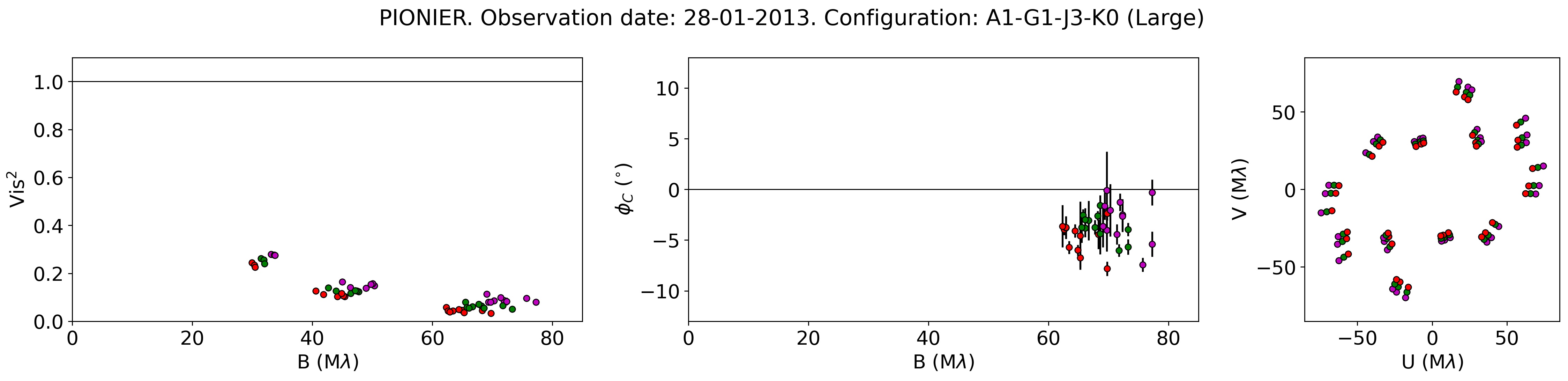}\\
    \includegraphics[width=\textwidth]{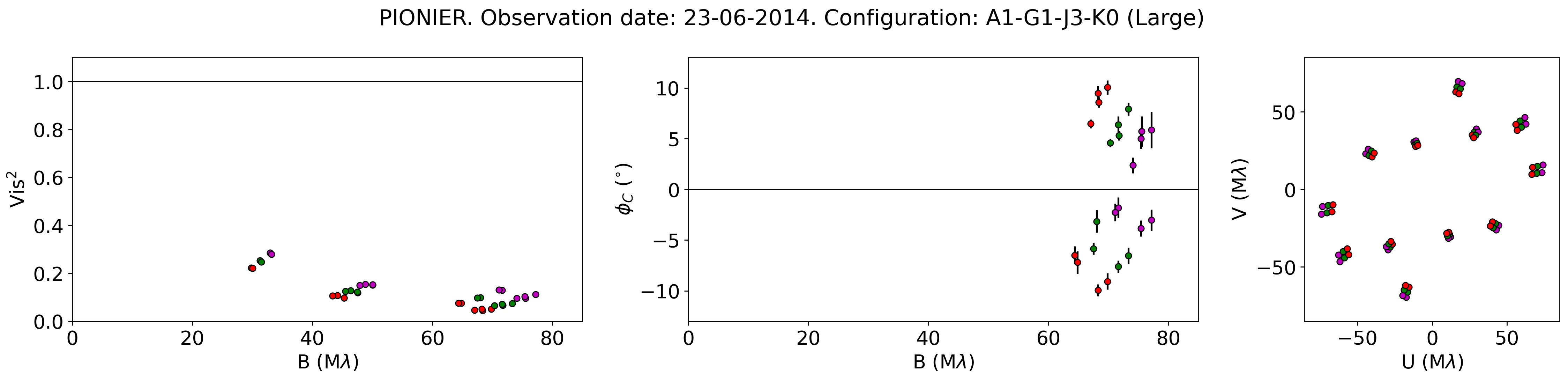}\\
    \includegraphics[width=\textwidth]{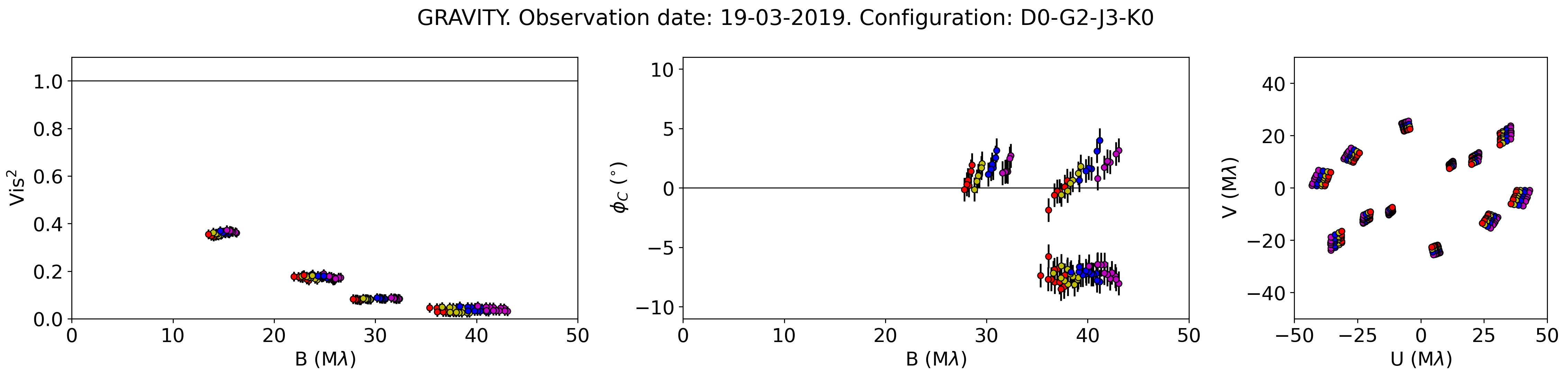}\\
    \includegraphics[width=\textwidth]{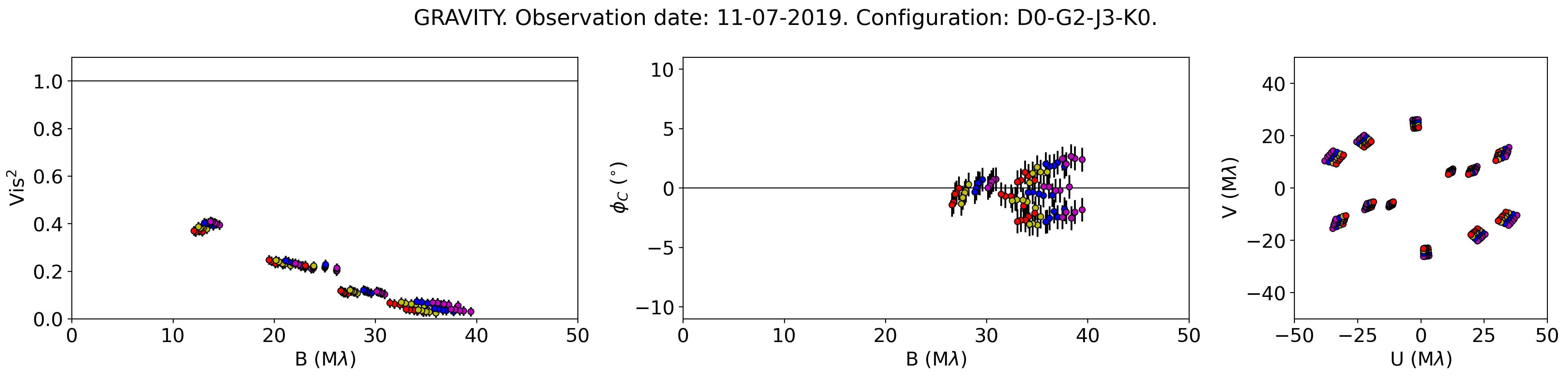}\\
    \caption{\object{HD\,98922} PIONIER (first two rows) and GRAVITY (last two rows) data, squared visibilities (left panel), closure phases (central panel), and \textit{(u,$\varv$)} plan coverage (right panel) for two epochs of the complete data set shown in Figs.~\ref{fig:PIONIER-Data+Mod} and~\ref{fig:GRAVITY-FT-Data}. Colors refer to the different spectral channels.}
    \label{fig:Continuum-data_intro}
\end{figure*}

\section{Results derived from the continuum interferometric data}\label{sec:continuum_analysis}

The reduced data are illustrated in Fig.~\ref{fig:Continuum-data_intro} for a few selected epochs and are shown in full in Figs.~\ref{fig:PIONIER-Data+Mod} and~\ref{fig:GRAVITY-FT-Data}. The plots show the calibrated squared visibilities, the closure phases, and the \textit{(u,$\varv$)} coverage. 
PIONIER and GRAVITY spatially resolves the H\,band and K\,band continuum emission of \object{HD\,98922} at all baselines and epochs, with visibilities ranging between 0 and 0.8. Clear closure-phase signals up to 20$^{\circ}$ with PIONIER and 40$^{\circ}$ with GRAVITY are detected for all epochs. The data therefore clearly suggest asymmetries in the brightness distribution of \object{HD\,98922} at spatial scales probed by the VLTI. 
Importantly, for the configurations with a similar \textit{(u,$\varv$)} coverage, we detect significant variations in the closure phases across the epochs. On the contrary, we observe that the change in the corresponding visibilities is not very strong, typically below V$^2$\,$\sim$\,0.05. 

\subsection{Modeling methodology}\label{subsec:continuum_methodology}

The observational results show that, for the epochs using VLTI configurations with a similar \textit{(u,$\varv$)} plane coverage, we see clear variations in the interferometric quantities and in particular in the closure phase signal. This indicates a noncentrosymmetric brightness distribution that is temporally variable. 
We adopt a classical approach in long-baseline interferometry based on the parametric fit of geometrical models to the squared visibilities and closure phase signals. As HD\,98922 is known from previous works to host a circumstellar disk (e.g., \cite{Kluska2020}), we focus our methodology on the analysis of a disk-like parametric model.

\subsubsection{Choice of the model}
A geometrical disk model with an azimuthal modulation is one possible solution to account for the asymmetric brightness distribution revealed by the nonzero closure phases \citep{Lazareff2017}; we adopt such a model here. We used chromatic geometric models that consist of a point-like central star, a scattered light component (called halo), and a circumstellar environment to fit the continuum interferometric quantities. The star is assumed to be unresolved and the halo to be fully resolved 
with a visibility of zero. 
The circumstellar emission is modeled through an azimuthally modulated wireframe \citep{Lazareff2017} with a radial brightness distribution given by
\begin{equation}\label{eq:flux-dist-lazareff}
\mathcal{F}(r) = \frac{1}{2\pi}\delta(r-{\rm a}_r)\cdot \Big( 1+\sum_{j=1}^m (c_j\cos j\phi + s_j\sin j\phi) \Big)
,\end{equation}
where $\phi$ is the polar angle. The order of the azimuthal modulation is taken in our case to be $m$\,=\,1. The wireframe is convolved by an ellipsoid kernel that regulates the width of the ring-like emission, and whose visibility is given by Eq.~9 of \cite{Lazareff2017}. Hence, the model can describe from infinitesimally thin rings to very wide rings tending to ellipsoids. It is described by ten parameters: the star flux contribution F$_s$;  the halo flux contribution F$_h$; the spectral index of the circumstellar disk k$_c$; the radius of the wireframe a$_r$ and of the Kernel a$_k$; the inclination $i$; the position angle (PA); the weighted contribution of a Gaussian or Lorentzian distribution $f$Lor; and the azimuthal modulation parameters $c_1$ and $s_1$. More details on the relation between the geometrical, physical, and fitted parameters is provided in Sect.~3.6 of \cite{Lazareff2017}.\\
The total complex visibility of the system at spatial frequencies $(u,\varv)$ and at wavelength $\lambda$ is therefore described by a linear combination of the three components as 
\begin{equation}\label{eq:Visibility-continuum}
    V(u,\varv,\lambda) = \frac{F_{\rm s}\,(\lambda/\lambda_0)^{k_{\rm s}} + F_{\rm c}\,(\lambda/\lambda_0)^{k_{\rm c}}\,V_{\rm c}(u,\varv)}{(F_{\rm s} + F_{\rm h})\,(\lambda/\lambda_0)^{k_{\rm s}} + F_{\rm c}\,(\lambda/\lambda_0)^{k_{\rm c}}},
\end{equation}
where $V_{\rm c}$ is the complex visibility of the circumstellar environment, and $F_{\rm s}$, $F_{\rm h}$, and $F_{\rm c}$ are the specific fractional flux contributions of the star, of the halo, and of the circumstellar environment, respectively, at $\lambda_{0}$\,=\,$\lambda_{K,0}$\,=\,2.15\,$\mu$m
(the wavelength of the central spectral channel of the GRAVITY FT), or at $\lambda_{0}$\,=\,$\lambda_{H,0}$\,=\,1.68\,$\mu$m (for PIONIER). The parameter $k_{\rm s}$\,=\,$d \, \mathrm{log} \, F_{\lambda, s} / d \, \mathrm{log} \, \lambda$ is the spectral index of the star assumed to be a black body at $T_{\mathrm{eff}}$=10500\,K. This translates into a spectral index of $k_{\rm s}$\,=\,-3.645 at $\lambda_{K,0}$ and -3.523 at $\lambda_{H,0}$. The parameter $k_{\rm c}$\,=\,$d \, \mathrm{log} \, F_{\lambda, c} / d \, \mathrm{log} \, \lambda$ is the spectral index of the circumstellar environment.\\
At this point, we must reiterate an important convention of the azimuthal modulation relevant for the correct interpretation of our results. The azimuthal modulation, which is parametrized with the variables c1 and s1, has a PA of the peak emission given by the argument of the complex number $c1$\,+\,j.$s1$. The origin of the angular position of the azimuthal modulation is the PA of the disk (cf. Table~\ref{tab:Continuum_param}) measured from north to east, with east to the left. As an example, an azimuthal modulation described by c1=1 and s1=1 in a disk with a PA\,=\,45$^{\circ}$ will show a visual rendering where the azimuthal modulation peak emission appears at 90$^{\circ}$ towards east.\\
The model fitting consists in an initial minimization procedure with \texttt{scipy.optimize.minimize} using a sequential least squares programming method to obtain an initial guess of the free parameters, followed by a procedure based on a Markov chain Monte Carlo  (MCMC; \citealt{Foreman-Mackey2013}) numerical approach, which is robust against trapping in local minima. 
We also report, when applicable, the uncertainty on the reduced chi-square $\chi_r^2$ by computing the error on the mean of the stochastic variable $T_i$ described in Eq.~8 in \cite{Ganci2021}. 
 
\subsubsection{Global fit modeling}

As the epochs are constituted by observations using different array configurations, different epochs probe different spatial frequencies of the source. This may lead to the detection of spurious variability in the model parameters, as shown in our test analysis in Appendix~\ref{sec:appendix_variability}. Therefore, instead of fitting each single epoch with our ten-parameter model and searching for individual variability trends, we implement a so-called {global fit} of the data. In this approach, we take advantage of the large temporal baseline of our data set {to decipher which of the parameters tested could be }{}the dominant parameter causing the variability visible in the data of Fig.~\ref{fig:Continuum-data_intro}. The global fit is obtained by forcing all the model parameters to be nonvariable over time, except for the parameter ---or set of parameters--- describing the feature to be tested for variability.

\subsection{Temporal variability}\label{subsubsec:variab}

Using the global fit approach on the continuum data, we explore which among the seven model parameters (star flux contribution F$_s$, ring inclination $i$, PA, characteristic size (a$_r$,a$_k$), and azimuthal modulation (c$_1$,s$_1$)) is most prone to cause the temporal variability of the data visible in Fig.~\ref{fig:Continuum-data_intro}. The parameters describing the halo contribution F$_h$, the spectral index k$_c$, and the weight parameter $f$Lor were not tested for temporal variability. We also report for comparison the analysis result for a fully nonvariable system model. 
The results are shown in Table~\ref{tab:GlobalFit_chi2}.\\
For the GRAVITY data set, we observe from the $\chi_r^2$ analysis that the 
time-variable azimuthal modulation model\,\#6 leads to the smallest $\chi_r^2$ value. The nonvariable system (\#1) gives the poorest fit. 
This is less marked for the PIONIER data set for which the $\chi_r^2$ values are larger. However, the trend in terms of decreasing $\chi_r^2$ suggests an analogous behavior for both the PIONIER and GRAVITY data sets. 
This analysis indicates that, among the tested models, the one with a time-variable disk azimuthal modulation (model\,\#6) best describes our data, and that such asymmetry is probably the dominant effect in the variability of the system, as opposed to other geometrical effects. 
The parameter values for the best-fit result of model\,\#6  are shown in Table~\ref{tab:Continuum_param} and the corresponding marginal posterior distribution is presented in Appendices~\ref{fig:PF_corner} and \ref{fig:GF_corner}. The fit results for each individual epoch of our PIONIER and GRAVITY data sets are shown in Appendix~\ref{tab:ObsLogPIONIER}.

\begin{table}[t]
\centering
\caption{Results of the global fit for different parameters tested for variability (Models \#2 to \#6). Model \#1 corresponds to a nonvariable system. 
The uncertainty on the $\chi_r^2$ is reported.}
\begin{tabular}{llllll} 
\hline
\hline
\# &Model & Var. par. & Nfp & $\chi_r^2$\,(P)& $\chi_r^2$\,(G) \\
\hline
1 & SHRM1 & --                    & 10    & 17.2$\pm$0.8
& 21.0$\pm$0.9
\\
2 & SHRM1 & \textit{i}            & 9+Ne  & 14.9$\pm$0.7
& 14.7$\pm$0.5
\\
3 & SHRM1 & F$_s$                 & 9+Ne  & 13.5$\pm$0.7
& 11.1$\pm$0.4
\\
4 & SHRM1 & PA                    & 9+Ne  & 12.4$\pm$0.6
& 12$\pm$0.5
\\ 
5 & SHRM1 & a$_r$, a$_k$          & 8+2Ne & 12.0$\pm$0.6
& 7.7$\pm$0.3
\\
6 & SHRM1 & c$_1$, s$_1$          & 8+2Ne & 11.5$\pm$0.6
& 5.3$\pm$0.3
\\ \hline
\multicolumn{4}{l}{Nb of data points} & 3855 & 1803 \\
\hline
\end{tabular}\\
\vspace{1ex}
{\raggedright \footnotesize{ \textbf{Notes}.
The model nomenclature (column\,2) gives S = star, H = halo, RM1 = 1$^{st}$ order azimuthal modulated ring; 
Nfp gives the number of free parameters. Ne is the number of epochs, namely 20 for PIONIER (P) and 13 for GRAVITY (G)). \par}}
\label{tab:GlobalFit_chi2}
\end{table}

\begin{table}[b]
\centering
\caption{Nonvariable parameters of the azimuthal modulation global fit continuum model. 
F$_c$ is not a free parameter, but is obtained following $F_c$=1-$F_h$-$F_s$. The $\sigma_{MCMC}$ error estimates derived through the MCMC fitting procedure are given by the 16$^{th}$ and 84$^{th}$ percentiles of the samples in the MCMC marginalized distributions.}
\begin{tabular}{lcllll} 
\hline
\hline
& & \multicolumn{2}{c}{PIONIER} & \multicolumn{2}{c}{GRAVITY} \\
\hline
Parameter & Unit & Value & $3\sigma_{MCMC}$ & Value & $3\sigma_{MCMC}$\\
\hline
F$_s$      & \%  & 23.09 & 0.39 & 22.69 & 0.42 \\
F$_h$      & \%  & 3.92  & 0.78 & 10.36 & 0.57 \\
F$_c$      & \%  & 72.99 & 1.17 & 66.95 & 0.99 \\
k$_c$      &     & 3.47  & 0.18 & -0.39 & 0.33 \\
a$_r$      & mas & 0.93  & 0.06 & 2.02  & 0.06 \\
a$_k$      & mas & 1.60  & 0.03 & 1.73  & 0.06 \\
\textit{i} & deg & 0.67  & 2.01 & 34.15 & 1.08 \\
PA         & deg & 106.9 & 1.80 & 122.9 & 1.50 \\
$f$Lor     &     & 1.00  & 0.03 & 1.00  & 0.03 \\
$\chi^2_r$ &     & 11.52 &      & 5.26  &      \\ \hline
\end{tabular}
\label{tab:Continuum_param}
\end{table}

\subsection{Disk azimuthal asymmetry}\label{subsubsec:disk-asymm}

\begin{figure*}
\centering
\includegraphics[width=\textwidth]{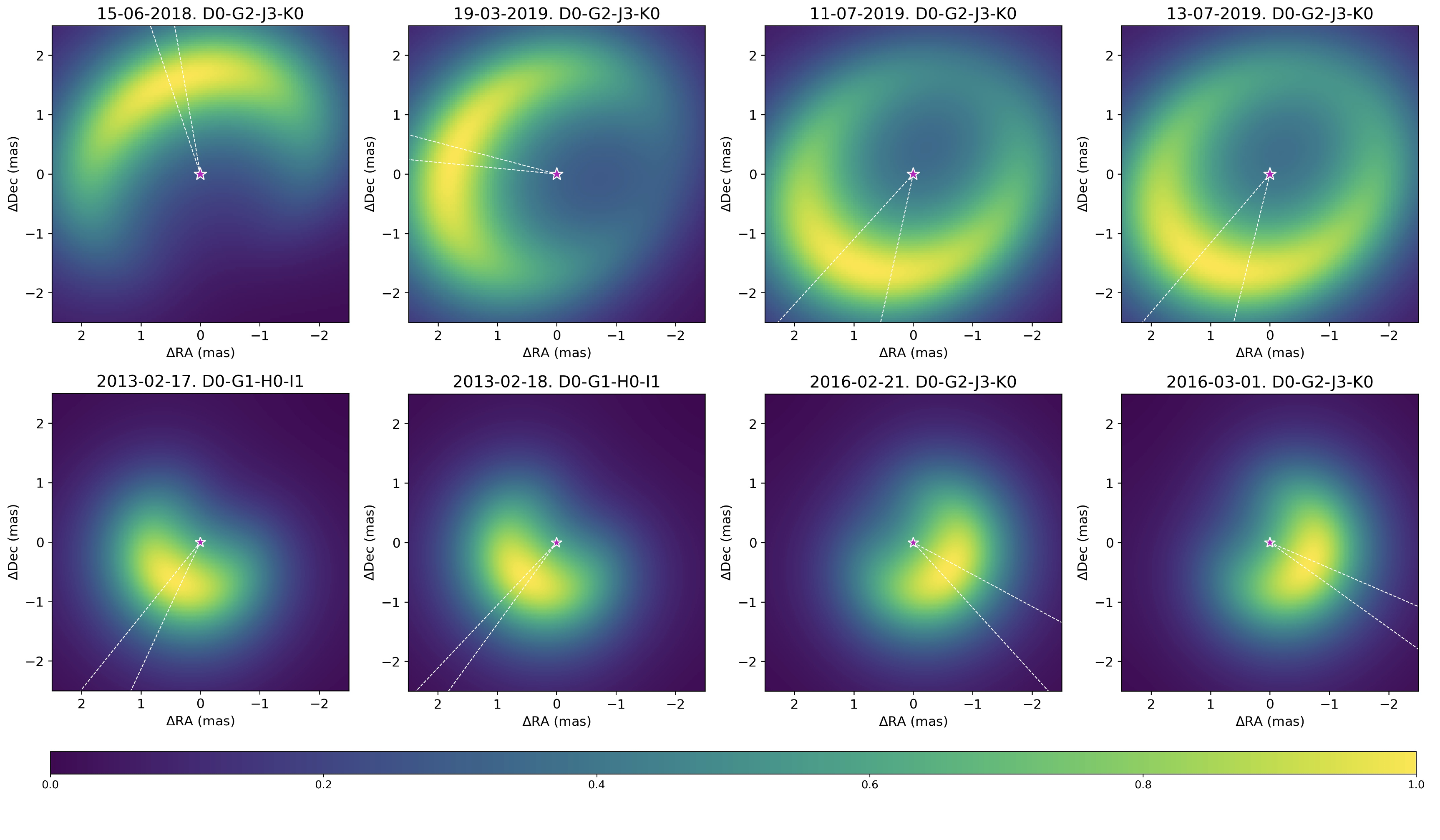}
\caption{Peak-normalized GRAVITY (top row) and PIONIER (bottom row) continuum model images. The dashed white lines represent the $\pm$3$\sigma$ uncertainty on the PA of the azimuthal modulation. 
The central object is not displayed but is marked with a star to enhance the circumstellar emission. North is up, east is to the left.
See Appendix~\ref{sec:visualisation} for the full data set.}
\label{fig:Continuum-imgs}
\end{figure*}

Our best-fit model exhibits a crescent-like asymmetric feature in the disk resulting from the azimuthal modulation that varies in PA through the epochs and revolves around the central star. 
In Fig.~\ref{fig:Continuum-imgs}, 
we present a subset of the continuum model images corresponding to our fitted models and from which the geometry, extent, and location of the asymmetry can be followed as a function of time. The complete time sequence is presented in Figs.~\ref{fig:PIONIER-Continuum-imgs_Full} and~\ref{fig:GRAVITY-Continuum+line-imgs_Full}. 
The fit of the variable azimuthal modulation appears quite robust in Fig.~\ref{fig:AzMod-chi2maps}, with one single global minimum identified in the $c_1$,$s_1$ diagrams. 
Figure~\ref{fig:Continuum-imgs} displays the azimuthal uncertainties in the form a white-line cone as derived from the 3$\sigma$ error bars on $c_1$ and $s_1$. 
The azimuthal uncertainty is generally smaller (up to $\sim$10-20$^{\circ}$, depending on the configuration) in the K-band than in the H-band. In addition, the azimuthal position is most poorly constrained with the small configuration (e.g., P2-P5-P6 with PIONIER) because the closure phase signal is marginal for the shortest baselines 
For epochs only separated by a few days at most and taken with the same array configuration, the azimuthal locations of the asymmetric feature are consistent in most cases within the error bars (e.g., P7-P8-P9-P10-P11 and P19-P20 with PIONIER, or G7-G8 with GRAVITY). 
We also generally observe that the continuum emission appears azimuthally more compact in the H-band than in the K-band. 
The brightness contrast between the crescent-like feature and the corresponding centro-symmetric position in the disk is estimated from Figs.~\ref{fig:PIONIER-Continuum-imgs_Full} and ~\ref{fig:GRAVITY-Continuum+line-imgs_Full}.
The contrast is found to be $\sim$4 in the K-band (ranging from $\sim$1.6 to 10 across the epochs, with $\sigma$$\sim$2.8) and $\sim$2.4 in the H-band (ranging from $\sim$1.3 to 3.3, with $\sigma$\,$\sim$0.7). No specific trend is found in the temporal evolution of the contrast. 
The dynamical properties of the inner disk feature revealed by our observations are further discussed in Sect.~\ref{ssec:DustPeriod-Discussion}.

\subsection{Orbital period of the crescent-like feature}
\label{ssec:DustPeriod-Discussion}

\begin{figure}[t]
    \centering
    \includegraphics[width=\columnwidth]{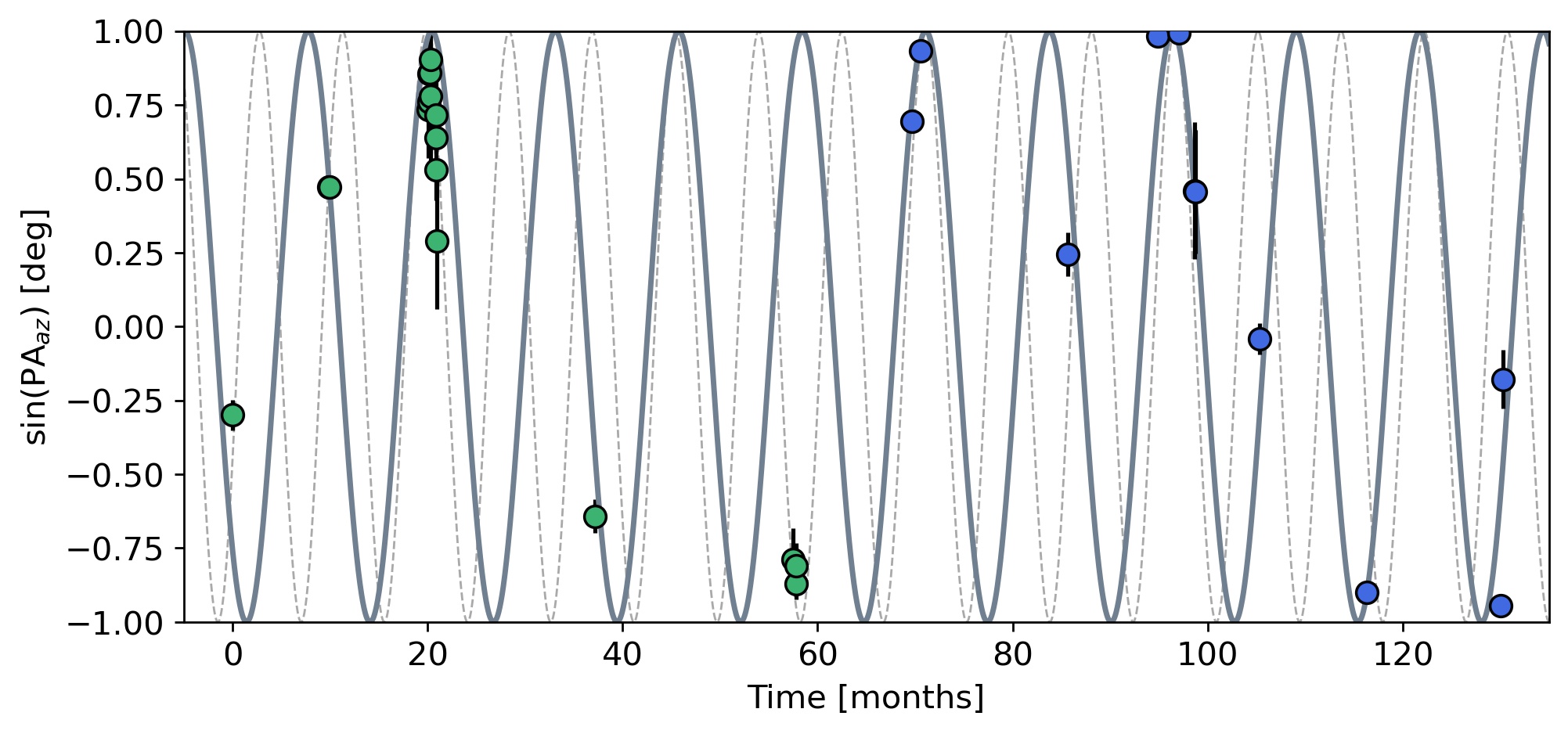}\\
    \caption{Sine of the azimuthal modulation PAs as a function of time. The markers represent the sine of the variable dusty feature PAs depicted in Fig.~\ref{fig:Continuum-imgs}, green for the PIONIER data and blue for the GRAVITY one. The full gray line represents the best fit ($12.6\pm0.1\,$months) of the uniform circular motion PA expressed as a sine function. The dashed gray line represents the fit solution corresponding to a period of $8.50\pm0.25\,$months.}
    \label{fig:DustPeriod}
\end{figure}

The revealed azimuthal asymmetry in the inner disk of HD\,98922 shows orbital motion around the central star. 
Assuming this is the same asymmetric feature that is monitored with the VLTI over the 11 year period in both the H and K bands, we attempt to investigate the orbital properties of the emission feature. We show in Fig.~\ref{fig:DustPeriod} the distribution of the time-variable PA of the emission feature. The small-configuration epochs P2, P5, P6, G6, and G10 are not taken into account as the small baselines provide 
marginally weaker constraints on the azimuthal modulation because of the low level of closure phase signal detected. 
The time origin corresponds to the first PIONIER observation of June 2011. The gray line is a sine function corresponding to a uniform circular motion fitted to the data to derive a period estimate. 
The best sine fit gives a period of 12.6$\pm$0.1 months. The $\chi^2_r$ value is large (111$\pm$51), which is due to the group of epochs at $\sim$60\,months that deviate from the best fit. 
If these epochs were found to correspond to outliers and were removed, the fit would give a $\chi^2_r$ value of 38$\pm$17, while the estimated period would remain the same (see Table~\ref{tab:Dust-Periods}). However, at this stage, there is no clear justification for the removal of these epochs.
 \\
Using Kepler's law, we estimate the central mass of our target star for a separation of the azimuthal asymmetry ranging from 0.6 to 1.3\,au based on the fitted ring annular radius a$_r$, as well as for a range from 1.2 to 1.7\,au based on the half-light radius $a$, where $a$\,=\,$(a^2_r + a^2_k)^{1/2}$ following \cite{Lazareff2017}. From the derived orbital period of the azimuthal feature, we estimate a central mass of $\sim$1--3\,M$_\odot$ depending on the separation reported above. This is significantly lower than the literature value of $\sim$6\,M$_\odot$, which was robustly established using UVES high-spectral-resolution observations \citep{Garatti2015}. For this higher mass, the estimated Keplerian orbit for a circular motion would be 9.2$\pm$0.4\,months, which is shorter than that found with our measurement. It is noteworthy that our sine fit also shows a local minimum for a similar period of around $\sim$8.5\,months, although with a poorer $\chi^2_r$ value of $\sim$350. We show this sine fit for comparison as a dashed line in Fig.~\ref{fig:DustPeriod}.\\
Some caveats are worth mentioning here. The coverage of the orbital motion is relatively sparse and a finer temporal sampling would allow us to estimate the period with greater confidence. Furthermore, it should be noted that we have implemented here the simplest case of a circular orbit, because our model\,\#6 of Table~\ref{tab:GlobalFit_chi2} foresees a nonvariable ring radius $a_r$. Accounting for possible eccentricity of the orbit of the crescent-like feature may impact the derived orbital period, which is not explored in this work. The orbital motion is further discussed in Sect.~\ref{ssec:Dynamics}.

 \begin{table}
    \centering
    \caption{\centering Orbital period estimates of the dust azimuthal asymmetry}
    \begin{tabular}{cccc}
   \hline
   \hline
   Period & $\chi_r^2$ & M$_\star$ & Method\\
   \hline
   [months] &  & [M$_\odot$] & \\
   \hline
   $12.6\pm0.1$ & $111\pm 51$ & -- & (1) \\
   $12.6\pm0.1$ & $38\pm 17$ & 0.9$\pm$0.5 & (2) \\
   $12.6\pm0.1$ & $38\pm 17$ & 3.0$\pm$0.8 & (3) \\
    \hline
    \end{tabular}\\
    \vspace{2ex}
    {\raggedright \footnotesize{\textbf{Notes:} (1) period estimated from the fit of the time-variable PA of the continuum azimuthal feature; (2) same as (1) but with the group of epochs at 60 months removed. The mass of the central star is derived by assuming that the feature has a separation ranging from 0.6 to 1.3\,au as given by the ring radius a$_r$; (3) same as (1) but with the mass of the central star derived by assuming that the feature has a separation ranging from 1.2 to 1.7\,au as given by the half-light radius $a$. 
    \par}}
    \label{tab:Dust-Periods}
 \end{table}

\subsection{Fractional flux and disk characteristic size}\label{subsubsec:frac-flux}

\begin{figure}[t]
    \centering
    \includegraphics[width=\columnwidth]{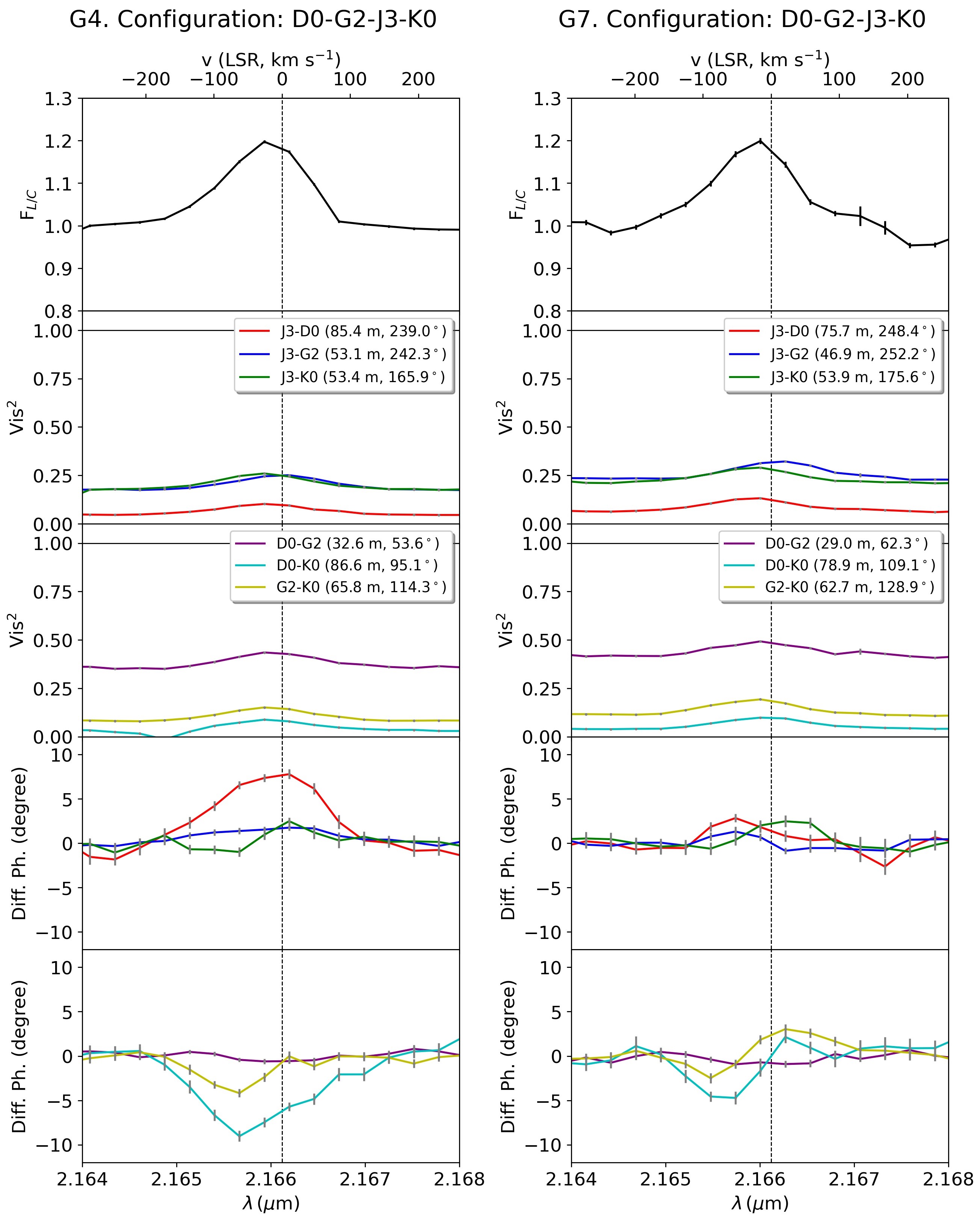}
    \caption{\object{HD\,98922} GRAVITY SC data for two different epochs (one per column) with similar \textit{(u,$\varv$)} plan coverage. From top to bottom: Wavelength-calibrated and continuum-normalized spectrum, total squared visibilities (two panels), and total differential phases (two panels). Colors refer to the different baselines.}
    \label{fig:Gas-data_intro}
\end{figure}

In H-band, we find fractional flux ratios F$_s$, F$_h$, and F$_c$ comparable to those found by \cite{Lazareff2017}. 
In the K-band, we measure a stellar flux contribution of $\sim$\,23\% (see Table~\ref{tab:Continuum_param}). 
Different values for the K-band fractional stellar contribution based on photometry and SED analysis are reported in the literature, ranging from $\sim$15\,\% \citep{Kraus2008,Hales2014,Perraut2019} to $\sim$23\,\% \citep{Garatti2015}. However, when comparing these values, we must keep in mind the different distance estimates used in these previous studies. 
We also find a stronger halo contribution than \cite{Perraut2019}, which can be explained by the fact that these latter authors constrained the circumstellar emission of HD\,98922 in the K-band only with the astrometric configuration, whereas the short baselines of the small configuration are usually needed to better constrain the halo. \\
The radius of the ring disk emission in K-band is found to be a$_r$=2.02$\pm$0.1\,mas (or 1.31$\pm$0.07\,au), in good agreement with the estimates of \cite{Kraus2008} and \cite{Garatti2015}, of namely 2.2\,mas and 1.6\,mas, respectively. 
Our result is slightly larger than the estimate of a$_r$=1.8\,mas of \cite{Perraut2019}, who only used two snapshots with the astrometric configuration.
In the H-band, we find the circumstellar emission to be more compact than in K, with a characteristic radius of a$_r$=0.93$\pm$0.1\,mas (or 0.60$\pm$0.07\,au), which is in line with the estimate of 0.87\,mas by \cite{Lazareff2017}. \cite{Kluska2020} report a larger half-flux radius of 2.1\,mas based on image reconstruction, which evidences the impact of different modeling approaches. The K-to-H size ratio is further discussed in Sect.~\ref{sec:Discussion}. 
The ratio a$_k$/a$_r$ being close to or larger than unity suggests a wide, smooth ring emission as opposed to a sharp edge. 
The disk is known to have a low inclination, which is therefore more difficult to accurately constrain in the small angle range. This applies to the PA as well (Fig.~\ref{fig:GF_chi2maps}).

\section{Results on the Br$\gamma$-line interferometric data}\label{sec:brg_analysis}

The GRAVITY line spectrum, the calibrated squared visibilities, and the differential phases are shown in Figs.~\ref{fig:Gas-data_intro} and \ref{fig:GRAVITY-SC-Data} for 12 of the 13 epochs after zooming into the spectral region between 2.164 and 2.168\,$\mu$m. No high-spectral-resolution data could be acquired on June 15, 2018 (epoch G3). 
From the high-spectral-resolution K\,band data, \object{HD\,98922} shows a slightly blueshifted ($\approx - 23.5\,$km/s) single-peaked Br$\gamma$ emission line in all epochs. Considering the 3\,\AA{} spectral resolution, we can consider the peak's position of the continuum-normalized line ($21659.7^{+0.8}_{-1.3}\,$\AA{}) to be constant through the epochs at our spectral resolution. The normalized peak flux varies between 1.17 and 1.25 depending on the epoch, while the line width measured at the peak's 10\% flux level ranges from 14 to 15\,\AA{}.
The total squared visibilities in the Br$\gamma$ region vary between $\approx 0$ and $\approx 0.4$ depending on the epoch and baseline configurations (medium or large), while they reach $\approx 0.7$ for the small configuration. 
Finally, the differential phase signals vary between $-15^{\circ}$ and $25^{\circ}$ and have significantly different shapes (flat, single-peaked, double-peaked, or \textit{S}-shape) for the different epochs and baselines.

\subsection{Modeling methodology}\label{subsec:method-brg}

\begin{figure*}[t]
    \centering
    \includegraphics[width=0.49\textwidth]{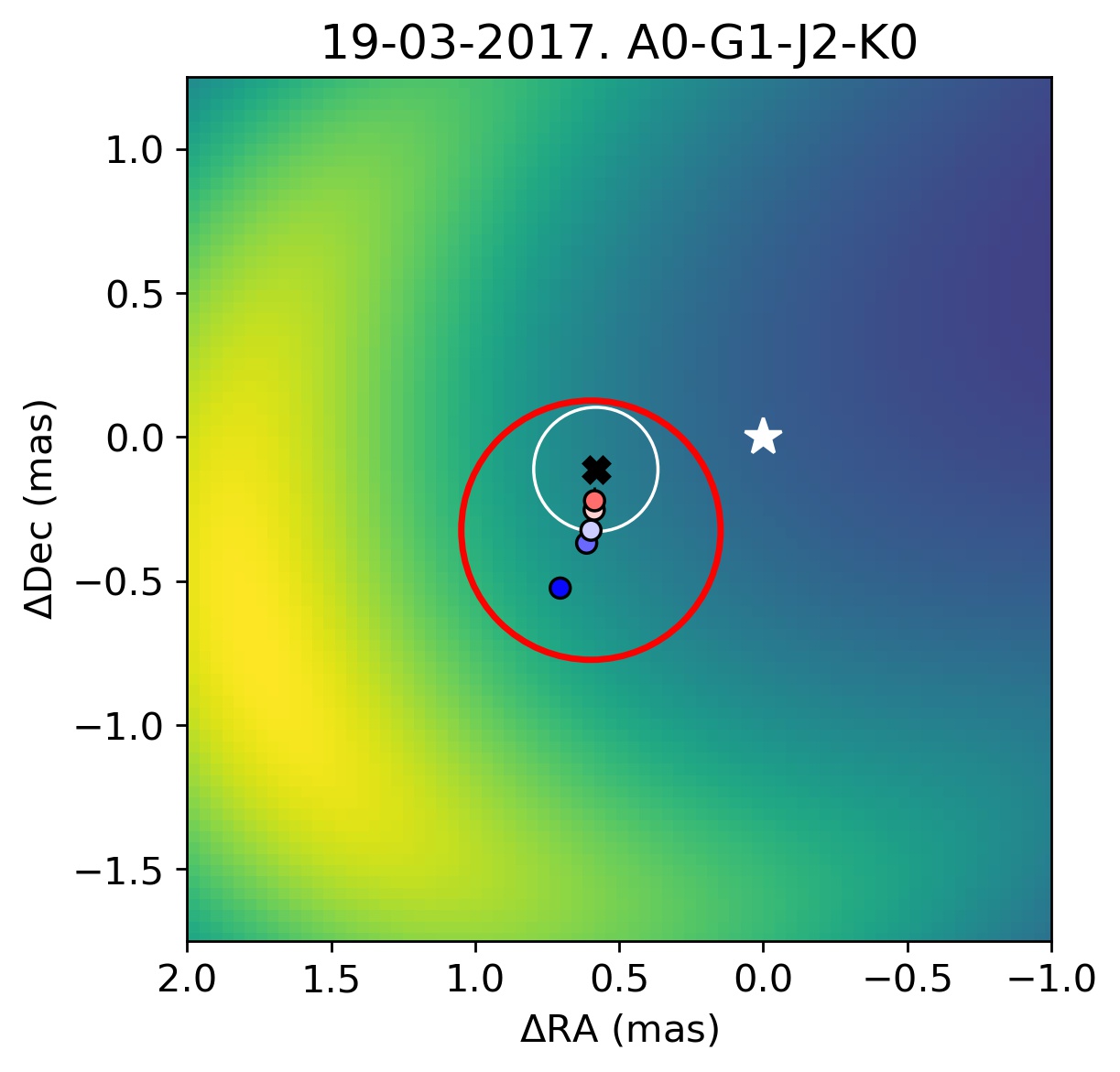}
    \includegraphics[width=0.49\textwidth]{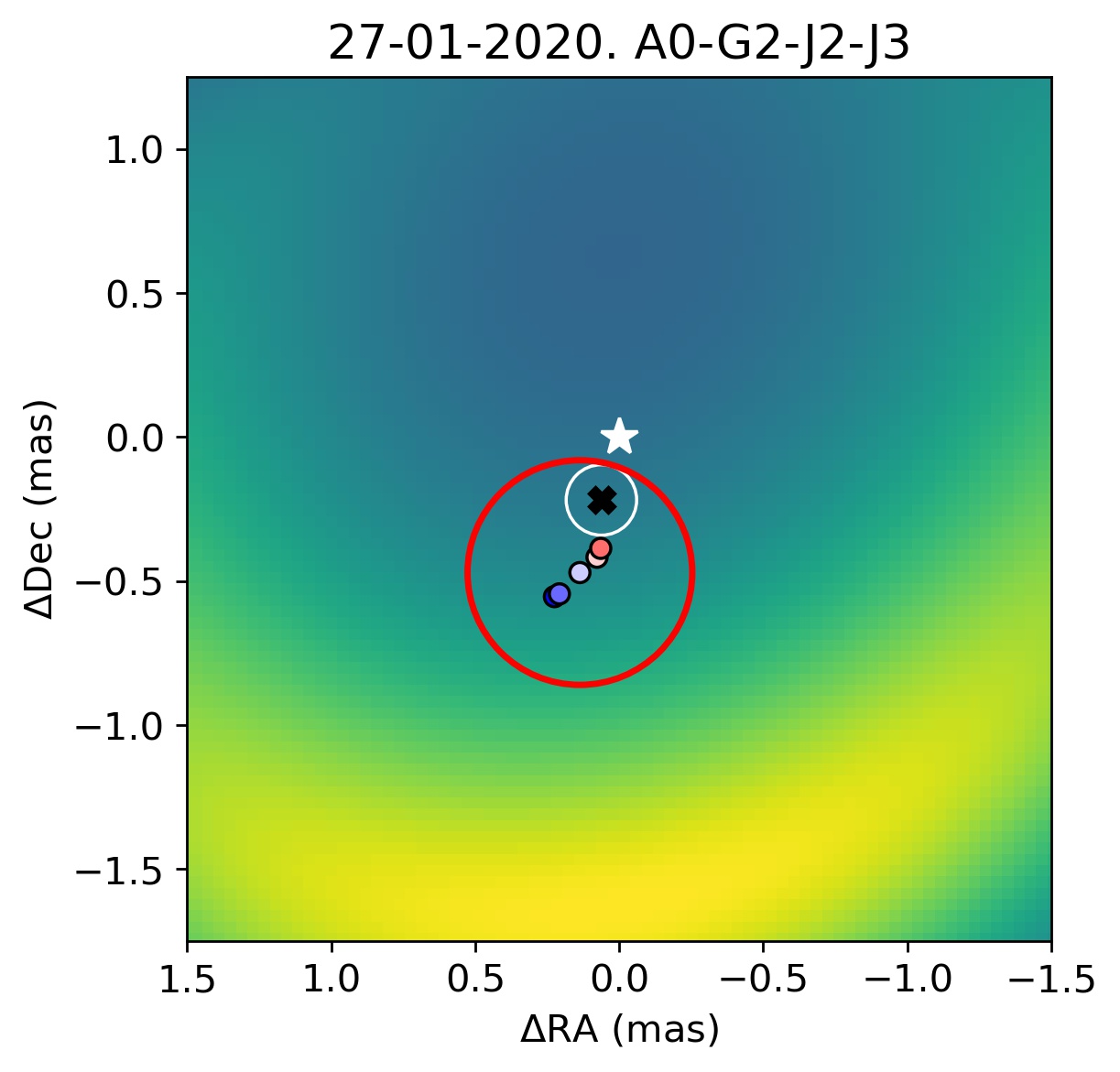}
    \caption{Spatial and kinematic properties of the Br$\gamma$-line-emitting gas region overlaid to the continuum model for two epochs. The complete sequence is shown in Appendix~\ref{apx:brg-visualisation}. The star is centered on the origin. The black cross and the white circle around it show the position of the continuum photocenter and its uncertainty estimated from the continuum modeling. The red circle represents the extent of the gas-emitting region estimated at the peak of the line emission and centered at about $\sim$\,0\,km\,s$^{-1}$. The blue to red colored filled dots show the gas photocenter positions for the five spectral channels across the Br$\gamma$ line corresponding to velocities ranging from -100\,km\,s$^{-1}$ to +100\,km\,s$^{-1}$, as color-coded in Fig.~\ref{fig:GRAVITY-SC-Data}.}
    \label{fig:BrG_PhCenters_small-sample}
\end{figure*}

To  estimate  the  Br$\gamma$ gas  region  size, kinematics, and displacement with respect to the continuum emission from  the GRAVITY SC  visibilities and differential phases, we extrapolated  the  pure-line  contribution (marked with subscript $L$)  from  the interferometric observables. 
Following \cite{Weigelt2011}, the pure-line interferometric quantities characterizing the gas-emitting region, the visibility $V_{\rm L}$ and differential phase $\phi_{\rm L}$, are related as follows:
\begin{eqnarray}\label{eq:PL-Visibility}
{F^2_{\rm L}V^2_{\rm L} = F^2_{\rm {tot}}V^2_{\rm {tot}} + F^2_{\rm {cont}}V^2_{\rm {cont}} - 2 F_{\rm {tot}}V_{\rm {tot}}F_{\rm {cont}}V_{\rm {cont}}\cdot \rm{cos}\,\phi_{\rm {tot}}}
,\end{eqnarray}
\begin{equation}\label{eq:Diff-phase}
\rm{sin\,\phi_{\rm {L}} = \rm{sin}\,\phi_{\rm {tot}} \frac{F_{\rm {tot}}V_{\rm {tot}}}{F_{\rm L}V_{\rm L}}}.
\end{equation}
The quantities reported in Eq.~\ref{eq:PL-Visibility} and ~\ref{eq:Diff-phase} refers to values inside the Br$\gamma$-line spectral region. As the continuum quantities $F_{\rm {cont}}$ and $V_{\rm {cont}}$ {within the line} are not directly measurable, they are estimated from the continuum near to the line region. However, hot Herbig stars exhibit a strong Br$\gamma$ photospheric absorption feature, which needs to be accounted for in order to retrieve the correct pure-line quantities. In the case where photospheric absorption is present, knowledge from a model for the continuum emission is required. One can show that
\begin{align}
V_{\rm L} & = \left(\frac{(V_{\rm tot}\,F_{\rm L/C}^{\prime})^2 + 
\left(\frac{F_{\rm s}^{\prime}}{F_{\rm cont}^{\prime}}(\alpha-1)+V_{\rm cont}^{\prime}\right)^2}
{\left (F_{\rm L/C}^{\prime} - \frac{\alpha+\beta+\gamma}{1+\beta+\gamma}\right )^2} \right. \nonumber \\ 
& \left. \frac{- 2\,(V_{\rm tot}\,F_{\rm L/C}^{\prime})
\left(\frac{F_{\rm s}^{\prime}}{F_{\rm cont}^{\prime}}(\alpha-1)+V_{\rm cont}^{\prime}\right)
\cos(\phi_{\rm tot})}
{\left(F_{\rm L/C}^{\prime} - \frac{\alpha+\beta+\gamma}{1+\beta+\gamma}\right)^2} \right)^{1/2},
\label{eq:VL}
\end{align}
\\
where the superscript ($^\prime$)  indicates that the quantities $V_{\rm cont}^\prime$ and $F_{\rm cont}^\prime$ are estimated outside the emission line spectral region, as opposed to Eq.~\ref{eq:PL-Visibility}. The line-to-continuum ratio $F_{\rm L/C}^\prime$ is also normalized to the nearby continuum value. The parameters $\beta$ and $\gamma$ are the disk-to-star $\beta$=$F_{\rm c}$/$F_{\rm s}$ and halo-to-star $\gamma$=$F_{\rm h}$/$F_{\rm s}$ flux ratios. 
\\
The parameter $\alpha$ describes the star photospheric absorption, with 0<$\alpha$\,$\leqslant$1, so that 
\begin{eqnarray}\label{eq:PL-Flux}
F_{\rm L} = F_{\rm tot} - F_{\rm cont}^\prime + F_{\rm s}\,(1 - \alpha)
,\end{eqnarray}
where $\alpha$ is equal to 1 when there is no absorption, and equal to zero when 100\% of the star flux is absorbed (see Appendix \ref{apx:spctr-corr} for a description of the photospheric absorption model).\\
Similarly, Eq.~\ref{eq:Diff-phase} is written in terms of known quantities as
\begin{equation}\label{eq:Diff-phase2}
\sin\,\phi_{\rm L} = \sin\,\phi_{\rm tot}
\frac{V_{\rm tot}}{V_{\rm L}}
\frac{1}{\left(1-\frac{\alpha+\beta+\gamma}{1+\beta+\gamma}\,
\frac{1}{F_{\rm L/C}^\prime}\right)}
.\end{equation}\\
For a compact and marginally resolved gas emission component, we derive the photocenter displacement along each baseline from the pure-line differential phases as in \cite{Lachaume2003}:
\begin{equation}\label{eq:Photo-shift}
\mathbf{p} = \frac{- \phi_{\rm L}}{2 \pi} \cdot \frac{\lambda}{\mathbf{B}},
\end{equation}
where $\mathbf{p}$ is the projection on the baseline $\mathbf{B}$ of the 2D photocenter vector with its origin on the continuum photocenter of the system. 

\subsection{Properties of the line-emitting region}\label{subsec:results-brg}

We exploit the high-spectral resolution data of GRAVITY in the Br$\gamma$-line region to constrain the spatial scale and the kinematics of the hot-gas component. We used 10 out of 13 epochs for this analysis: in addition to epoch G3, for which no high-spectral resolution data could be acquired, epochs G6 and G10 are not considered, as no differential phase signal was detected using the small configuration. \\
From the pure-line visibilities, we estimated the characteristic size of the gas-emitting region at the peak wavelength of 2.1662\,$\mu$m using a simple Gaussian disk model, as well as a ring model with 20\% radial thickness for comparison. We obtain a Gaussian HWHM radius of 0.47\,mas with a temporal standard deviation of 0.04\,mas. Similarly, we obtain a radius of 0.50$\pm$0.05\,mas  for the thick ring. In both cases, the gas-emitting component model is found to be consistent with a face-on orientation, as we find an inclination of 5$^\circ$$\pm$5$^\circ$. 
These values translate into a physical radius of $\sim$0.3\,au at a distance of 651\,pc, in agreement with the results of \cite{Garatti2015}. \\
As the estimation of the characteristic size of the hot-gas region has been studied in previous works, we focus on the determination of the location of the compact hot-gas component in relation to the star, and further explore whether there is a correlation between the locations of the gas and the dust feature observed in the continuum. 
This can be investigated through an analysis of the interferometric differential phase, which provides precise information about the spatial location of the photocenter of the gas emission component on angular scales that surpass the nominal resolution of the interferometer. \\
The detection of a differential phase signal as reported in Fig.~\ref{fig:GRAVITY-SC-Data} indicates that the photocenter positions of the continuum and gas emission components are not coincident. 
From the continuum-corrected (or pure-line) differential phases (Eq.~\ref{eq:Diff-phase2}), we calculated the deprojected photocenter shifts of the hot-gas component with respect to the continuum photocenter for few spectral channels around the 0\,km/s velocity (Eq.~\ref{eq:Photo-shift}) following the formalism of \cite{Alex2023}. 
The location of the continuum photocenter may differ from the position of the central star if the dust emission is noncentrosymmetric. This is indeed the case for HD\,98922, because our modeling of the continuum shows a strongly asymmetric time-variable brightness distribution of the inner circumstellar disk.  
Therefore, the location of the continuum photocenter is the parameter that most  affects the position of the Br$\gamma$-line photocenter with respect to the central star. 
Figure~\ref{fig:BrG_PhCenters} shows the spatial location of the gas emission photocenter with respect to the continuum photocenter, and therefore with respect to the central star. For most of the epochs, the bulk of the Br$\gamma$ emission appears offset ---considering the retrieved characteristic size of the gas emission component--- with respect to the stellar position by up to 0.5\,mas. We also observe that the location of the compact gaseous component varies with time and that it is, on first order, consistently found in an area between the central star and the peak of the dusty feature, following its orbital motion. 
Looking closer at Fig.~\ref{fig:BrG_PhCenters_small-sample}, we see that, for each epoch, the positions of the photocenter across the Br$\gamma$ line are not spatially colocated but are distributed in a profile that qualitatively resembles a pattern of Keplerian motion. 

\section{Physical properties of the disk}
\label{sec:RT_modeling}

In an earlier work, \cite{Hales2014} proposed a model for the disk of HD\,98922 accounting for a distance of 507\,pc and exploiting photometry data up to 160\,$\mu$m. These authors proposed a disk model with an inner radius located at 1.5\,au. We revisit this model in light of our new interferometric data.

\subsection{Disk structure}

\begin{figure*}[t]
    \centering
    \includegraphics[width=\textwidth]{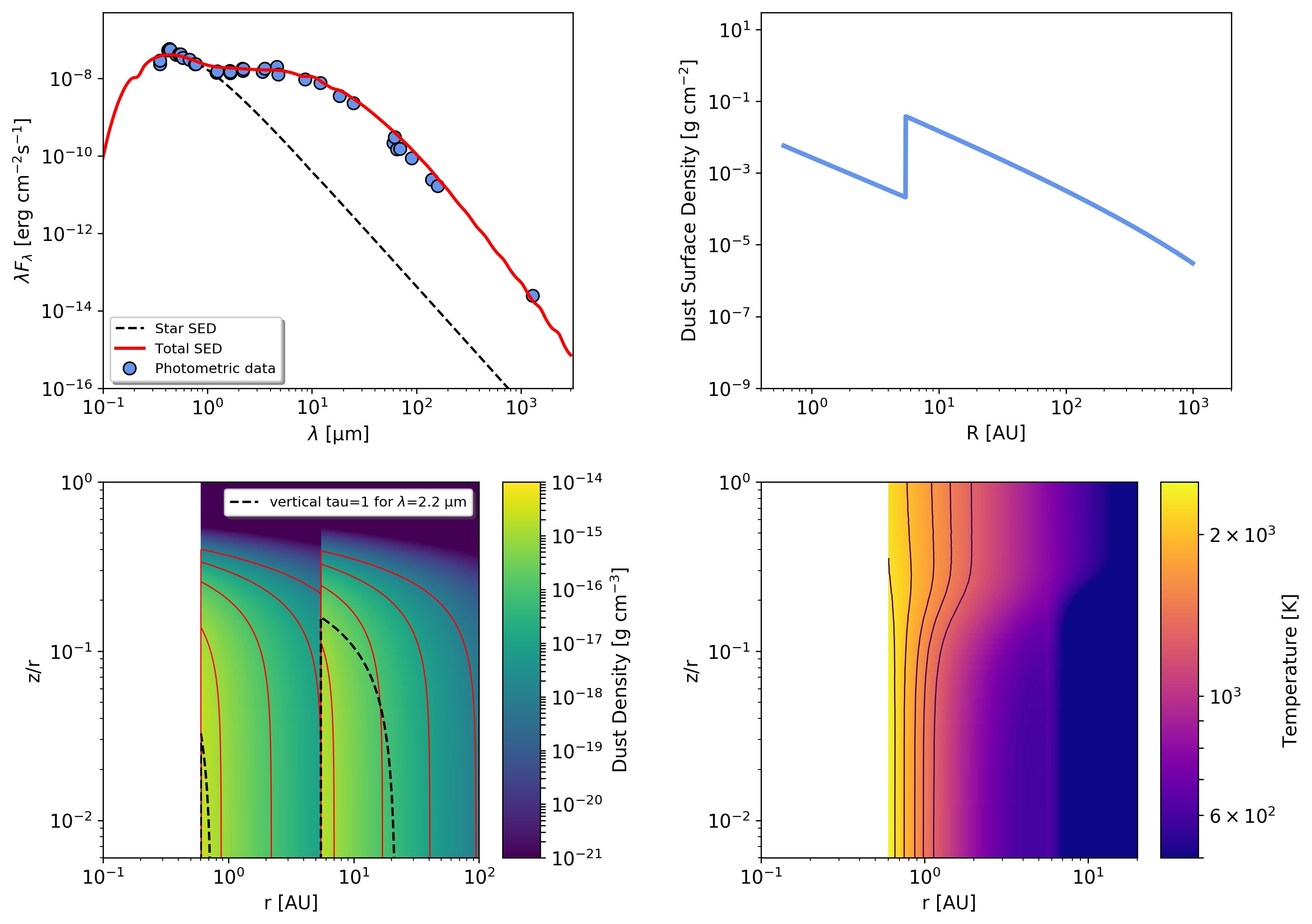}
    \caption{Radiative transfer modeling corresponding to Table~\ref{tab:RT-BestModel}. The top left panel shows the SED for Model\,CS, with the blue dashed line representing the stellar black-body function and the red line showing the modeled total emission. The black bars represent the photometric data. The top right panel shows the derived dust surface density profile as a function of the distance from the star. The bottom left plot shows the dust density structure, where the black dashed line represents the $\tau$\,$=$\,1 surface at 2.2\,$\mu$m, and the red lines represent, from left to right and for both components, the density contours at $10^{-15}$, $10^{-16}$, $10^{-17}$, and $10^{-18}$\,g\,cm$^{-3}$, respectively. The bottom right plot shows the dust temperature structure, where the black lines represent, from left to right, the isothermal contours at 2300, 2000, 1700, 1500, and 1300\,K, respectively.
    }
    \label{fig:RT-BestModel}
\end{figure*}

We used the radiative transfer code MCMax \citep{Min2009} to constrain the disk density and temperature structure based on archival broadband photometry data and on the spatial structure evidenced by our interferometric measurements. We started from a one-component model based on \cite{Hales2014}. 
The inner disk radius is 1.5\,au, the outer radius is 320\,au, the flaring index $\gamma$=1, and the scale-height is 15\,au at a radius of 100\,au. 
We implement a grain population based on DIANA standard dust grains \citep{Woitke2016} composed of 75\% amorphous silicates (e.g., Mg$_{0.7}$Fe$_{0.3}$SiO$_3$), 25\% porosity, by volume, and initially no amorphous carbon. The grain size $a$ ranges from 1\,$\mu$m to 2200\,$\mu$m with a distribution of d$n(a)$\,$\propto$\,$a^{-3.5}$, as large grains appear to dominate in this source \citep{Bouwman2001,VanBoekel2003}. 
The surface density profile is based on a modified version from \cite{Li2003} with a power-law exponent of $p_\Sigma = -1.5$, a gas-to-dust ratio of $100$, and a dust disk mass of $2\times10^{-5}\,$M$_\odot$. The extinction is $A_V$\,=\,0.5\,mag \citep{Guzman2021}. The parameters of the central star are taken from Table~\ref{tab:StellarParameters}.\\
With the revised stellar parameters, the stellar luminosity increases by a factor $\sim$3 in comparison to \cite{Hales2014} and the photospheric emission contribution results in overestimation of the K-band NIR flux in the SED by a factor $\sim$1.2. On the other hand, our GRAVITY and PIONIER observations show that dust is present at separations of less than 1.5\,au. As the SED fitting process is a degenerated problem, we attempt to better constrain the inner disk structure based on the consideration  that a dust component at $\lesssim$1\,au, producing $\sim$75\%-80\% NIR excess and resolved with the VLTI has to be accounted for. 
Simply moving the inner radius to $\sim$0.6\,au in the one-component model described above is not possible because this leads to further overestimation of the NIR flux in the SED. We therefore revise this disk model and propose a two-component disk model in which the inner component has a lower surface density in comparison to the outer component and an inner radius at 0.6\,au. \\
The mass of the outer component is set to 2$\times$10$^{-5}$\,M$_\odot$ as in \cite{Hales2014}. This value can be compared to the rough mass estimate obtained from the archival ALMA photometry point at 1.3\,mm (ID:\,2015.1.01600.S/PI Panic), which was not included in the  work of these latter authors. 
We used the CARTA visualization tool and estimate a flux density at 1.3\,mm of $F_{\nu}$=10.4\,mJy  for the unresolved HD\,98922
source, with a typical conservative uncertainty of 10\% . 
We compute the dust disk mass $M_{\rm d}$ following \cite{Beckwith1990}. 
Assuming optically thin emission, we use the relation
\begin{eqnarray}
F_\nu \approx \kappa_\nu\frac{2k\overline T \nu^2}{c^2 d^2}M_{\rm d}\label{dmass}
,\end{eqnarray}
where $\overline T$ is the dust mean temperature, $\kappa_\nu$  the dust absorption coefficient at 1.3\,mm, $d$ the distance, $k$ the Boltzmann constant, and $c$ is the speed of light. 
A large uncertainty subsists on the dust opacity value $\kappa_\nu$ and on the mean temperature, and therefore the mass value is only a crude approximation for first-order estimates. If we consider standard values of $\overline T$\,=50\,K, and $\kappa_\nu$(1.3\,mm)\,=\,0.02\,cm$^2$/g \citep{Beckwith1990}, we obtain a disk dust mass of $M_d$\,$\sim$\,0.013\,$M_\odot$. 
However, \cite{Woitke2016} underline the strong dependence of $\kappa_\nu$ on the fraction of amorphous carbon and grain size distribution, and report values as high as $\kappa_\nu$(1.3\,mm)\,$\sim$\,5-10\,cm$^2$/g. The latter value results in a lower mass of $M_d$\,$\sim$5\,$\times$\,10$^{-5}$\,M$_\odot$, which is comparable to that found by \cite{Hales2014}. Because of the uncertain estimate of the disk dust mass, we choose to adopt the value reported by the latter authors 
for the outer disk component. 
For the disk composition, we considered two approaches: in the first case, we included a 25\% fraction of carbon grains (model CS) to quench the strength of the silicate emission feature \citep{VanBoekel2003}; in the second case, we consider an alternative case of an inner component only composed of quantum-heated particles (model Q) strongly coupled to the gas (e.g., \citealt{Kluska2018,Ganci2021}). \\
\begin{table}[t]
    \centering
    \caption{\object{HD\,98922} RT models. The stellar parameters are from \cite{Guzman2021}.}
    \begin{tabular}{lllll}
    \hline
    \hline
    \multicolumn{5}{c}{Star}\vspace{0.1cm} \\
    \hline
    \multicolumn{2}{l}{Param.} & \multicolumn{1}{l}{Unit} & \multicolumn{1}{l}{Value} & \multicolumn{1}{l}{Range} \\ \hline
    \multicolumn{2}{l}{T$_\star$ } & \multicolumn{1}{l}{K} & \multicolumn{1}{l}{10500} & \multicolumn{1}{l}{Fixed} \\
    \multicolumn{2}{l}{R$_\star$} & \multicolumn{1}{l}{R$_\odot$} & \multicolumn{1}{l}{11.45} & \multicolumn{1}{l}{Fixed} \\
    \multicolumn{2}{l}{M$_\star$} & \multicolumn{1}{l}{M$_\odot$} & \multicolumn{1}{l}{7.0} & \multicolumn{1}{l}{Fixed} \\
    \multicolumn{2}{l}{d} & \multicolumn{1}{l}{pc} & \multicolumn{1}{l}{650.9} & \multicolumn{1}{l}{Fixed} \\
    \multicolumn{2}{l}{A$_V$} & \multicolumn{1}{l}{mag} & \multicolumn{1}{l}{0.5} & \multicolumn{1}{l}{Fixed} \\    \hline
    \multicolumn{5}{c}{Disk Inner Component} \vspace{0.1cm} \\
    \hline
    Param. & Unit & Model CS & Model Q & Range\\
    \hline
    R$_{\rm in}$ & au &  0.6 & 0.6 & Fixed \\
    R$_{t}$ & au &  5.5 & 3.5 & [2.5 - 7.5]\\
    M$_{\rm d,in}$ & M$_\odot$ & 6$\times$10$^{-9}$ & $2\times10^{-12}$ & $[10^{-13, -7}]$\\
    H$_{100\,\rm au}$ & au &  10.0 & 10.0 & Fixed \\
    $\gamma$ &  & 1.0 & 1.0 & Fixed \\
    $a_{\rm min}$ & $\mu$m &  1.0 & 0.006 & Fixed \\
    $a_{\rm max}$ & $\mu$m &  2200 & 0.006 & Fixed \\
    $p_\Sigma$ &  & -1.5 & -1.5 & Fixed \\
    Carbon & \% & 25 & 0 & [0-35]\\
    \hline
    \multicolumn{5}{c}{Disk Outer Component} \vspace{0.1cm} \\
    \hline
    \multicolumn{2}{l}{Param.} & \multicolumn{1}{l}{Unit} & \multicolumn{1}{l}{Value} & \multicolumn{1}{l}{Range} \\ \hline
    \multicolumn{2}{l}{R$_{\rm out}$} & \multicolumn{1}{l}{au} & \multicolumn{1}{l}{320} & \multicolumn{1}{l}{Fixed}\\
    \multicolumn{2}{l}{M$_{\rm d,out}$} & \multicolumn{1}{l}{M$_\odot$} & \multicolumn{1}{l}{2$\times$10$^{-5}$} & \multicolumn{1}{l}{Fixed}\\
    \multicolumn{2}{l}{H$_{100\,\rm au}$} & \multicolumn{1}{l}{au} & \multicolumn{1}{l}{10.0} & \multicolumn{1}{l}{Fixed}\\
    \multicolumn{2}{l}{$\gamma$} & \multicolumn{1}{l}{-} & \multicolumn{1}{l}{1.0} & \multicolumn{1}{l}{Fixed}\\
    \multicolumn{2}{l}{$a_{\rm min}$} & \multicolumn{1}{l}{$\mu$m} & \multicolumn{1}{l}{1.0} & \multicolumn{1}{l}{Fixed}\\
    \multicolumn{2}{l}{$a_{\rm max}$} & \multicolumn{1}{l}{$\mu$m} & \multicolumn{1}{l}{2200} & \multicolumn{1}{l}{Fixed}\\
    \multicolumn{2}{l}{$p_\Sigma$} & \multicolumn{1}{l}{-} & \multicolumn{1}{l}{-1.5} & \multicolumn{1}{l}{Fixed}\\
    \multicolumn{2}{l}{Carbon} & \multicolumn{1}{l}{\%} & \multicolumn{1}{l}{25} & \multicolumn{1}{l}{[0-35]}\\
    \hline
    \end{tabular}\\
    \label{tab:RT-BestModel}
\end{table}
\\
We conducted a grid search for the inner disk dust mass M$_{\rm d,in}$ and for the transition radius between the low- and high-surface-density disk R$_t$ (cf. Table~\ref{tab:RT-BestModel_FluxContr}) to match the SED as well as the NIR excess (F$_c$+F$_h$) estimated by interferometry (cf. Table~\ref{tab:Continuum_param}). 
The other disk parameters were kept identical for the inner and outer components. 
 Table~\ref{tab:RT-BestModel} shows our best result for the models CS and Q, respectively, for which our radiative transfer simulation produces the SED displayed in the top left panel of Fig.~\ref{fig:RT-BestModel}. For better readability, only the case of model CS is shown because the result for model Q is almost identical. Visually, these models provide the best overlap between the model and the data, except in the FIR region where the photometry data are more severely overestimated by the model. The FIR emission traces the disk at large radii not immediately relevant in our study. 
For the model CS, a transition radius R$_t$ at $\sim$\,5.5\,au and an inner disk dust mass of $\sim$\,6$\times$10$^{-9}$\,$M_\odot$ are derived from the radiative transfer modeling (cf. Table~\ref{tab:RT-BestModel}). We note that our two best models CS and Q may still contain some level of degeneracy, because other parameters such as the scale height, the index of the surface density power law, or the flaring index may influence the NIR excess. However, we limited ourselves to  the inner disk properties that can be directly constrained by the NIR flux ratios and spatial confinement of the emission derived from our interferometric measurements. \\

\begin{table}[b]
    \centering
    \caption{\centering RT results for Model CS. The sign $^{(*)}$ indicates that the relative flux contributions of the disk and halo components are added. LDC and HDC stand for low-density and high-density component, respectively.}
    \begin{tabular}{l|cccc|cccc}
    \hline
    \hline
    R$_t$, M$_{\rm d,in}$ & \multicolumn{4}{|c|}{5.5\,au, 6$\times$10$^{-9}$\,M$_\odot$} & \multicolumn{4}{c}{3.5\,au, 3$\times$10$^{-9}$\,M$_\odot$}  \\ \hline\hline
Flux & \multicolumn{2}{|c}{SED} & \multicolumn{2}{c|}{Interf.} & \multicolumn{2}{|c}{SED} & \multicolumn{2}{c}{Interf.}  \\ 
      (\%) & H & K & H & K & H & K & H & K \\    \hline
    Star        & 30    & 16 & 22 & 23 & 28 & 14 & 22 & 23 \\
    LDC  & 70    & 75 & 78$^{(*)}$ & 77$^{(*)}$ & 41 & 44 & 78$^{(*)}$& 77$^{(*)}$\\
    HDC  & 0     & 9  & -- & -- & 31 & 42 & -- & -- \\
    \hline
    \end{tabular}\\
    \label{tab:RT-BestModel_FluxContr}
\end{table}

\subsection{Transition radius}

For the adopted two-component disk model for HD\,98922, the low- to high-density transition radius R$_t$ is found to be located at $\sim$5.5\,au, with a lower-limit at 2.5\,au below which the NIR excess is systematically overestimated in the SED. 
Despite the degeneracy between R$_t$ and M$_{\rm d,in}$ for the determination of the NIR excess flux in the SED fit, 
our interferometric measurements can set constrains on the inner radius, relative flux contribution, and compact spatial extent of the low-density component that dominates the NIR excess. 
Table~\ref{tab:RT-BestModel_FluxContr} illustrates this point by comparing the relative flux contributions of the inner low- and outer high-density components inferred from radiative transfer to the estimates from the interferometric measurements, and this for two values of R$_t$, both giving a good fit of the SED. For smaller values of R$_t$, the NIR flux contribution of the outer high-density component increases, while that of the inner low-density component decreases. Qualitatively, this would translate for our geometrical models into wider rings, or into a higher contribution of the halo component, which we only find to be less than 10\% in both bands. For instance, in the case R$_t$=3.5\,au, the innermost low-density component probed by GRAVITY and PIONIER contributes only up to $\sim$40\%. 
However, it should be noted that the proposed argument suffers from some degree of uncertainty: as the modeling of our interferometric observables relies on single-ring geometrical models rather than on radiative transfer images, the accuracy on the determination of R$_t$ remains limited. As the detailed interferometric modeling of radiative transfer images goes beyond the scope of this paper, we limit ourselves to propose that the structured disk of HD\,98922 shows a dust density transition located no closer than $\sim$4-6\,au to the central star. Estimating an upper value for R$_t$ is not feasible using only the presented data because the NIR excess (set on the first order by R$_t$, M$_{\rm d,in}$ and M$_{\rm d,out}$) becomes a degenerated quantity for excessively large values of R$_t$.

\subsection{Surface density profile}

The dust surface density profile we implement has a sharp discontinuity at R$_t$, which does not correspond to a realistic case, given that a continuous transition profile would be more physically meaningful. Such sharp transitions are nonetheless proposed to differentiate between volatile-rich and volatile-free regions, as for instance in the case of the water snow line \citep{Hayashi1981,Lecar2006}. More recent works have reported structured inner disks with regions of differentiated surface density \citep{Tatulli2011,Matter2016}. We derive a low- to high-density transition of two orders of magnitude at about 5\,au (Fig.~\ref{fig:RT-BestModel}), with a dust surface density $\Sigma_d$ ranging from $\sim$\,2$\times$10$^{-4}$\,g\,cm$^{-2}$ to $\sim$\,0.006\,g\,cm$^{-2}$ in the low-density region. We emphasize that we cannot exclude that a gap is present in the disk of HD\,98922 beyond $\sim$5\,au, but mid-infrared (MIR) interferometric data are necessary to explore this hypothesis. However, we also note that HD\,98922 is classified as a group II source in the classification of \cite{Meeus2001} and based on the N-to-K size ratio \citep{Perraut2019}, suggesting a preferentially flat, gap-free disk.

\subsection{Location and temperature of the dust}

The dust emission in the H and K bands are expected to originate in a region close to the disk inner rim. In our case, the bulk of the emission is located at different radii for the two bands. We find the characteristic size
 of the K band is larger than that of the H band by a factor $\sim$1.4 to 2 when considering either the ring annular radius a$_r$ or the half-light radius $a$ parameters. This trend is seen in previous works when comparing the results of \cite{Lazareff2017} and \cite{Perraut2019}: out of 21 common sources, 17 show a larger K band by a factor ranging from 1.1 to $\sim$3. At first glance, the difference in resolution between GRAVITY (1.75\,mas) and PIONIER (1.22\,mas) by a factor 1.4 seems capable of explaining the fact that PIONIER resolves a more compact emission, similarly to what observed for HD\,163296 \citep{Sanchez2021}. A more physical, albeit simplistic, argument can be formulated based on the Wien's temperature of the dust in a temperature-gradient disk (i.e., $T(r)$\,=\,$T_0\,(r/r_0)^{-q}$) emitting as a blackbody. For a typical value of $q$\,=\,0.75 for a flat group II disk, we derive a size ratio of $r_K$/$r_H$\,$\sim$\,$(1350/1750)^{1/0.75}$, or 1.41, in agreement with our findings.\\
\\
One challenge posed by our results is the presence of H-band-emitting dust as close as $\sim$0.6\,au to the star. In the scenario of a passively irradiated disk with an optically thin inner cavity \citep{Muzerolle2004,Monnier2005}, the dust temperature at this location, assuming black-body grain emitters ($\epsilon$=1), is $T_g$\,$\sim$\,2200\,K, which is above the sublimation temperature of standard silicates. This is also seen in Fig.~\ref{fig:RT-BestModel} for the temperature structure of our carbon-rich disk model. The problem of very hot dust grains inside the theoretical sublimation radius has been observed in other sources such as \object{Z\,CMa}, \object{V1685\,Cyg}, \object{MWC\,297}, and \object{HD\,190073} \citep{Monnier2005,Hone2017,Setterholm2018}, invoking the presence of refractory dust (e.g., corundum or iron), the sublimation temperature of which critically depends on the disk density \citep{Kama2009}. Alternatively, \citet{Benisty2010} proposed the possibility that a small fraction of highly refractory graphite grains with high sublimation temperature ($T_s$>2000\,K) contribute to the NIR excess very close to HD\,163296, and this might apply to HD\,98922 as well. Clearly, further radiative transfer modeling is required to tackle this question by considering other values of $a_{\rm min}$ and $p_\Sigma$, as well as grain composition. We note that adding amorphous carbon to our initial pure-silicate model was essential to decrease the dust temperature at 0.6\,au from $T>$3000\,K to $T$\,$\sim$\,2300\,K. \\
Alternatively, for Herbig stars with  masses similar to or larger than that of HD\,98922, the undersized inner disk is explained by the presence of optically thick gas in the inner cavity, which is found to be favored in systems with accretion rates of $\dot{\rm M}_{\rm acc} \gtrsim 10^{-8}\,$M$_{\odot}\,$yr$^{-1}$ \citep{Muzerolle2004}, as for HD\,98922.\\
Another possible explanation for the detection of excess emission very close to the star is the presence in the inner region of a fraction of quantum heated particles (QHPs), which can be stochastically heated by the strong UV radiation field from the central star, reaching temperatures higher than the equilibrium temperature and producing NIR continuum emission. This scenario was proposed for \object{HD\,100453}, \object{HD\,179218}, and \object{HD\,141569} \citep{Klarmann2017,Kluska2018,Ganci2021}. Polycyclic aromatic hydrocarbons (PAHs), as an example of QHPs, are detected in HD\,98922 \citep{Geer2007,Acke2010}. Interestingly, the 6.2-to-11.3\,$\mu$m feature ratio is estimated from \cite{Seok2017}\,\footnote{The ratio is uncertain due to incomplete wavelength coverage} to be $I_{\rm 6.2}/I_{\rm 11.3}$\,$\sim$\,3--4, with ratios larger than unity pointing at predominantly unshielded ionized PAH species in the disk inner regions that are directly exposed to the intense UV radiation field of the star \citep{Maaskant2014}. We tested a radiative transfer model with a PAH-based inner component  (Model Q, Table~\ref{tab:RT-BestModel}), which allows a good fit of the SED with a very small mass of QHP grains (2$\times$10$^{-12}$\,M$_\odot$) and a transition radius of 3.5\,au. This value of R$_t$ is estimated as the balance between the increasing contribution of the outer high-density component (HDC) to the NIR SED (for R$_t$<3.5\,au) and the increasing contribution of the inner low-density component (LDC) to the SED at $\lambda$\,$\sim$1.5\,$\mu$m (for R$_t$>3.5\,au and due to the blue spectral index of the QHP emission). However, in the case of Model Q, the flux fractional contribution of the LDC is estimated to be only $\sim$\,5\% in the K-band and 10\% in the H-band.\\
Ultimately, a mixture of refractory grains and QHPs could explain the presence of continuum emission as close as 0.6\,au and detected in the H-band. The presence of predominantly ionized PAH species close to the star is therefore not to be discarded and requires further work.

\section{Discussion}\label{sec:Discussion}

\subsection{The origin of the time-variable inner disk asymmetry}
\label{ssec:AsymmetryNature-Discussion}
Our analysis suggests a crescent-like asymmetric dust feature in the inner 1 au region of \object{HD\,98922}   for which orbital motion is detected. 
Comparable azimuthal asymmetries are observed in other disks both at small and large scales: \cite{Varga2021} and \cite{Sanchez2021} characterized a time-variable crescent-like feature  in
the MIR and NIR in the innermost disk of HD\,163296, whereas \cite{Ibrahim2023} directly imaged the complex asymmetric and time-variable inner rim of HD\,190073  at similar wavelengths. Using ALMA, \cite{Casassus2015} and \cite{VanDerMarel2013} reveal a strong azimuthal asymmetry at separations of $>$\,50\,au in HD\,142527 and Oph\,IRS\,48, respectively. Infrared and submillimeter observations trace micron-sized and millimeter-sized dust grains, respectively, and vortices or efficient dust traps are tentatively invoked for these sources. \\
Regarding HD\,98922, it is remarkable that \cite{Garatti2015} also detect large-scale disk asymmetry at $\sim$40\,au from the central star. Imaged with SINFONI in the K-band, the emission is associated with scattered light, and the authors exclude the possibility that the arc-shaped structure is due to the presence of a possible stellar companion as close as $\sim$\,20\,au to the central source and with a mass of $\ge$\,$0.5$\,$\rm M_\odot$. Here, we explore a few hypotheses as to the nature and dynamical evolution of the asymmetric structure in the innermost disk of HD\,98922.
  
\subsubsection{Disk hydrodynamic instabilities}

Inclination projection effects as in the case of FS\,CMa \citep{Hofmann2022,Kluska2020} are very unlikely to be able to explain the azimuthal asymmetry in HD\,98922. The strength of the closure phase signal ---despite the low disk inclination ($\lesssim$30$^{\circ}$)--- and the detected orbital motion suggest a physical effect in the disk, possibly resulting from hydrodynamic instabilities. \\
Local perturbations in the inner disk may explain the variable azimuthal asymmetry under the assumption of strong dust--gas coupling. The Stokes parameter quantifies the degree of coupling and is given by \cite{Birnstiel2010}: 
\begin{eqnarray}\label{eq:stokes}
{\rm St} = \frac{\rho_{\rm s}.a}{\Sigma_{\rm g}}\frac{\pi}{2}.
\end{eqnarray}
The condition for dust--gas coupling in the Epstein regime, namely St$\lesssim$1, is fulfilled for grain sizes of $a$\,$\lesssim$\,2$\Sigma_{\rm g}$/$\pi \rho_{\rm s}$. Considering a gas surface density at 1\,au of 0.6\,g\,cm$^{-2}$ (see Fig.~\ref{fig:RT-BestModel}) and a bulk density of silicate-rich grains of $\rho_{\rm s}$\,$\sim$\,3\,g.cm$^{-3}$ \citep{Pollack1994}, we obtain a grain size of $a$\,$\lesssim$\,1300\,$\mu$m, which suggests efficient dust--gas coupling for most grain sizes assumed in our model CS (see Table~\ref{tab:RT-BestModel}). Under this hypothesis, a vortex resulting from disk instabilities could result in the observed over-density. The vortex might be formed through Rossby wave instability (RWI, \citealt{Meheut2010}), which could be induced by the presence of a Jupiter-mass planetary companion. RWI-induced vortices occur preferentially in inviscid disks (typically $\alpha$\,$\lesssim$\,10$^{-4}$) associated to a low-turbulence environment, which allows the vortex to survive for thousands of orbits. It is therefore highly unlikely that we trace within the time baseline of our observations two unrelated over-densities that may have developed in different regions of the disk.\\
\\
Regarding the morphology of the crescent-like  feature, both \cite{Meheut2010} and \cite{Barge2017} show the formation of a vortex on one side of the disk, although the latter authors explore this aspect in the outer disk, at 60\,au. The length/width aspect ratio of the simulated vortex in \cite{Meheut2010} and \cite{Varga2021} appears to be large, with the vortex covering at least one-third of the circumference. 
From our interferometric K-band observations, we model a crescent-like structure with an azimuthal span comparable to these authors (cf. Fig.~\ref{fig:Continuum-imgs}). 
In \cite{Barge2017}, the vortex appears azimuthally more compact.  The latter authors include dust grains in their simulations and show that the contrast of the over-density is lowest for the smallest grains, with a value of $\sim$\,10 for a size of 42\,µm in the {dust}-density map. These latter authors report a fainter contrast of $\sim$\,1.3 in the {gas}-density map for which no dust is included. Similarly, these levels of contrast compare well with the values derived from our geometrical modeling (see Sect.~\ref{subsubsec:disk-asymm}). \\
\\
An azimuthally asymmetric feature can also arise from density waves due to magnetorotational turbulence \citep{Flock2017}. The observational signatures developed in this scenario also include a vortex visible in the form of an arc that develops at the edge of the dead zone, with a brightness contrast of comparable magnitude to that in the case of RWI. \\
It is noteworthy that, while the flux properties of the crescent-like feature are derived from a pure geometrical modeling of our interferometric data, the radiative transfer simulations predict an inner disk essentially optically thin at 2.2\,$\mu$m due to the low surface density. It would therefore be interesting to include a nonaxisymmetric model  in future radiative transfer simulations in order to better constrain the influence of the grain properties and size on the azimuthal NIR contrast of the crescent-like feature.

\subsubsection{Disk warp induced by a close companion}\label{sec:warp}

An alternative scenario that can explain our observations is the presence of a close companion inside the cavity of what then becomes a circumbinary (CB) disk. \cite{Ragusa2017} show from 3D SPH gas and dust simulations that a crescent-like asymmetry develops at the CB disk inner edge only (and not further out in the disk) due to the dynamics of the central pair. The contrast of the nonaxisymmetric feature depends on the mass ratio of the  components $q$, with $q$\,$\geqslant$\,0.05 needed to produce a sufficiently contrasted over-density. This would translate in the case of HD\,98922 into a companion mass of at least $\sim$0.3\,M$_\odot$ at a separation of $\lesssim$\,1\,mas. \\
\cite{Baines2006} report a spectro-astrometric companion candidate with a separation of between 0.5$^{\prime\prime}$ and 3$^{\prime\prime}$ ---that is, wider than 325\,au---, which is not reported by \cite{Garufi2022} from the SPHERE total intensity images. Furthermore, the RUWE parameter reported from Gaia measurements is 0.94, which means that a single-star model provides, in principle, a good fit to the astrometric observations. 
Finally, a comparison of the photospheric lines obtained in 2005 and 2013 with the spectrograph/spectroplarimeter ESPaDOnS\footnote{https://www.cfht.hawaii.edu/Instruments/Spectroscopy/Espadons/} between 0.3 and 1~µm does not reveal any clear sign of variability in the profile or position of the line down to $\sim$10\,km\,s$^{-1}$ (Alecian et al., private communication). With the realistic hypothesis that the period of such a companion does not coincide with the time lapse between the two ESPaDOnS epochs, we propose 
that no companion more massive than 0.3\,M$_\odot$ at 0.1\,au or 0.6\,M$_\odot$ at 0.5\,au is seen in the inner cavity. Of course, the low inclination of the system and the limited number of epochs mean that this preliminary conclusion requires further strengthening.\\
While the close binarity status of HD\,98922 ---in particular for companions lighter than 0.3\,M$_\odot$--- cannot be totally excluded in the existing literature, we do not favor the idea that the presence of an inner stellar companion is inducing the crescent-like asymmetry.

\subsubsection{Dynamics of the asymmetric feature}
\label{ssec:Dynamics}

Thanks to our 11 year temporal baseline, we suggest that we are tracing the same azimuthal asymmetric feature possibly interpreted as a vortex in the inner disk at $\sim$1.0\,au. 
One hypothesis resulting from our work is that the bright over-density could show a sub-Keplerian motion with a period of $\sim$1\,year. We emphasize however that this hypothesis requires further assessment in the future, since our disk data modeling only accounts for circular motion. 
Gas in protoplanetary disks is expected to orbit at sub-Keplerian rotation due to a radial pressure gradient with a deviation from Keplerian rotation of $\eta$\,$\propto$\,$c_s^2$/v$_K^2$. Even assuming gas temperatures of a few thousand Kelvin at $\sim$1.0\,au, the deviation factor is $\eta$\,$\sim$\,10$^{-3}$, and would therefore not explain the difference we find between Keplerian and sub-Keplerian motion. 
Although the scenario of a close binary inside the disk cavity is not favored (Sect.~\ref{sec:warp}), this is a case where more significant sub-Keplerian orbital velocity could arise. \cite{Ragusa2020} suggest that disk eccentricity due to the presence of a companion can result in the nesting of ellipses with a different eccentricity leading to a slowly precessing over-density through which the disk material passes. The caveat, in this case, is that the precession timescale of the over-density is of the order of $\sim$100-1000 orbits, which should have resulted in a very steady feature considering our temporal baseline. \\
A more qualitative discussion concerns the structures formed in the disk by the presence of a planetary-mass companion. It is well known that the perturbing planet develops a pattern of wound spirals on either side of it (e.g., \citealt{Kley1999}). The spiral arms are corotating with the planet, or, in other words, they are stationary in the reference frame of the planet. If the crescent-like feature in HD\,98922 were associated with such a companion, this would lie at $\sim$1.8\,au to match the orbital period of 12 months, considering the literature value of 6\,M$_\odot$ for the mass of the central star.\\
We underline that the question of the dynamics of asymmetric features in the innermost disk regions will become
an important question for future interferometric observations. It is remarkable to note that in their recent imaging campaign of HD\,190073, \cite{Ibrahim2023} present evidence that the detected sub-AU structure rotates two times slower than Keplerian, pointing at dynamics effects from the outer disk.

\subsection{Kinematics of the hydrogen hot gas: Wind or companion?}

\label{ssec:BrG-Discussion}

The origin of the hydrogen Br$\gamma$-line in the system is debated: previous works suggest various scenarios such as a stellar wind, an X-wind, or magnetospheric accretion \citep{Kraus2008}, with the latter case being favored by these authors.
On the other hand, strong variability over a timescale of a  few days is instead detected in the blueshifted absorption lobe of the Na\,I\,D and Balmer lines, which typically probes the inner dust-free cavity region \citep{Aarnio2017}. 
The blueshifted component can be modeled with a disk wind extending from 0.17\,au to 0.42-0.85\,au and a wind mass-loss rate of 10$^{-7}$\,M$_{\odot}$\,yr$^{-1}$. Similarly, a disk-wind model extending from $\sim$\,0.1\,au to $\sim$\,1\,au with a wind mass-loss rate of 2$\times$\,10$^{-7}$\,M$_{\odot}$ along with an asymmetric continuum disk model can be successfully fitted to VLTI/AMBER Br$\gamma$-line interferometric data \citep{Garatti2015}, which is in agreement with our continuum results.\\
\\
Another interpretation involves the presence of an accreting very low-mass or planetary-mass companion. 
It is interesting to note that not only does the location of the compact Br$\gamma$-line emitting region change through the epochs, but so does the line luminosity. 
In the scenario of a young embedded planet, variability of the line luminosity resulting from changes in the accretion rates can be seen during the orbital motion of the planet \citep{Szulagyi2020}. 
\\
By integrating the spectral line region through the observed wavelength-calibrated and continuum-normalized spectrum (Fig.~\ref{fig:GRAVITY-SC-Data}), we calculated the observed Br$\gamma$-line equivalent width ($W_{\rm Br\gamma}^{\rm obs}$) 
for each epoch,
from which the line equivalent width corrected for photospheric absorption and veiling ($W_{\rm Br\gamma}^{\rm corr}$) was derived (Table~\ref{tab:LineQuant}). 
Here, we discard the first two epochs G1 and G2 where the line is strongly affected by the continuum normalization and is therefore not highly reliable. 
We derive the flux density of the line $F_{\rm Br\gamma}$ in erg/sec/cm$^2$ by multiplying the spectral flux density of the nearby continuum,
\begin{equation}
    F_{\rm cont,\lambda} = F_{0,K} 10^{-0.4(m_K-0.1A_V)}\\
,\end{equation}
with the corrected equivalent width $W_{\rm Br\gamma}^{\rm corr}$. 
The K band zero-point flux is $F_{0,K}$\,=\,4.28$\times$10$^{-11}$\,erg/s/cm$^2$/\AA\,\citep{Rodrigo2020}. Finally, we compute the line luminosity as $L_{\rm Br\gamma}$\,=\,4$\pi$d$^2$$F_{\rm Br\gamma}$.
\\
In Table~\ref{tab:LineQuant}, we see that in the four-month period from March 2019 (G4) to July 2019 (G8), the line luminosity $L_{\rm Br\gamma}$ is relatively constant at $\sim$\, 3.3$\times$10$^{-2}$\,L$_\odot$, while after 6 months (G9) it increases by $\sim$\,10\% and does not vary significantly for 11 months (G11), after which it increases again by $\sim$\,5\% (G12) and again by $\sim$\,4\% in the following week (G13), for a total increase of $\sim$20\%. \\
We note here that the values listed in Table~\ref{tab:LineQuant} do not follow the known linear relationship between equivalent width and peak flux: $W = A \cdot F_{\rm peak} + B$. This could be due to the fact that our $W_{\rm Br\gamma}^{\rm obs}$ values are measured by integrating the spectral line region using the data themselves and not by integrating a Gaussian profile fitted to the data points. Additionally, the relative errors on $W$ and $F_{\rm peak}$ may differ depending on the value of $B$. In any case, the key insight captured by the data is that the variation in both peak intensity and in $W_{\rm Br\gamma}^{\rm obs}$ is a positive one.\\
Under the hypothesis of an accreting planetary-mass companion, we use the empirical relation of Table~3 from \cite{Szulagyi2020} that relates the Br$\gamma$-line luminosity to the mass of the accreting planet $M_{\rm p}$:
\begin{equation}
\log\left(\frac{L_{{\rm Br}\gamma}}{L_\odot}\right) = a\cdot M_{\rm p} + b
,\end{equation}
where the coefficients $a$ and $b$ depend on the opacity model. Considering the different opacity cases, we derive a mass $M_{\rm p}$ ranging between 10.3\,$M_{\rm J}$ and 12.6\,$M_{\rm J}$.\\
A number of caveats must nevertheless be considered. First, the value of $M_{\rm p}$ should be treated as a loose estimate because the simulation from \cite{Szulagyi2020} accounts for a model with a central stellar mass of 1\,M$_\odot$, a circumstellar environment extending from 2 to 12.4\,au, and a planet at 5.2\,au. Furthermore, in this scenario,  we make the implicit assumption that all the emission line flux is associated with the putative accreting companion, whereas a stellar component (e.g., in the form of a wind) may also contribute to the emission budget. In this scenario, the retrieved positions of the gas photocenter could result from the combined and weighted contributions from a stellar-driven component and a companion-driven component. Based on this argument, if it is a companion located further out at 1.8\,au that causes the azimuthal asymmetry (see 
Sect.~\ref{ssec:Dynamics}), a suitable weighting between the stellar-driven and companion-driven emission contributions could explain why the measured location of the combined photocenter is found between the star and the crescent-like feature. It is not the goal of this study to explore the case of a possible disk-embedded companion and its impact on the size, location, and dynamics of the Br$\gamma$-line-emitting region  in detail; nonetheless, we speculate that a young accreting planetary-mass companion is a possible scenario to consider, with further modeling required.

\begin{table}[t]
    \centering
    \small
    \caption{Properties of the Br$\gamma$-line emission. The relation between W$_{\rm Br\gamma}^{\rm obs}$ and W$_{\rm Br\gamma}^{\rm corr}$ follows Eq.~2 of \cite{Grant2022}.}
    \begin{tabular}{lccc}
    \hline
    \hline
    Epoch & W$_{\rm Br\gamma}^{\rm obs}$ & W$_{\rm Br\gamma}^{\rm corr}$ & L$_{\rm Br\gamma}$ \\ 
          & [\AA{}] & [\AA{}] & [$10^{-2}\,$L$_\odot$] \\ 
    \hline
        G4 & $-1.69\pm 0.06$  & $-2.90\pm 0.06$  & $3.34\pm 0.11$ \\ 
        G5 & $-1.56\pm 0.15$  & $-2.77\pm 0.15$  & $3.19\pm 0.19$ \\ 
        G6 & $-1.63\pm 0.13$  & $-2.84\pm 0.13$  & $3.27\pm 0.17$ \\ 
        G7 & $-1.59\pm 0.24$  & $-2.80\pm 0.24$  & $3.23\pm 0.29$ \\ 
        G8 & $-1.59\pm 0.07$  & $-2.80\pm 0.07$  & $3.23\pm 0.12$ \\ 
        G9 & $-1.93\pm 0.11$  & $-3.14\pm 0.11$  & $3.62\pm 0.16$ \\ 
        G10 & $-1.96\pm 0.15$ & $-3.17\pm 0.15$ & $3.65\pm 0.20$ \\ 
        G11 & $-1.97\pm 0.14$ & $-3.18\pm 0.14$ & $3.67\pm 0.19$ \\ 
        G12 & $-2.14\pm 0.10$ & $-3.35\pm 0.10$ & $3.86\pm 0.16$ \\ 
        G13 & $-2.32\pm 0.10$ & $-3.53\pm 0.10$ & $4.07\pm 0.16$ \\ 
         \hline
    \end{tabular}
    \label{tab:LineQuant}
\end{table}

\section{Summary}
\label{sec:Summary}

We present new multi-epoch VLTI/GRAVITY observations of \object{HD\,98922} which, coupled with VLTI/PIONIER archival data, form an 11 year observational-period data set for the system, accounting for a total of 33 different epochs between 2011 and 2022. This data set allows a unique interferometric study to test the potential time variability of the innermost circumstellar morphology. We can summarize the main conclusions of our work as follows:
\begin{itemize}
    \item The system, which is spatially resolved by both instruments at all epochs, shows temporal variability dominated by changes in the asymmetric  spatial distribution of brightness rather than in the flux ratio between the central star and its circumstellar environment. This is supported by the significant variability in the nonzero closure phase signals for observations obtained with comparable \textit{(u-$\varv$)} coverage, whereas little time variability is observed in the continuum squared visibilities.  
    \item Among the different modeling scenarios that we tested, the data are best explained by a crescent-like asymmetric dust feature radially extending from $\sim$\,0.6 to $\sim$\,2\,au, which dominates the NIR excess. The revealed disk feature revolves around the central star with an estimated period of $\sim$\,1\,year. This could point to sub-Keplerian orbital motion, which is also seen in the recently characterized system HD\,190073; however, a shorter period cannot be excluded. 
    \item The innermost location of the warm emitting dust traced by the interferometric data ($\sim$\,0.6\,au) is coupled to revised radiative transfer models that suggest a radially structured disk formed by an inner low-density component (optically thin at 2.2\,µm) and an outer high-density component with a transition between 3 to 5\,au. The models favor either an inner region with a mass of 2$\times$10$^{-9}$\,M$_\odot$
 made up of large, carbon-enriched  silicate  grains, or the presence of very small stochastically heated particles. The high dust temperature requires the presence of some level of refractory dust.    
    \item The origin of the azimuthal asymmetry is preferentially connected to hydrodynamic effects, either induced by disk instabilities generating a vortex or by an undetected low-mass (substellar or planetary-mass) companion embedded in the low-density region of the disk and launching a wound spiral. An asymmetry resulting in disk fragmentation is unlikely considering the low surface density of $\lesssim$\,10\,g\,cm$^{-2}$, and we do not find evidence for the presence of a close (0.1--0.5\,au) stellar companion inside the inner dust rim that could drive eccentricity in the central cavity resulting in a crescent-like structure. However, we do believe that our data cannot completely rule out this hypothesis, in particular regarding a stellar companion less massive than $\sim$\,0.3\,M$_\odot$. 
    \item The high-resolution GRAVITY data show a single-peaked Br$\gamma$ emission line with a luminosity increasing by 20\,\% in 3 years. The location of the compact Br$\gamma$-line-emitting region is offset with respect to the central star and appears to be consistently located inside the dust cavity and between the dusty feature and the star.    
    \item The interpretation of the interferometric observations of the hot, gaseous component in HD\,98922 leaves some open questions: while a  stellocentric magnetospheric accretion scenario is not favored, a wider asymmetric disk wind is a qualitatively plausible scenario, though more advanced numerical simulations would be needed to explain the time variability in its spatial distribution. We also speculate that a strongly accreting  substellar or planetary companion with a mass of larger than $\sim 10$\,M$_J$  could explain our measurement of the Br$\gamma$ emission line. However, this would trigger further questions regarding its dynamics and connection to the detected crescent-like structure in the inner disk.
\end{itemize}

\section*{Acknowledgements}

\begin{footnotesize}

The authors would like to thank H\'elo\"ise Meheut and Fran\c cois M\'enard for the fruitful discussions on the inner disk of HD\,98922. 
V.G. was supported for this research through a stipend from the International Max Planck Research School (IMPRS) for Astronomy and Astrophysics at the Universities of Bonn and Cologne, and from the Bonn-Cologne Graduate School of Physics and Astronomy (BCGS).
A.C.G. has been supported by PRIN-INAF MAIN-STREAM 2017 and PRIN-INAF 2019 (STRADE), P.
This work is based on observations made with ESO Telescopes at the La Silla Paranal Observatory under program IDs listed in Table~\ref{tab:ObsLogPIONIER} and Table~\ref{tab:ObsLogGRAVITY}.
This work has made use of data from the European Space Agency (ESA) mission {\it Gaia} (\url{https://www.cosmos.esa.int/gaia}), processed by the {\it Gaia} Data Processing and Analysis Consortium (DPAC, \url{https://www.cosmos.esa.int/web/gaia/dpac/consortium}). Funding for the DPAC has been provided by national institutions, in particular the institutions in the {\it Gaia} Multilateral Agreement.
We acknowledge the Gemini Observatory for the use of the IR spectrum model of the atmospheric transmission above Cerro Pachon.
This research has made use of the model atmosphere grid NeMo, provided by the Department of Astronomy of the University of Vienna, Austria (\url{http://www.univie.ac.at/nemo/}). NeMo was funded by the Austrian Science Fonds. 
This research has made use of the Jean-Marie Mariotti Center \texttt{Aspro}.
This research has made use of the Spanish Virtual Observatory (\url{https://svo.cab.inta-csic.es}) project funded by MCIN/AEI/10.13039/501100011033/ through grant PID2020-112949GB-I00.
This paper makes use of the following ALMA data: ADS/JAO.ALMA\#2015.1.01600.S. ALMA is a partnership of ESO (representing its member states), NSF (USA) and NINS (Japan), together with NRC (Canada), MOST and ASIAA (Taiwan), and KASI (Republic of Korea), in cooperation with the Republic of Chile. The Joint ALMA Observatory is operated by ESO, AUI/NRAO and NAOJ. The authors thank the anonymous referee for assessing the quality of this work.

\end{footnotesize}

\bibliographystyle{aa}
\bibliography{sample} 

\appendix

\section{Logs and observation data}\label{apx:dataset}

\begin{table*}
\centering
\begin{small}
\caption{\centering Observation logs of the VLTI/PIONIER \object{HD\,98922} observations}
\begin{tabular}{lccccccccc} 
\hline
\hline
ID & \multicolumn{1}{c}{Date} & \multicolumn{1}{c}{UT} & \multicolumn{1}{c}{Configuration} & \multicolumn{1}{c}{N} & \multicolumn{1}{c}{Calibrator} & \multicolumn{1}{c}{Seeing\,[$^{\prime\prime}$]} & \multicolumn{1}{c}{Airmass} & \multicolumn{1}{c}{$\tau_0$\,[\rm ms]} & \multicolumn{1}{c}{ID$_{obs}$} \\
\hline
P1  & 03-06-2011 & 00:21 & D0-G1-H0-I1 & 4 & \object{NDA} & 0.5-0.9 & 1.17-1.26 & 1.6-2.9 & 087.C-0458(C)  \\
P2  & 08-06-2011 & 23:28 & A1-B2-C1-D0 & 2 & \object{NDA} & 0.8-1.3 & 1.15-1.16 & 2.0-2.7 & 087.C-0458(B)  \\
P3  & 25-03-2012 & 04:50 & A1-G1-I1-K0 & 2 & \object{NDA} & 0.5-0.9 & 1.16-1.38 & 2.6-4.8 & 088.D-0828(B)  \\
P4  & 28-03-2012 & 05:22 & A1-G1-I1-K0 & 2 & \object{NDA} & 1.4-1.6 & 1.22-1.28 & 1.3-1.4 & 088.D-0185(A)  \\
P5  & 20-12-2012 & 06:47 & A1-B2-C1-D0 & 2 & \object{NDA} & 0.5-0.6 & 1.38-1.47 & 8.1-8.5 & 190.C-0963(C)  \\
P6  & 22-12-2012 & 09:13 & A1-B2-C1-D0 & 1 & \object{NDA} & 0.8-0.9 & 1.15-1.16 & 8.1-9.5  & 190.C-0963(C) \\
P7  & 26-01-2013 & 08:48 & A1-G1-J3-K0 & 1 & \object{NDA} & 1.2-1.5 & 1.16-1.51 & 1.1-1.2& 190.C-0963(A) \\
P8  & 27-01-2013 & 07:11 & A1-G1-J3-K0 & 2 & \object{NDA} & 0.9-1.0 & 1.14-1.15 & 2.7-2.8 & 190.C-0963(A) \\
P9  & 28-01-2013 & 08:39 & A1-G1-J3-K0 & 3 & \object{NDA} & 0.8-1.5 & 1.17-1.28 & 1.5-3.3 & 190.C-0963(A) \\
P10  & 30-01-2013 & 07:25 & A1-G1-J3-K0 & 1 & \object{NDA} & 1.6-1.3 & 1.30-1.80 & 1.5-1.6 & 190.C-0963(C) \\
P11  & 31-01-2013 & 03:39 & A1-G1-J3-K0 & 2 & \object{NDA} & 0.9-1.1 & 1.15-1.16 & 1.8-2.3 & 190.C-0963(A) \\
P12  & 01-02-2013 & 04:35 & A1-G1-J3-K0 & 3 & \object{NDA} & 0.8-1.1 & 1.18-1.34 & 2.0-3.4 & 190.C-0963(A) \\
P13  & 17-02-2013 & 01:47 & D0-G1-H0-I1 & 3 & \object{NDA} & 0.7-0.8 & 1.19-1.22 & 3.7-4.4 & 190.C-0963(B) \\
P14  & 18-02-2013 & 03:27 & D0-G1-H0-I1 & 4 & \object{NDA} & 0.5-0.9 & 1.14-1.52 & 3.2-5.4 & 190.C-0963(B) \\
P15  & 19-02-2013 & 08:20 & D0-G1-H0-I1 & 1 & \object{NDA} & 0.7-0.9 & 1.26-1.27 & 3.1-3.9 & 190.C-0963(B) \\
P16  & 20-02-2013 & 02:36 & D0-G1-H0-I1 & 2 & \object{NDA} & 0.7-1.0 & 1.41-1.49 & 3.8-5.2 & 190.C-0963(B) \\
P17  & 23-06-2014 & 22:57 & A1-G1-J3-K0 & 2 & \object{HD\,98895} & 0.5-0.8 & 1.17-1.19 & 1.9-2.6 & 093.C-0559(D) \\
P18  & 21-02-2016 & 06:08 & D0-G2-J3-K0 & 1 & \object{HD\,98895} & 1.4-1.9 & 1.13-1.14 & 0.9-1.4 & 096.C-0867(C) \\
P19  & 01-03-2016 & 05:44 & D0-G2-J3-K0 & 2 & \object{HD\,98895} & 1.0-1.7 & 1.14-1.16 & 1.7-2.9 & 096.C-0867(D) \\
P20  & 02-03-2016 & 06:18 & D0-G2-J3-K0 & 2 & \object{HD\,98895} & 0.8-1.0 & 1.16-1.20 & 1.9-2.3 & 096.C-0867(E) \\
\hline
\end{tabular}\label{tab:ObsLogPIONIER}
\tablefoot{The date format is day-month-year. N denotes the number files that have been recorded on the target. NDA: no data available; These archival data were retrieved already calibrated from the JMMC Optical interferometry DataBase (\texttt{http://oidb.jmmc.fr/index.html}).}
\end{small}
\end{table*}

\begin{table*}
\centering
\begin{small}
\caption{\centering Observation logs of the VLTI/GRAVITY \object{HD\,98922} observations}
\begin{tabular}{lccccccccc} 
\hline
\hline
ID & \multicolumn{1}{c}{Date} & \multicolumn{1}{c}{UT} & \multicolumn{1}{c}{Configuration} & \multicolumn{1}{c}{N} & \multicolumn{1}{c}{Calibrator} & \multicolumn{1}{c}{Seeing\,[$^{\prime\prime}$]} & \multicolumn{1}{c}{Airmass} & \multicolumn{1}{c}{$\tau_0$\,[\rm ms]} & \multicolumn{1}{c}{ID$_{obs}$} \\
\hline
G1  & 22-02-2017 & 07:13 & A0-G1-J2-K0 [astro.] & 14  & \object{HD\,103125} & 0.7-1.7 & 1.18-1.70 & NDA & 098.C-0765(C) \\
G2  & 19-03-2017 & 04:36 & A0-G1-J2-K0 [astro.] &  5  & \object{HD\,100825} & 0.7-0.9 & 1.14-1.17 & NDA & 098.D-0488(A) \\
G3  & 15-06-2018 & 00:09 & D0-G2-J3-K0 [medium] &  4  & \object{HD\,100825} & 0.8-1.2 & 1.21-1.30 & 2.3-3.2 & 0101.C-0311(A)  \\
G4  & 19-03-2019 & 05:45 & D0-G2-J3-K0 [medium] &  6  & \object{HD\,100825} & 0.5-0.6 & 1.20-1.27 & 3.5-5.9 & 0102.C-0408(D)  \\
G5  & 24-05-2019 & 01:45 & A0-G1-J2-J3 [large] &  6  & \object{HD\,103125} & 0.6-0.9 & 1.22-1.29 & 1.8-3.2 & 0103.C-0347(C)  \\
G6  & 04-06-2019 & 23:21 & A0-B2-C1-D0 [small] &  7  & \object{HD\,103125} & 0.8-1.1 & 1.14-1.16 & 2.8-4.1 & 0103.C-0347(B)  \\
G7  & 11-07-2019 & 23:22 & D0-G2-J3-K0 [medium] &  5  & \object{HD\,103125} & 0.8-1.1 & 1.33-1.42 & 2.0-3.0 & 0103.C-0347(A)  \\
G8  & 13-07-2019 & 22:56 & D0-G2-J3-K0 [medium] &  7  & \object{HD\,103125} & 0.3-0.5 & 1.29-1.41 & 1.3-1.4 & 0103.C-0347(A)  \\
G9  & 27-01-2020 & 05:56 & A0-G2-J2-J3 [$\ast$] &  5  & \object{HD\,103125} & 0.7-1.0 & 1.16-1.21 & 3.6-5.9 & 0104.C-0567(A)  \\
G10  & 04-02-2020 & 05:35 & A0-B2-C1-D0 [small] & 11  & \object{HD\,103125} & 0.7-1.1 & 1.14-1.20 & 2.1-4.9 & 0104.C-0567(C)  \\
G11  & 23-12-2020 & 06:31 & D0-G2-J3-K0 [medium] & 13  & \object{HD\,99311} & 0.7-1.9 & 1.20-1.47 & 2.1-4.6 & 106.212G.002  \\
G12  & 07-02-2022 & 07:43 & D0-G2-J3-K0 [medium] &  8  & \object{HD\,103125} & 0.5-1.2 & 1.16-1.24 & 4.4-12.6 & 108.228Z.002 \\
G13  & 14-02-2022 & 06:11 & A0-G1-J2-J3 [large] &  6  & \object{HD\,103125} & NDA & 1.14-1.15 & NDA & 108.228Z.001 \\
\hline
\end{tabular}\label{tab:ObsLogGRAVITY}
\tablefoot{Same as \ref{tab:ObsLogPIONIER}. The configuration names -- small, medium, large, astrometric -- are reported. The configuration A0-G2-J2-J3 [$\ast$], not offered on a regular basis, covers spatial frequencies comparable to the large configuration.}
\end{small}
\end{table*}

\newpage 

\begin{figure*}[!ht]
\centering
\includegraphics[width=0.73\textwidth]{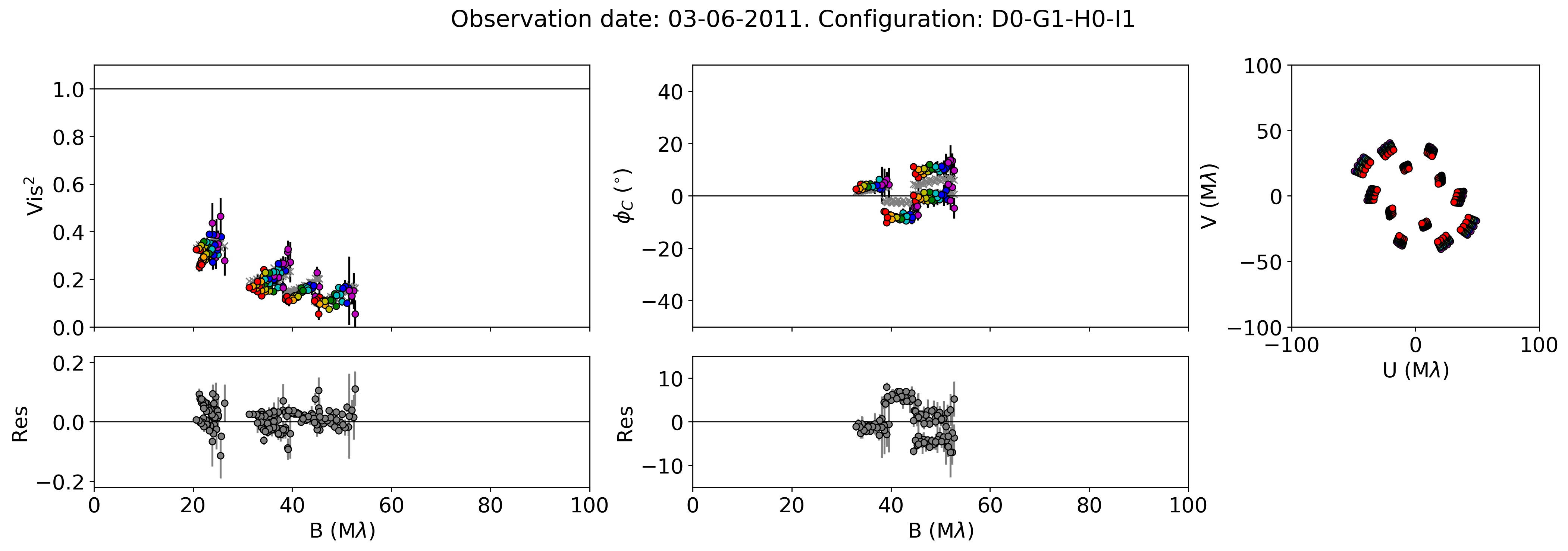} \\
\includegraphics[width=0.73\textwidth]{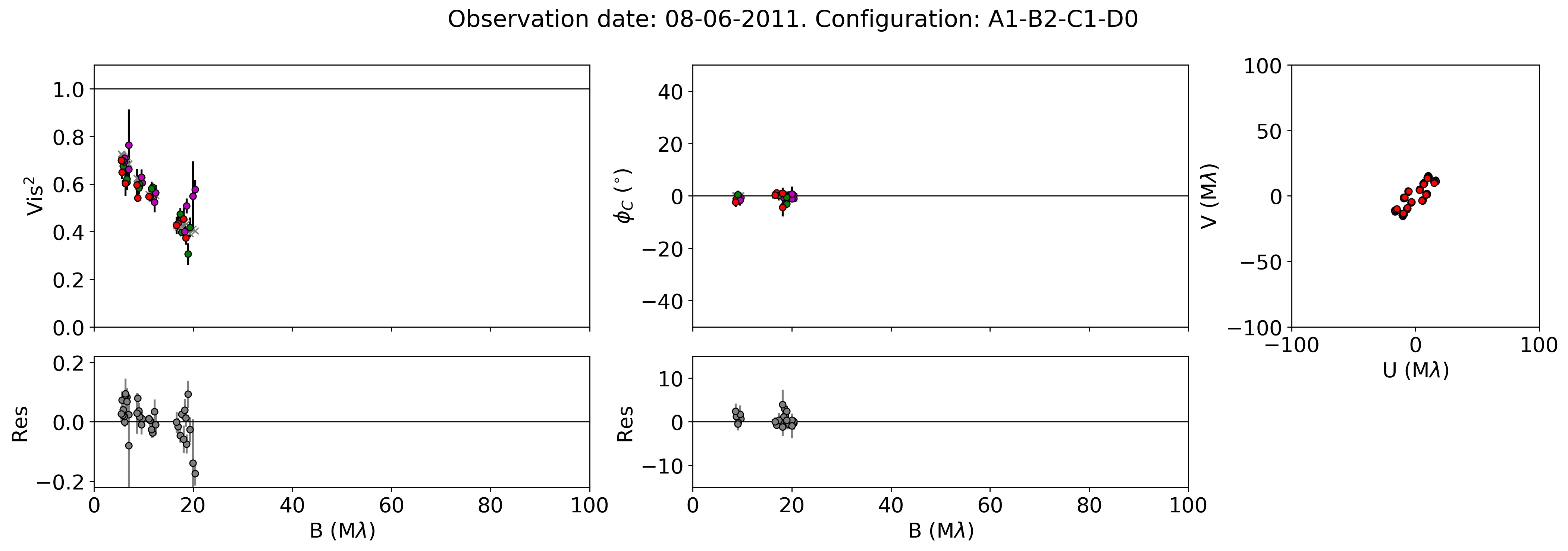} \\ 
\includegraphics[width=0.73\textwidth]{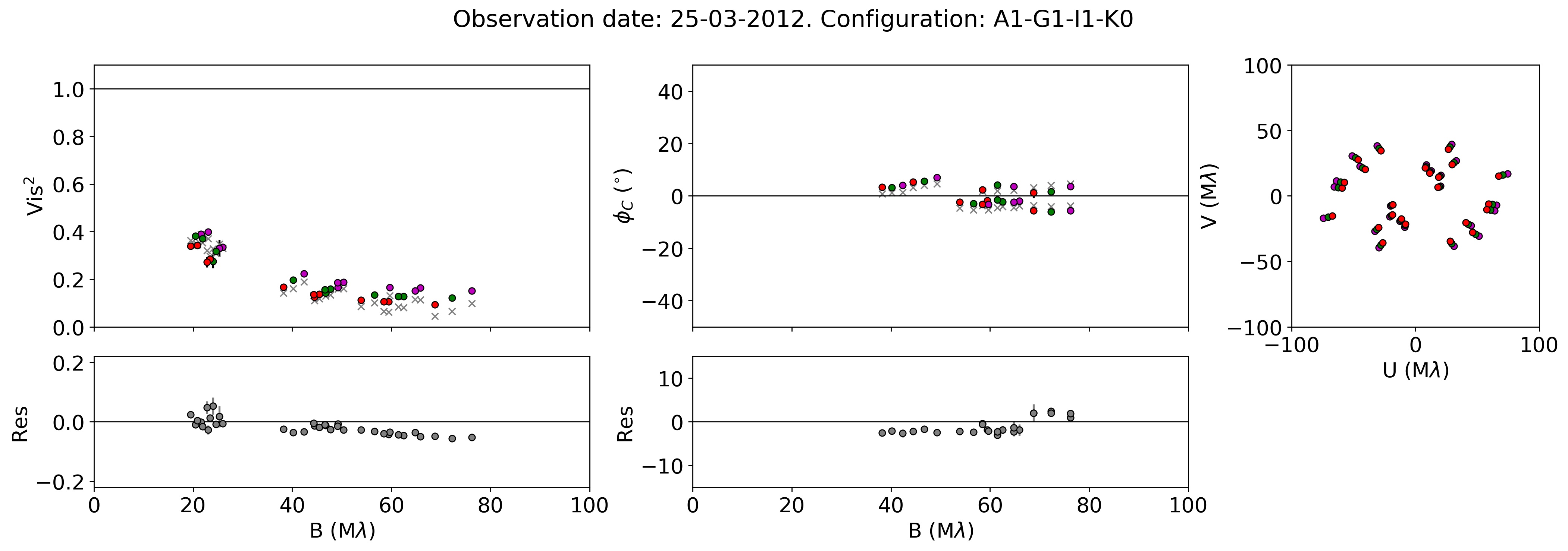} \\ 
\includegraphics[width=0.73\textwidth]{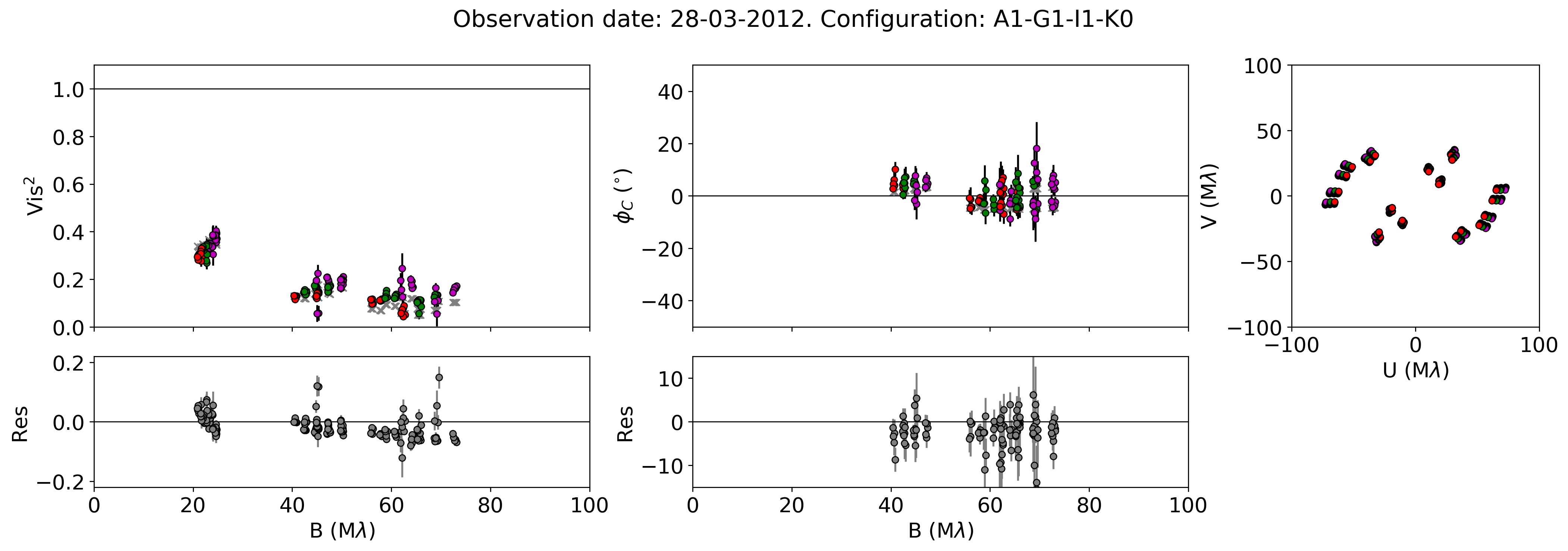}  \\ 
\includegraphics[width=0.73\textwidth]{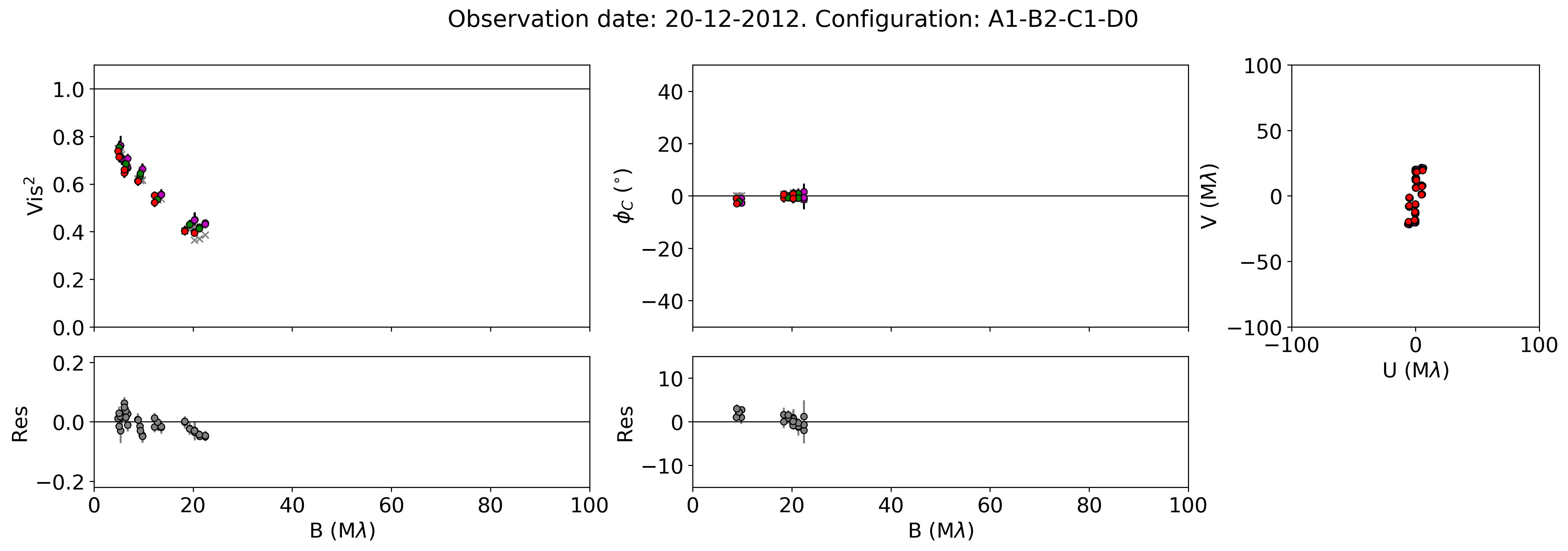} 
\caption{\object{HD\,98922} PIONIER data, squared visibilities, closure phases, and \textit{(u,$\varv$)} plan coverage for each epoch. Colors refer to the different PIONIER spectral channels. Gray crosses represent the model described in Sect.~\ref{subsec:continuum_methodology} and Table~\ref{tab:Continuum_param}. Gray circles in the bottom plots show the residuals.}
\label{fig:PIONIER-Data+Mod}
\end{figure*}

\addtocounter{figure}{-1}
\begin{figure*}[!ht]
\centering
\includegraphics[width=0.73\textwidth]{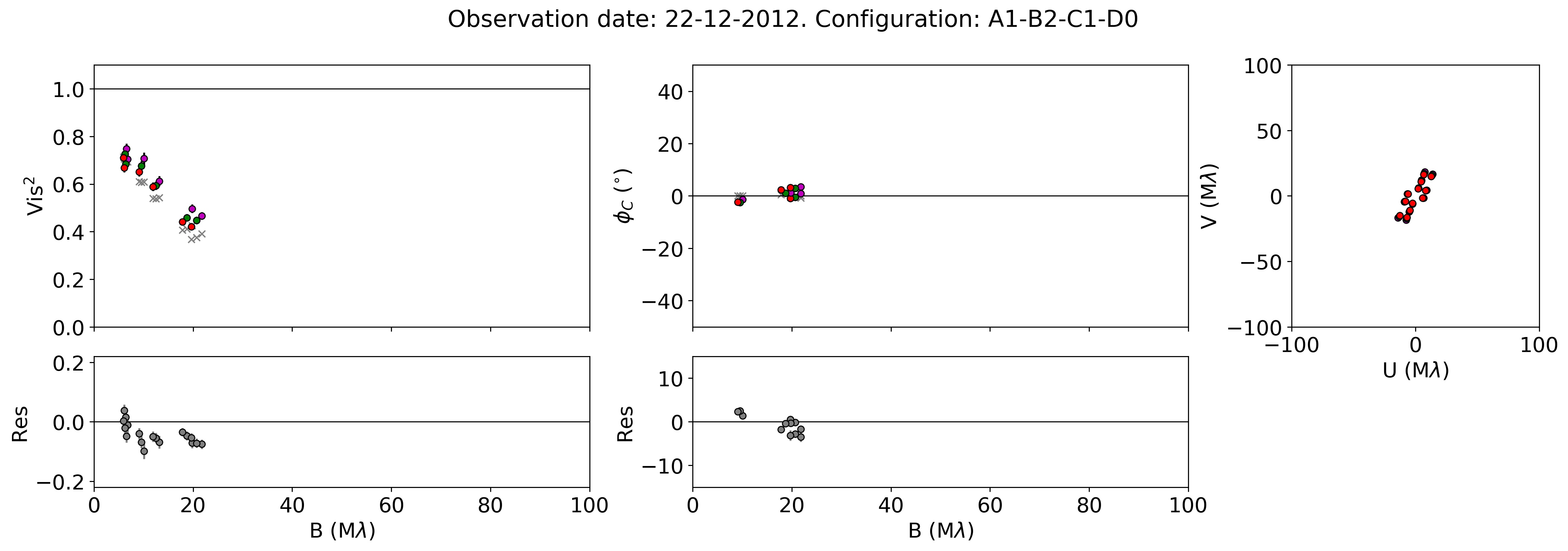}  \\ 
\includegraphics[width=0.73\textwidth]{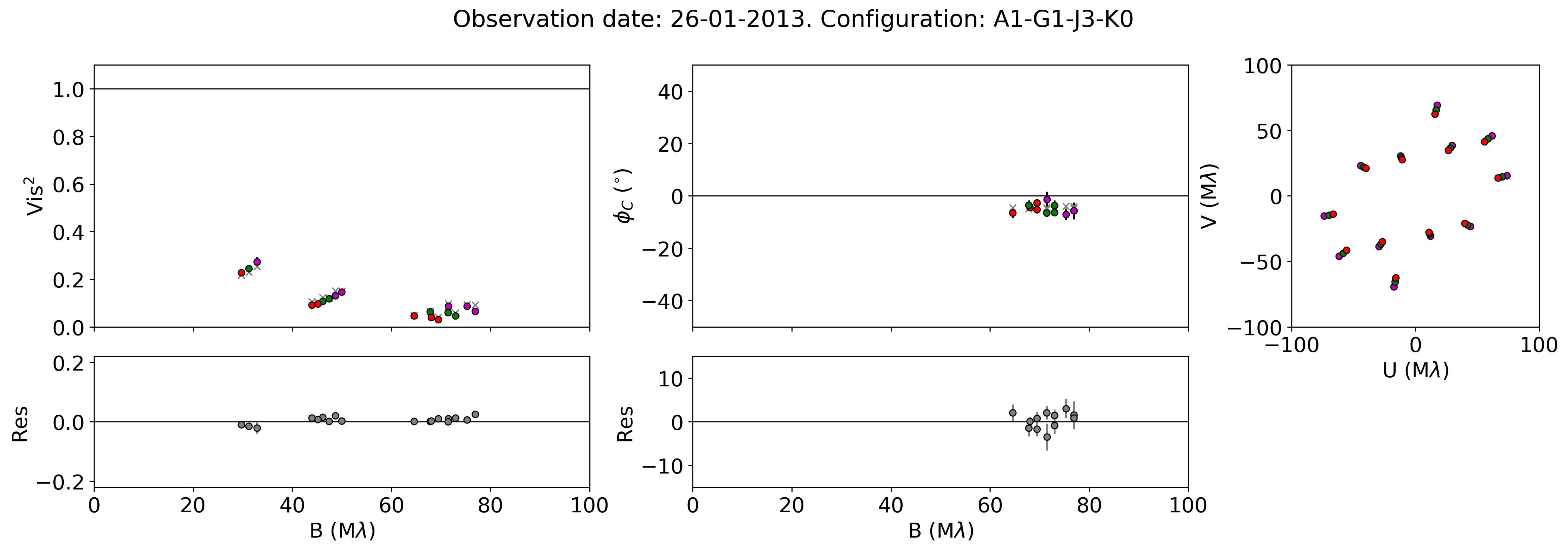}  
\includegraphics[width=0.73\textwidth]{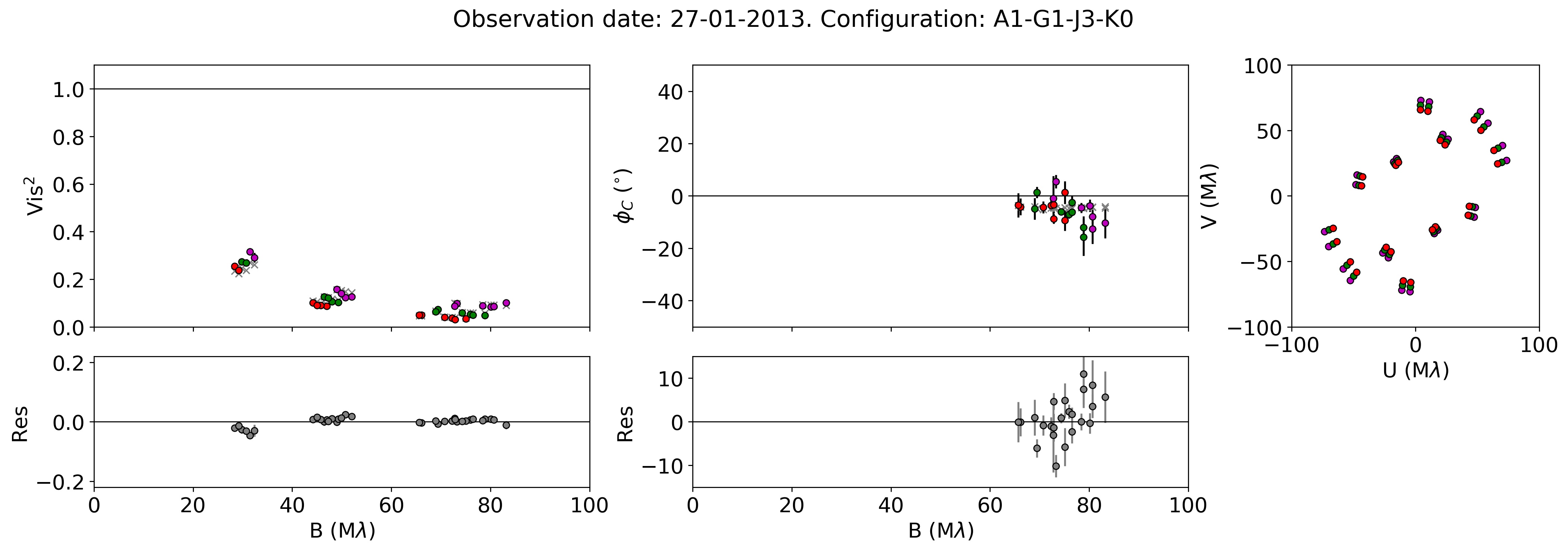}  \\ 
\includegraphics[width=0.73\textwidth]{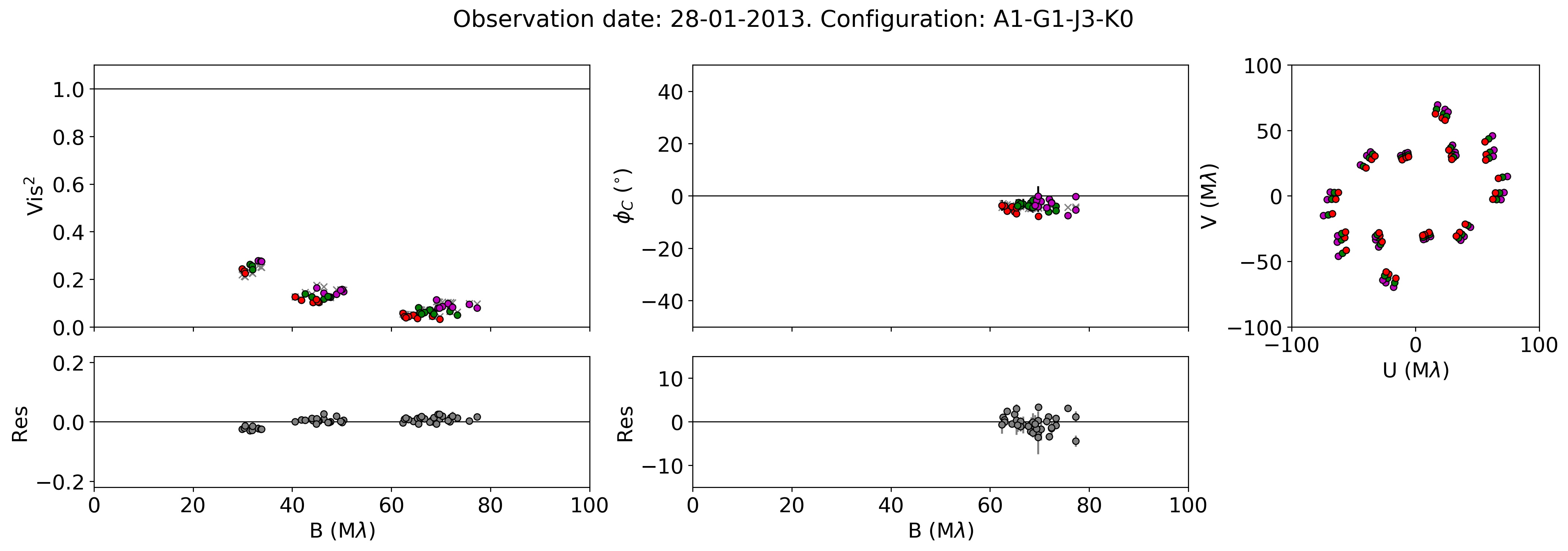} \\ 
\includegraphics[width=0.73\textwidth]{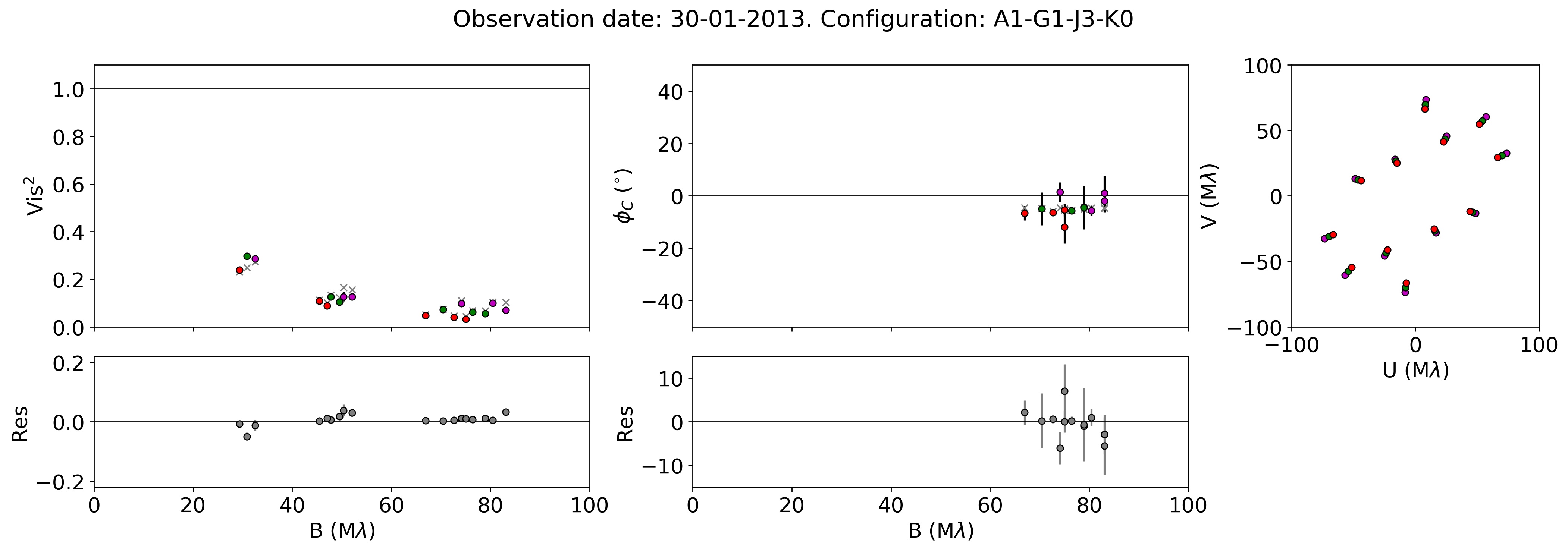}  
\caption{Continued.}
\end{figure*}

\addtocounter{figure}{-1}
\begin{figure*}[!ht]
\centering
\includegraphics[width=0.73\textwidth]{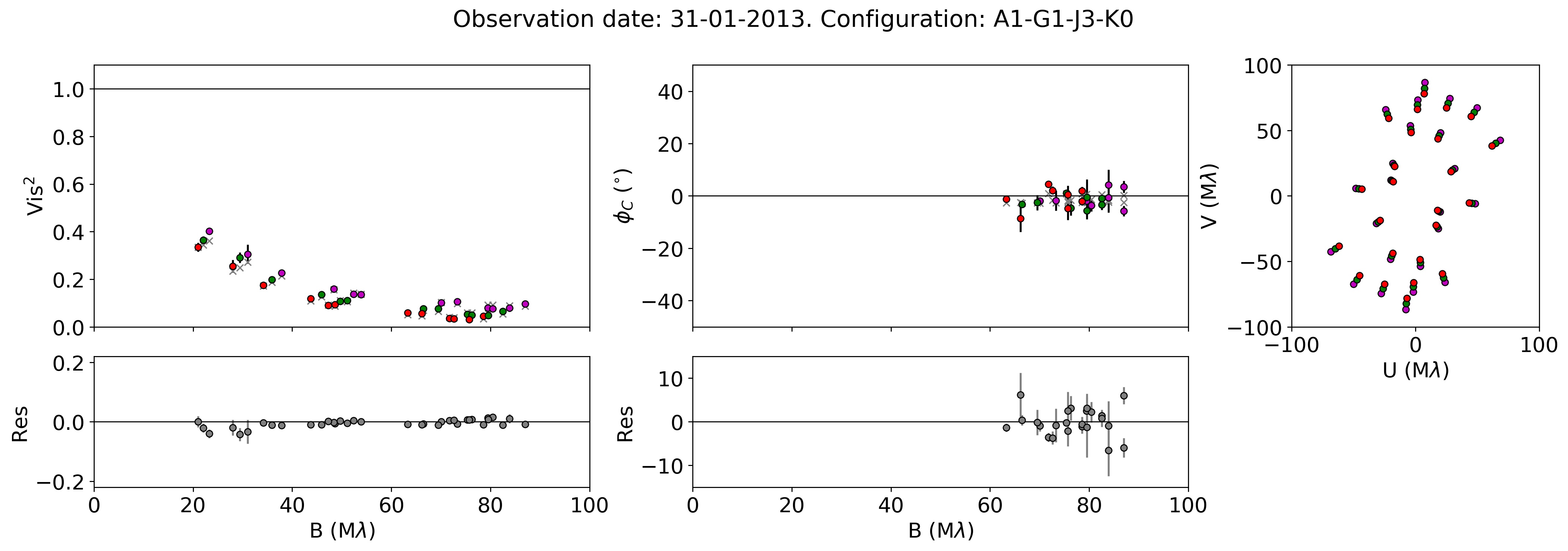} \\ 
\includegraphics[width=0.73\textwidth]{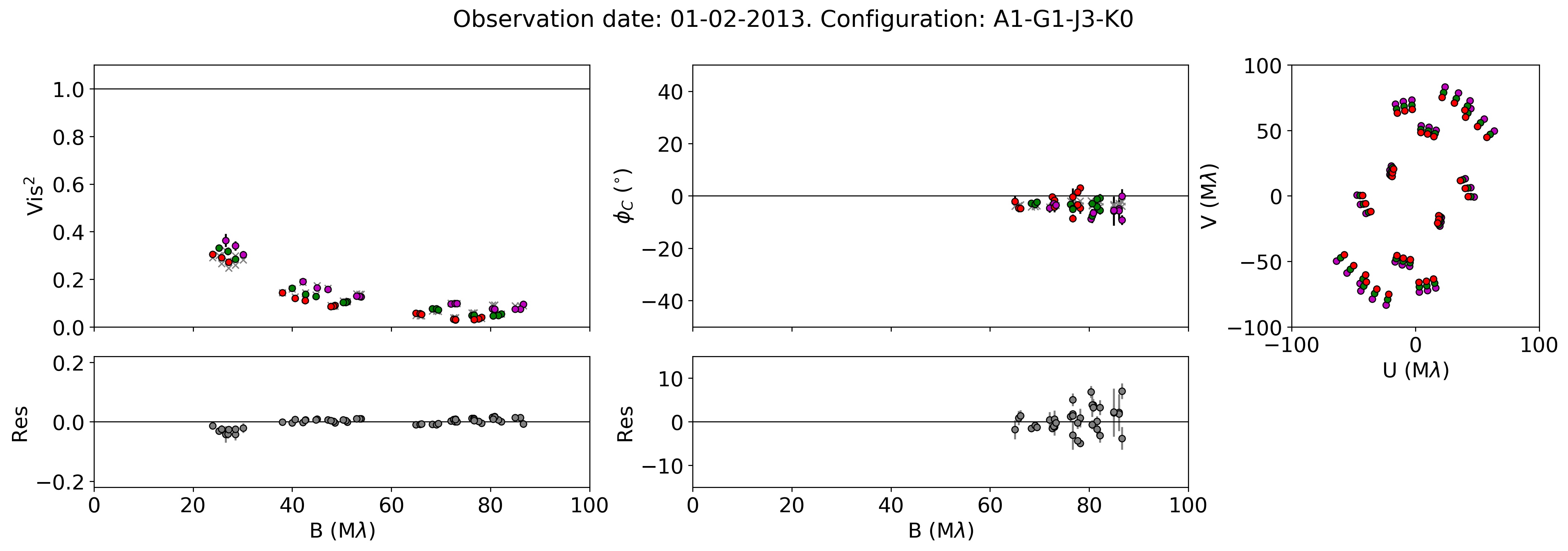}  \\ 
\includegraphics[width=0.73\textwidth]{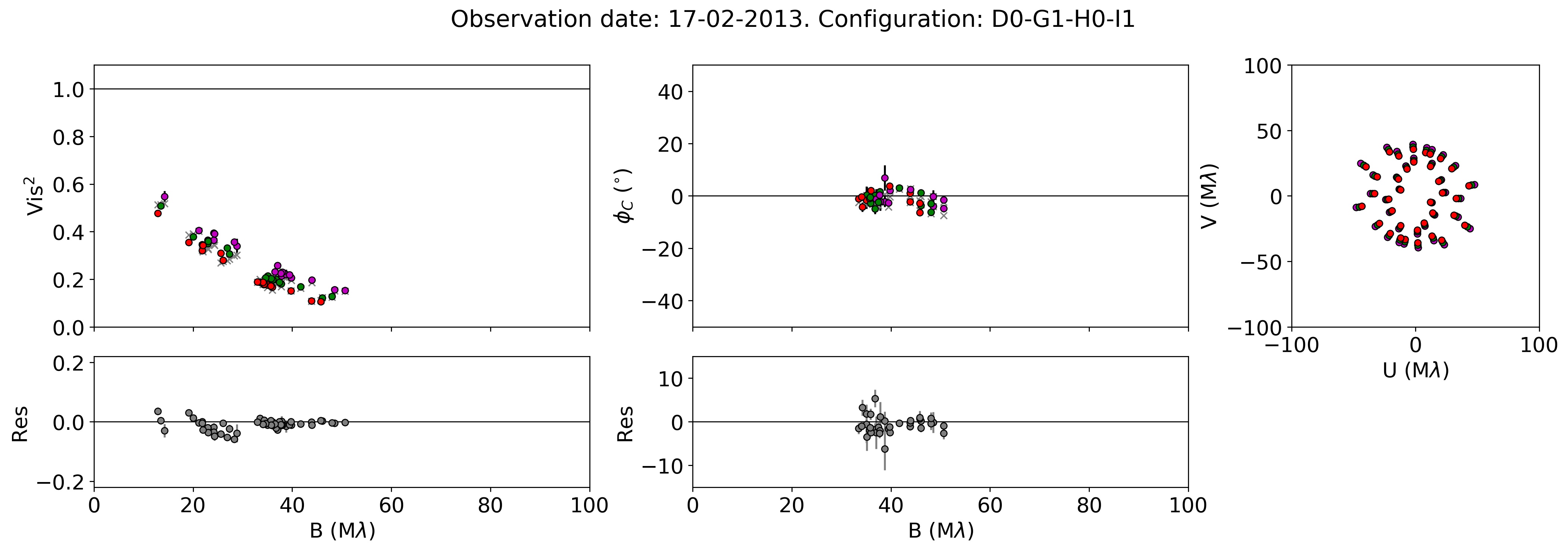} \\ 
\includegraphics[width=0.73\textwidth]{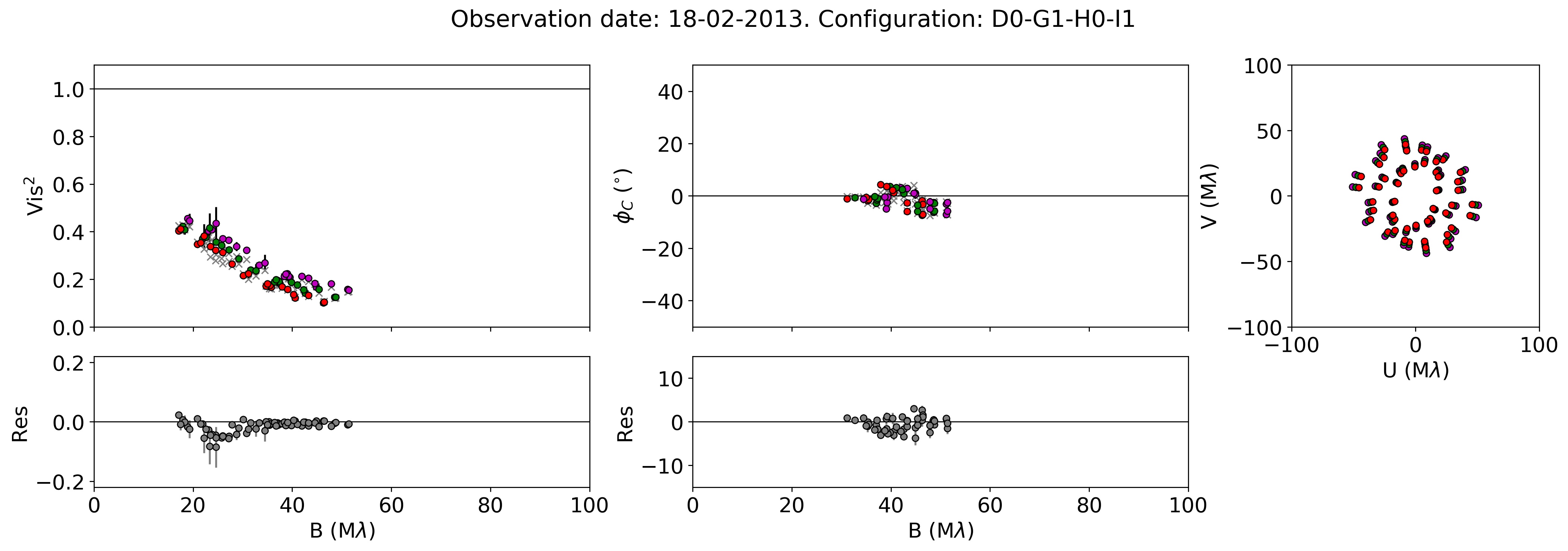} \\ 
\includegraphics[width=0.73\textwidth]{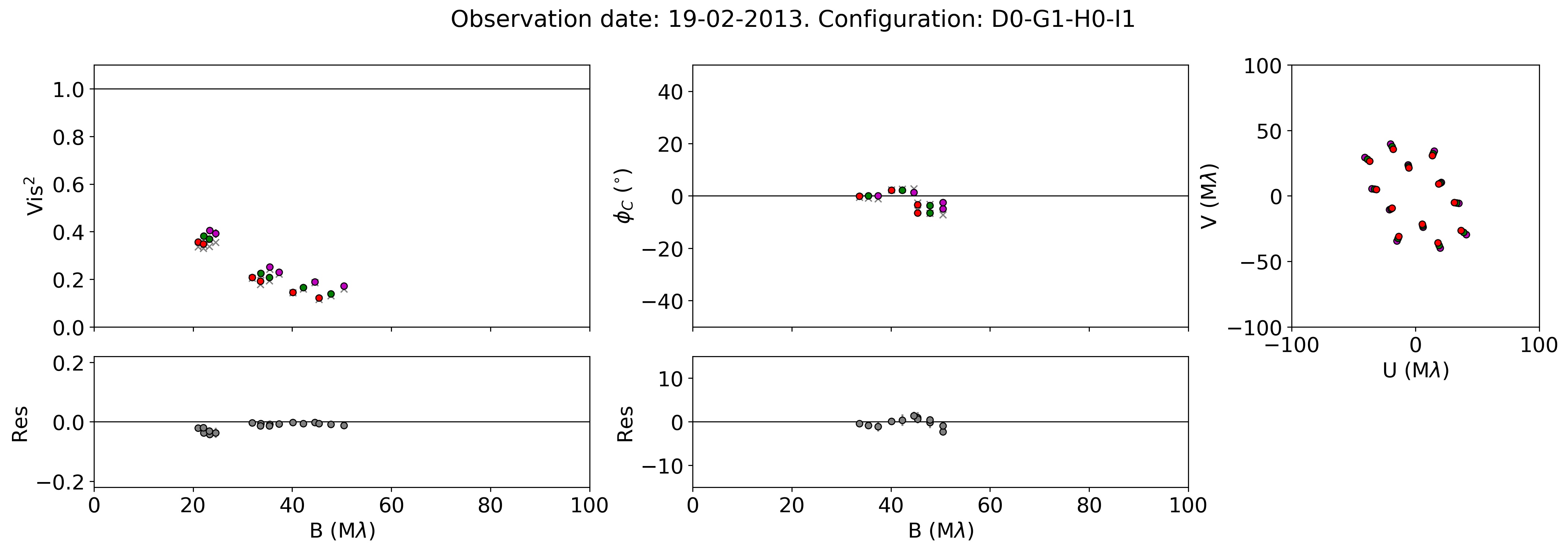}
\caption{Continued.}
\end{figure*}

\addtocounter{figure}{-1}
\begin{figure*}[!ht]
\centering
\includegraphics[width=0.73\textwidth]{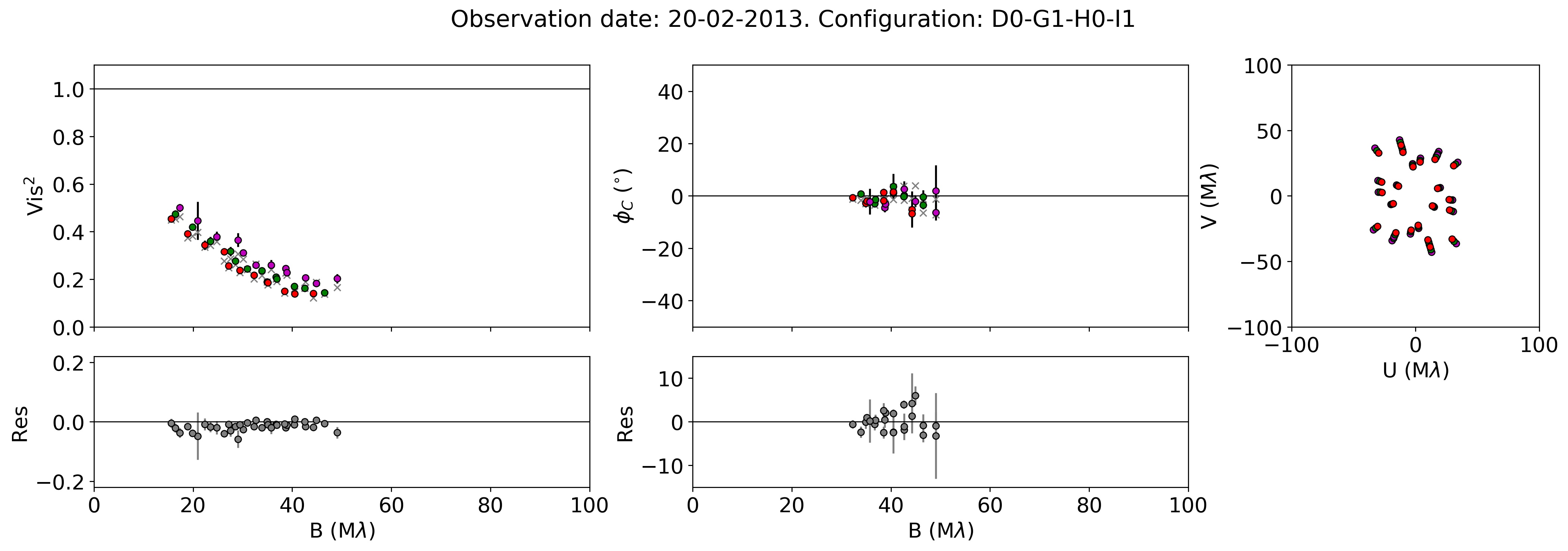} \\ 
\includegraphics[width=0.73\textwidth]{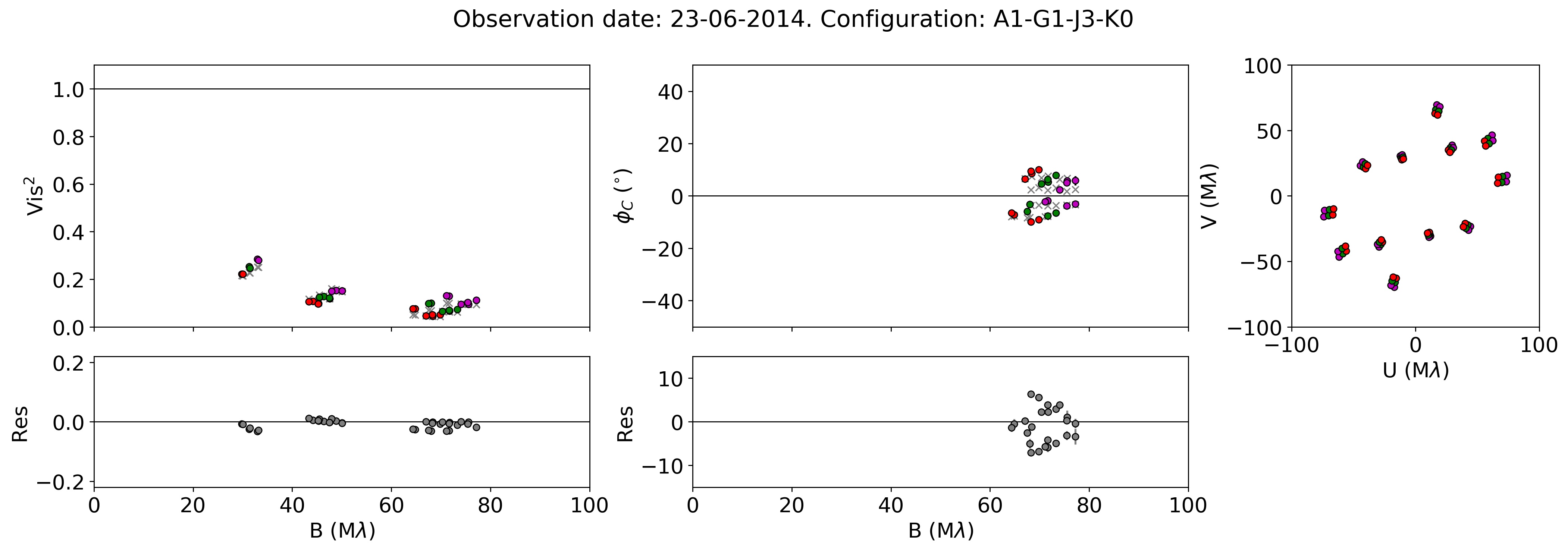} \\ 
\includegraphics[width=0.73\textwidth]{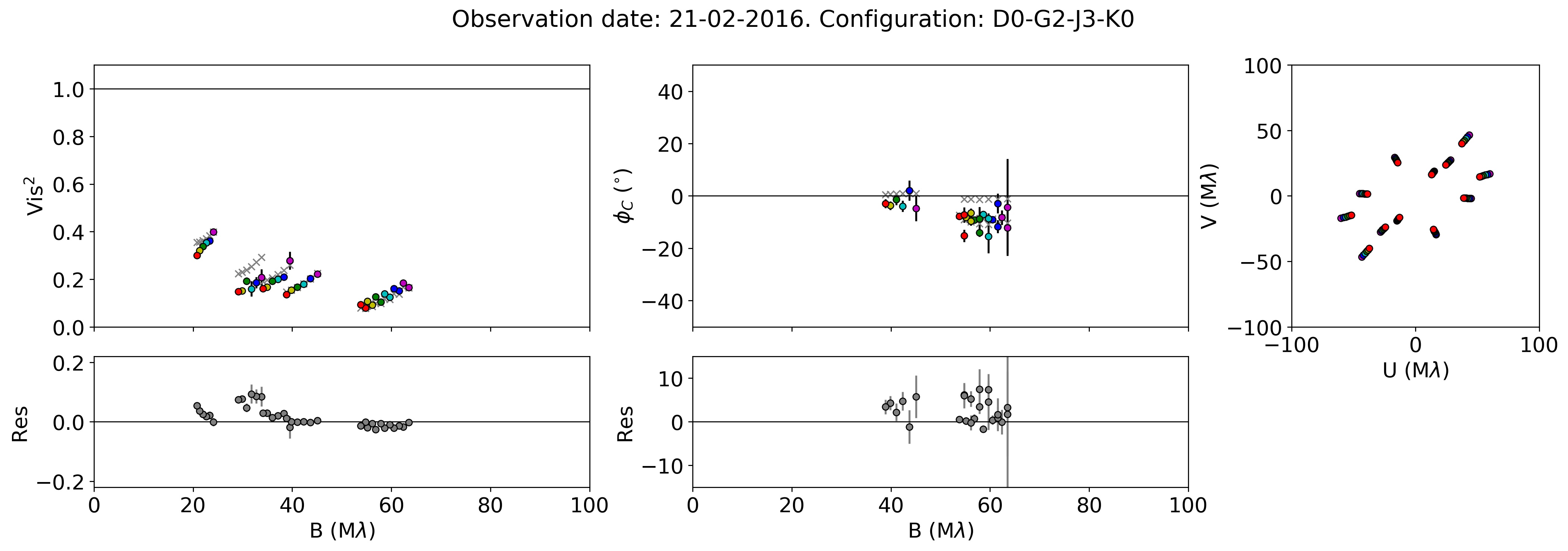} \\ 
\includegraphics[width=0.73\textwidth]{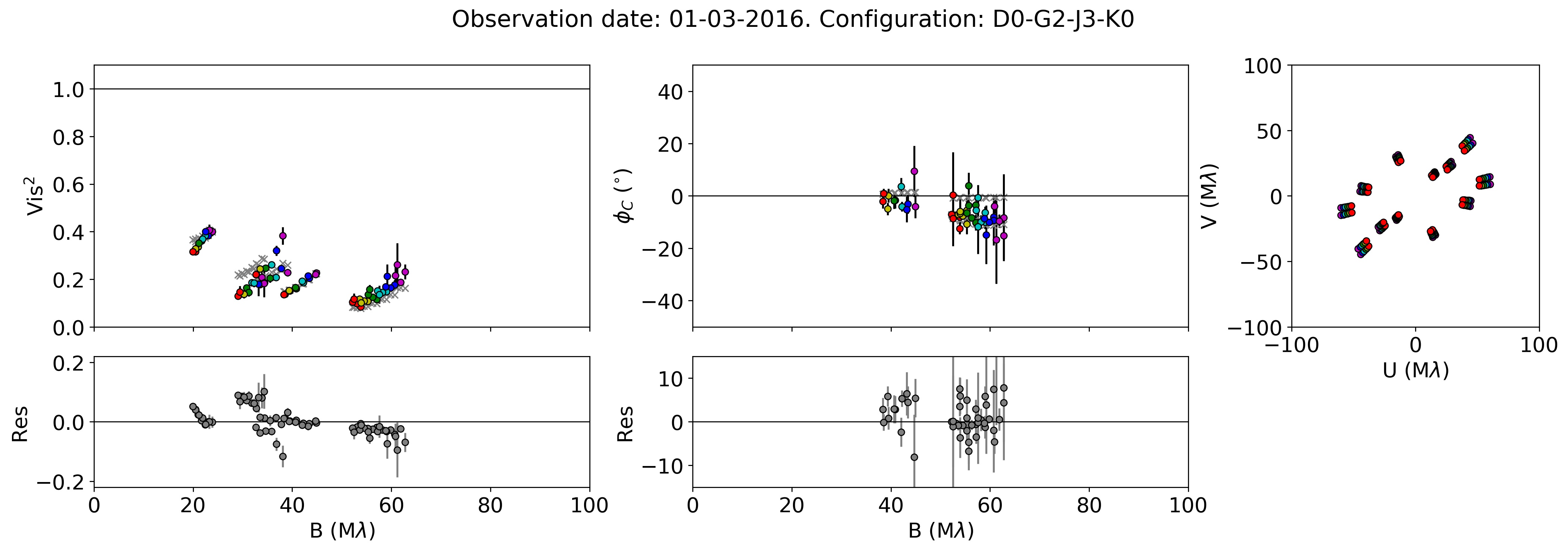} \\
\includegraphics[width=0.73\textwidth]{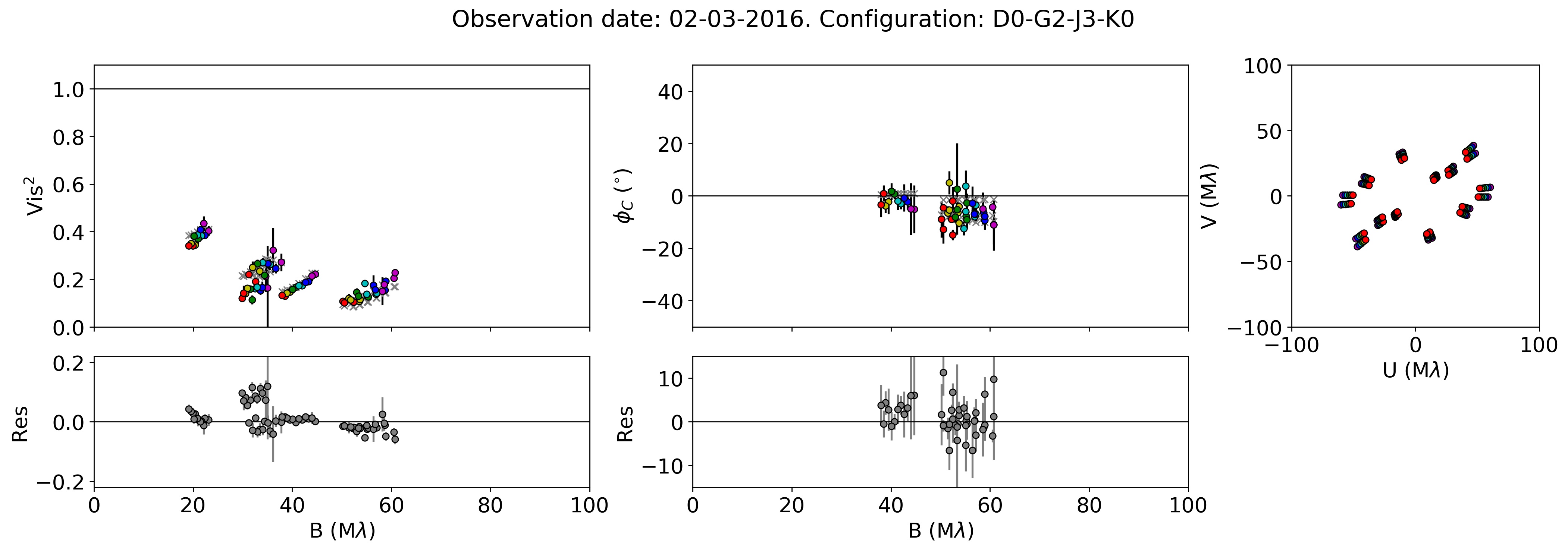}
\caption{Continued.}
\end{figure*}

\begin{figure*}[!ht]
\centering
\includegraphics[width=0.73\textwidth]{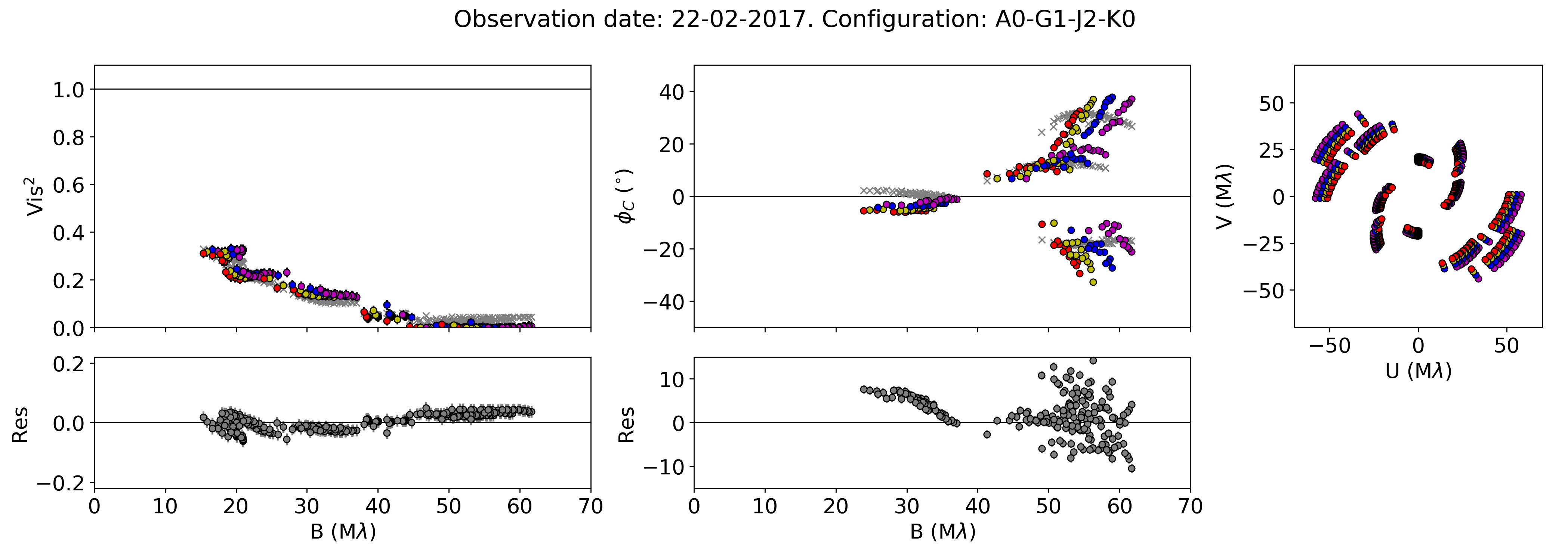} \\
\includegraphics[width=0.73\textwidth]{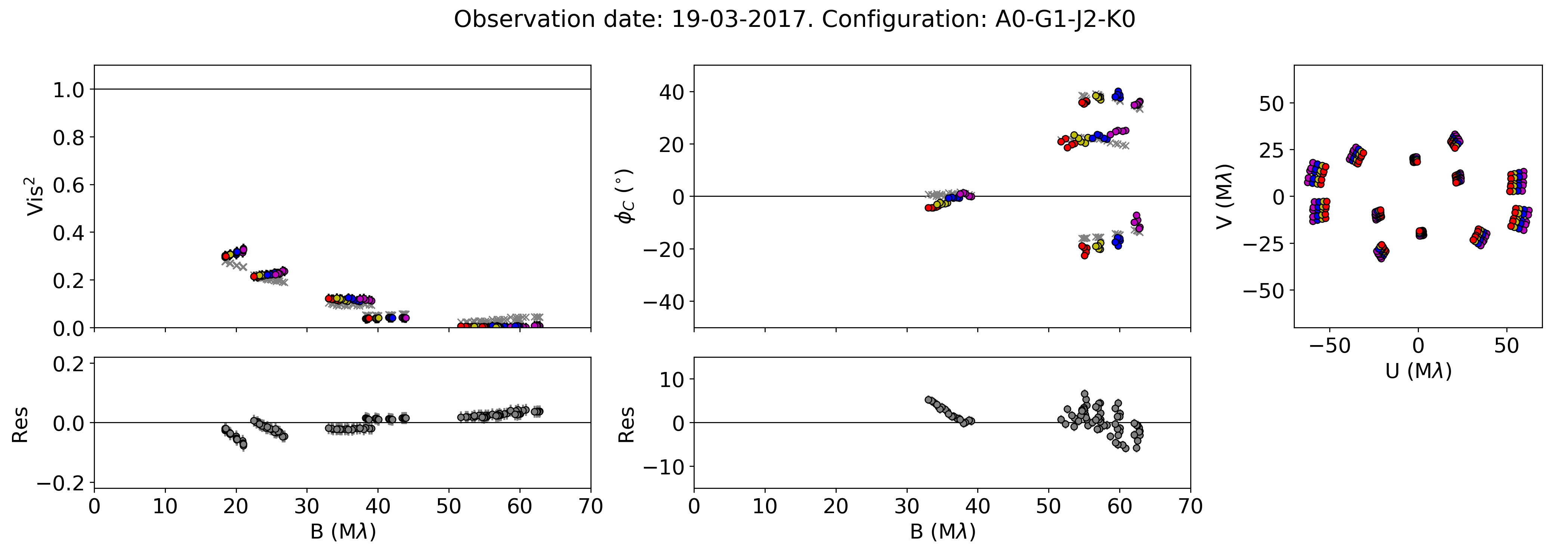} \\ 
\includegraphics[width=0.73\textwidth]{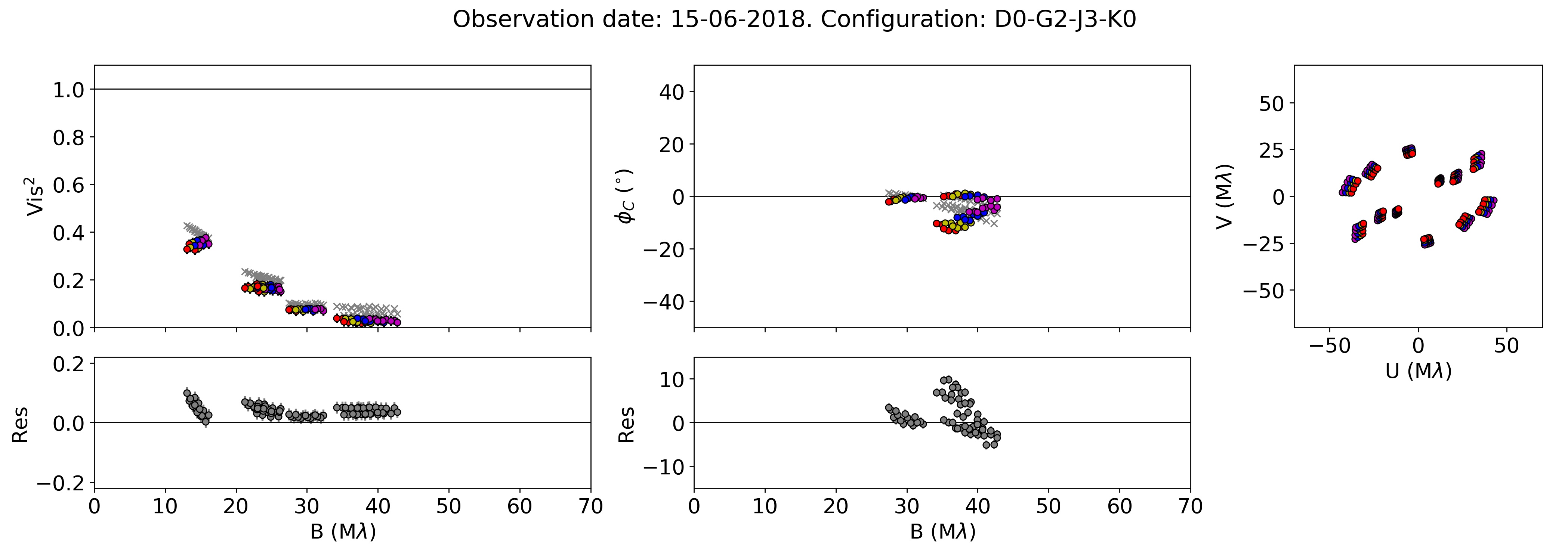} \\ 
\includegraphics[width=0.73\textwidth]{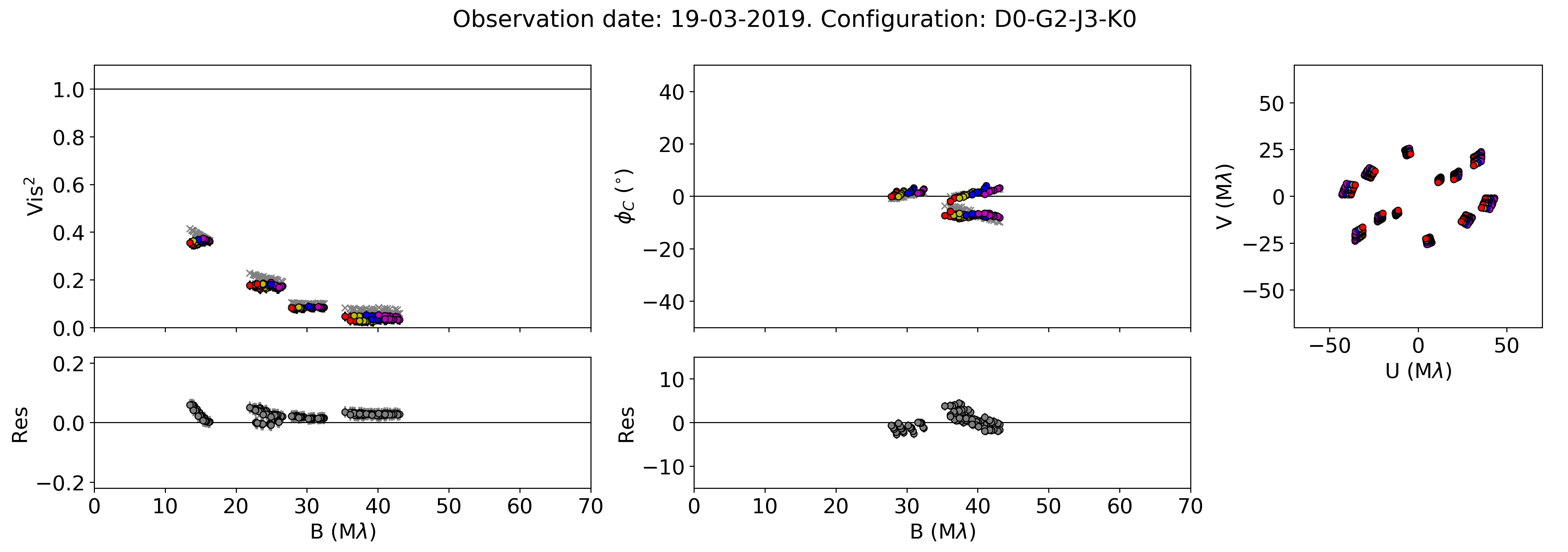} \\ 
\includegraphics[width=0.73\textwidth]{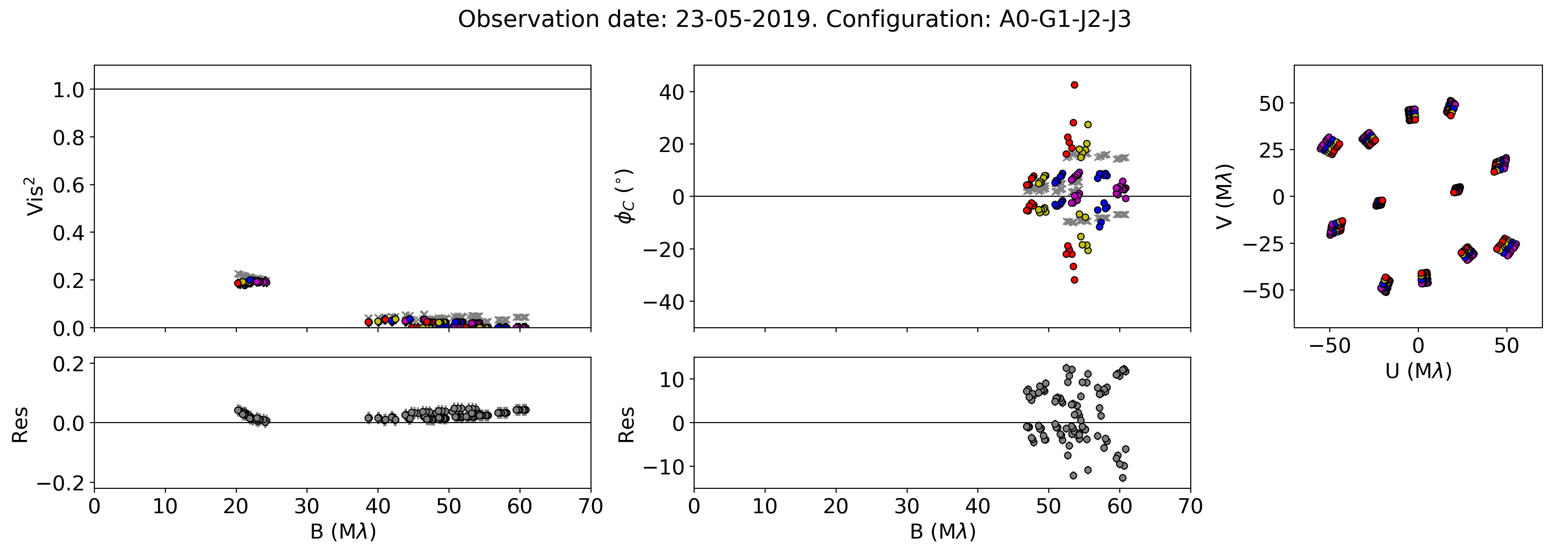}
\caption{\object{HD\,98922} GRAVITY FT data, squared visibilities, closure phases, and \textit{(u,$\varv$)} plan coverage for each epoch. Colors refer to the different GRAVITY spectral channels. Gray crosses represent the model described in Sect.~\ref{subsec:continuum_methodology}, and Table~\ref{tab:Continuum_param}. Gray circles in the bottom plots show the residuals of the model fitting.}
\label{fig:GRAVITY-FT-Data}
\end{figure*}

\addtocounter{figure}{-1}
\begin{figure*}[!ht]
\centering
\includegraphics[width=0.73\textwidth]{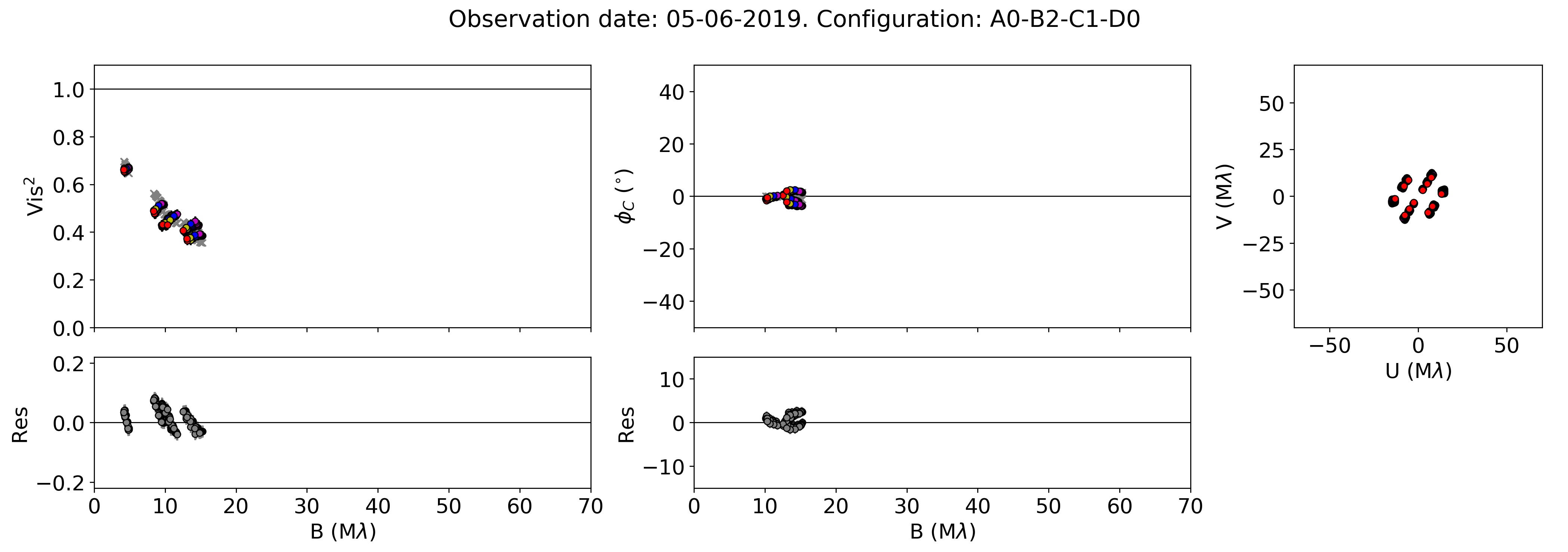} \\ 
\includegraphics[width=0.73\textwidth]{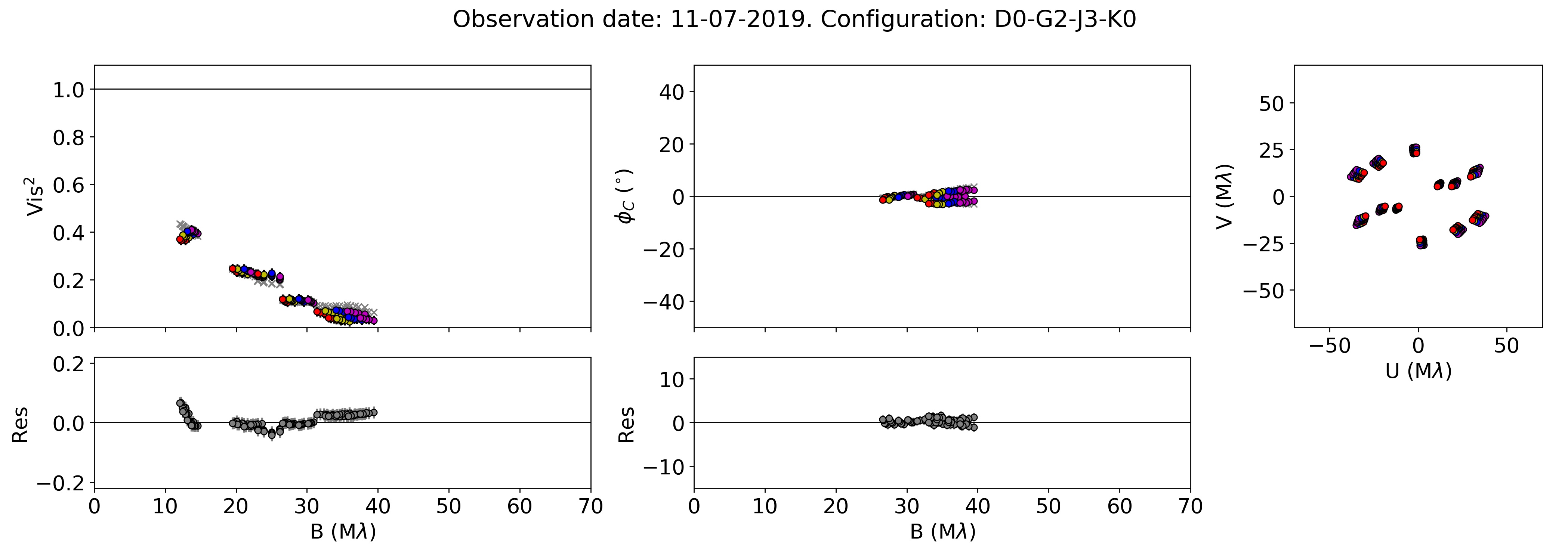} \\ 
\includegraphics[width=0.73\textwidth]{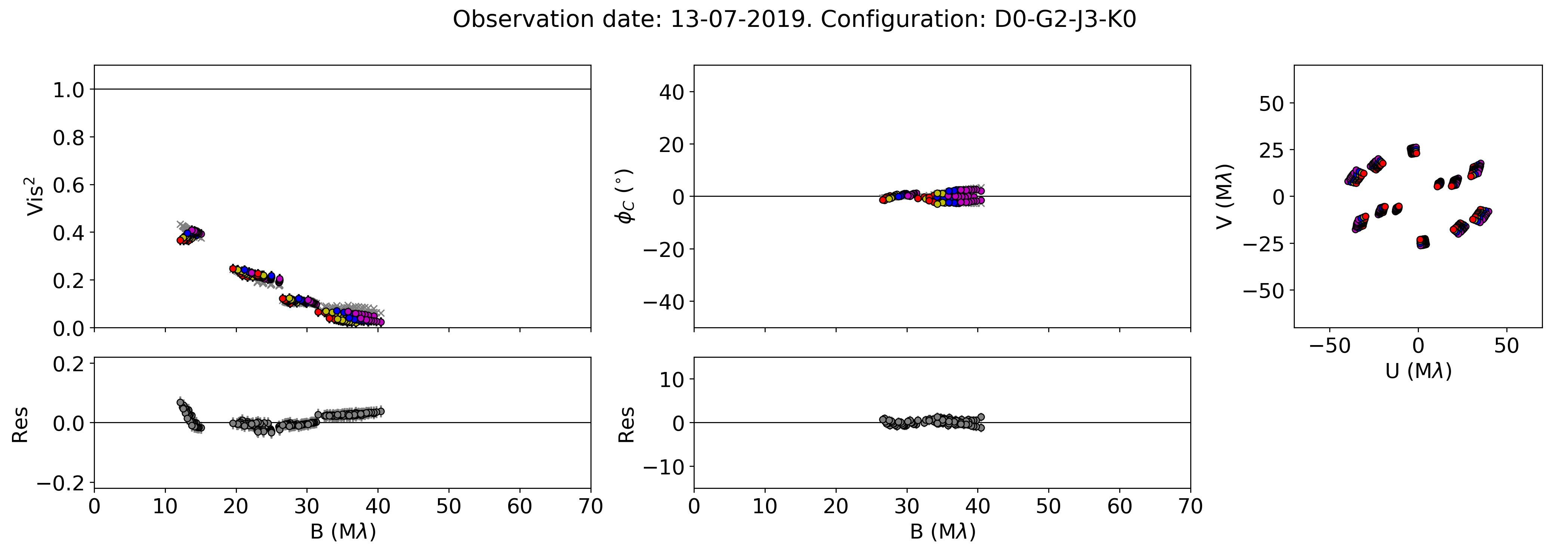} \\ 
\includegraphics[width=0.73\textwidth]{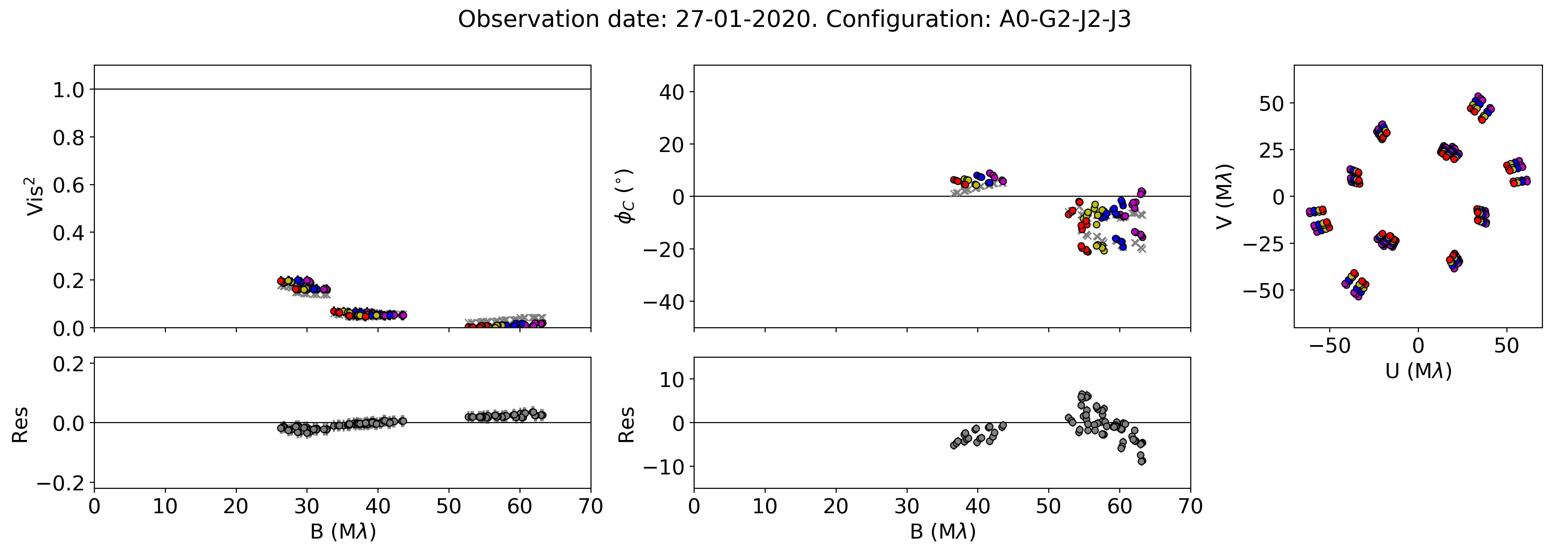} \\ 
\includegraphics[width=0.73\textwidth]{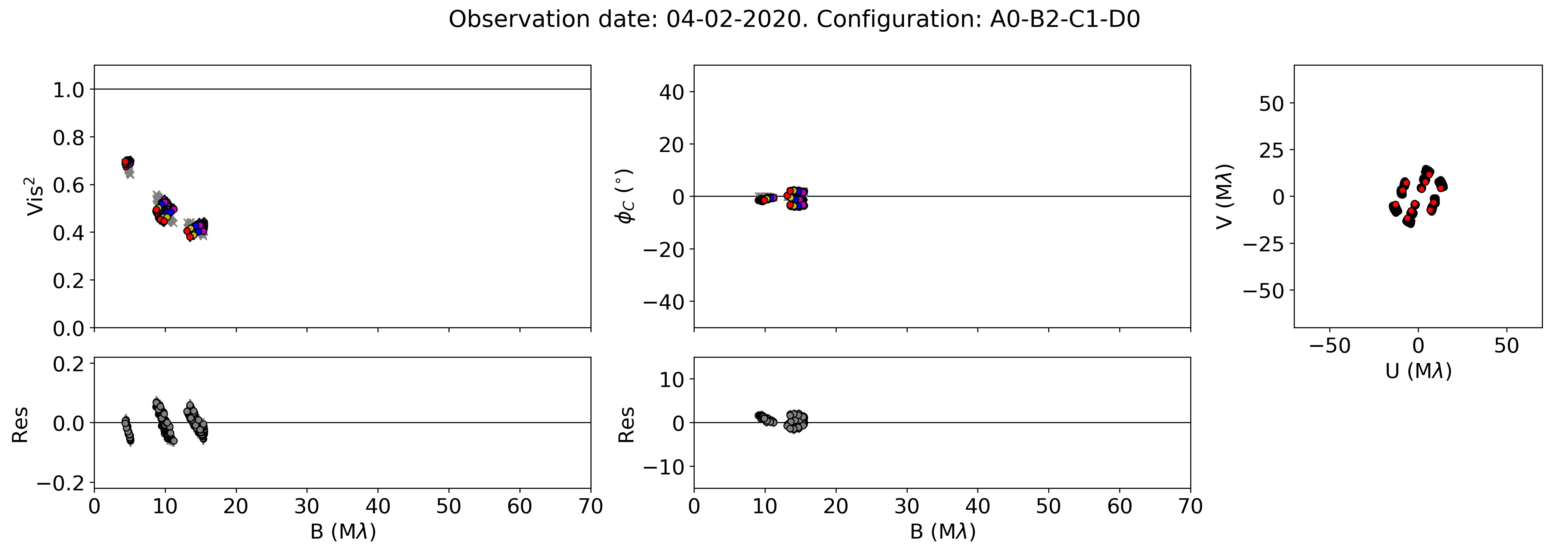} 
\caption{Continued.}
\end{figure*}

\addtocounter{figure}{-1}
\begin{figure*}[!ht]
\centering
 \includegraphics[width=0.73\textwidth]{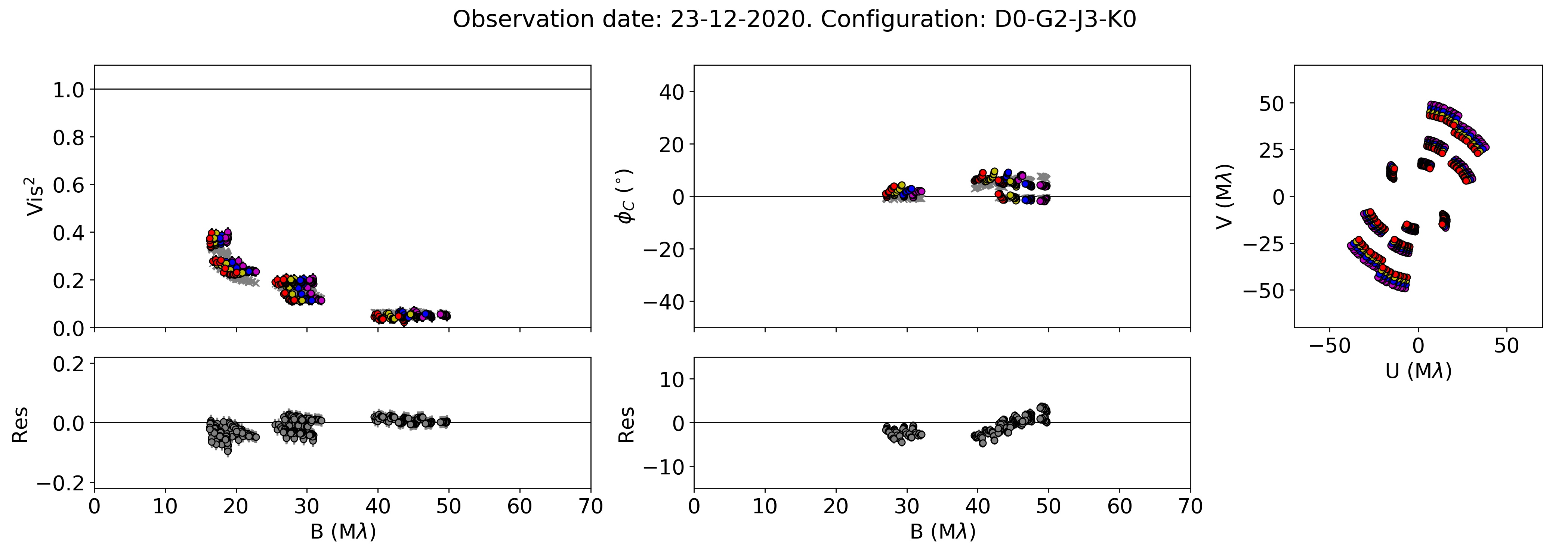} \\ 
 \includegraphics[width=0.73\textwidth]{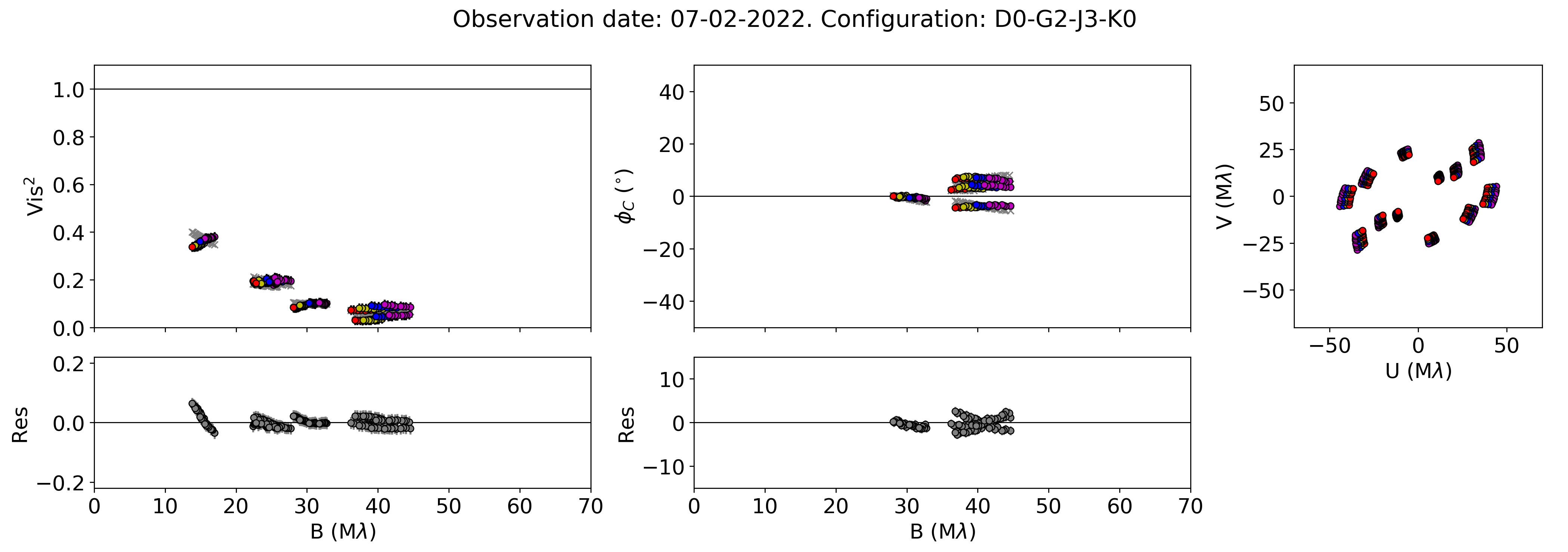} \\ 
\includegraphics[width=0.73\textwidth]{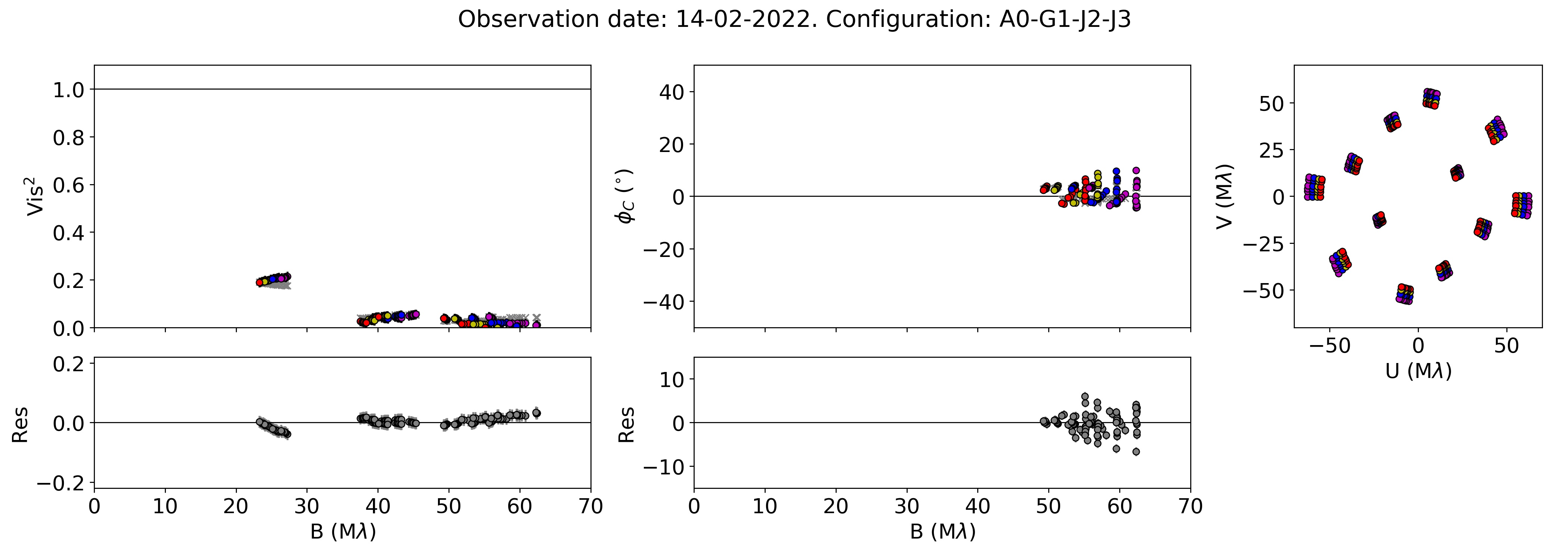} 
\caption{Continued.}
\end{figure*}

\begin{figure*}[!ht]
\includegraphics[width=\columnwidth]{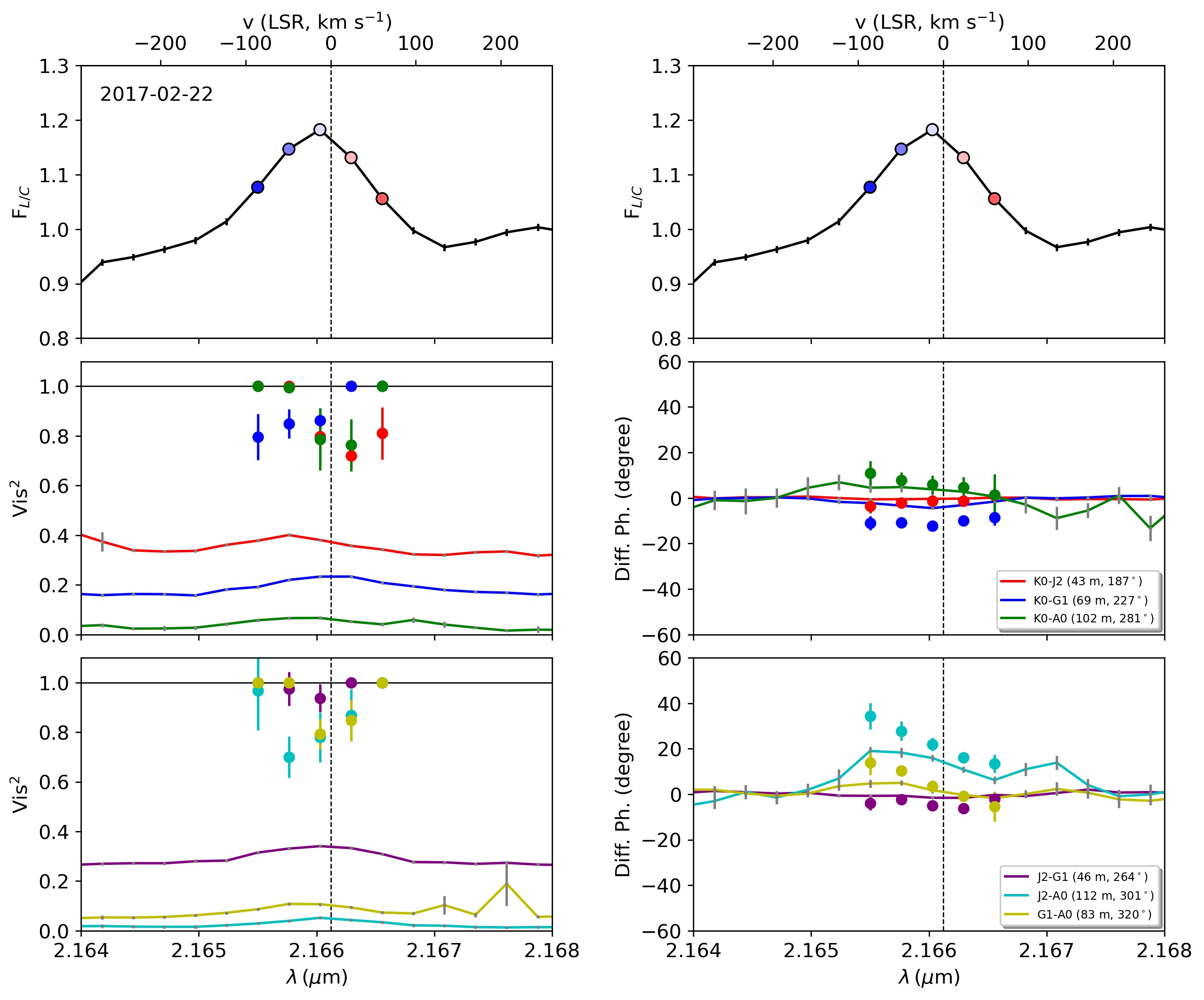}
\includegraphics[width=\columnwidth]{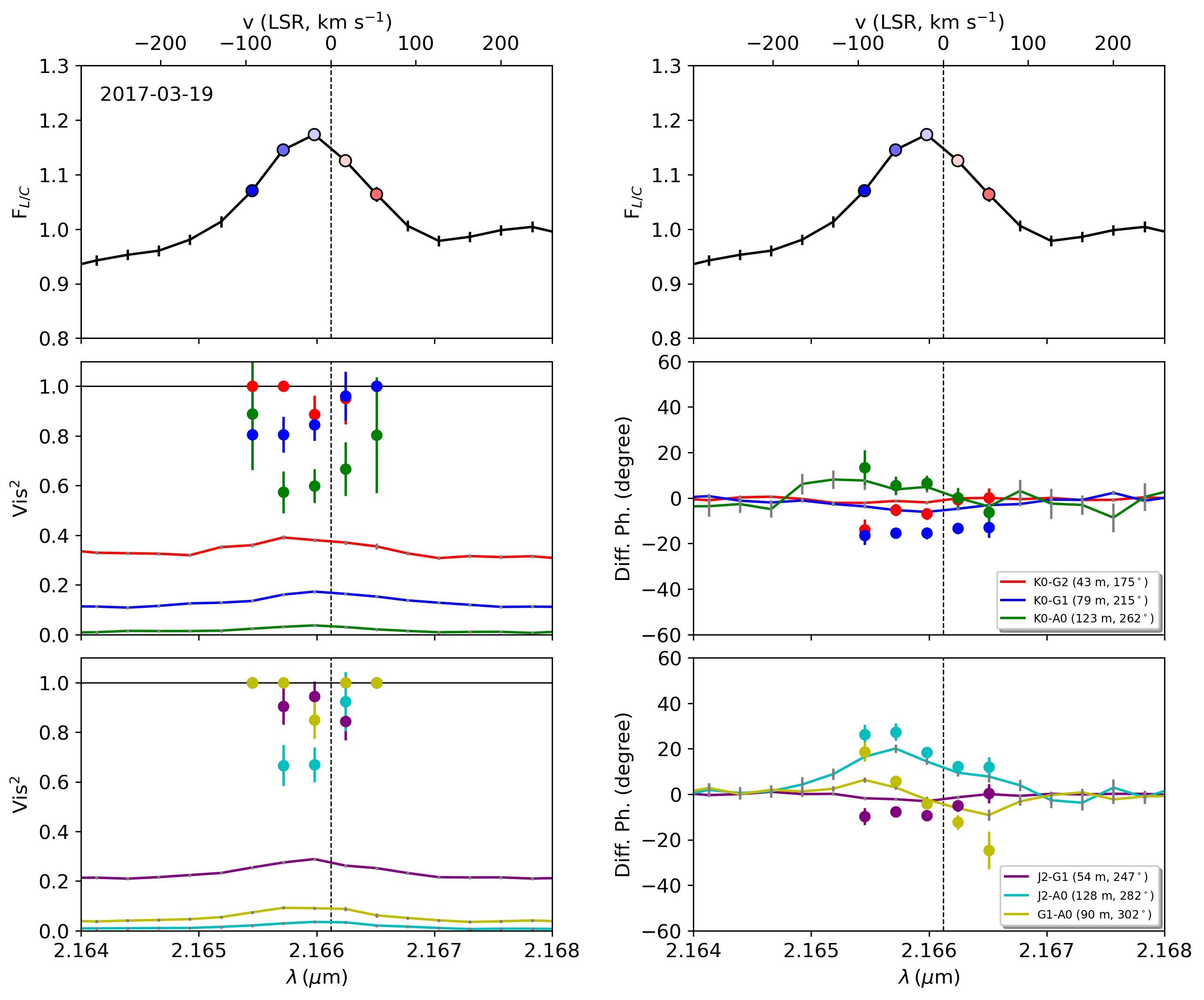}\\[0.5em]
\includegraphics[width=\columnwidth]{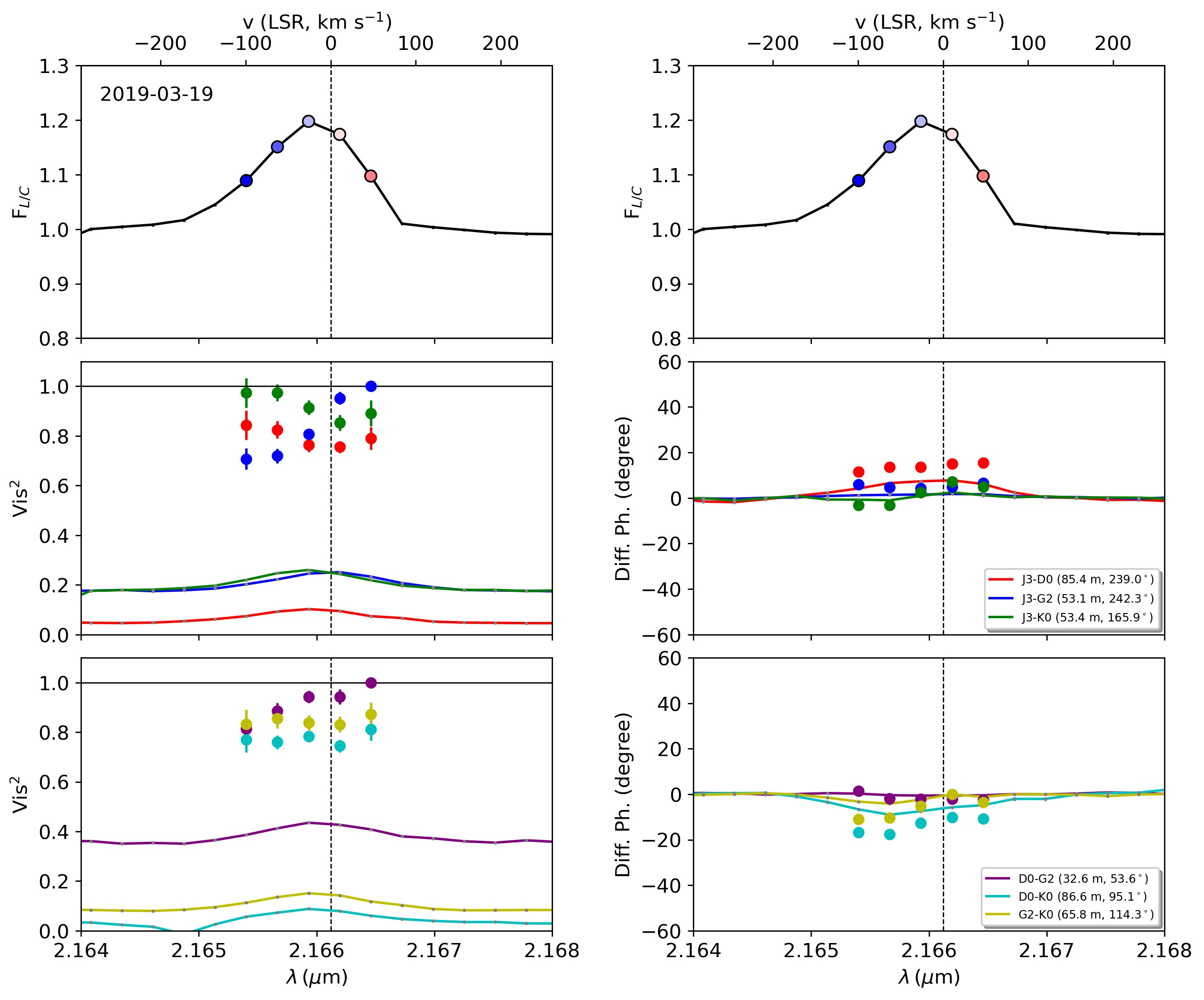}
\includegraphics[width=\columnwidth]{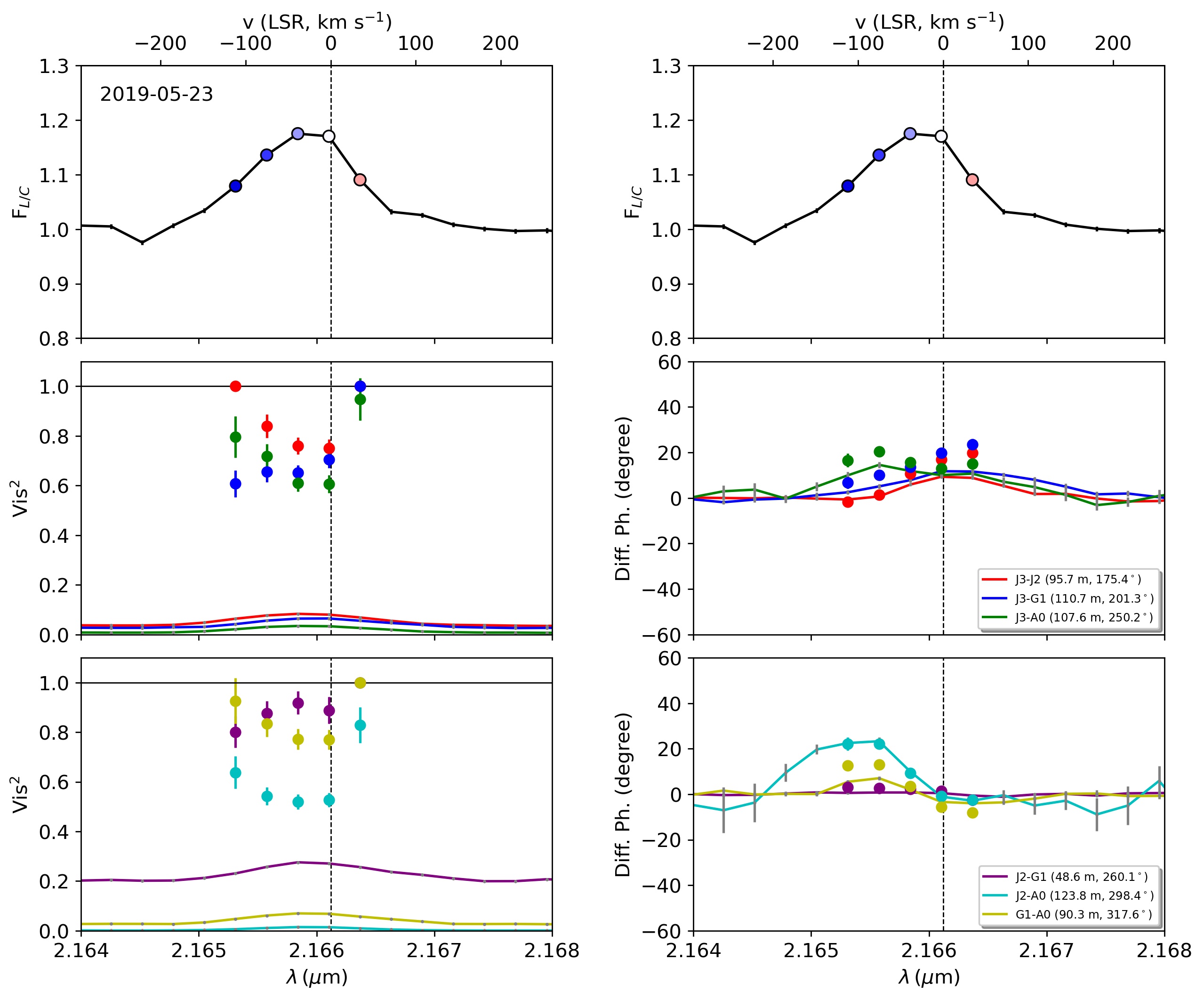}\\[0.5em]
\includegraphics[width=\columnwidth]{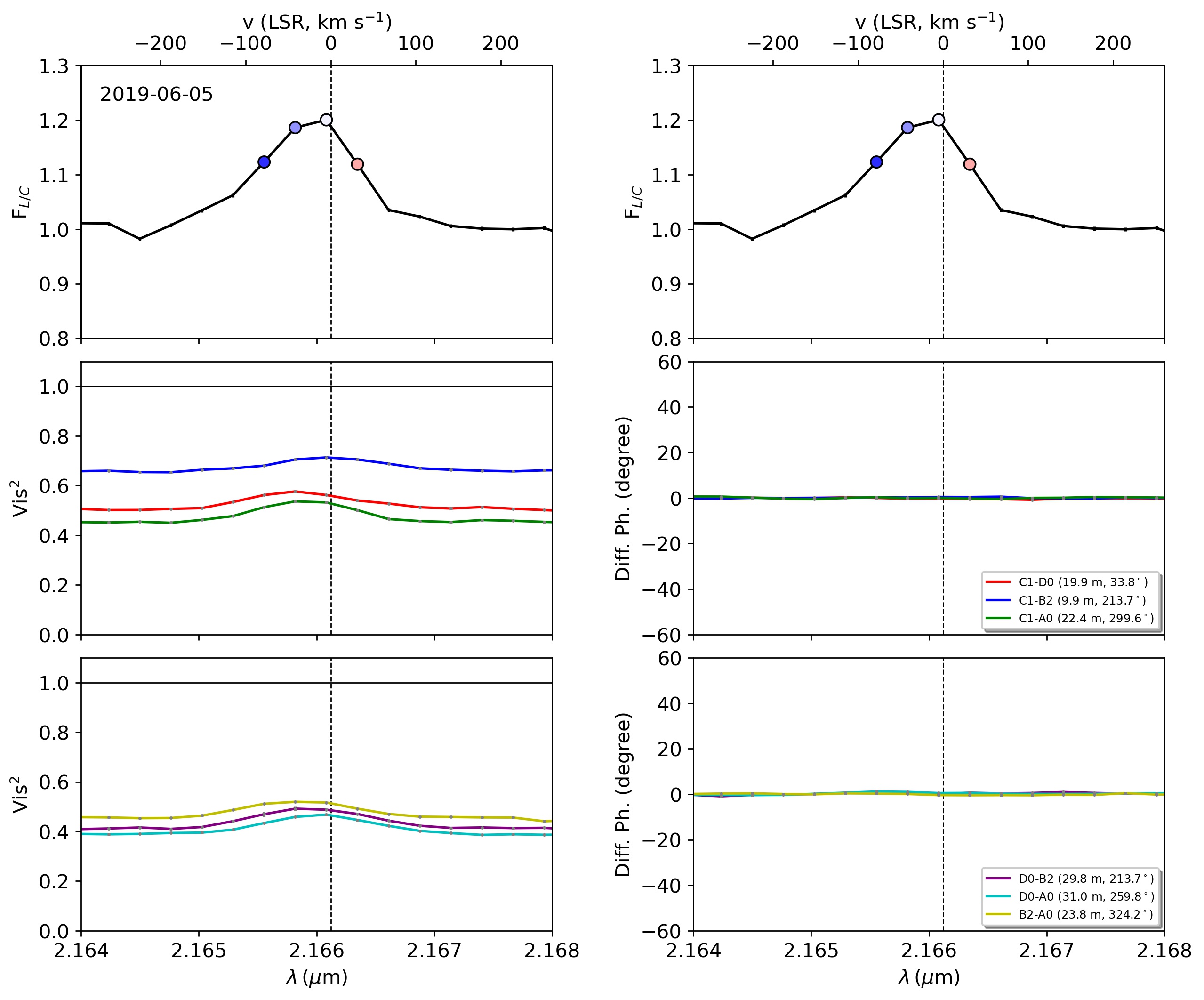}
\includegraphics[width=\columnwidth]{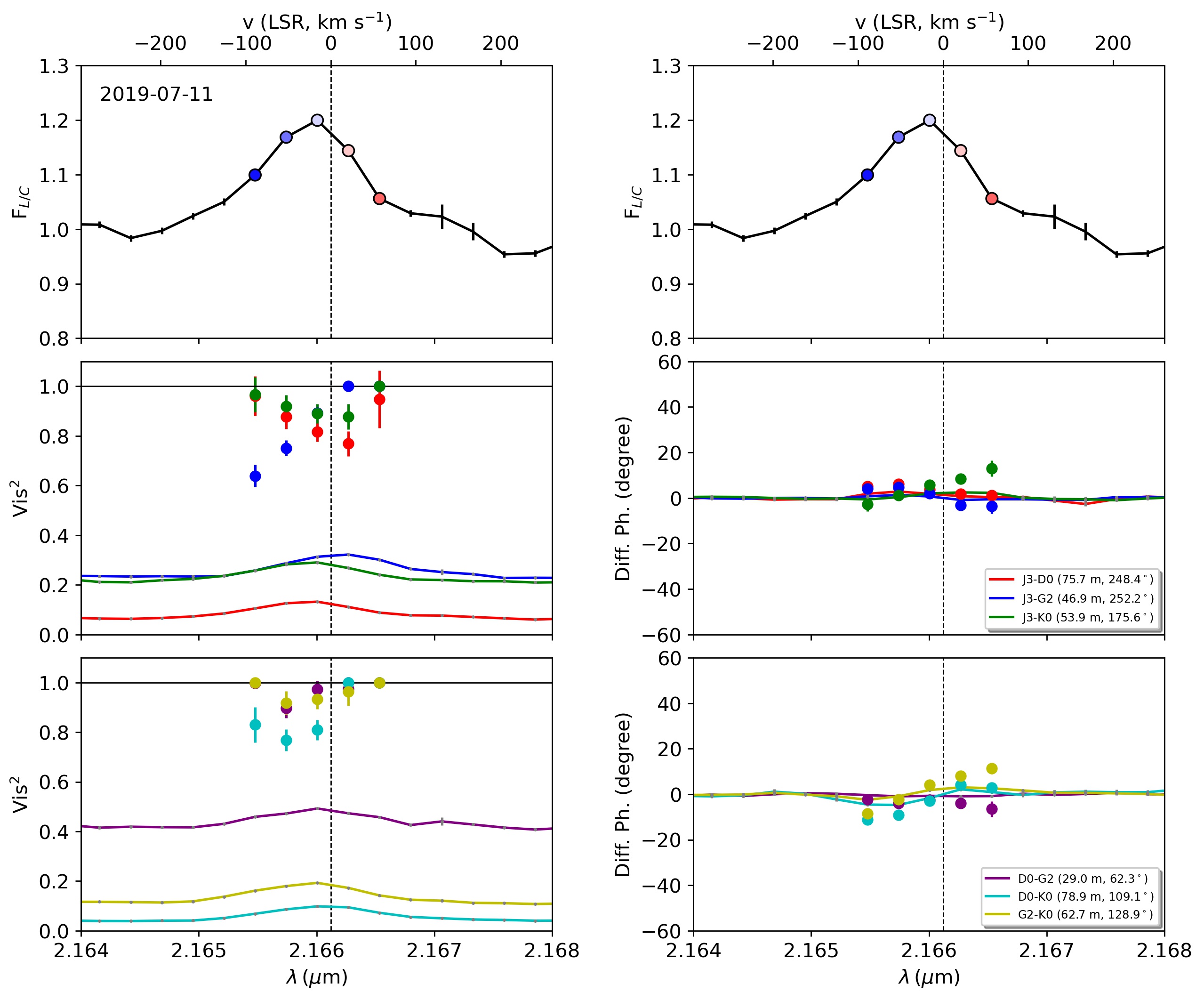}\\
\caption{\object{HD\,98922} GRAVITY SC data for the different epochs. For each epoch, top plots show the wavelength-calibrated and continuum-normalized spectrum, left plots show the total squared visibilities, and right plots show the total differential phases. Circles represent the pure-line quantities. Colors refer to the different baselines.}
\label{fig:GRAVITY-SC-Data}
\end{figure*}

\addtocounter{figure}{-1}
\begin{figure*}[!ht]
\includegraphics[width=\columnwidth]{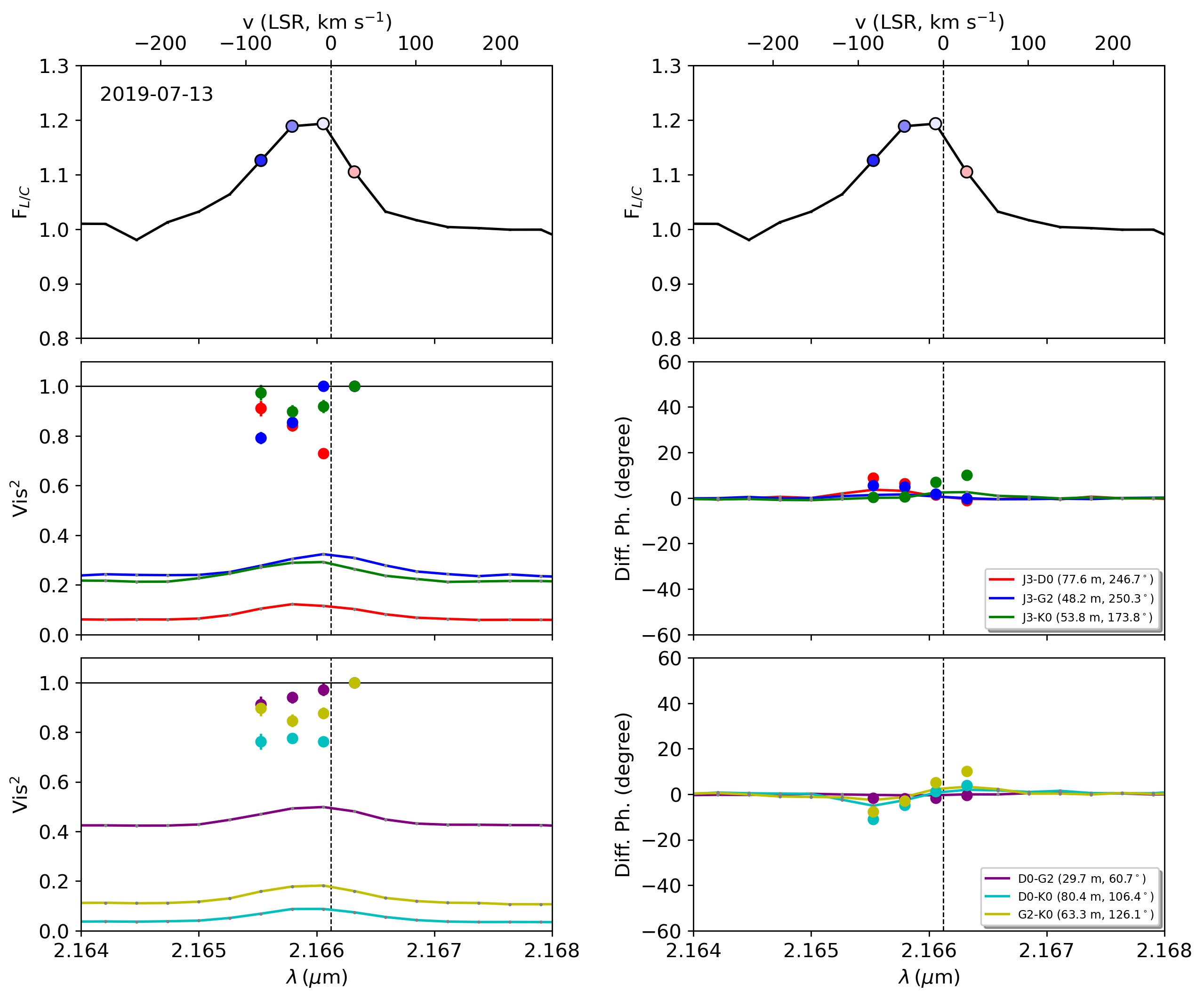}
\includegraphics[width=\columnwidth]{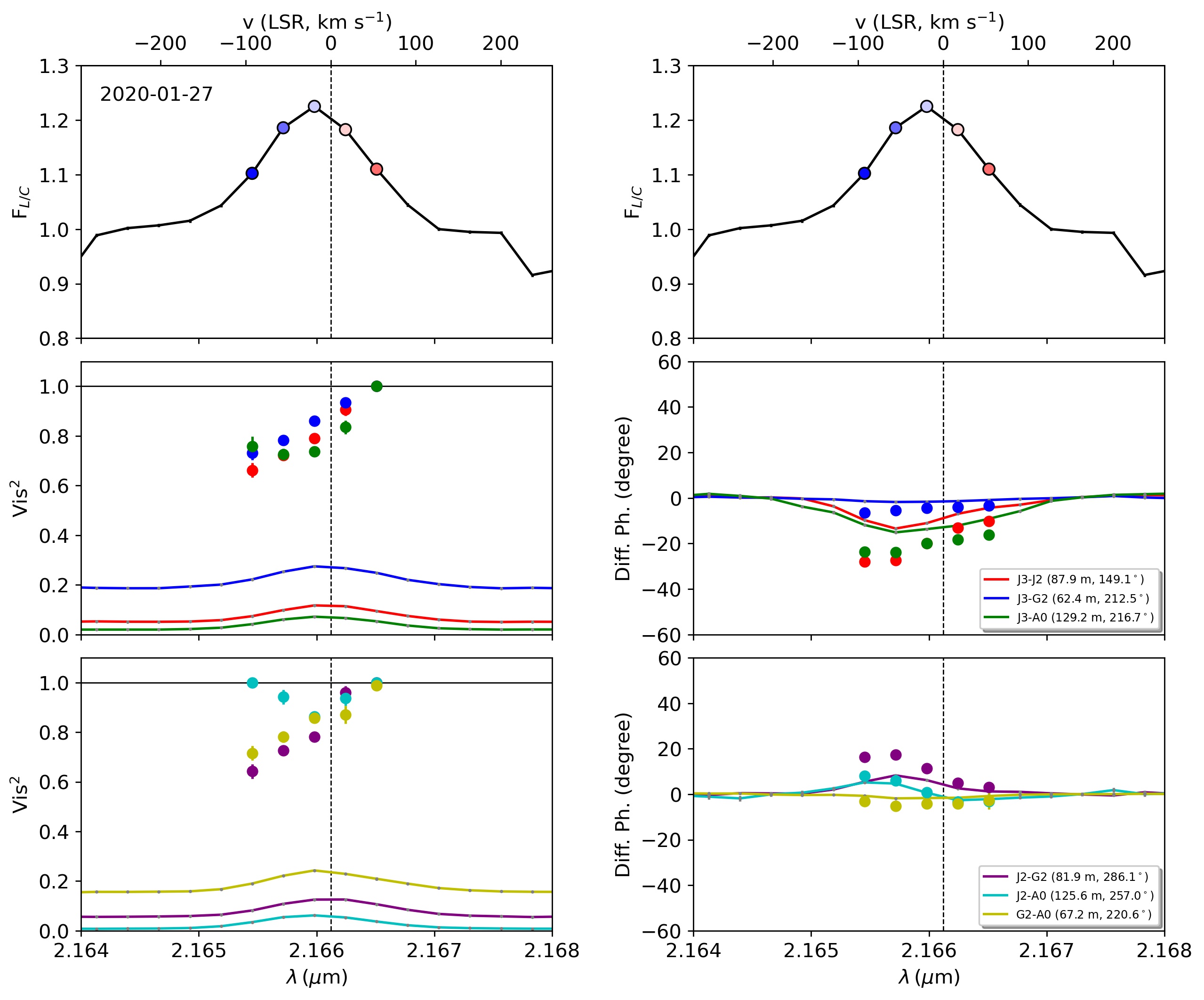}\\[0.5em]
\includegraphics[width=\columnwidth]{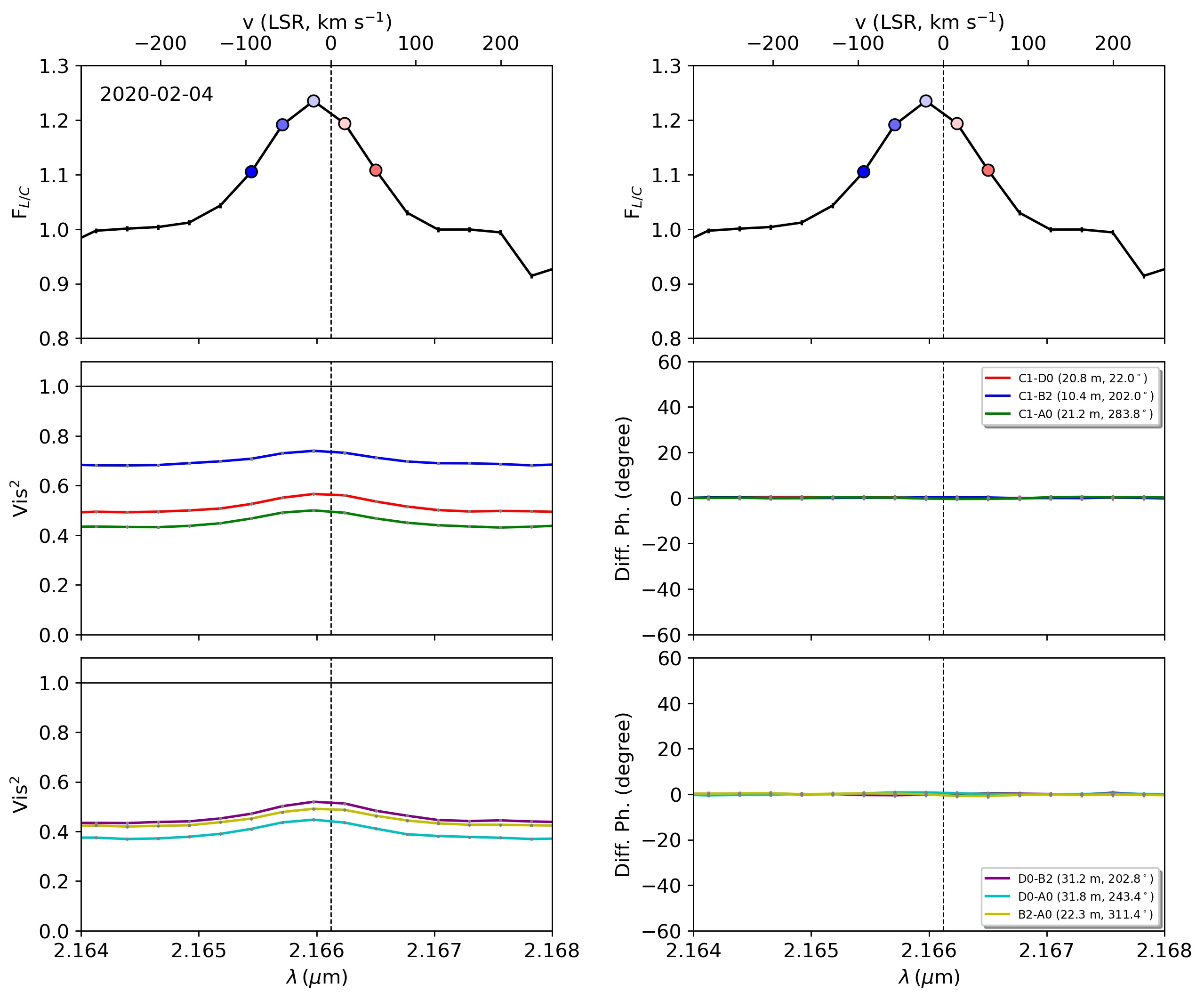}
\includegraphics[width=\columnwidth]{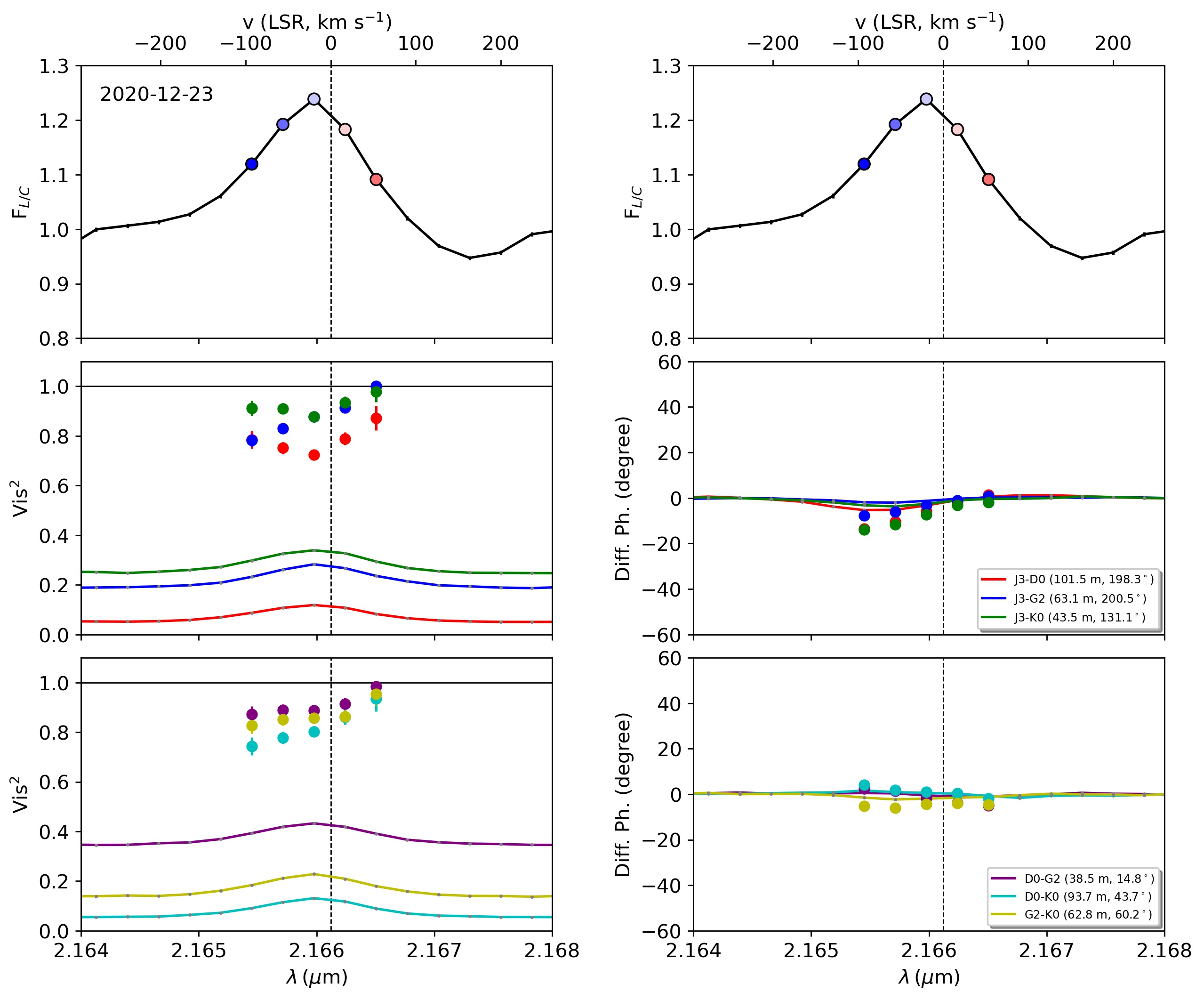}\\[0.5em]
\includegraphics[width=\columnwidth]{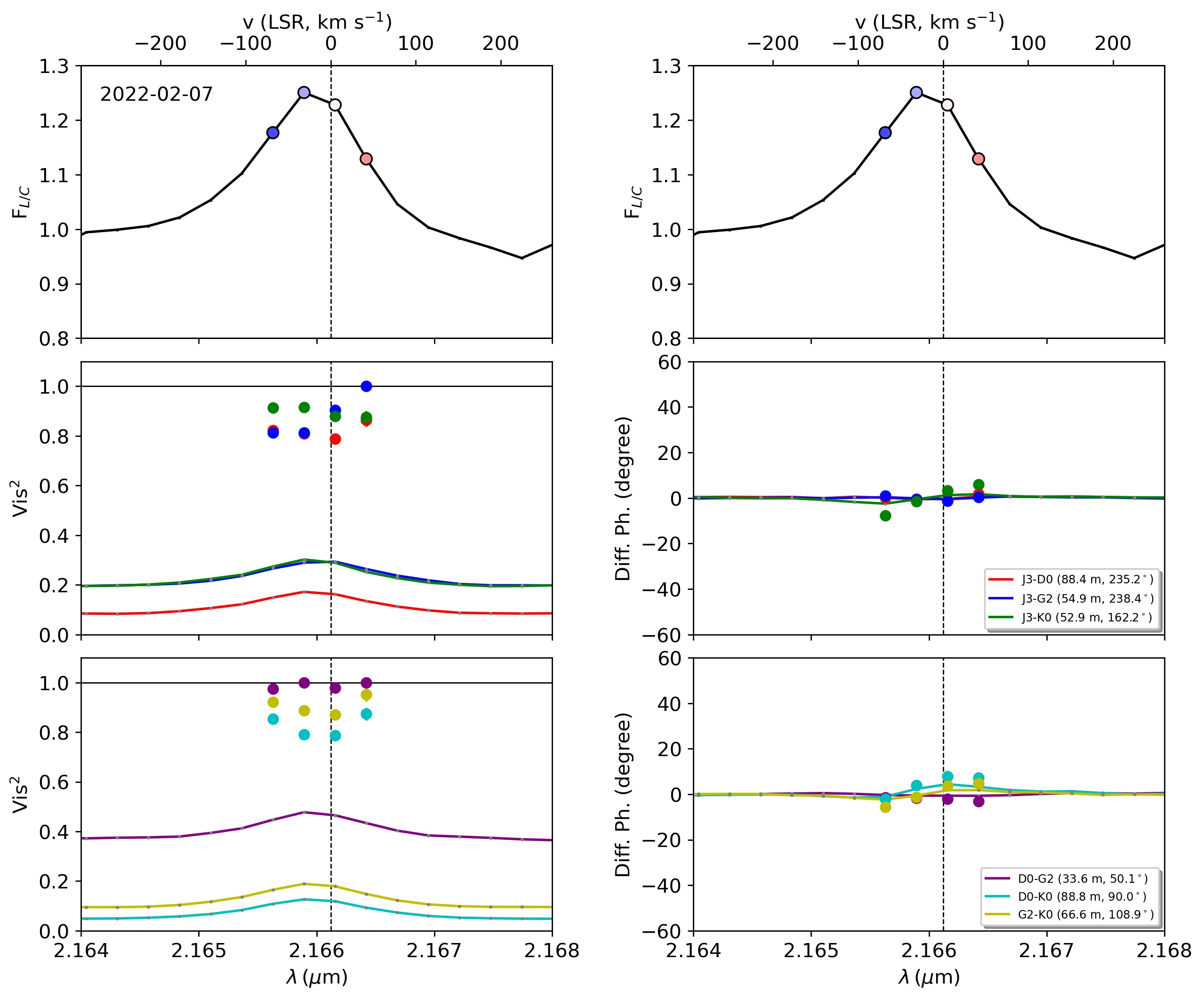}
\includegraphics[width=\columnwidth]{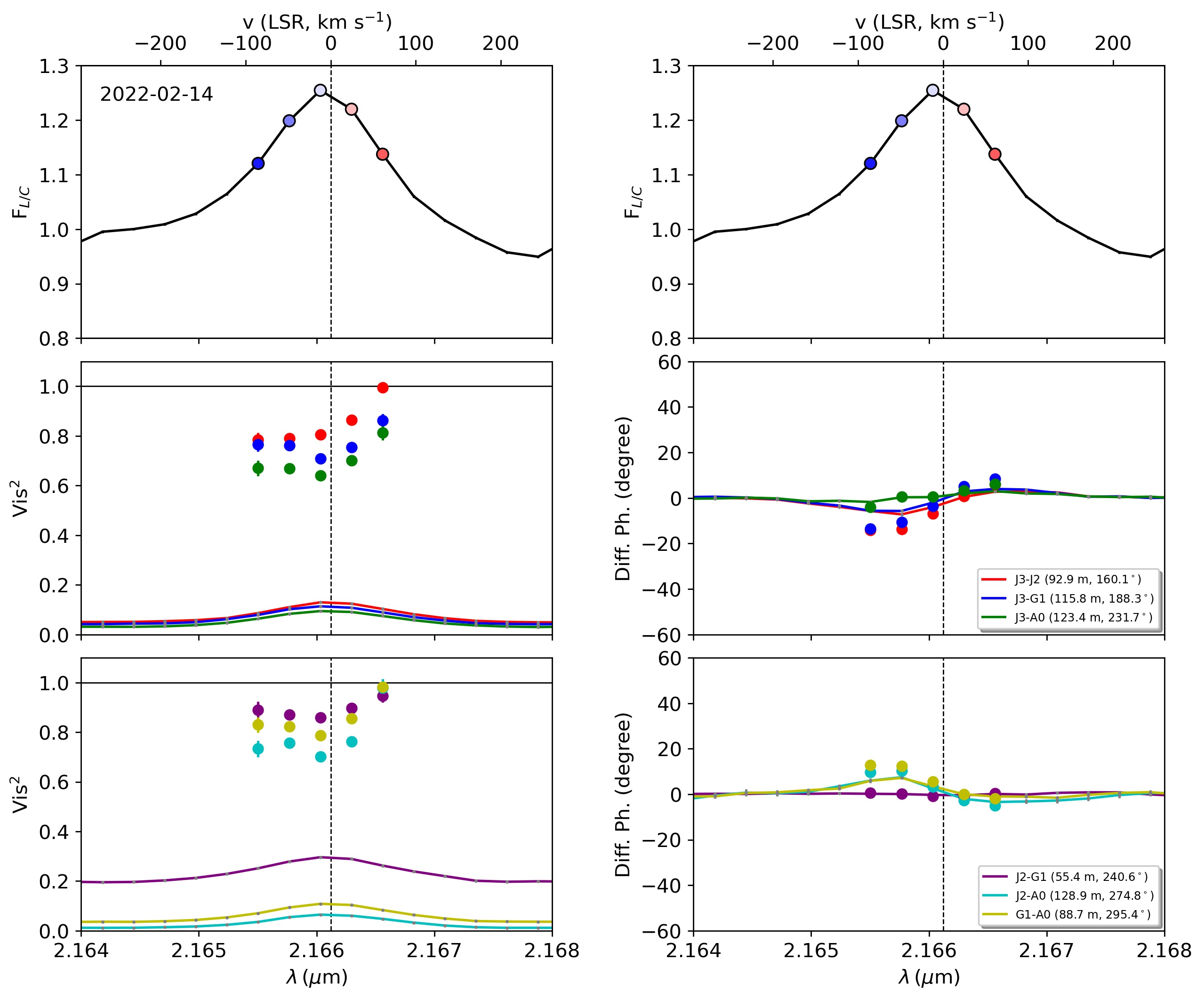}\\
\caption{Continued.}
\end{figure*}

\newpage

\begin{table*}
\centering
\caption{\centering Fit results per epoch for the model SHRM1 and combining the visibilities and the closure phases}
\begin{tabular}{l|cccccccccccccc} 
\hline
\hline
ID & P1 & P2 & P3 & P4 & P5 & P6 & P7 & P8 & P9 & P10 & P11 & P12 & P13 & P14 \\
$\chi^2_r$ & 13.53 & 3.48 & 39.36 & 13.64 & 2.64 & 16.25 & 16.24 & 6.97 & 16.56 & 6.97 & 6.68 & 12.36 & 4.74 & 8.67 \\ \hline
ID & P15 & P16 & P17 & P18 & P19 & P20 & {\it All epochs} & -- & -- & -- & -- & -- & -- & -- \\
$\chi^2_r$ & 5.07 & 3.29 & 36.03 & 10.82 & 12.81 & 9.09 & {\it 11.5} & -- & -- & -- & -- & -- & -- & --\\ \hline\hline
ID & G1 & G2 & G3 & G4 & G5 & G6 & G7 & G8 & G9 & G10 & G11 & G12 & G13 & {\it All epochs} \\
$\chi^2_r$ & 11.41 & 5.0 & 9.96& 2.53 & 24.29 & 2.89 & 1.19 & 1.04 & 5.54 & 2.42 & 2.79 & 1.14 & 2.36 & {\it 5.30} \\
\hline
\end{tabular}\label{tab:Single-Epoch_chi2}
\end{table*}


\clearpage

\section{Global fit MCMC posterior distribution functions and azimuthal modulation parameters' $\chi_r^2$ maps}
\label{apx:GF_AzMod-C2Maps}

\begin{figure*}[ht]
    \centering
    \includegraphics[width=\textwidth]{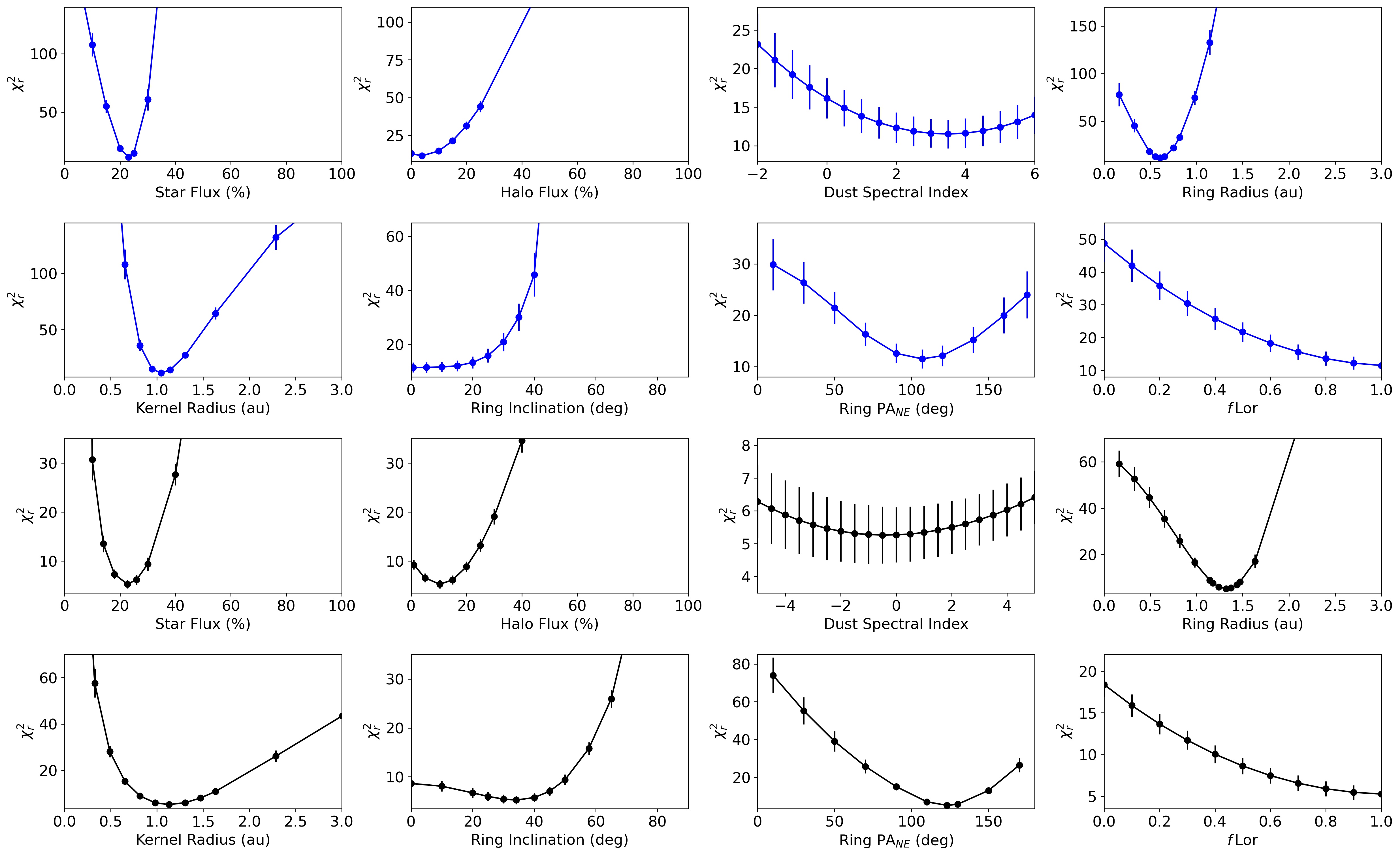}\\
    \caption{Reduced chi-square $\chi_r^2$ curves of each nontime-variable parameter from the azimuthal modulation global fit model. The first two rows, where the curves are blue, refer to the PIONIER data model, while the last two rows, where the curves are black, refer to the GRAVITY data model.}
    \label{fig:GF_chi2maps}
\end{figure*}

\begin{figure*}[ht]
    \centering
    \includegraphics[width=\textwidth]{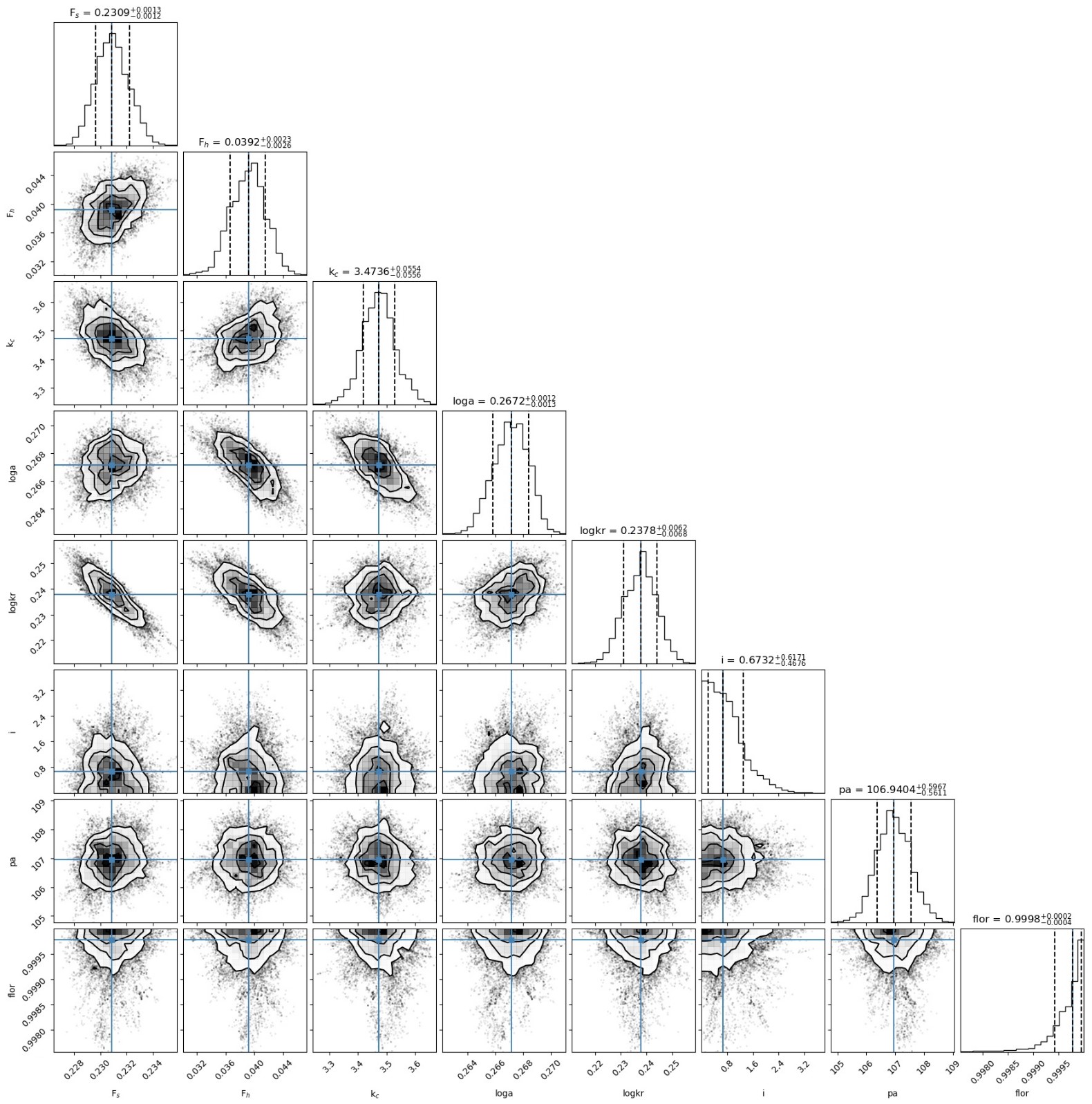}
    \caption{Global fit MCMC posterior distribution functions of the fitted parameters for the PIONIER data set. In the one-dimensional histograms, the blue line identifies the median of the distribution, the dashed black lines identify the 16$^{th}$ and the 84$^{th}$ percentiles.}
    \label{fig:PF_corner}
\end{figure*}

\begin{figure*}[ht]
    \centering
    \includegraphics[width=\textwidth]{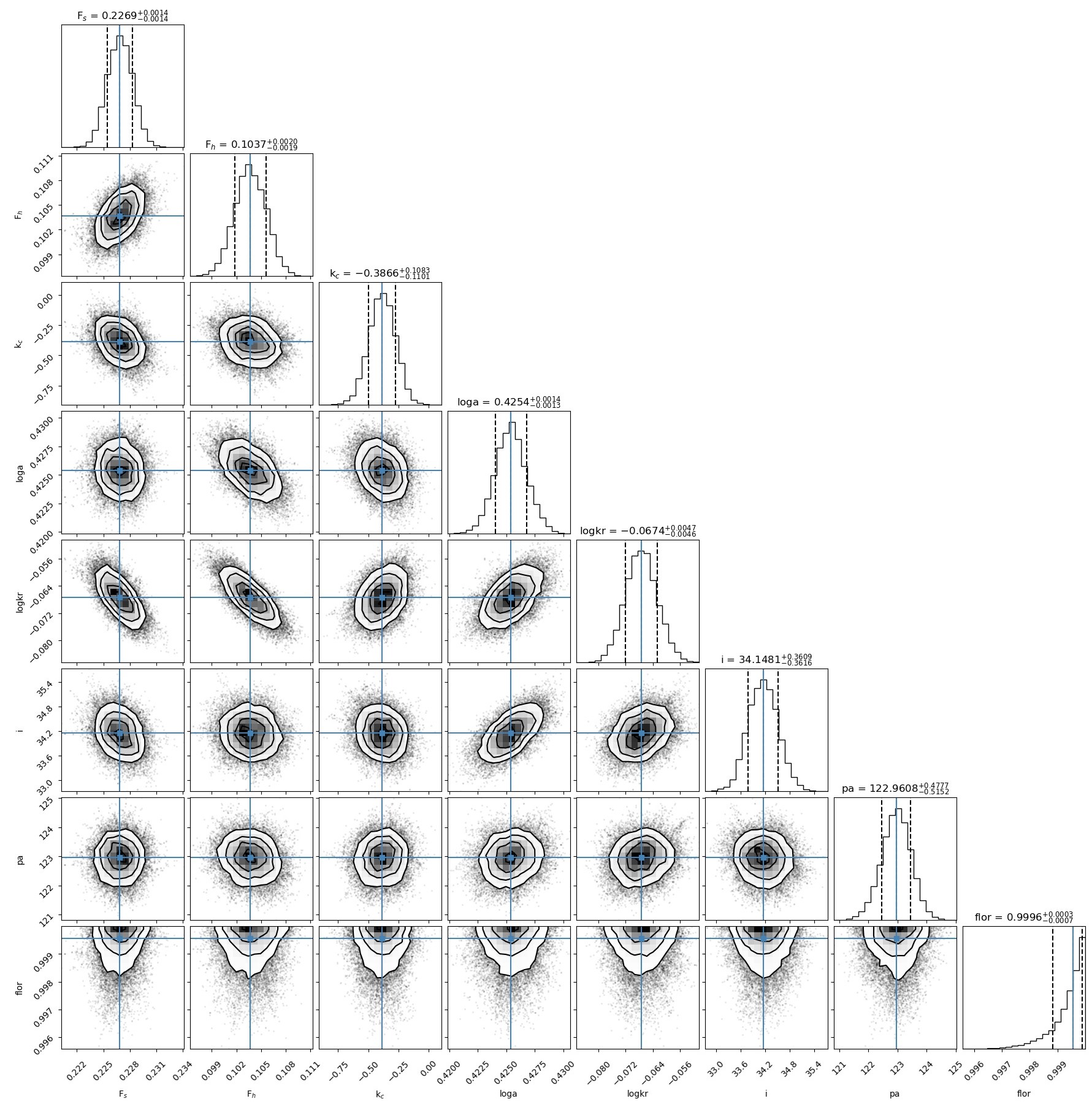}
    \caption{Global fit MCMC posterior distribution functions of the fitted parameters for the GRAVITY data set. In the one-dimensional histograms, the blue line identifies the median of the distribution, the dashed black lines identify the 16$^{th}$ and the 84$^{th}$ percentiles.}
    \label{fig:GF_corner}
\end{figure*}

\begin{figure*}[!ht]
\centering
    \includegraphics[scale=0.25]{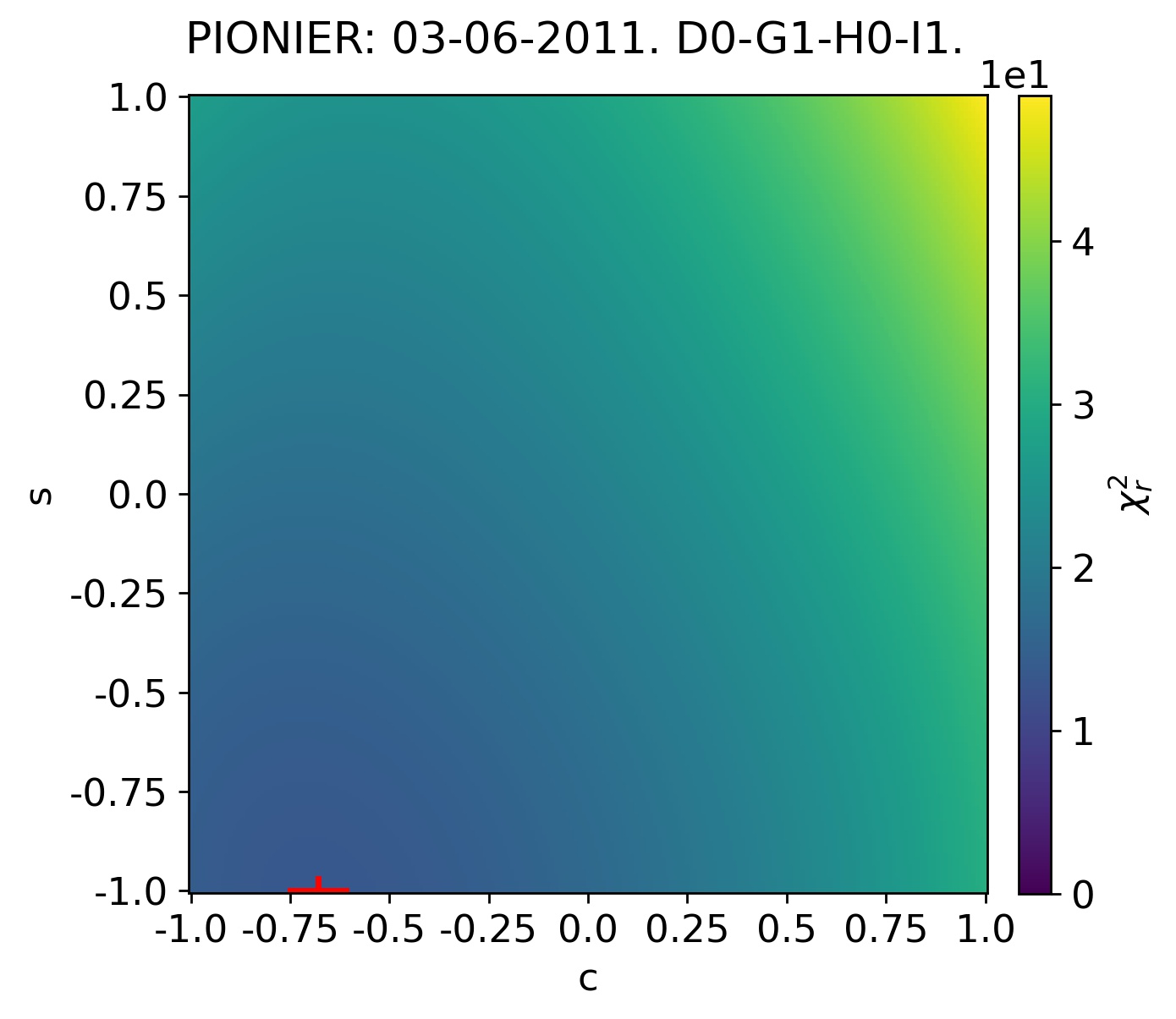}
    \includegraphics[scale=0.25]{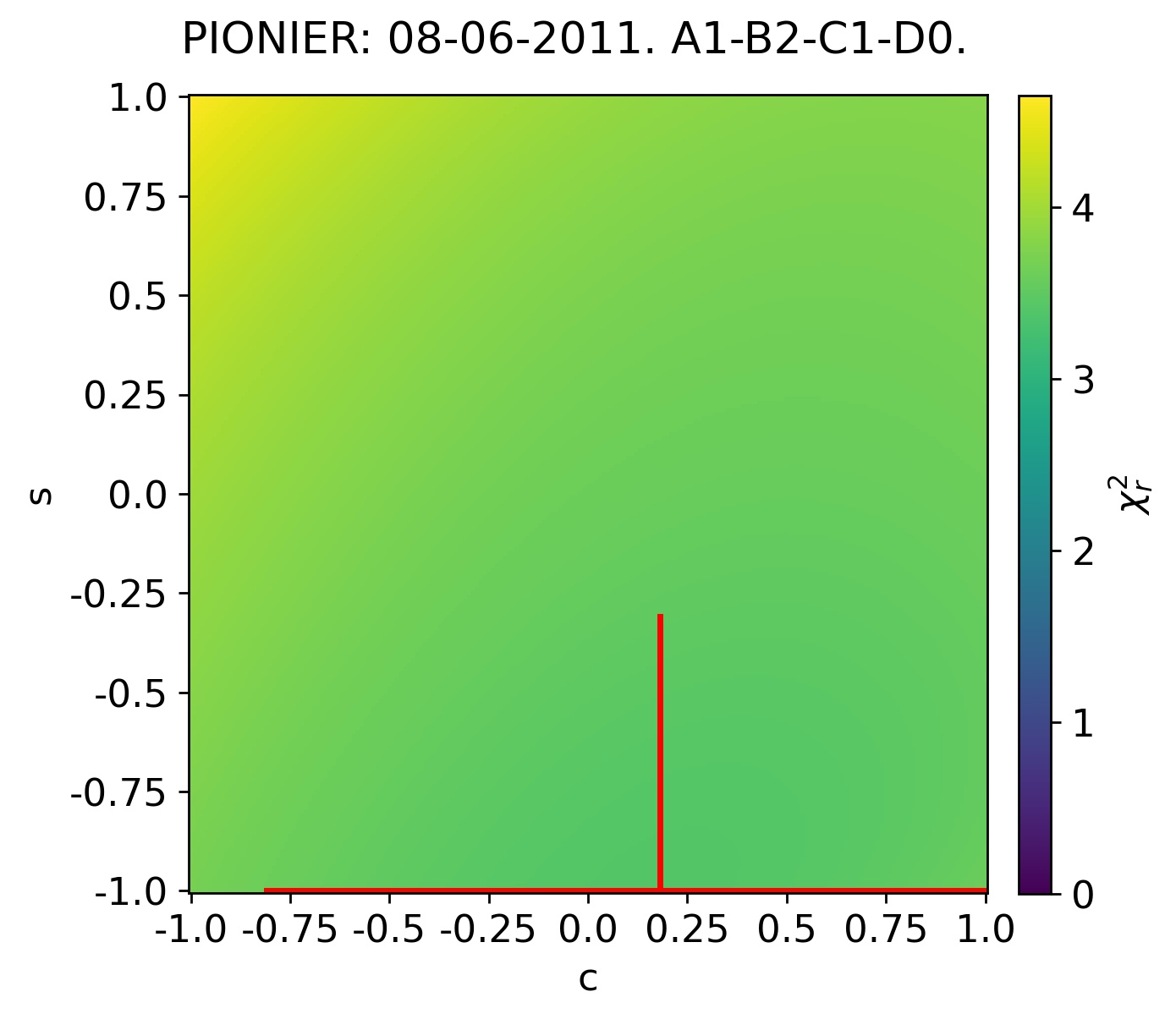}
    \includegraphics[scale=0.25]{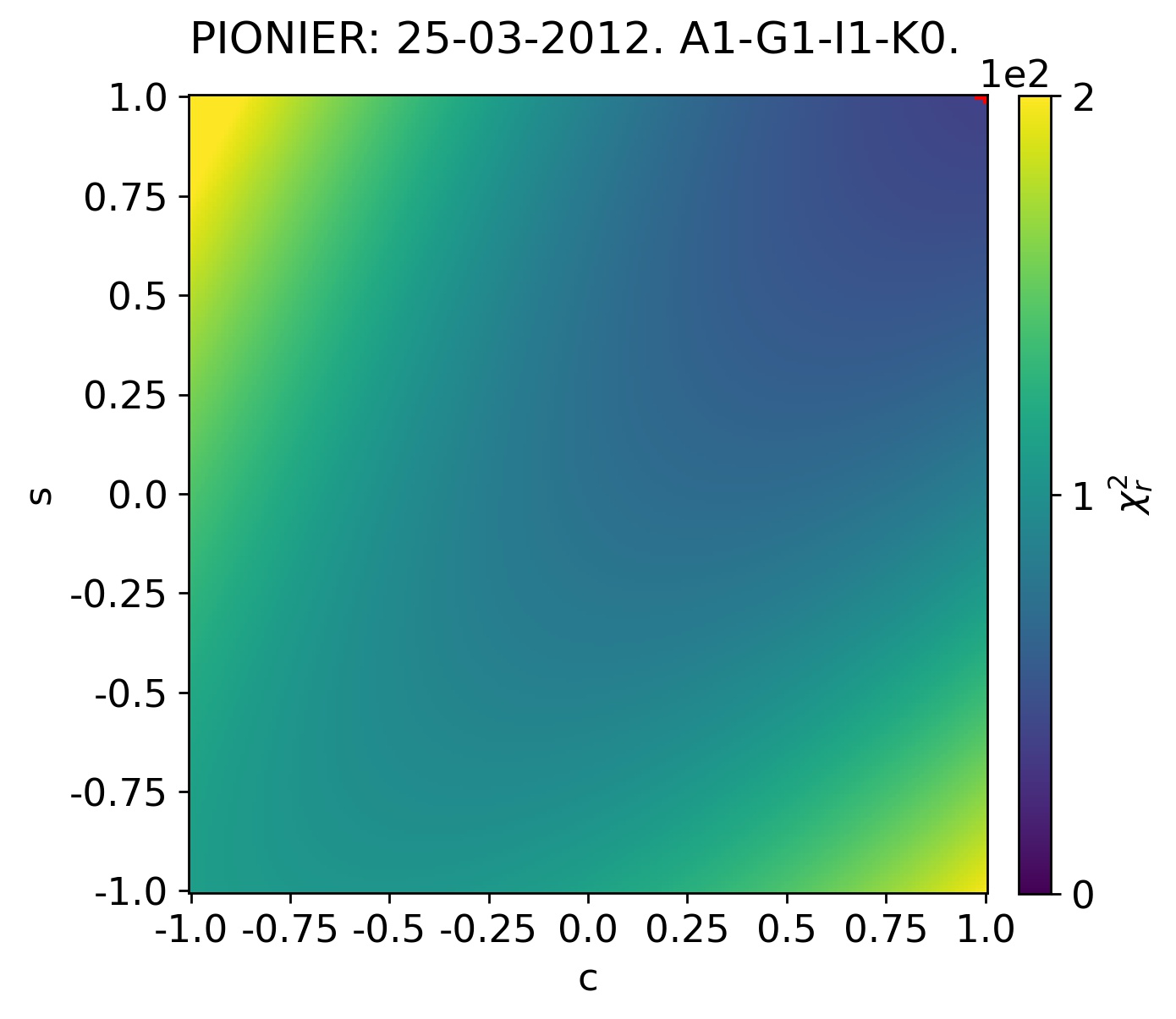}
    \includegraphics[scale=0.25]{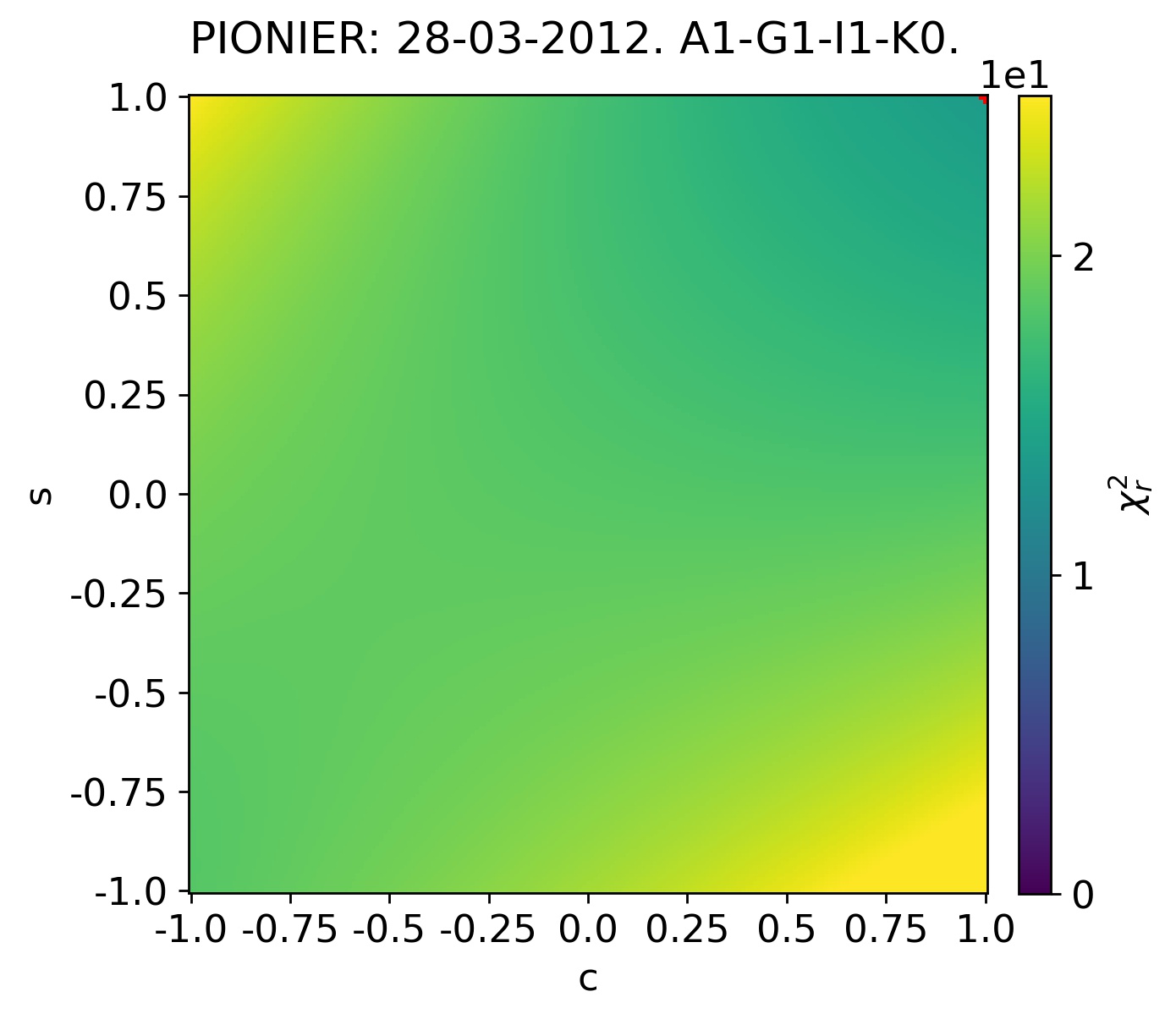}
    \includegraphics[scale=0.25]{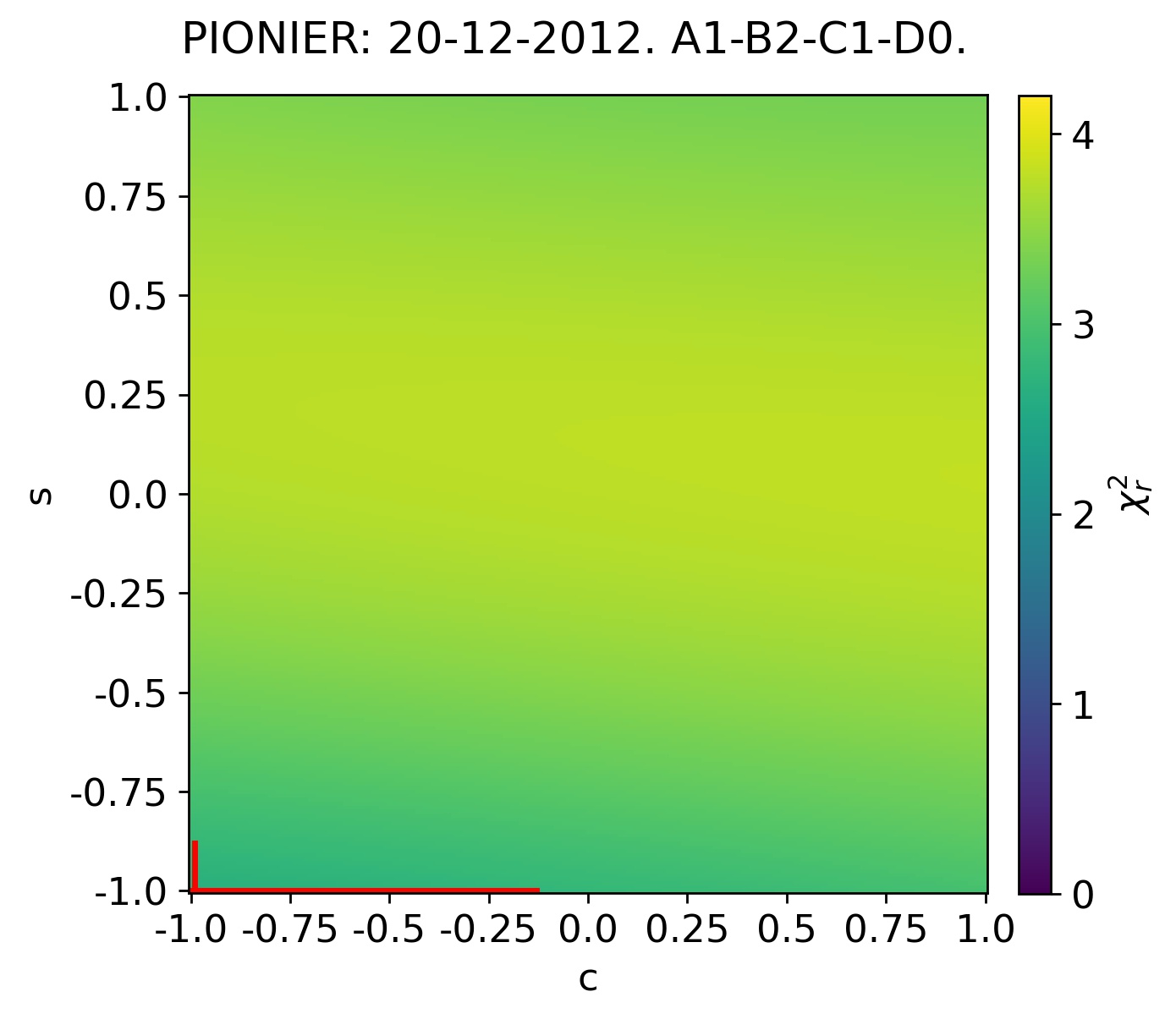}\\
    \includegraphics[scale=0.25]{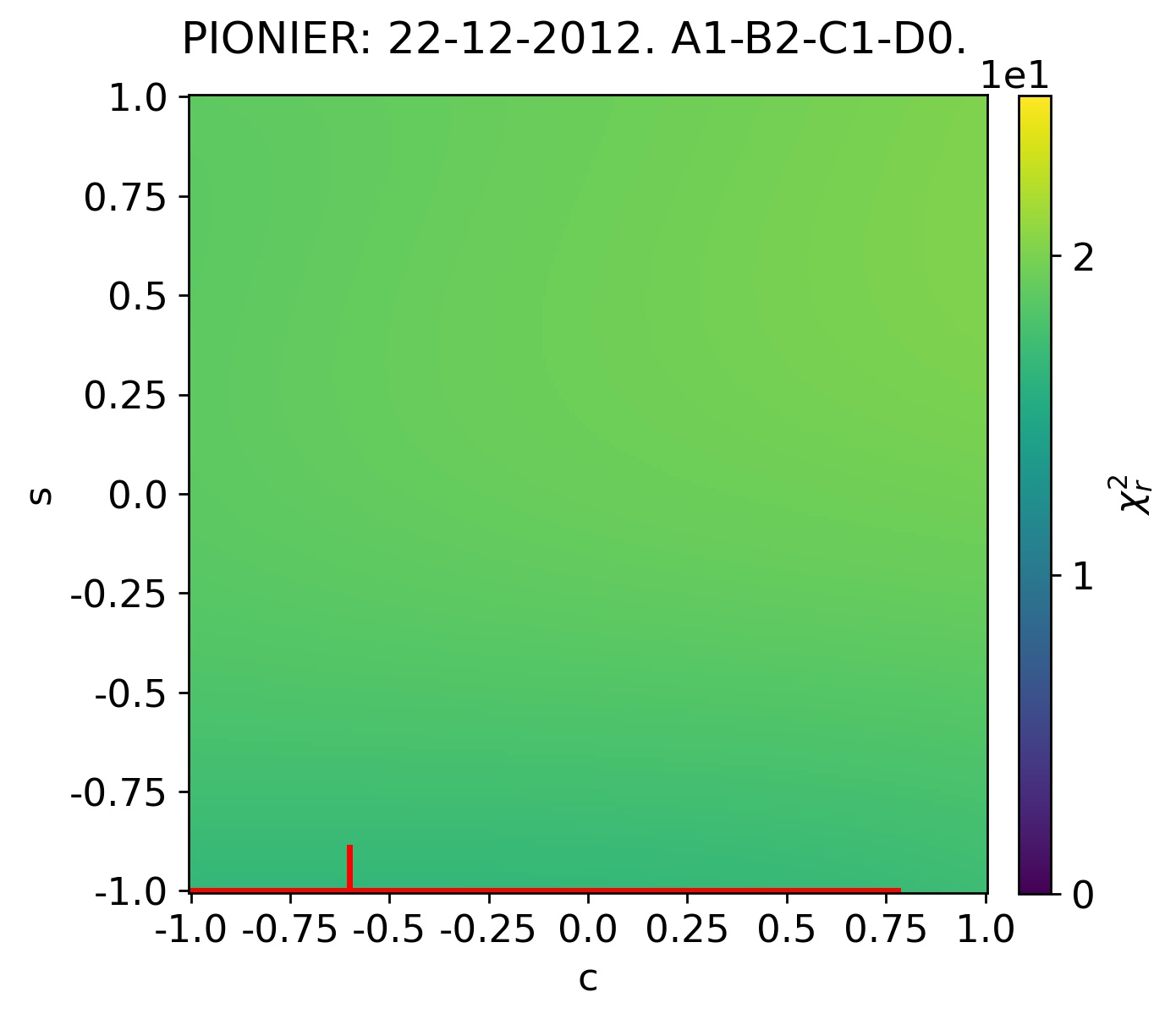}
    \includegraphics[scale=0.25]{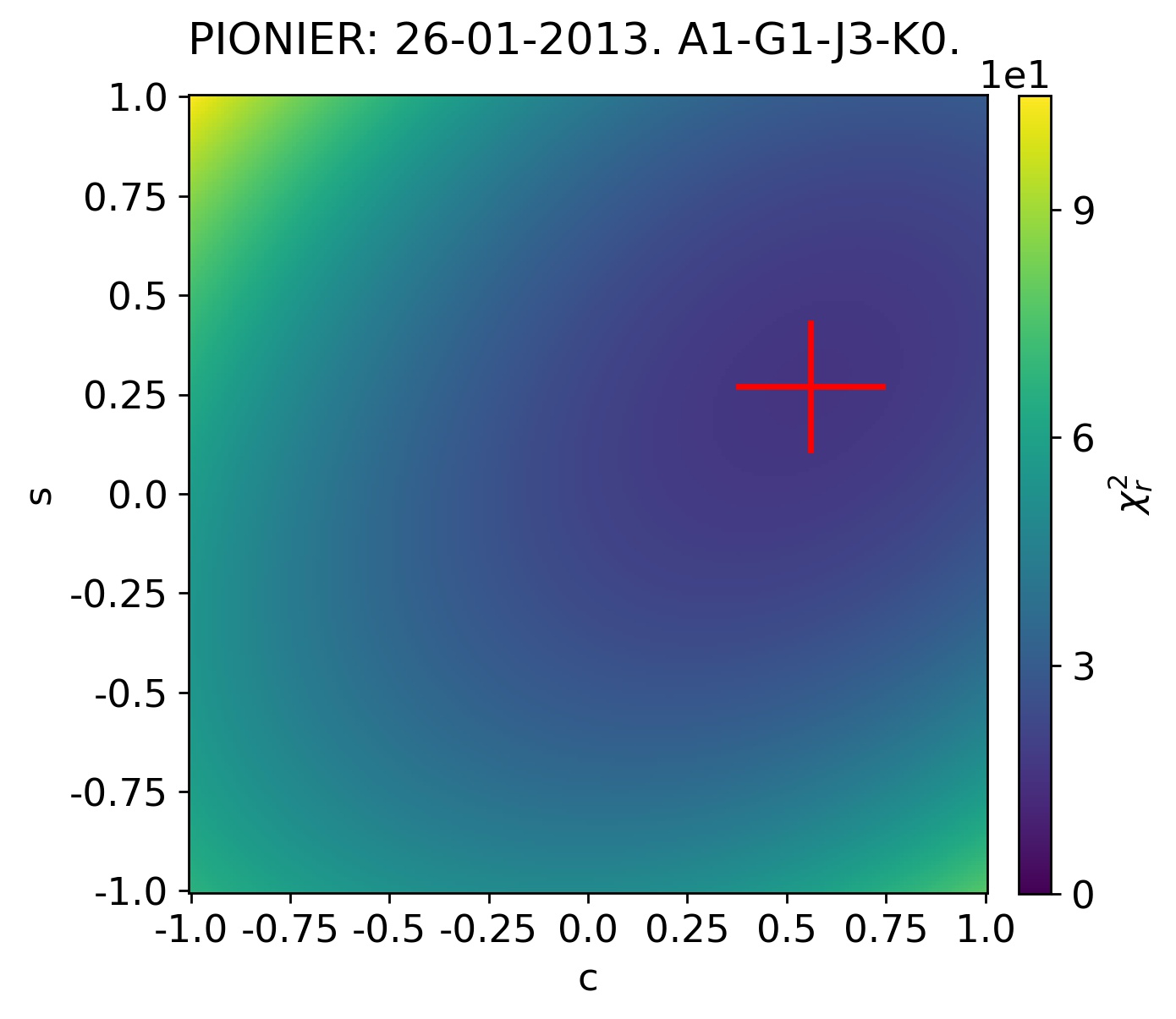}
    \includegraphics[scale=0.25]{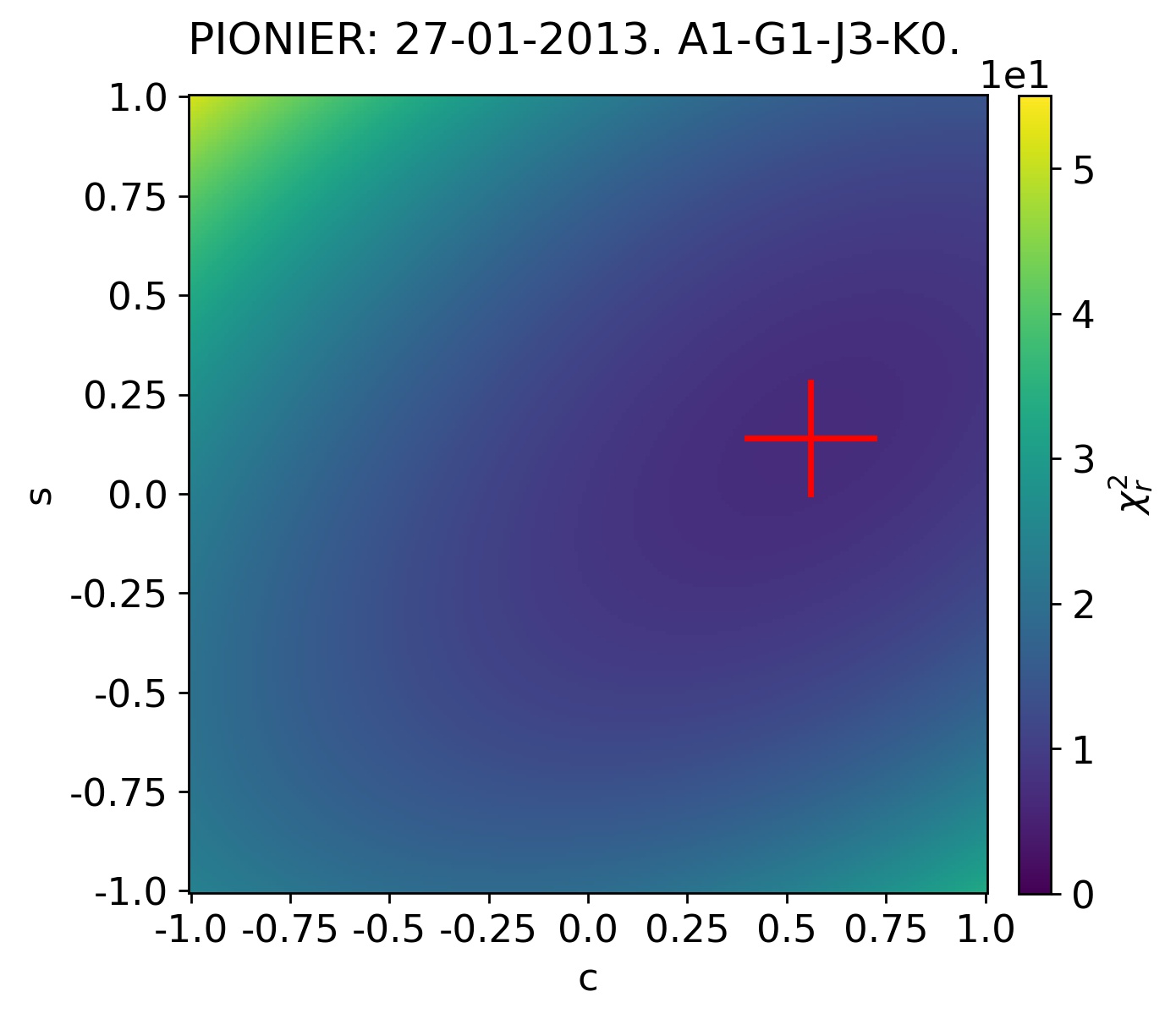}
    \includegraphics[scale=0.25]{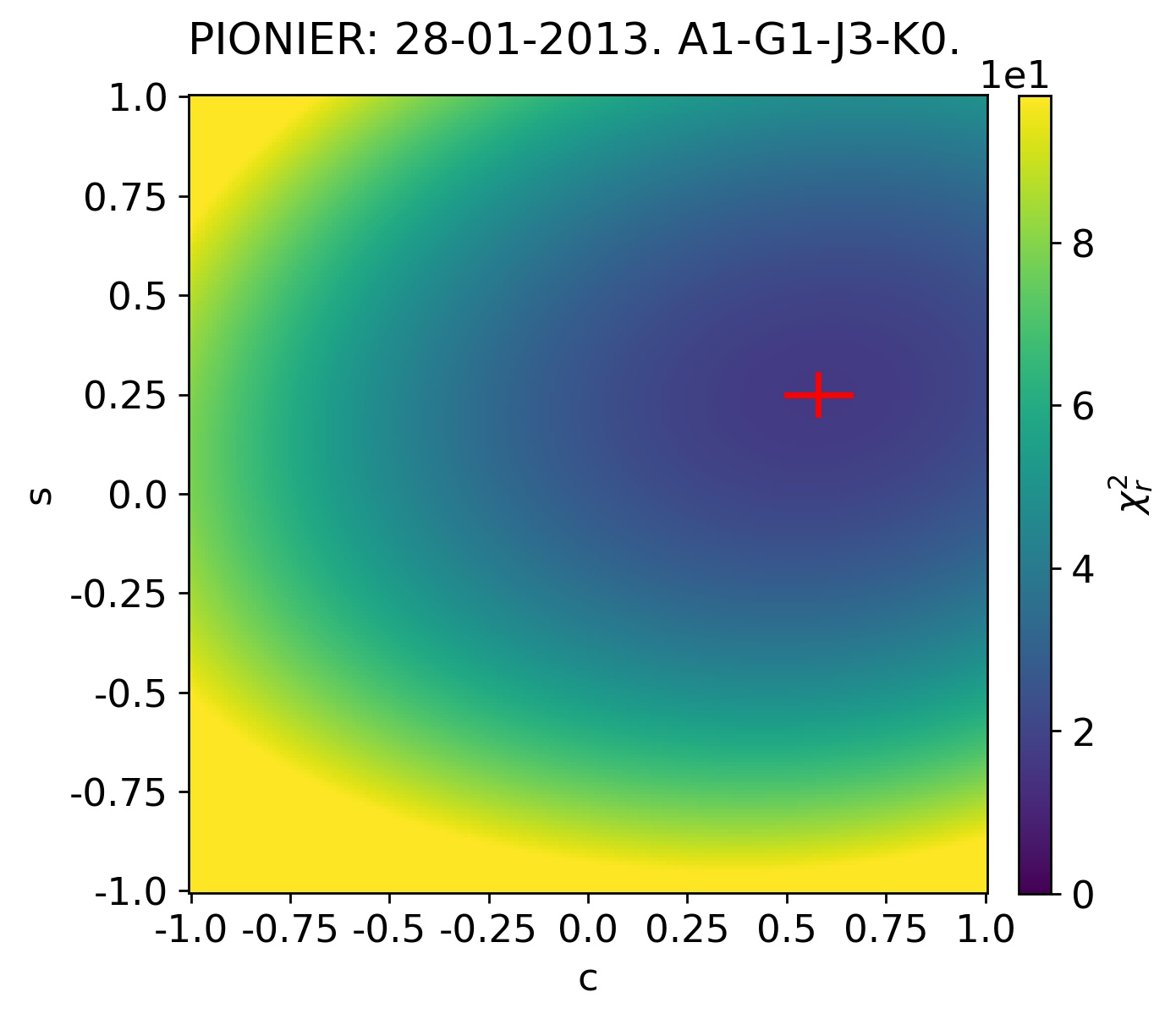}
    \includegraphics[scale=0.25]{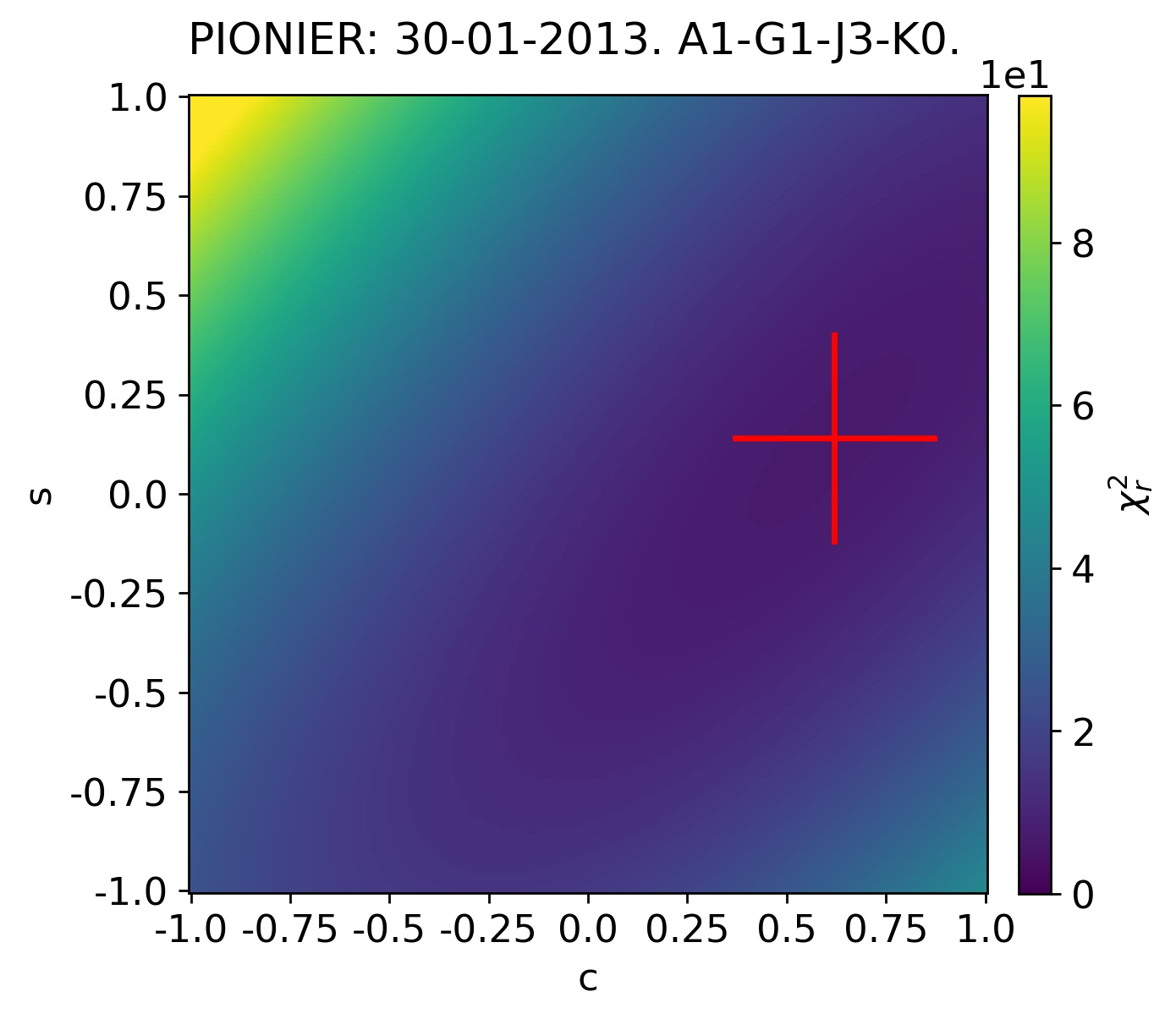}\\
    \includegraphics[scale=0.25]{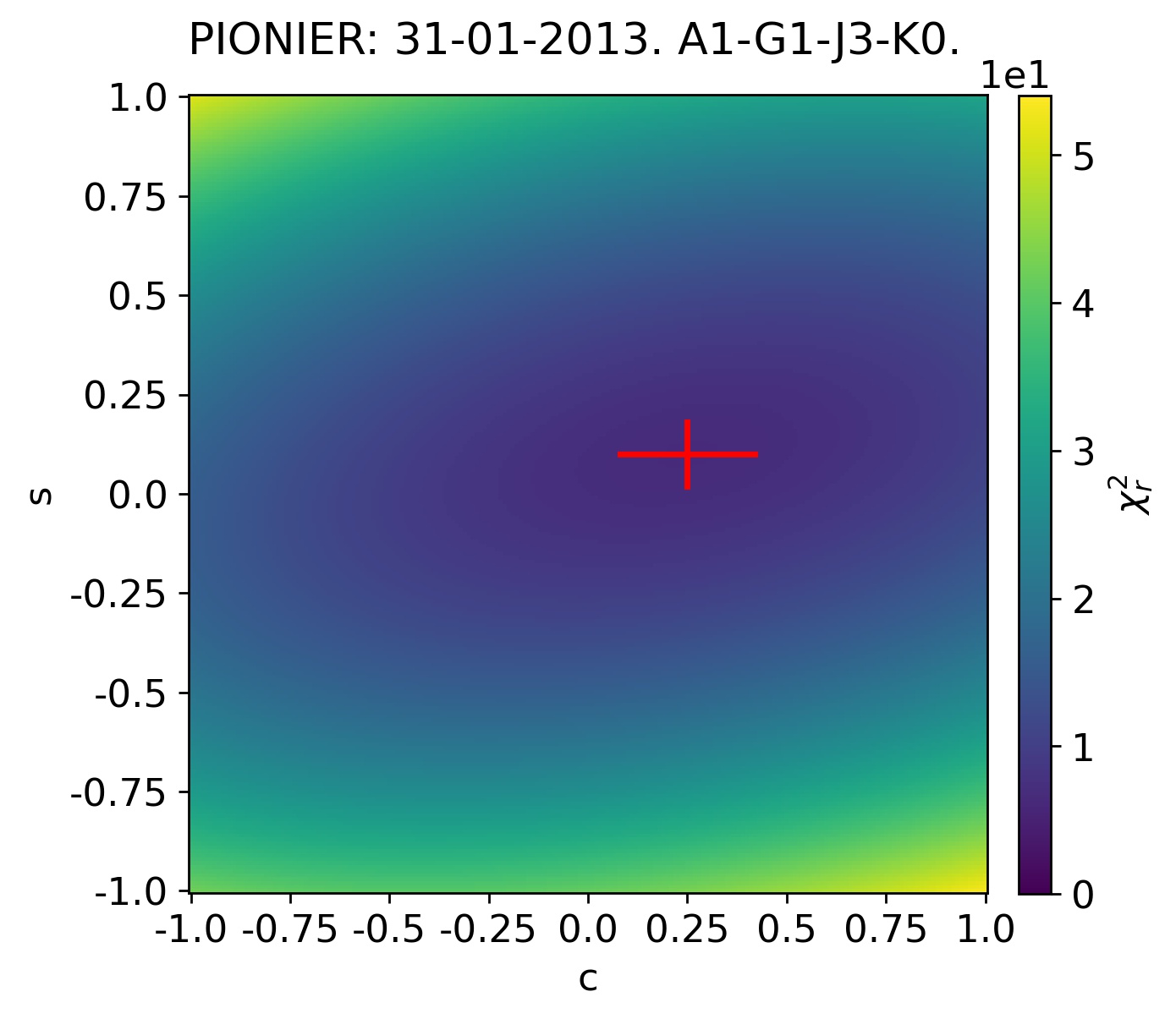}
    \includegraphics[scale=0.25]{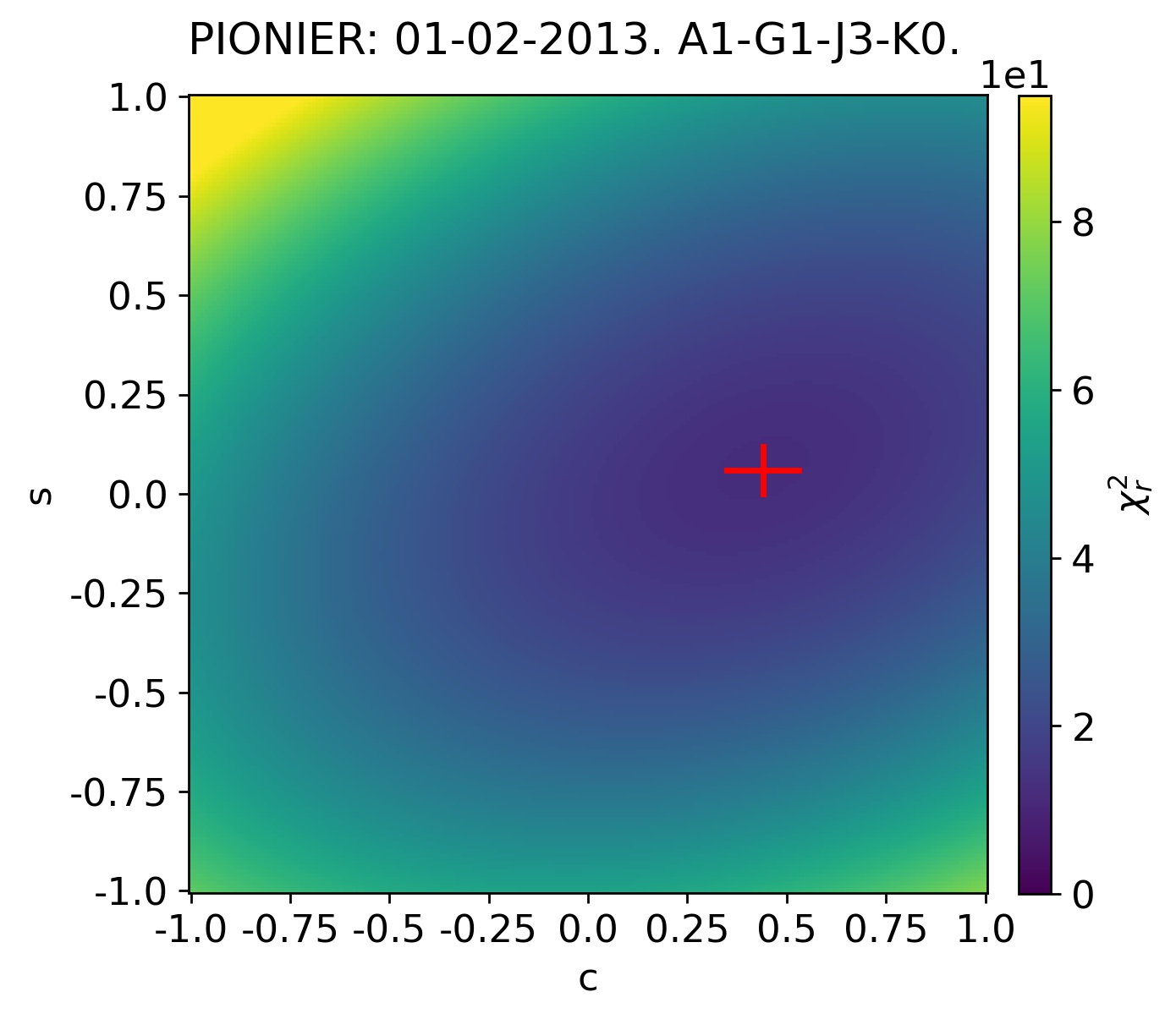}
    \includegraphics[scale=0.25]{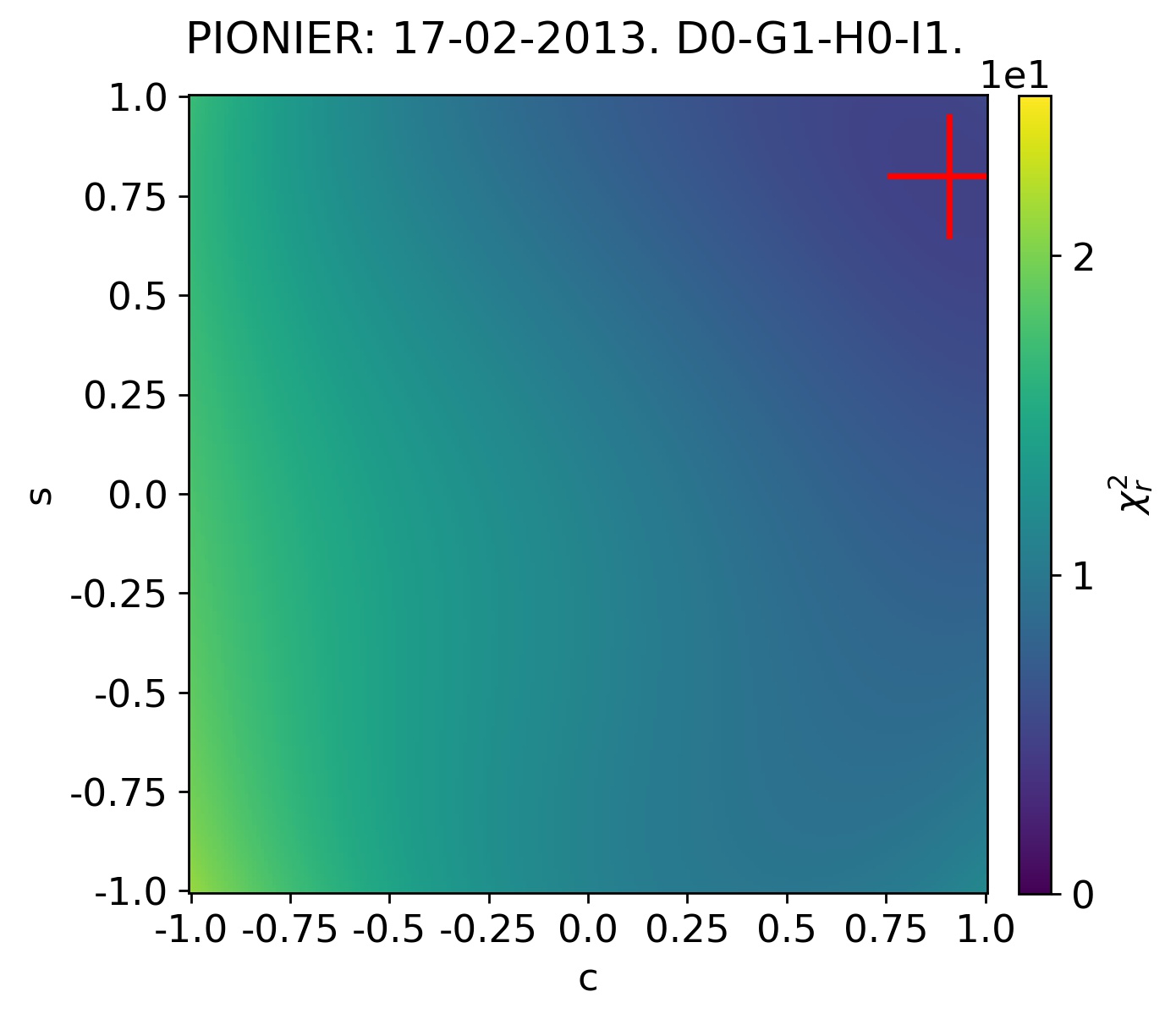}
    \includegraphics[scale=0.25]{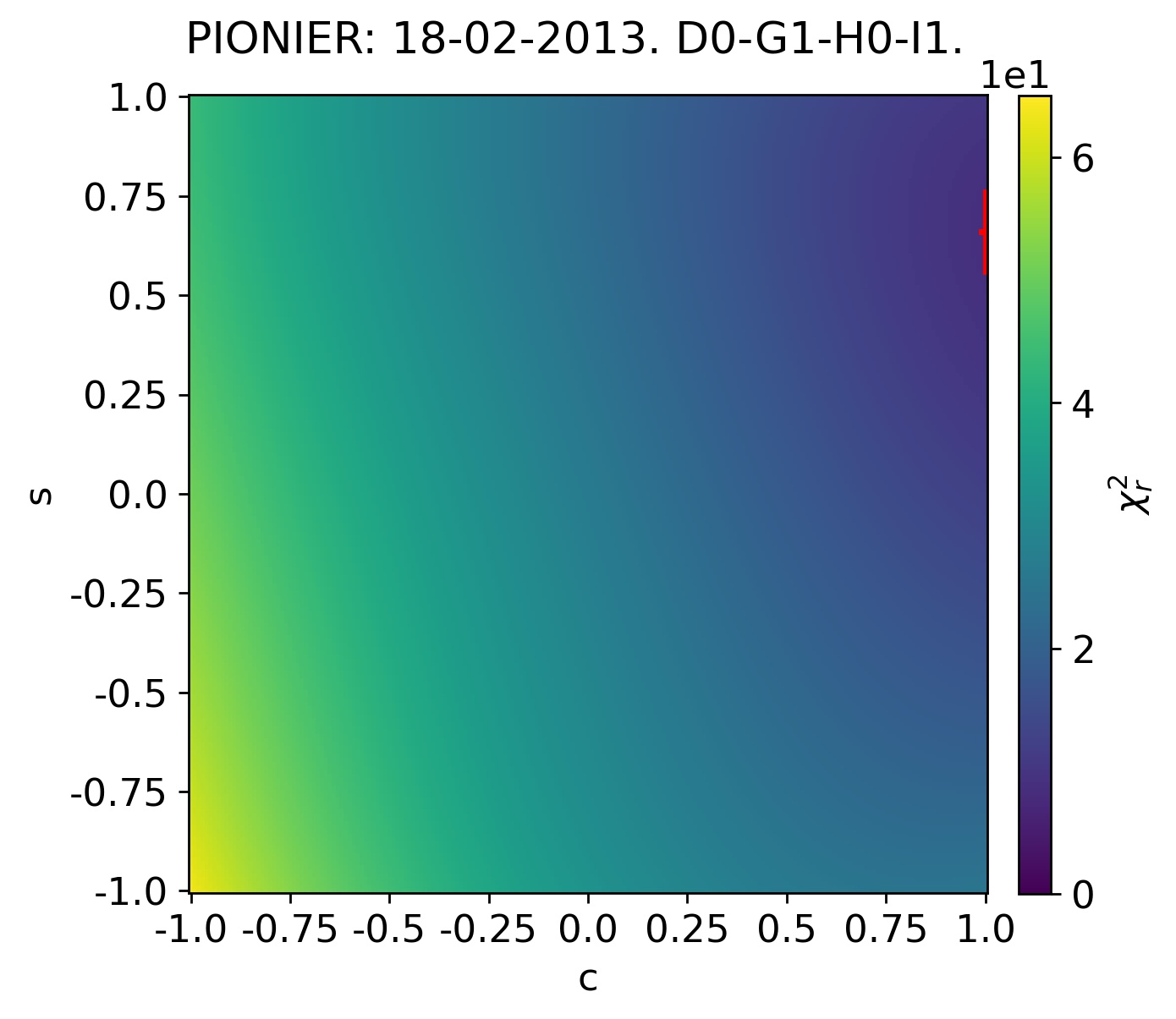}
    \includegraphics[scale=0.25]{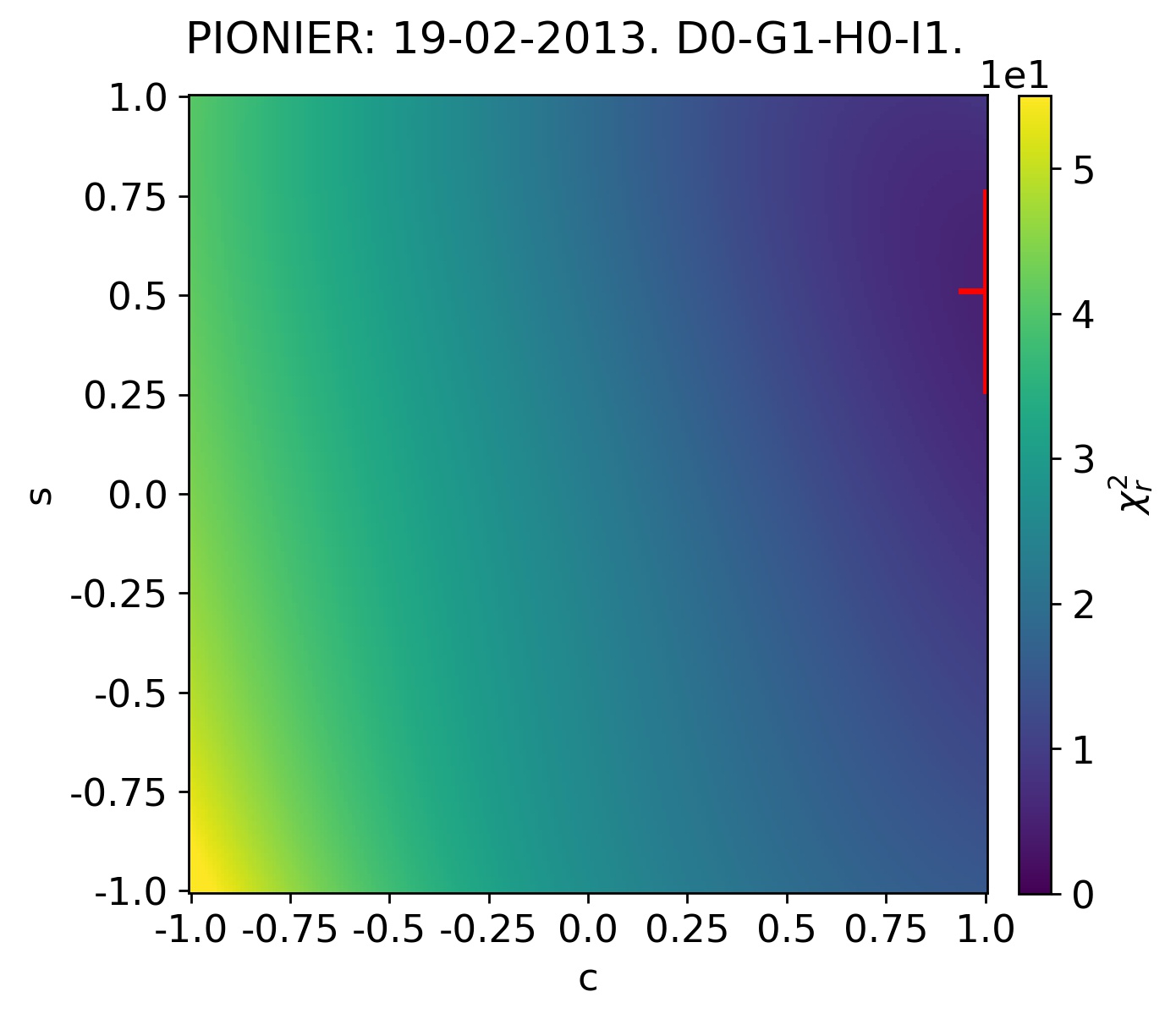}\\
    \includegraphics[scale=0.25]{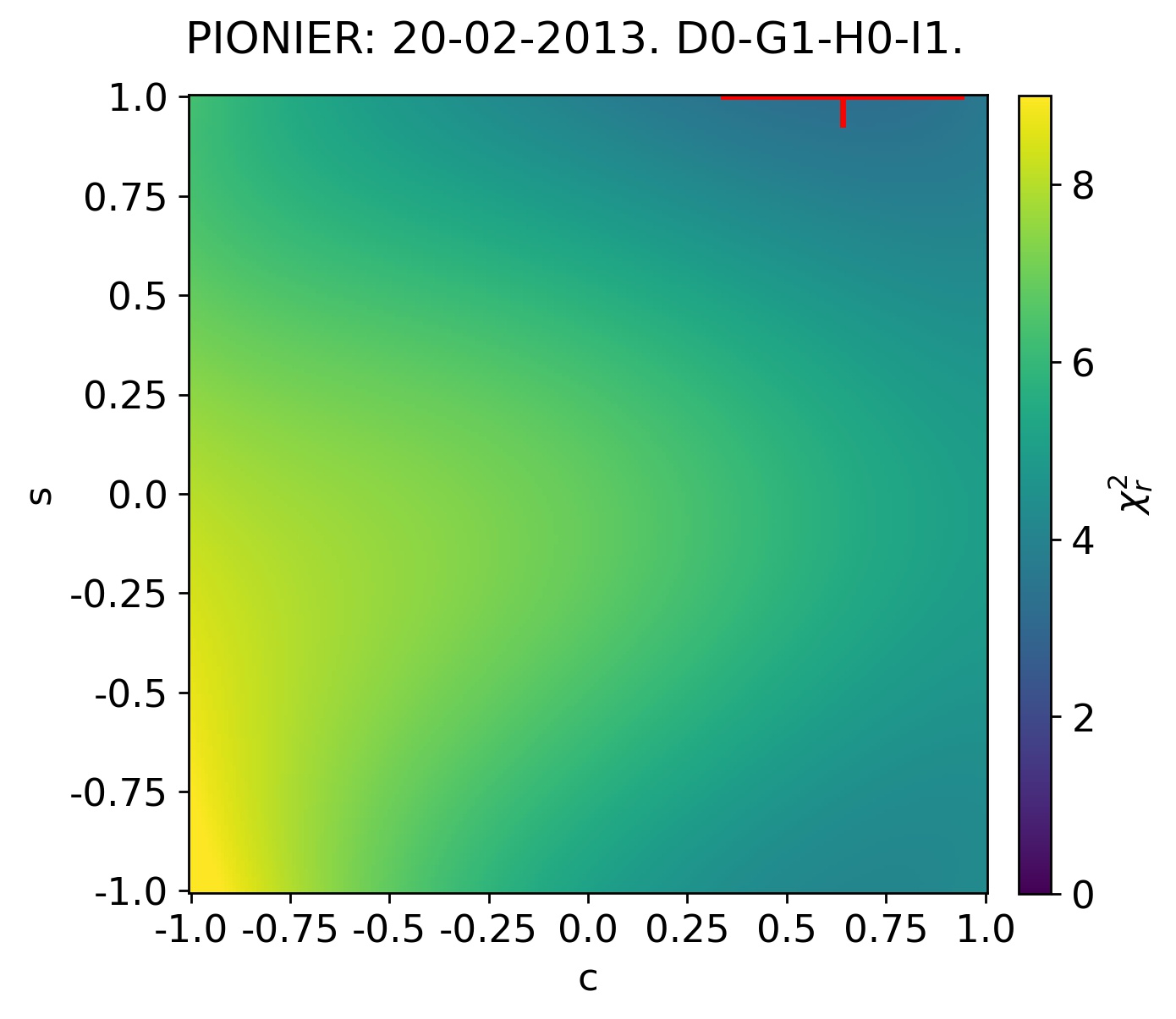}
    \includegraphics[scale=0.25]{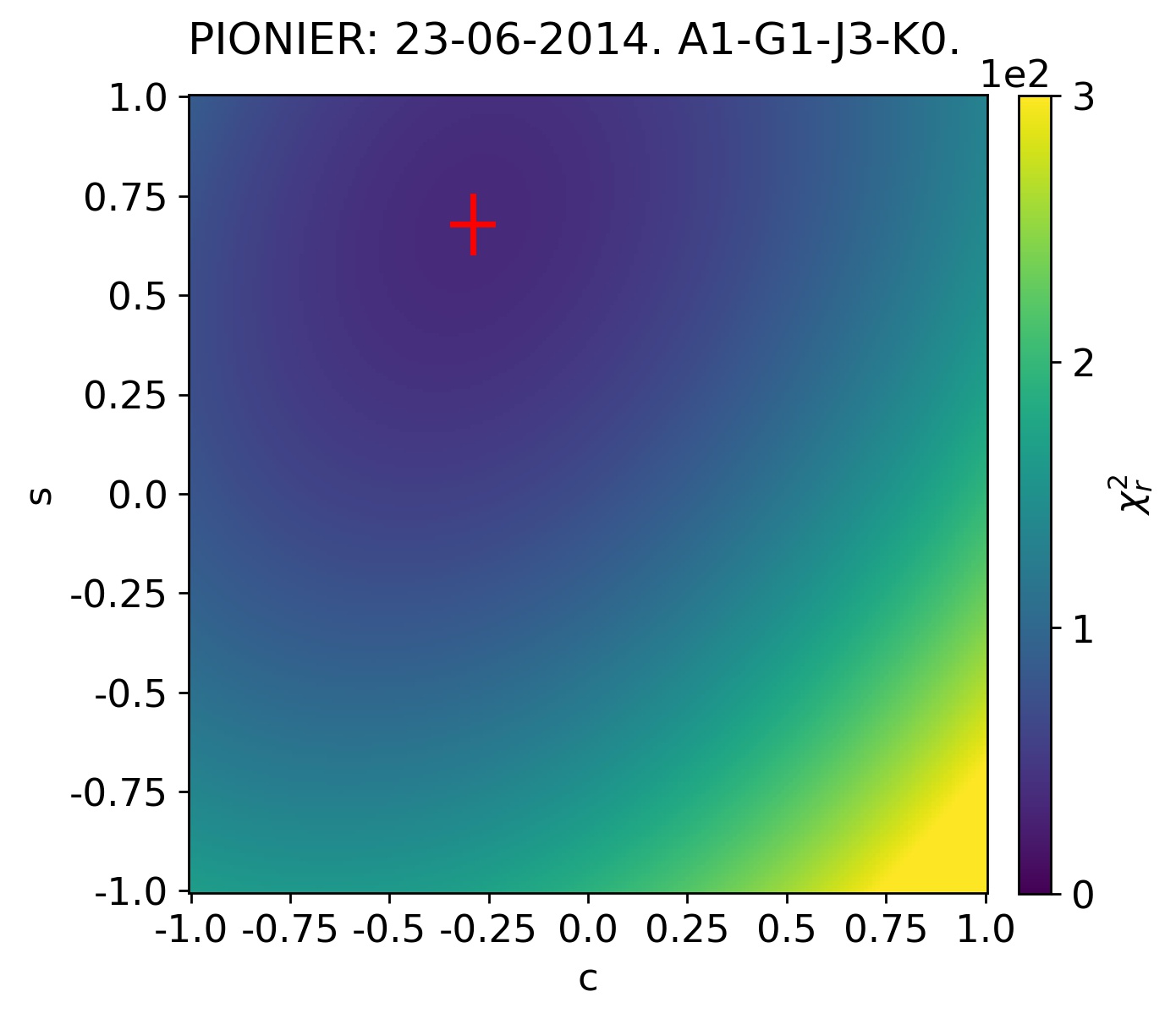}
    \includegraphics[scale=0.25]{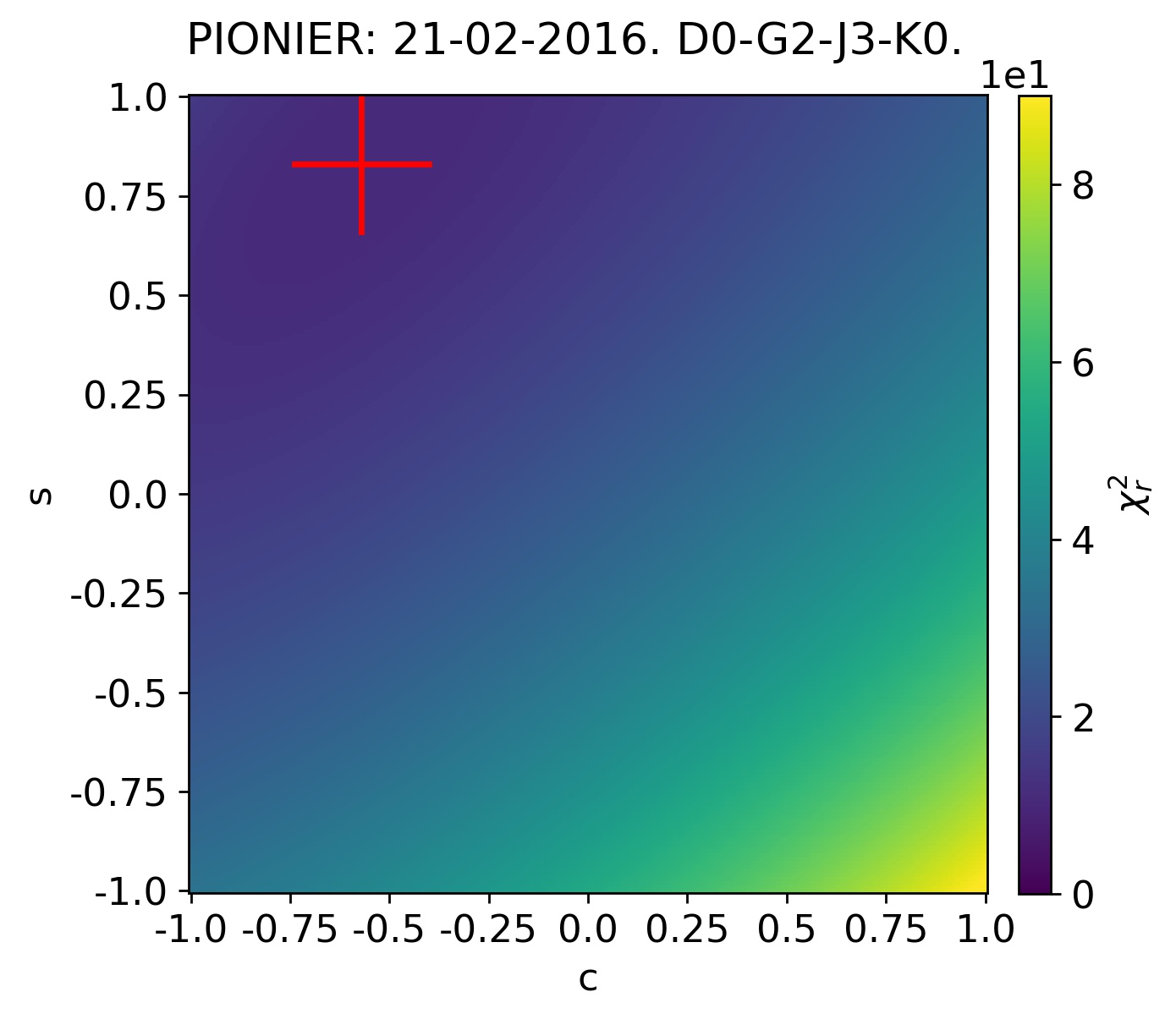}
    \includegraphics[scale=0.25]{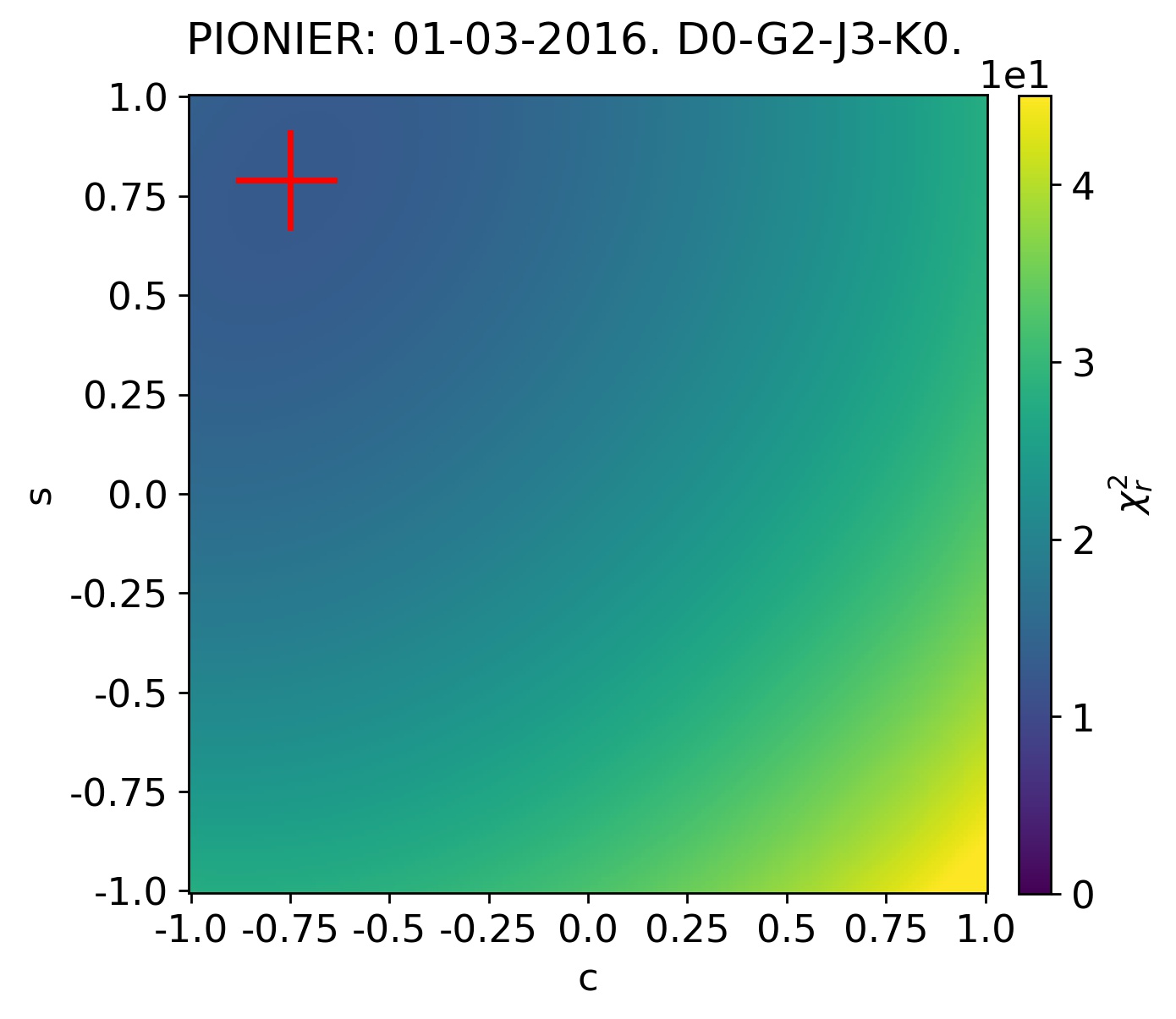}
    \includegraphics[scale=0.25]{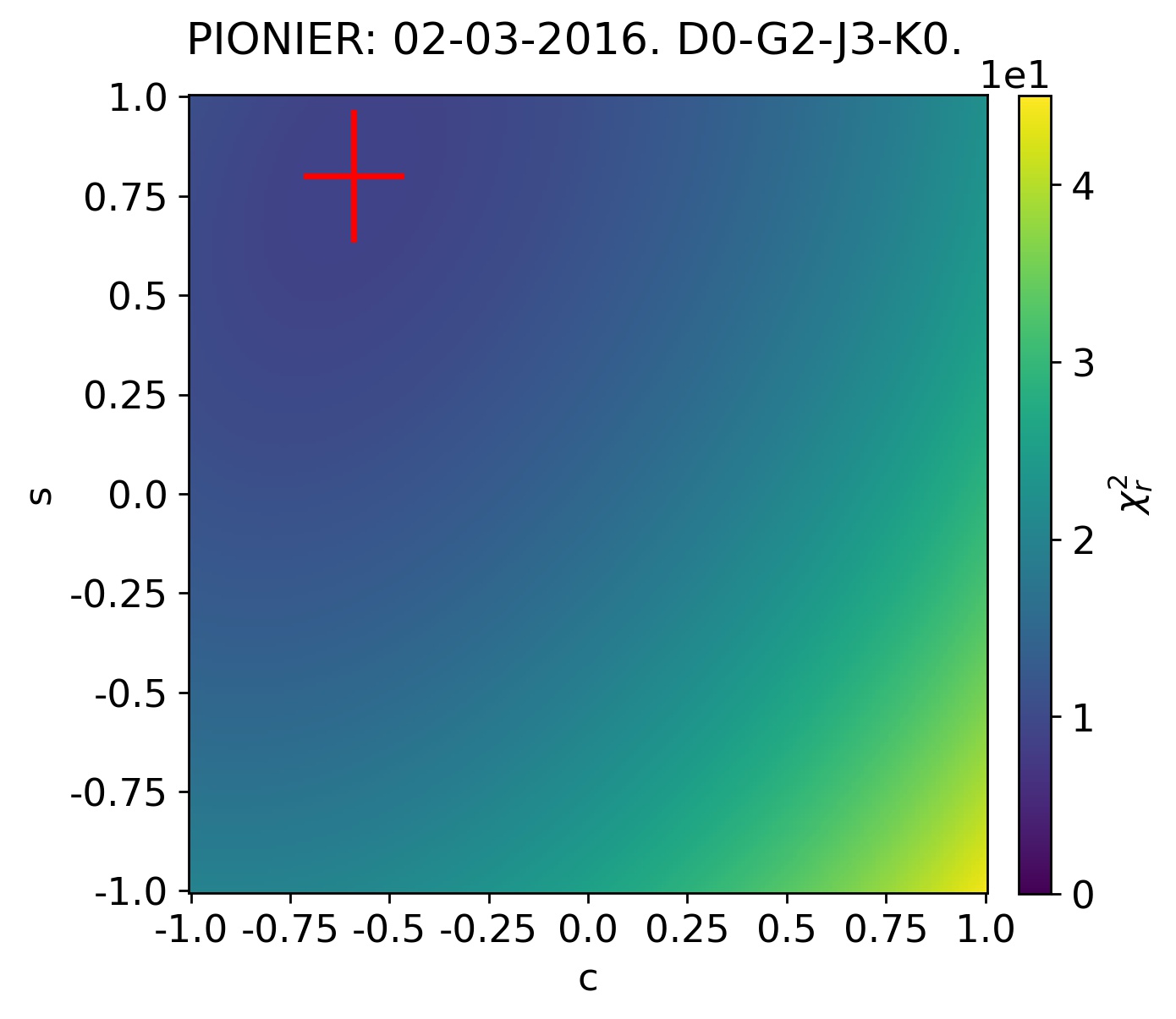}\\
    \includegraphics[scale=0.25]{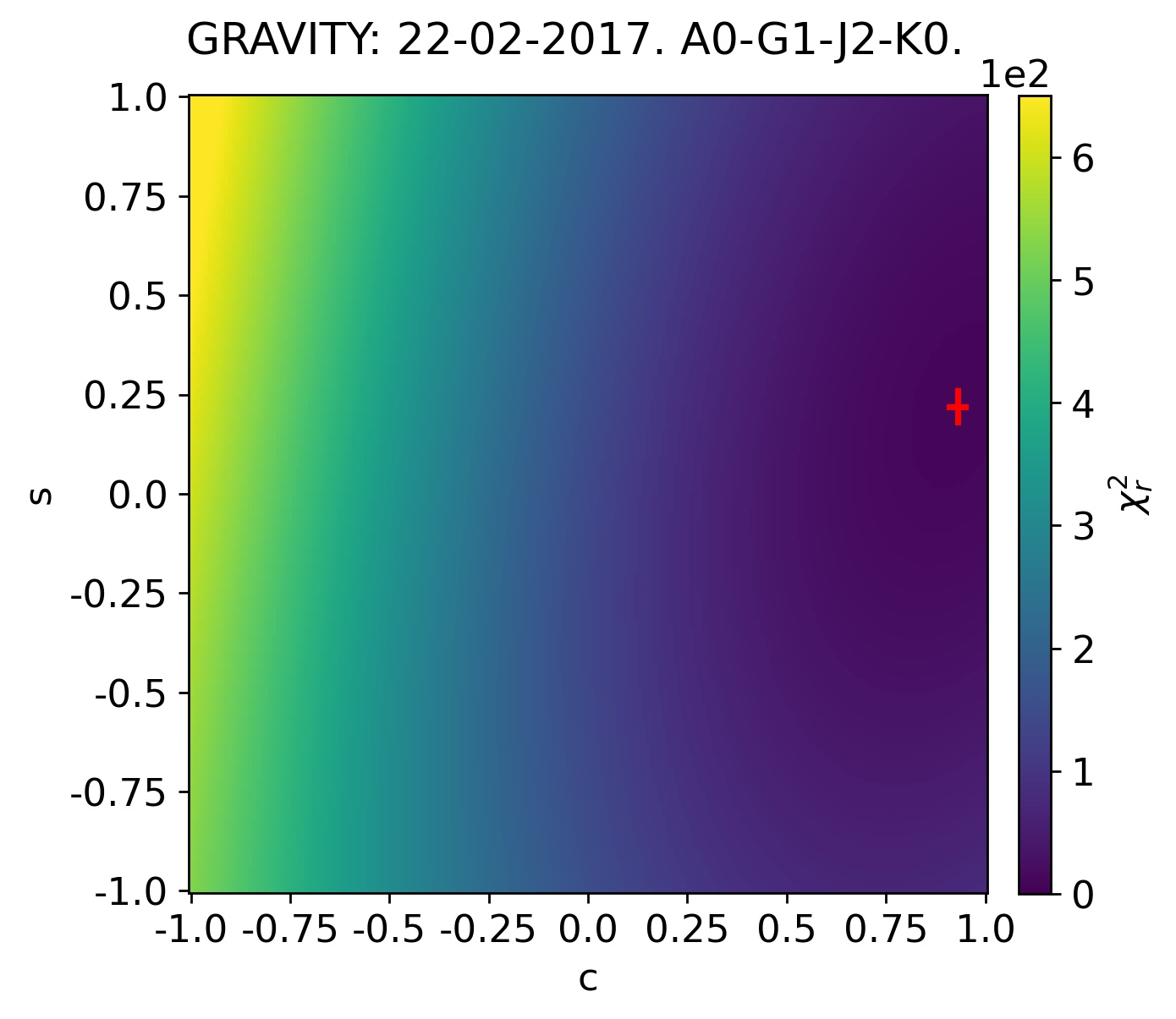}
    \includegraphics[scale=0.25]{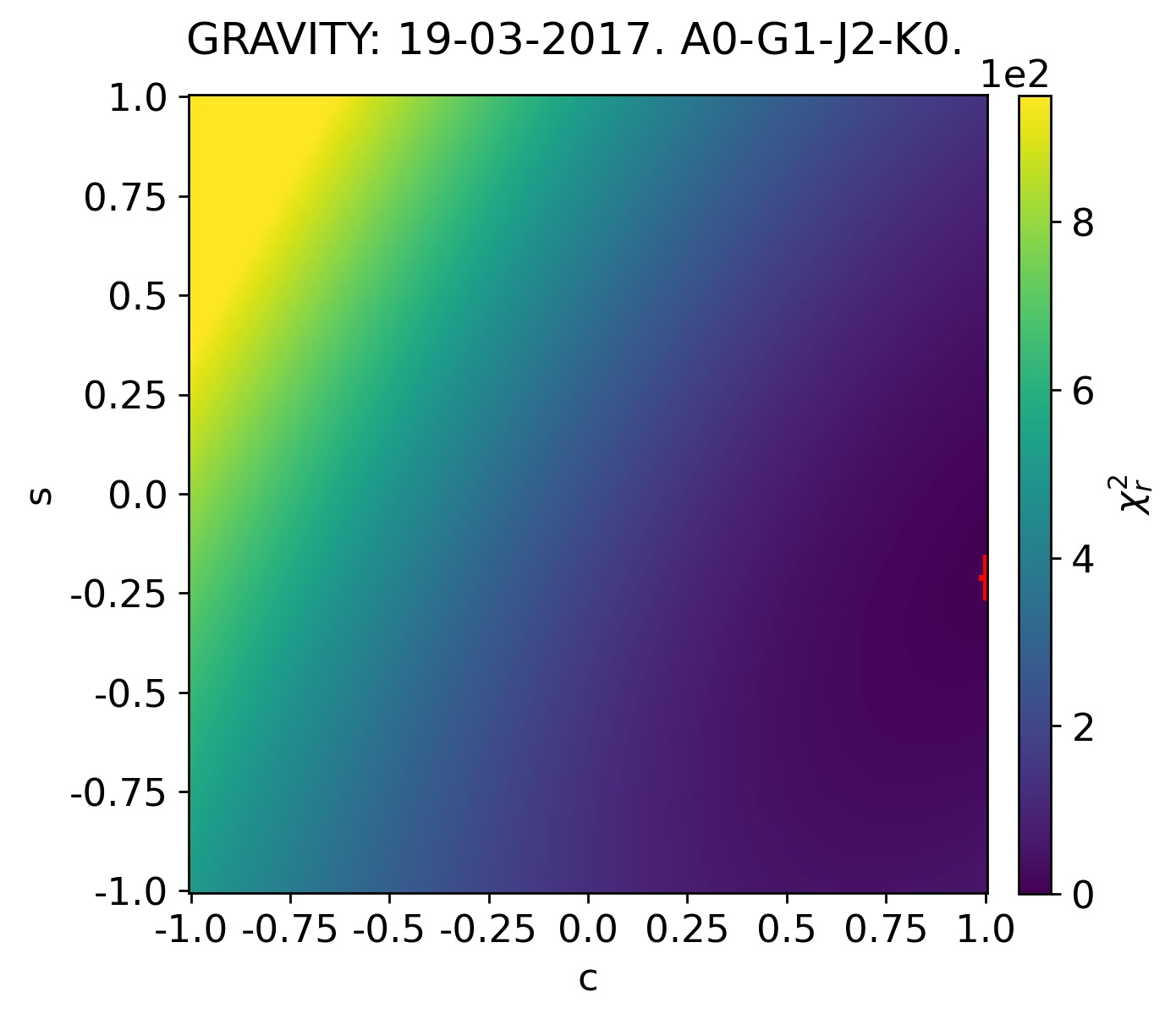}
    \includegraphics[scale=0.25]{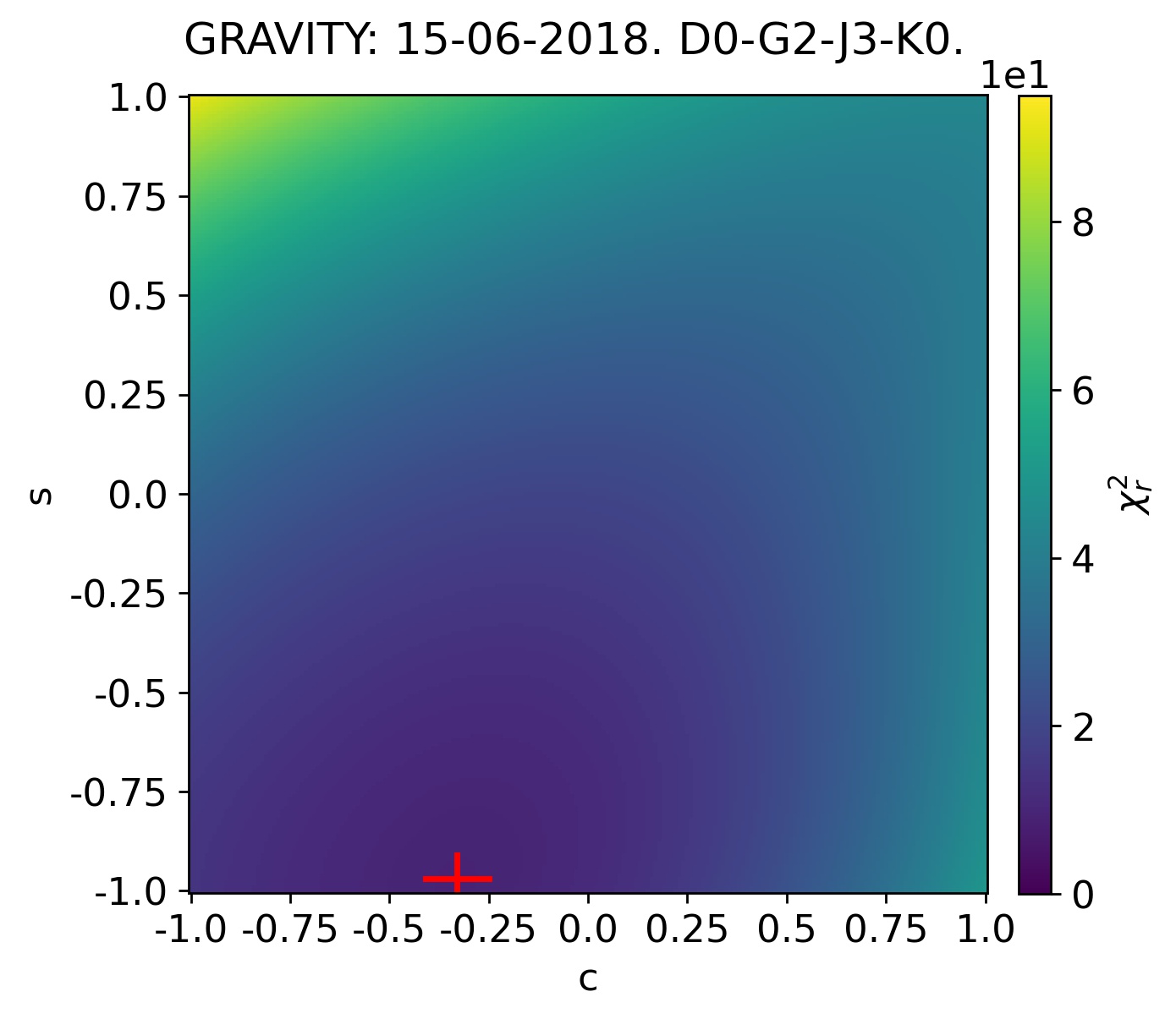}
    \includegraphics[scale=0.25]{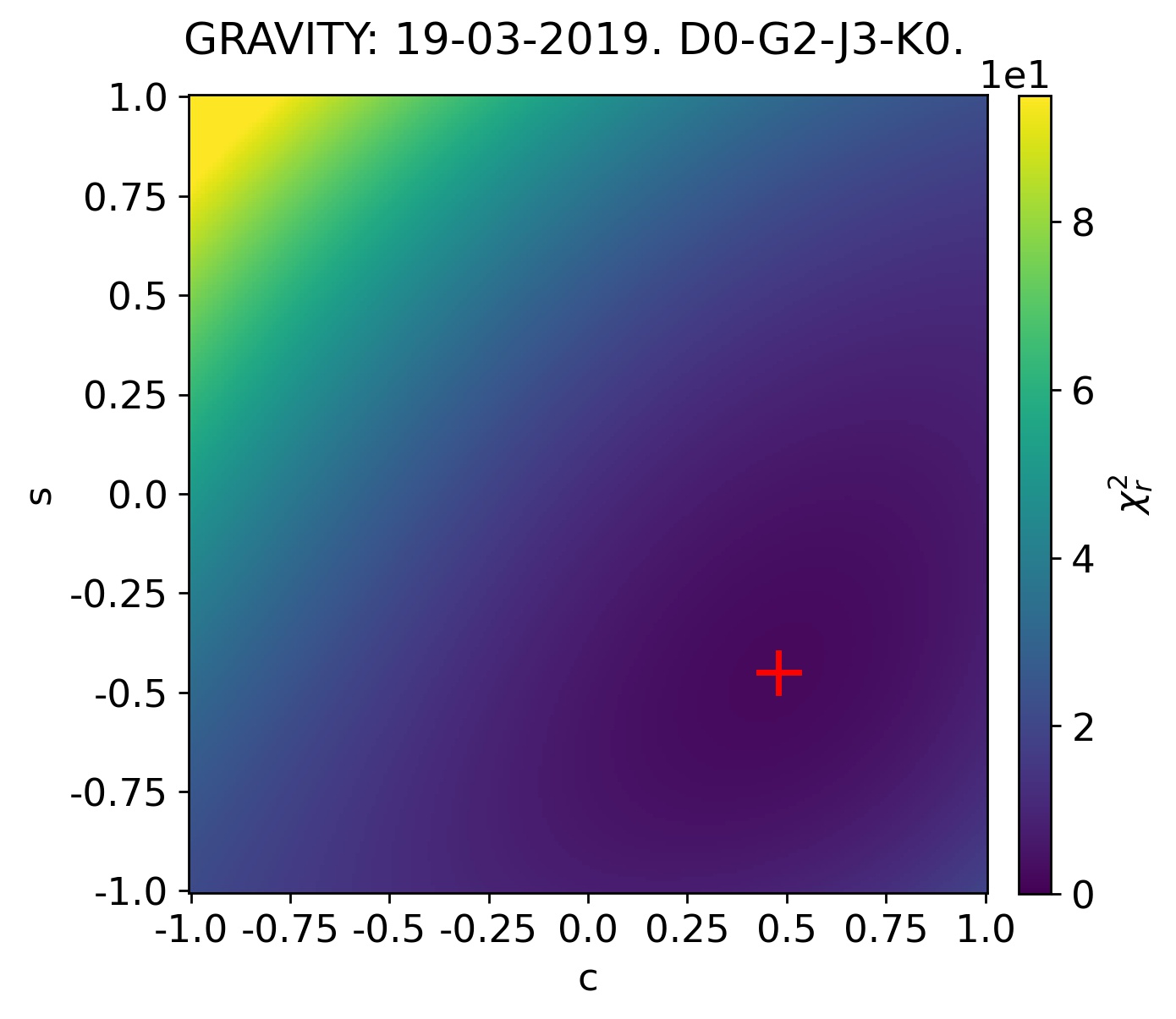}
    \includegraphics[scale=0.25]{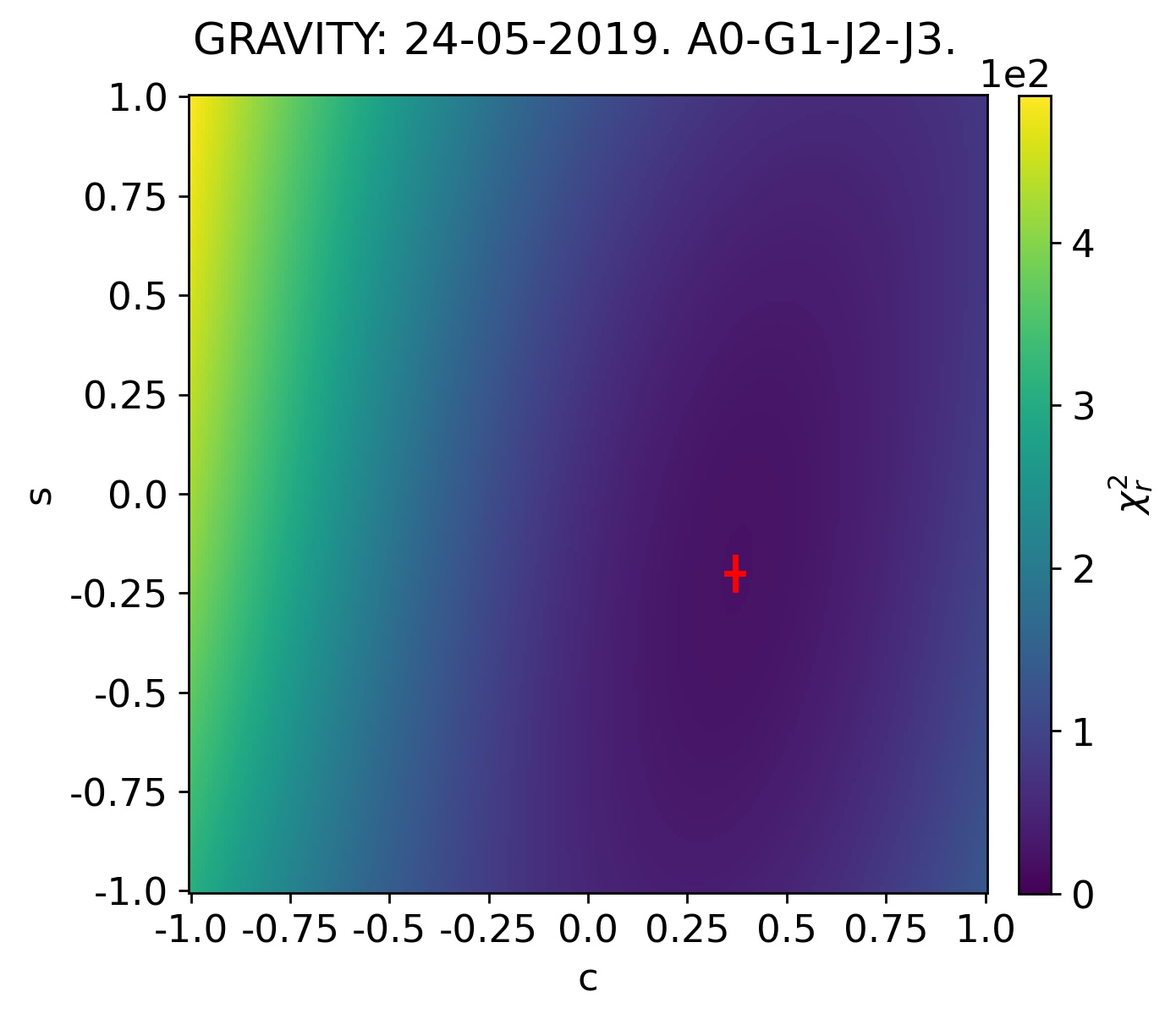}\\
    \includegraphics[scale=0.25]{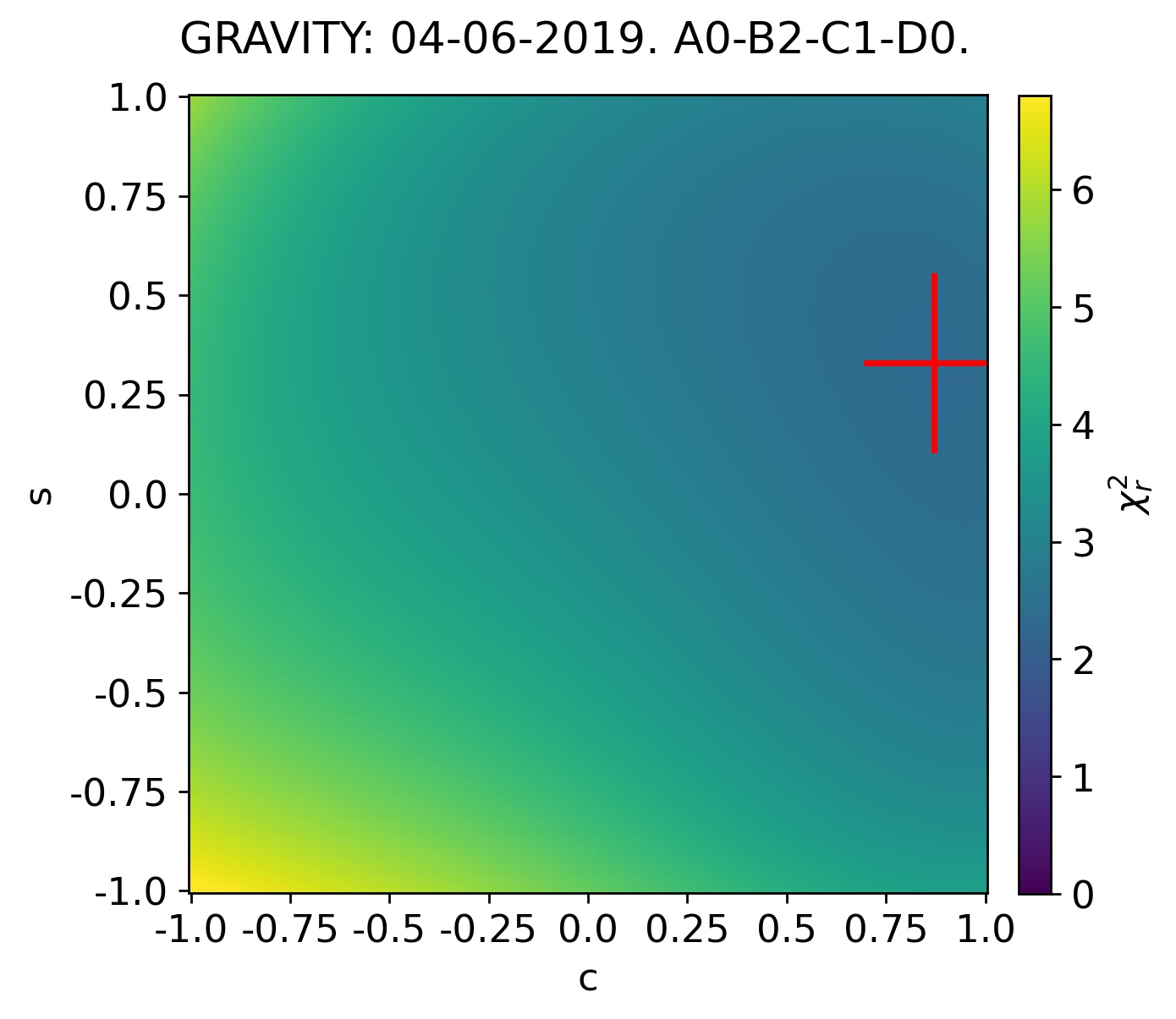}
    \includegraphics[scale=0.25]{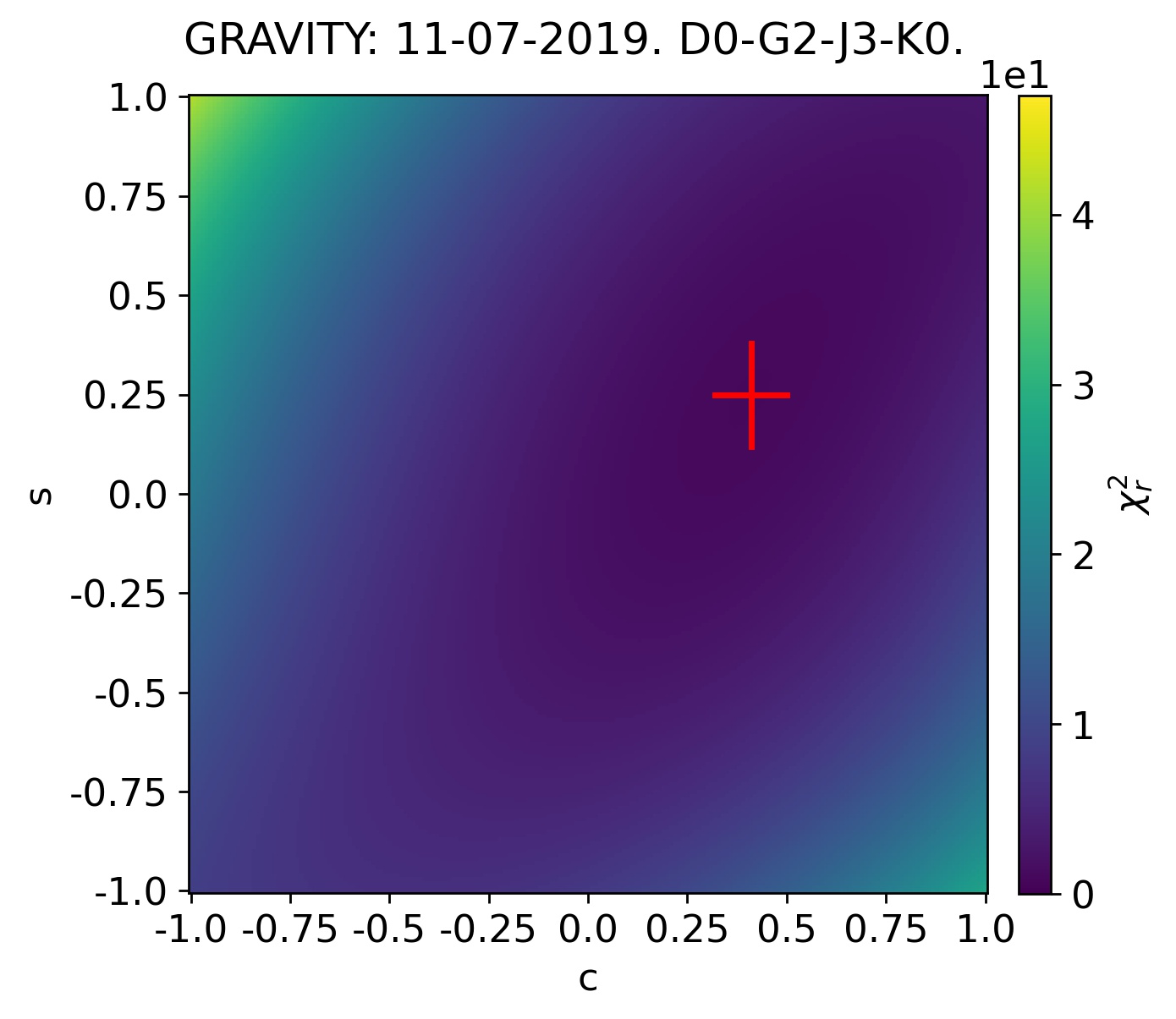}
    \includegraphics[scale=0.25]{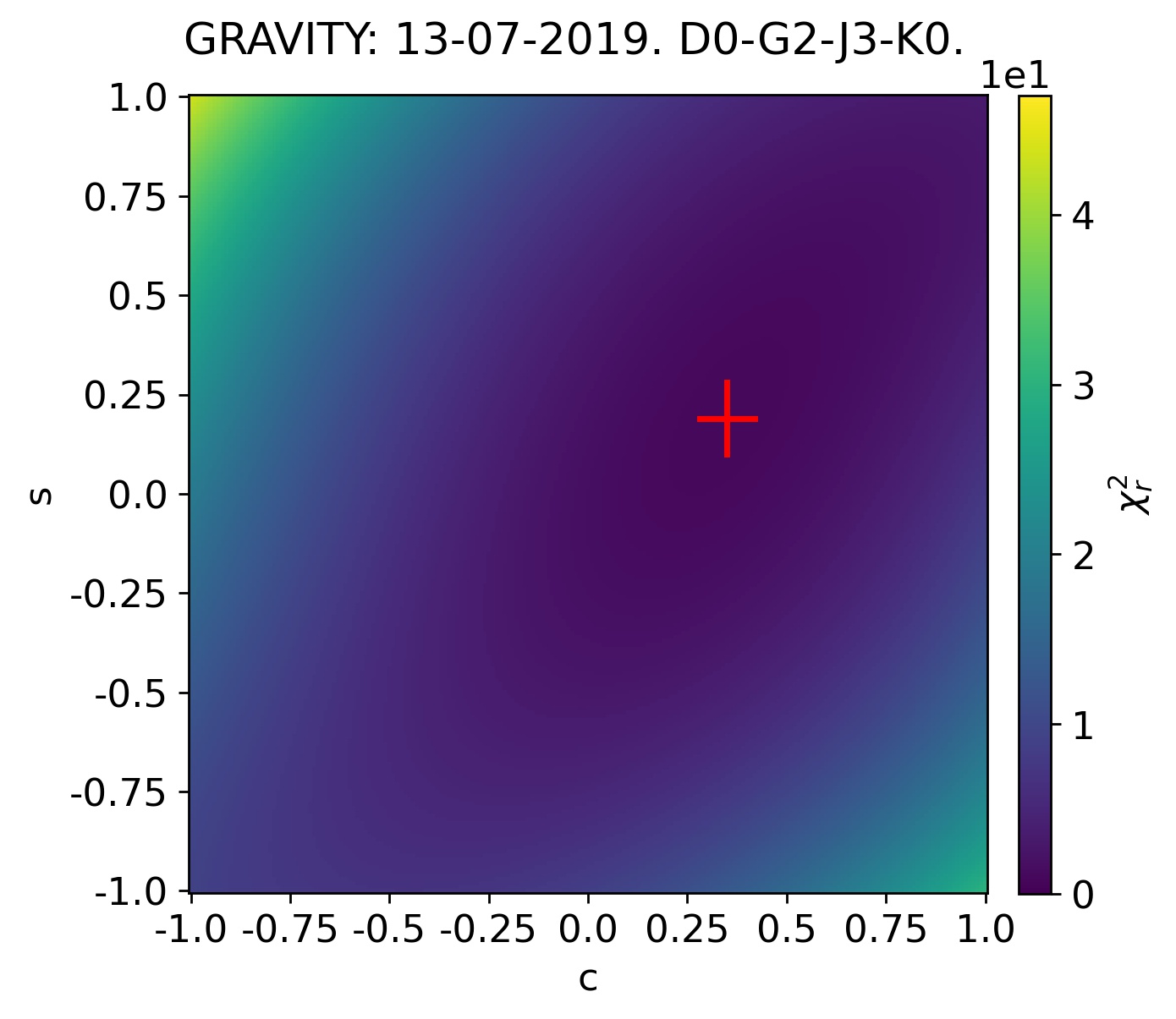}
    \includegraphics[scale=0.25]{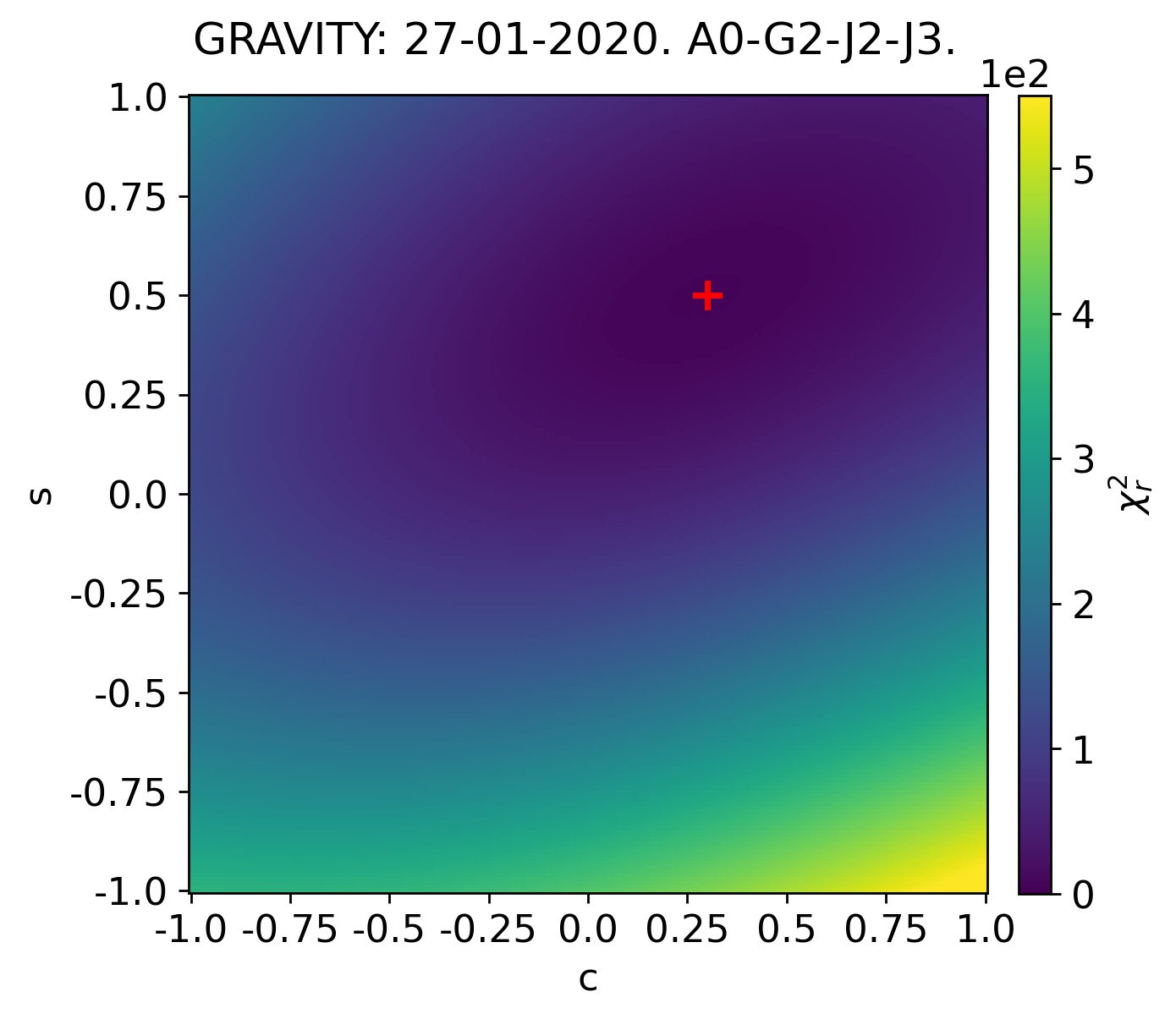}
    \includegraphics[scale=0.25]{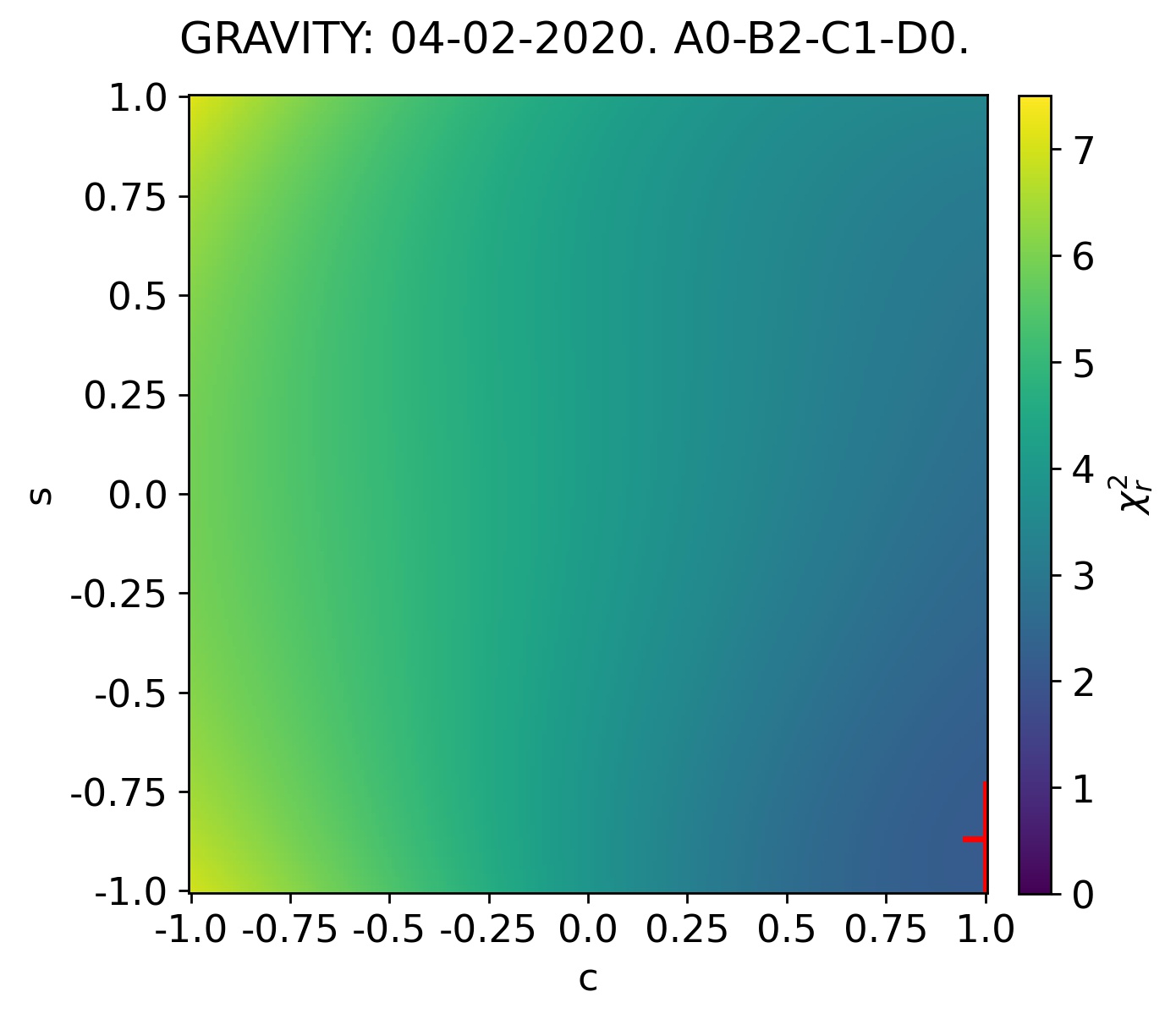}\\
    \includegraphics[scale=0.25]{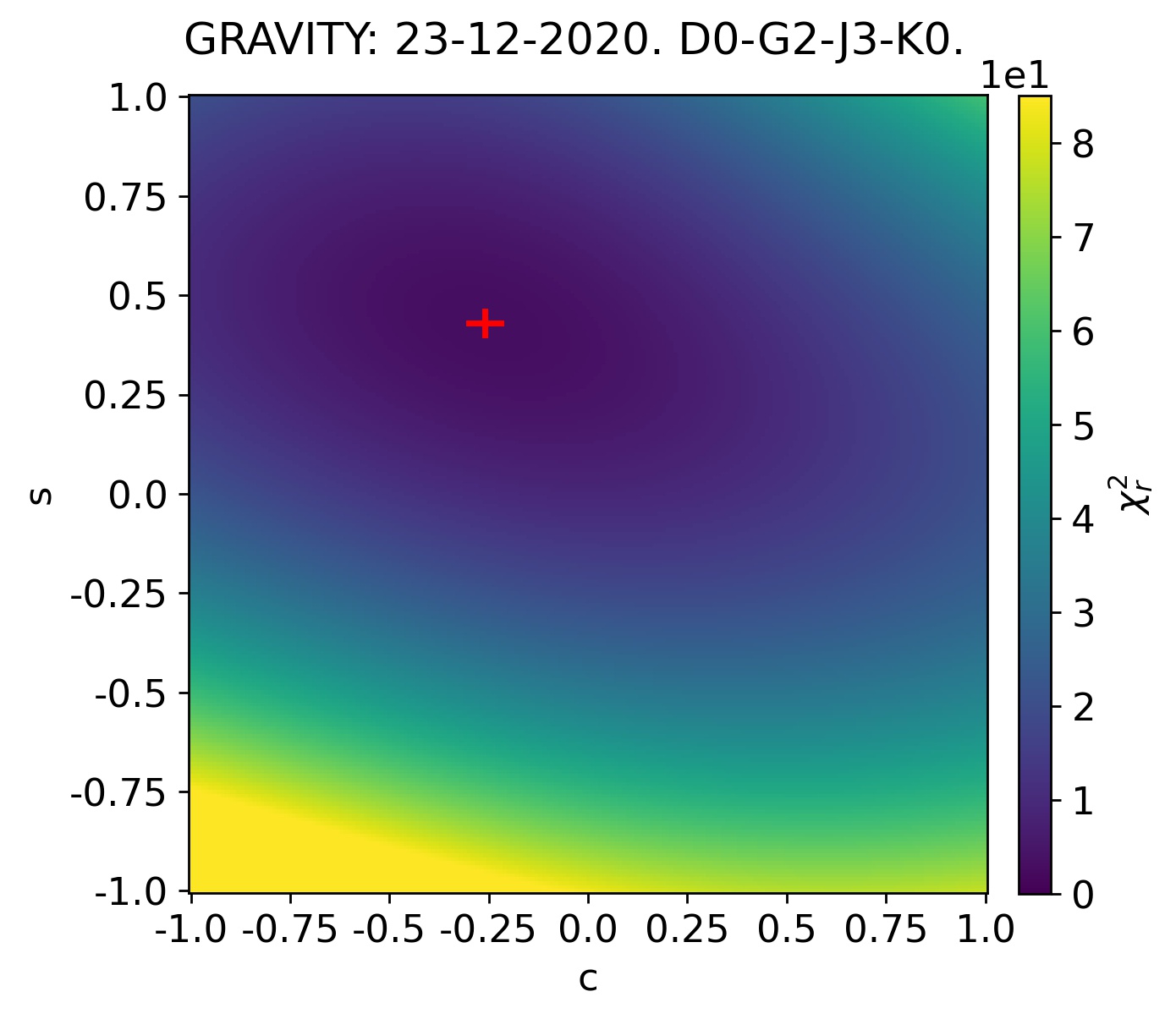}
    \includegraphics[scale=0.25]{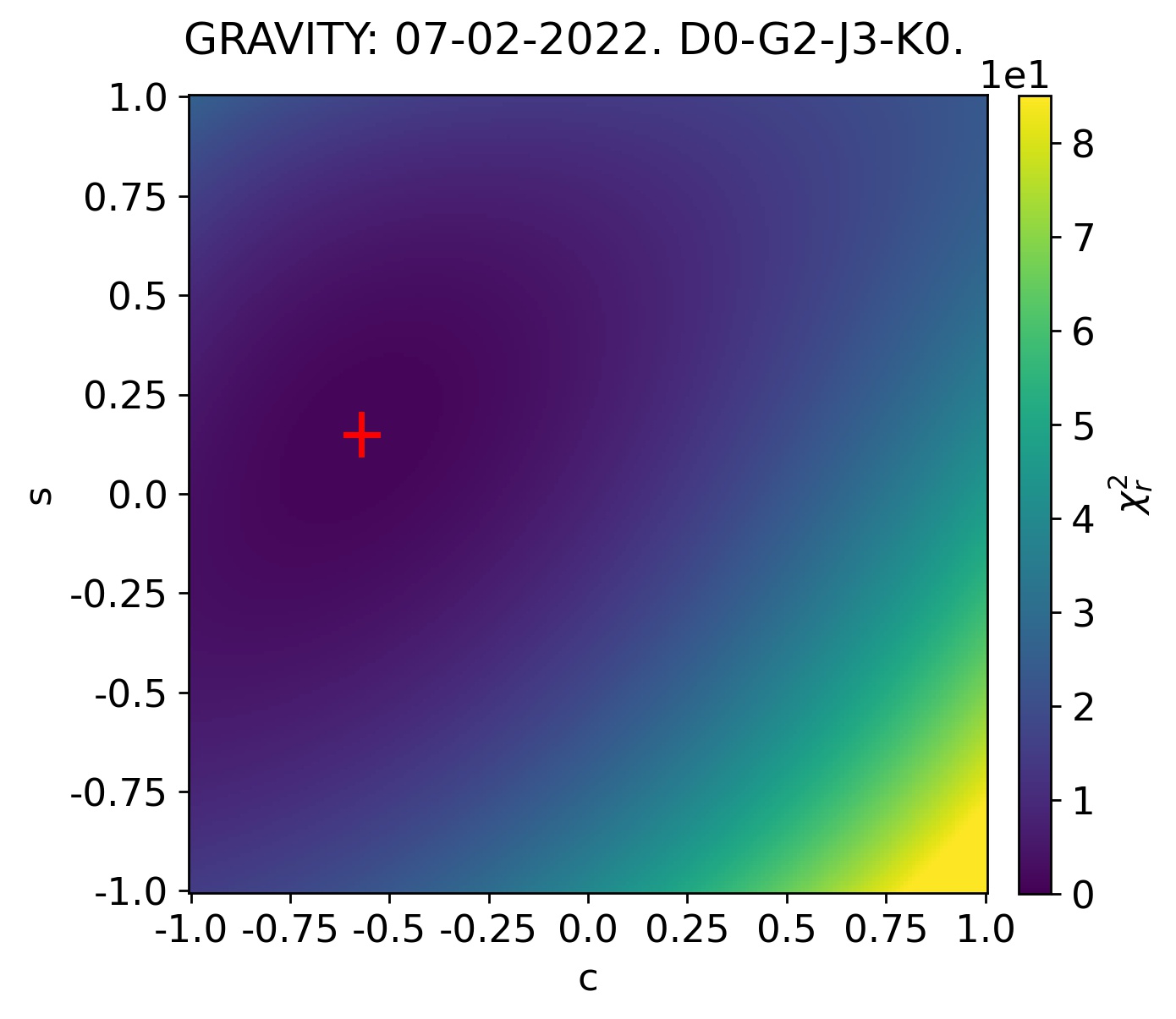}
    \includegraphics[scale=0.25]{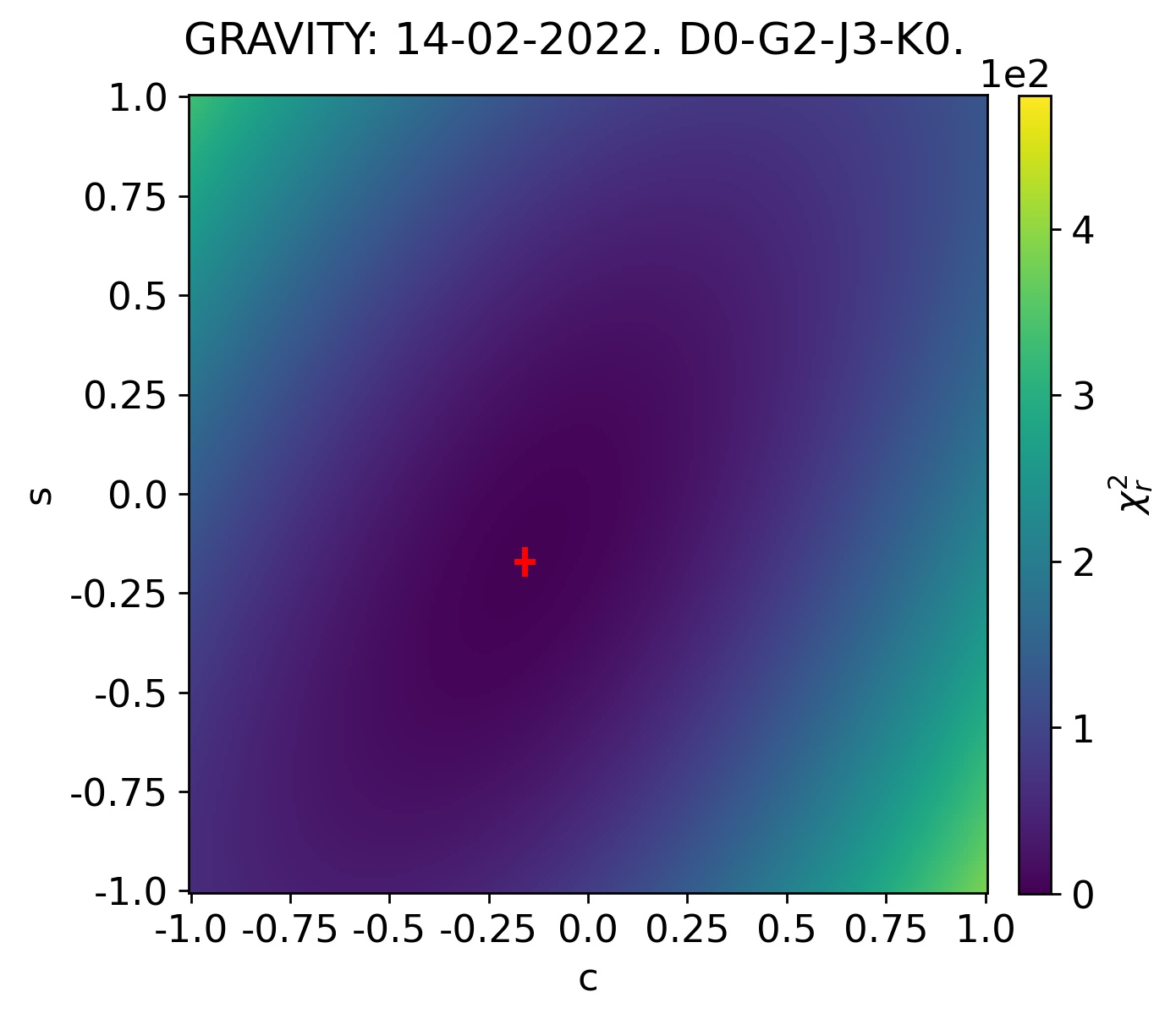}\\
    \caption{Global fit azimuthal modulation parameters $\chi^2_r$ maps. The red lines represent the 3\,$\sigma$ error bars of the parameters derived through the MCMC fitting procedure. Their intersections give the smallest $\chi^2_r$ for a given epoch.}
    \label{fig:AzMod-chi2maps}
\end{figure*}

\clearpage

\section{Spectrum wavelength calibration and star photospheric absorption model}
\label{apx:spctr-corr}

\begin{figure}[!ht]
    \centering
    \includegraphics[width=\columnwidth]{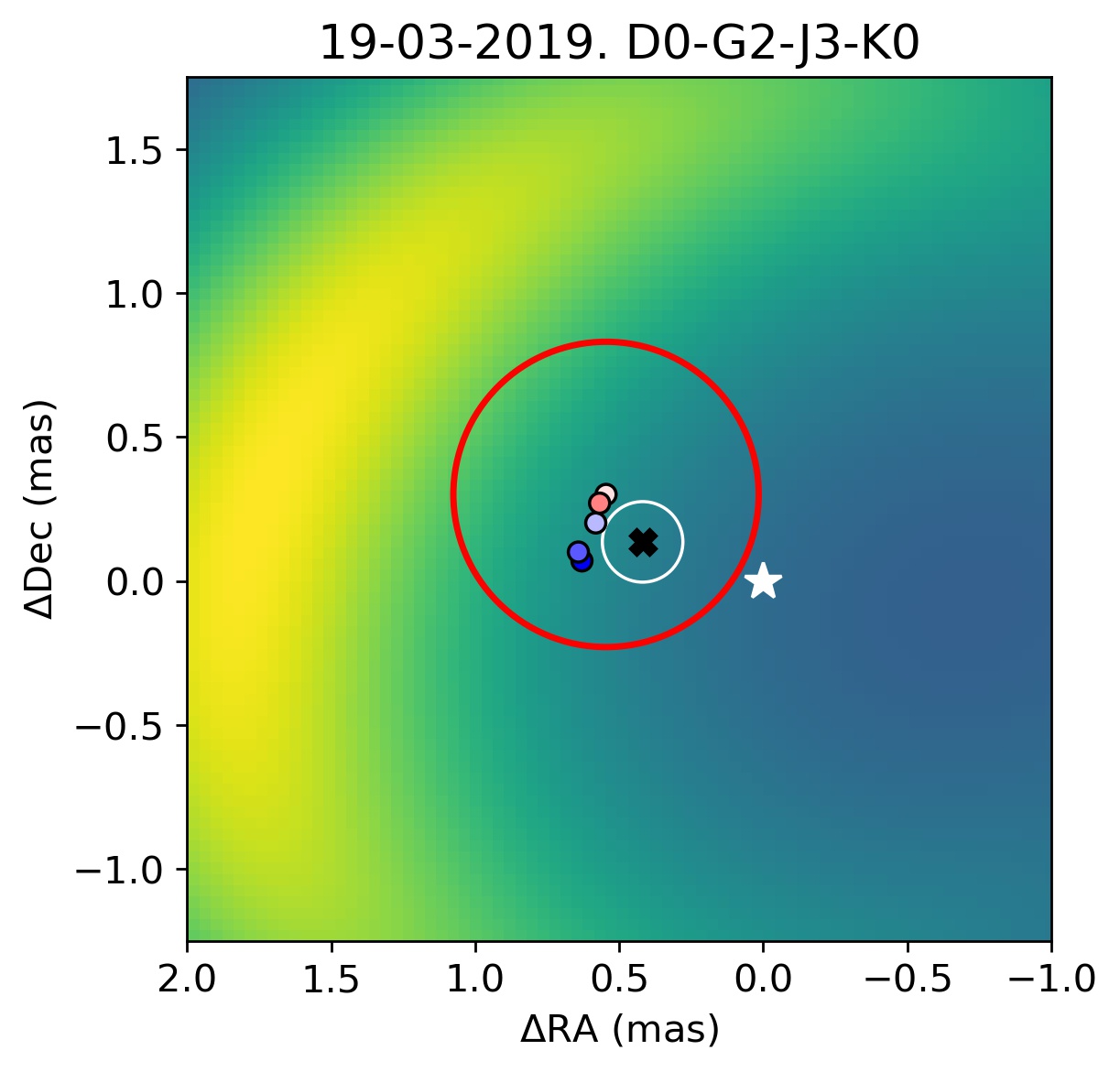}\\
    \includegraphics[width=\columnwidth]{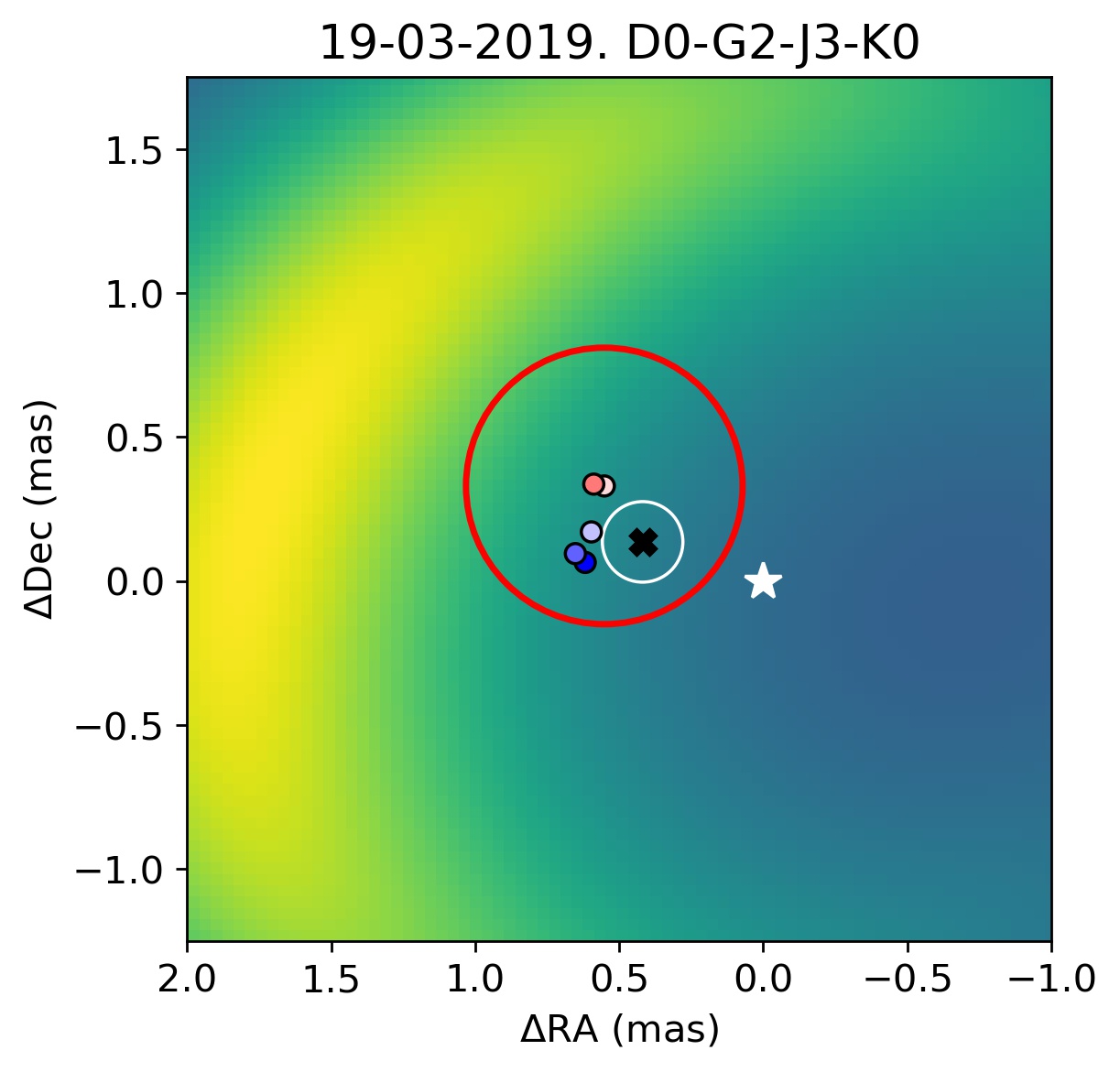}
    \caption{Difference between pure-line photocenters shift obtained with data corrected with the GRAVITY data reduction software (\citealt{Lapeyrere2014}, top panel), and with the \cite{Alex2023} high-precision spectral calibration (bottom panel). Same markers as in Fig.~\ref{fig:BrG_PhCenters_small-sample}.}
    \label{fig:SPChCorr}
\end{figure}

The continuum normalization of the GRAVITY science object spectrum of each epoch (average of all the observations blocks files that come from the four different ATs for that epoch) was done by fitting the slope of the raw spectrum and dividing the latter by the resulting fit.
The first step of the wavelength calibration was done by comparing the observed wavelength positions of the telluric lines in the \object{HD\,98922} spectrum for each epoch with respect to the positions of the telluric lines present in the IR spectrum of the atmospheric transmission above Cerro Pachon. We used
pre-computed ATRAN models \citep{Lord1992} available from the Gemini Observatory website accounting for $4.3\,$mm water vapor column and an airmass of 1.5.
The atmospheric transmission spectrum was convolved by a Gaussian with FHWM of $6\,$\AA{} to have the same resolution as GRAVITY. The correction results in a blueshift between 0 and $5\,$\AA{} depending on the epoch.
The \object{HD\,98922} spectrum was then corrected for the star radial velocity $(-4.9\,$km/s$)$ and its proper motion with respect to the local standard of rest, with a value that depends on the epoch of observations (between $\approx -35$ and $\approx 2\,$km/s).
The same corrections (telluric lines calibration, radial velocity correction, and local standard of rest correction) were also applied on the SC visibilities and differential phases for each epoch.
The stellar atmospheric Br$\gamma$ absorption was taken into account through a model selected from the Vienna New Model Grid of Stellar Atmospheres \citep{Heiter2002}\footnote{available on the NeMo webpage (Ch. St\"utz and E. Paunzen, \url{ http://www.univie.ac.at/nemo/})} that best represents the star (see Table~\ref{tab:StellarParameters}). The model accounts for a star effective temperature of $10500$ K, a surface gravity logarithm of 3.4, a metallicity of $[$Fe/H$] = -0.5$, and a microturbulence of 2.0 km/s, a common value for a Herbig star. Finally, we included in the model the rotation broadening effect due to the rotation of the star ($v\,sin\,i = 39.0\,$km/s) and we convolved the final model by a Gaussian with FHWM of $6\,$\AA{} to have the same resolution as GRAVITY using SPECTRUM \citep{Gray1994}.\\
\\
A systematic spectral shift of the line features was recently found in the GRAVITY SC data. The shift is seen when comparing data from an individual telescope with respect to the other telescopes, but also globally with regard to the known wavelength positions of the telluric lines in the K-band. It is thought that the shift is caused by the old grism of the science spectrometer which was upgraded in October 2019. GRAVITY data taken after this month should therefore no longer be affected. To check if our results are affected by the shift, we corrected the shift in the 19 March 2019 data following \citealt{Alex2023}, and compared the results with the ones obtained without the correction. For a description of the correction we refer to Sect.~3.2 of \citealt{Alex2023}. Figure~\ref{fig:SPChCorr} shows the pure-line photocenter shift obtained with the original data (top panel) and that obtained with the corrected ones (bottom panel). We note a slight change in the shift when comparing photocenters of the same spectral channel. However, in this work we are not primarily interested in the relative position of the photocenters for each spectral channel, but rather in the overall location of the gas with respect to the star and the dusty feature. This choice is also based on the fact that our data were obtained with the ATs, meaning that we have only 4 or 5 spectral channels across the line. Data obtained with the UTs have instead around 12 spectral channels across the line, making these data more suitable for high-precision spectral analysis. From this point of view, the differences between the results obtained with the corrected and the uncorrected ATs data is negligible. Therefore, we decided to not apply this high-precision spectral calibration to our data.

\clearpage

\section{Photometric data}

\begin{table*}[!ht]
    \centering
    \caption{\centering \object{HD\,98922} photometry data}
    \begin{tabular}{llll}
    \hline
    \hline
    $\lambda$ & F$_\lambda$ & $\Delta\,$F$_\lambda$ & Reference \\
    ($\mu$m) & (erg\,cm$^{-2}$\,s$^{-1}$\,\AA{}) & (erg\,cm$^{-2}$\,s$^{-1}$\,\AA{}) & \\
    \hline
    0.35 & 6.73$\times 10^{-12}$ & 6.73$\times 10^{-13}$ &\cite{Malfait1998} \\
    0.35 & 8.30$\times 10^{-12}$ & 1.44$\times 10^{-13}$ & \cite{Meyers2015} \\
    0.42 & 1.27$\times 10^{-11}$ & 6.79$\times 10^{-14}$ & \cite{ESA1997} \\
    0.50 & 8.05$\times 10^{-12}$ & 2.35$\times 10^{-14}$ & \cite{Gaia2018} \\
    0.43 & 1.33$\times 10^{-11}$ & 1.33$\times 10^{-12}$ &\cite{Malfait1998} \\
    0.44 & 1.25$\times 10^{-11}$ & 1.25$\times 10^{-12}$ & \cite{Hog2000} \\
    0.53 & 7.82$\times 10^{-12}$ & 7.42$\times 10^{-14}$ & \cite{ESA1997} \\
    0.55 & 7.17$\times 10^{-12}$ & 3.30$\times 10^{-13}$ &\cite{Malfait1998} \\
    0.55 & 7.39$\times 10^{-12}$ & 7.39$\times 10^{-13}$ & \cite{Hog2000} \\
    0.55 & 7.54$\times 10^{-12}$ & 7.54$\times 10^{-13}$ & \cite{Hauck1998} \\
    0.58 & 5.80$\times 10^{-12}$ & 1.77$\times 10^{-14}$ & \cite{Gaia2020} \\
    0.67 & 4.56$\times 10^{-12}$ & 1.99$\times 10^{-14}$ & \cite{Gaia2018} \\
    0.76 & 3.04$\times 10^{-12}$ & 1.55$\times 10^{-14}$ & \cite{Gaia2020} \\
    0.77 & 3.01$\times 10^{-12}$ & 1.51$\times 10^{-14}$ & \cite{Gaia2018} \\
    1.22 & 1.15$\times 10^{-12}$ & 1.15$\times 10^{-13}$ &\cite{Malfait1998} \\
    1.24 & 1.24$\times 10^{-12}$ & 2.29$\times 10^{-14}$ & \cite{Cutri2003} \\
    1.63 & 9.53$\times 10^{-13}$ & 2.60$\times 10^{-14}$ & \cite{Cutri2003} \\
    1.65 & 8.26$\times 10^{-13}$ & 8.26$\times 10^{-14}$ &\cite{Malfait1998} \\
    1.66 & 9.03$\times 10^{-13}$ & 2.41$\times 10^{-14}$ & \cite{Cutri2003} \\
    2.16 & 8.32$\times 10^{-13}$ & 2.76$\times 10^{-14}$ & \cite{Cutri2003} \\
    2.18 & 7.12$\times 10^{-13}$ & 7.12$\times 10^{-14}$ &\cite{Malfait1998} \\
    2.19 & 7.94$\times 10^{-13}$ & 2.50$\times 10^{-14}$ & \cite{Cutri2003} \\
    3.35 & 4.41$\times 10^{-13}$ & 6.41$\times 10^{-14}$ & \cite{Cutri2012} \\
    3.55 & 4.99$\times 10^{-13}$ & 4.99$\times 10^{-14}$ &\cite{Malfait1998} \\
    4.60 & 4.35$\times 10^{-13}$ & 5.67$\times 10^{-15}$ & \cite{Cutri2012} \\
    4.77 & 2.68$\times 10^{-13}$ & 2.68$\times 10^{-14}$ &\cite{Malfait1998} \\
    8.61 & 1.08$\times 10^{-13}$ & 4.04$\times 10^{-16}$ & \cite{Ishihara2010} \\
    12.0 & 6.34$\times 10^{-14}$ & 6.34$\times 10^{-15}$ & \cite{Beichman1988} \\
    18.4 & 1.91$\times 10^{-14}$ & 8.86$\times 10^{-17}$ & \cite{Ishihara2010} \\
    25.0 & 9.25$\times 10^{-15}$ & 9.25$\times 10^{-16}$ & \cite{Beichman1988} \\
    60.0 & 3.56$\times 10^{-16}$ & 3.56$\times 10^{-17}$ & \cite{Beichman1988} \\
    61.9 & 4.85$\times 10^{-16}$ & 4.39$\times 10^{-17}$ & \cite{Helou1988}\\
    65.0 & 2.30$\times 10^{-16}$ & 1.70$\times 10^{-17}$ & \cite{Yamamura2010} \\
    70.0 & 2.19$\times 10^{-16}$ & 1.10$\times 10^{-17}$ & \cite{Hales2014} \\
    90.0 & 9.60$\times 10^{-17}$ & 5.29$\times 10^{-18}$ & \cite{Yamamura2010} \\
    140.0 & 1.73$\times 10^{-17}$ & 7.98$\times 10^{-18}$ & \cite{Yamamura2010} \\
    160.0 & 1.03$\times 10^{-17}$ & 5.14$\times 10^{-19}$ & \cite{Hales2014} \\
    1290.0 & 1.88$\times 10^{-21}$ & 1.88$\times 10^{-22}$ & ALMA Archive data\\
    \hline
    \end{tabular}
    \label{tab:PhotometricData}
\end{table*}

\clearpage

\section{Interferometric variability and UV coverage}\label{sec:appendix_variability}

\subsection{Variability in the visibilities and closure phases}

\begin{figure*}
    \centering
    \includegraphics[width=\columnwidth]{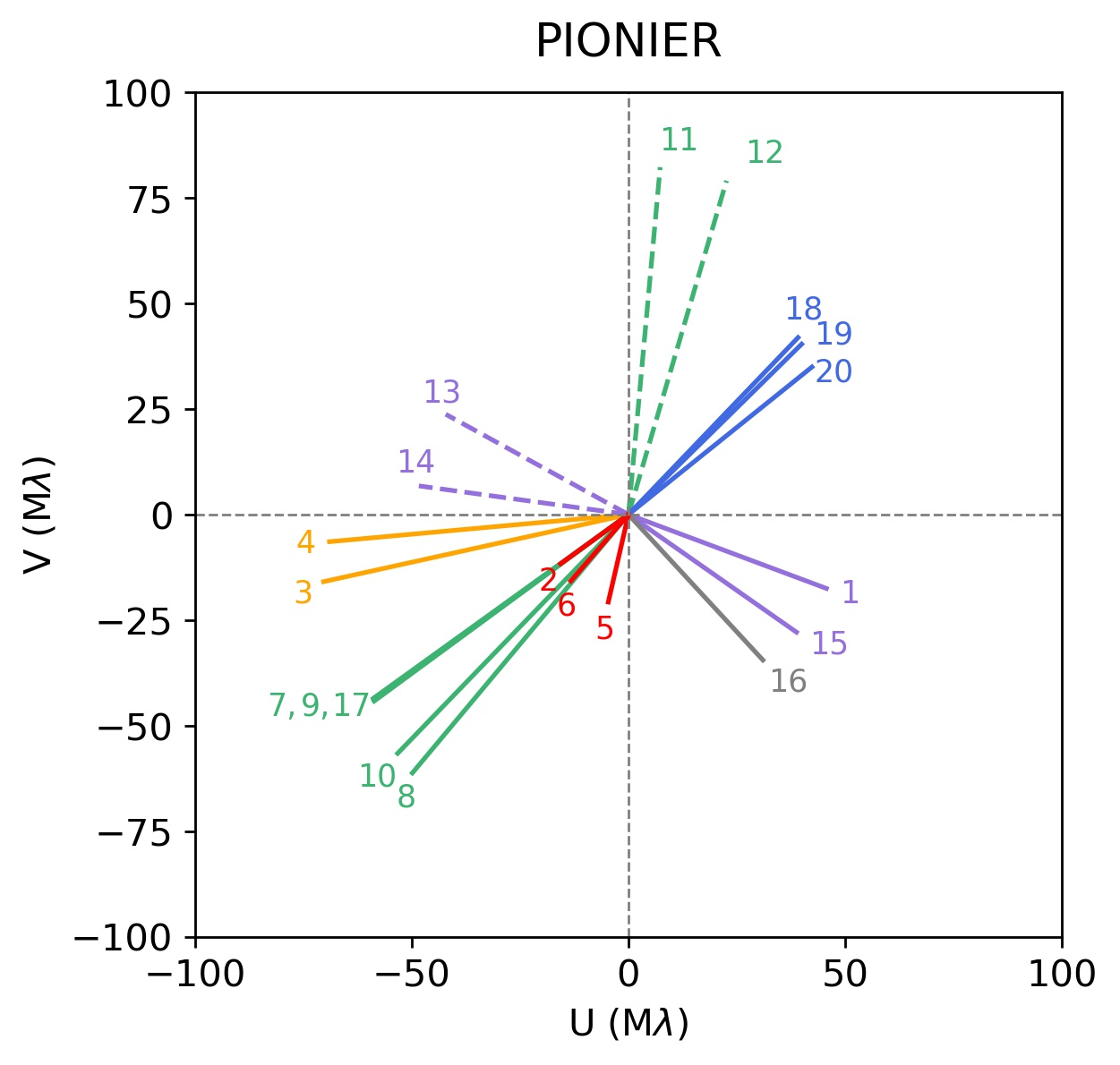}
    \includegraphics[width=\columnwidth]{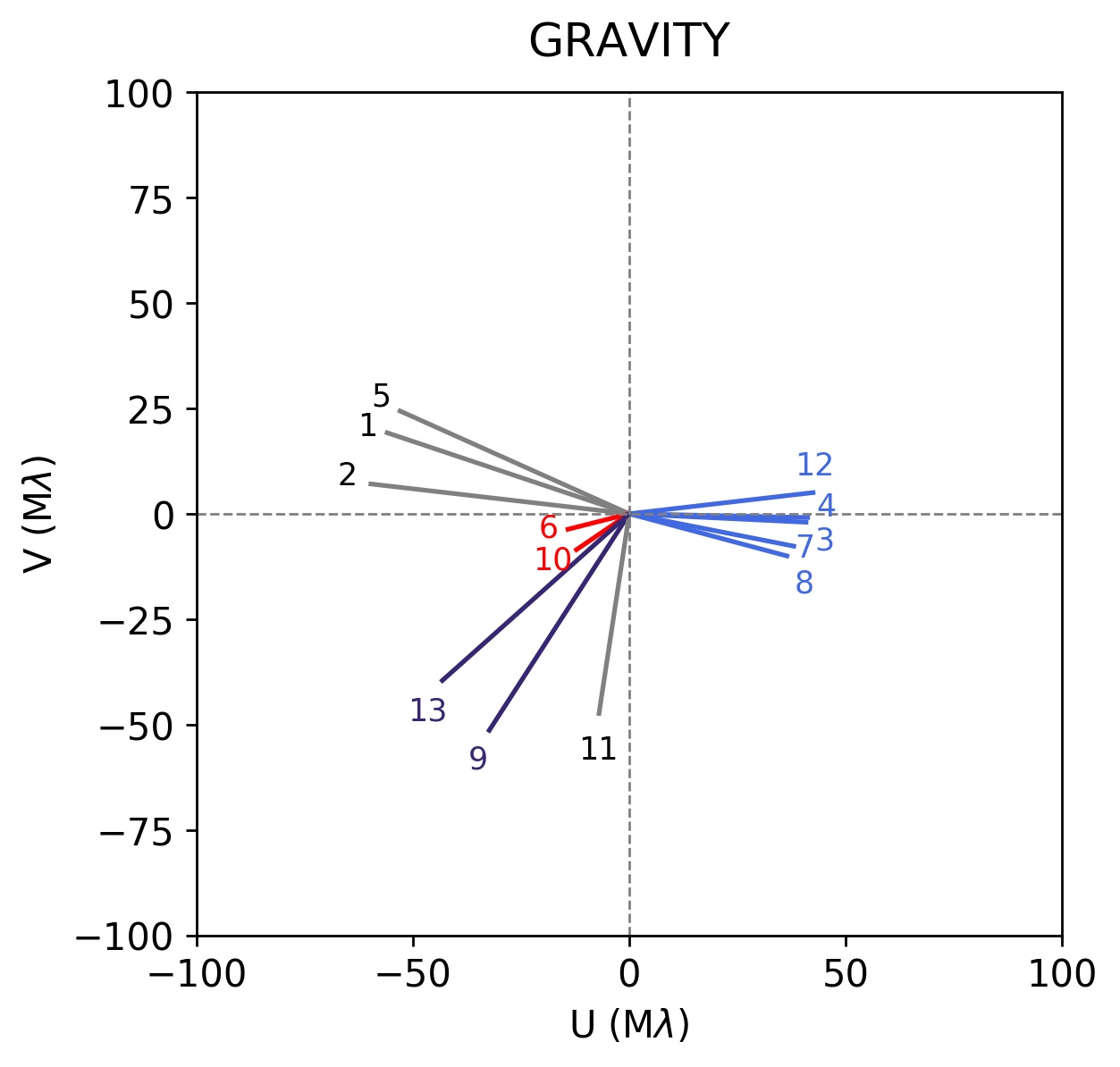}\\[0.5em]
    \includegraphics[width=\columnwidth]{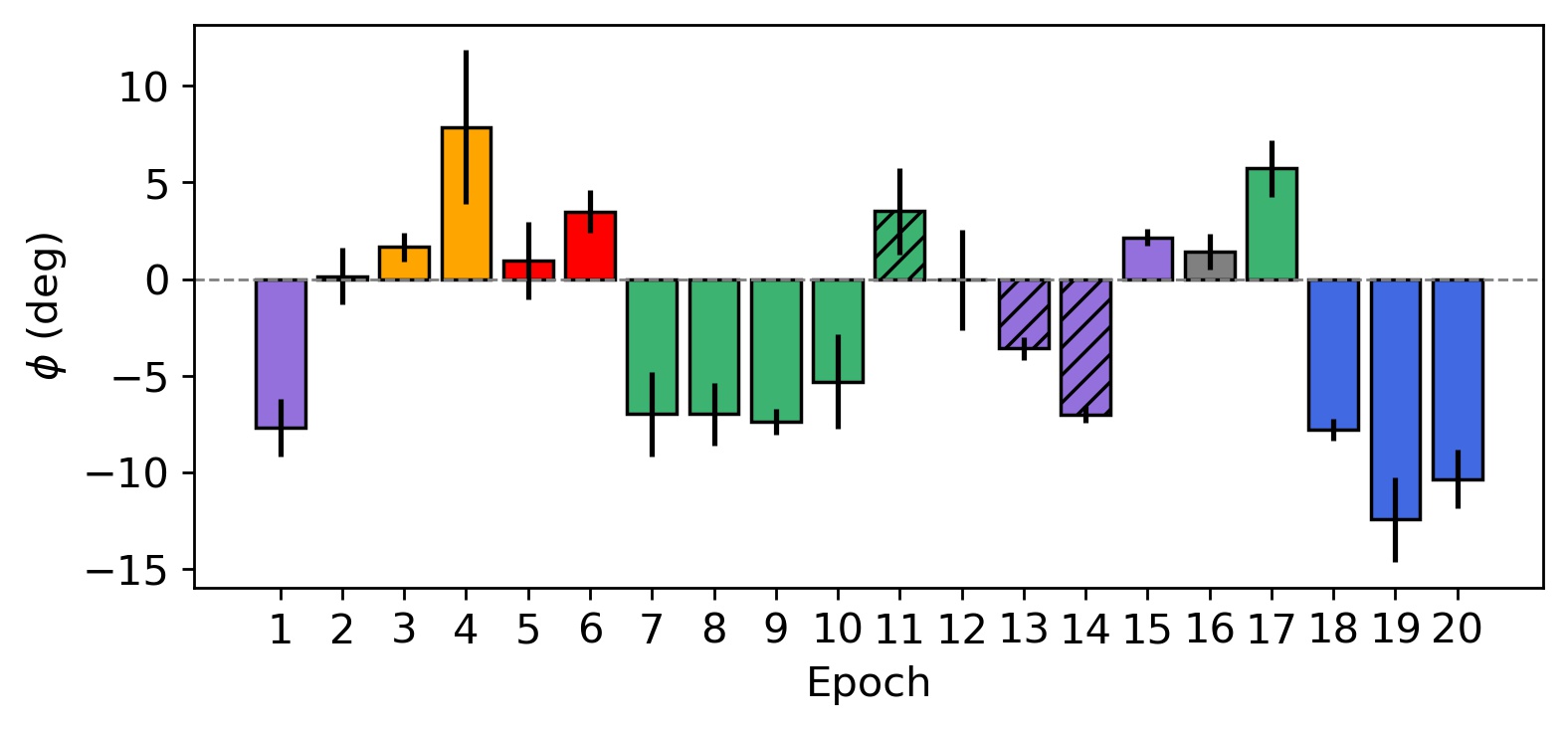}
    \includegraphics[width=\columnwidth]{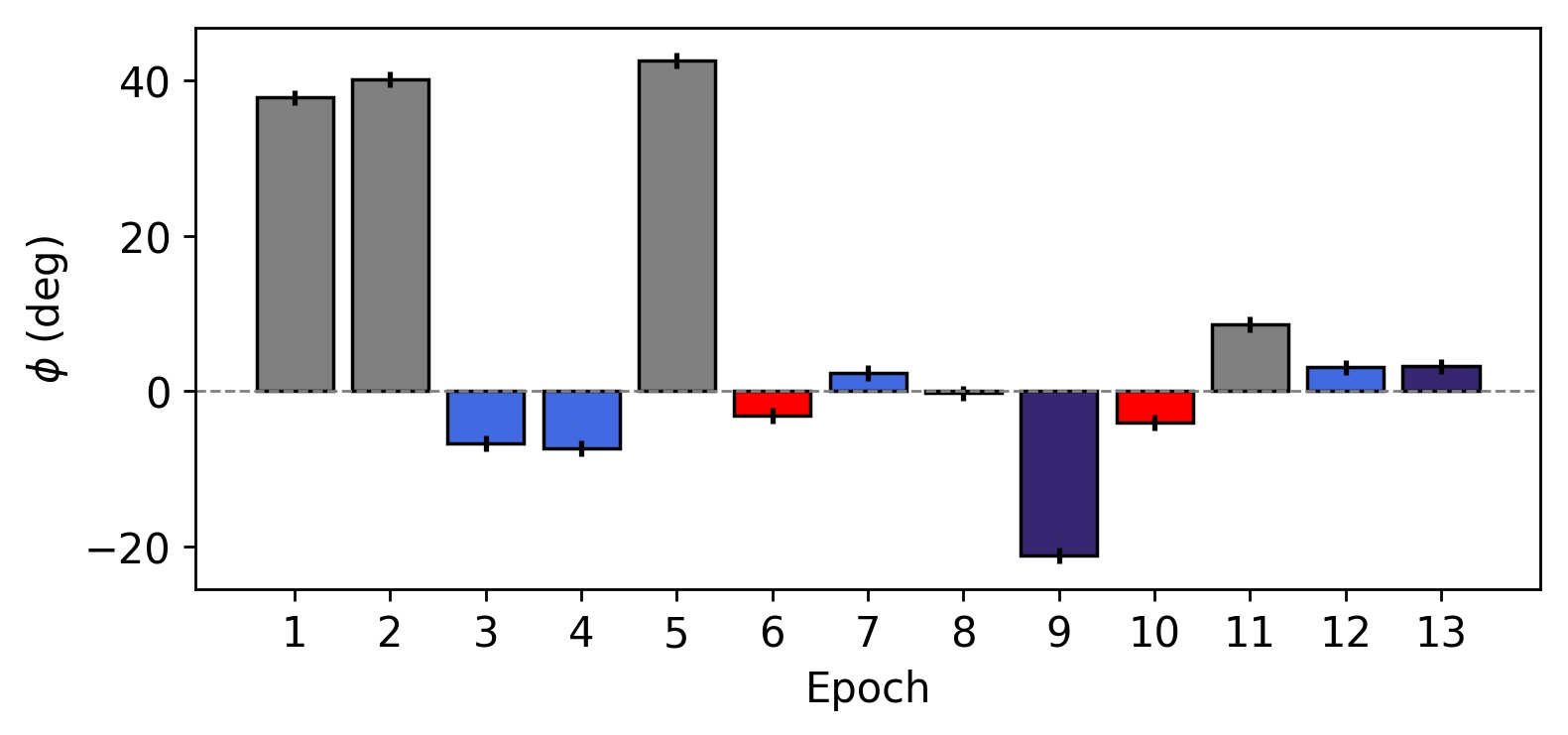}\\[0.5em]
    \includegraphics[width=\columnwidth]{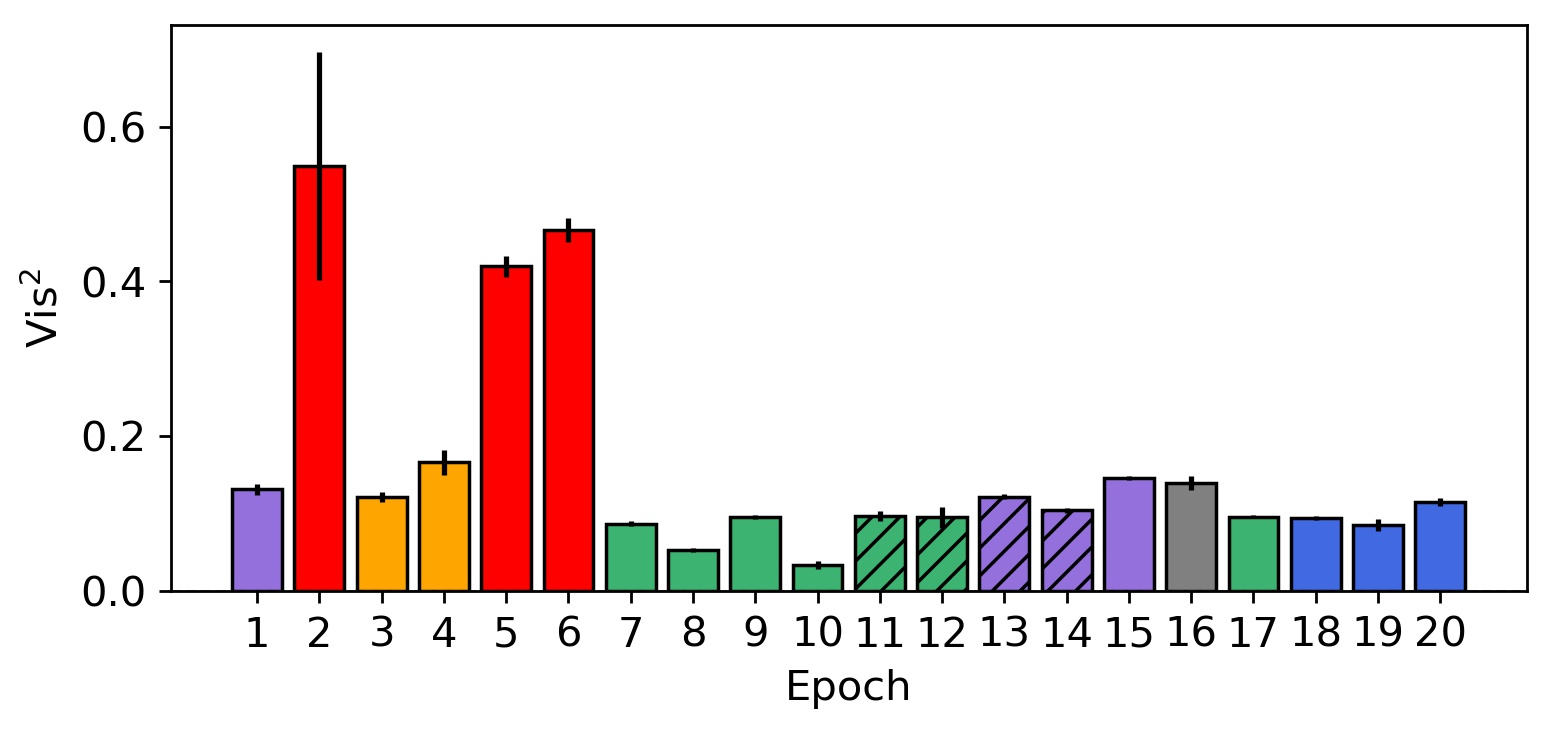}
    \includegraphics[width=\columnwidth]{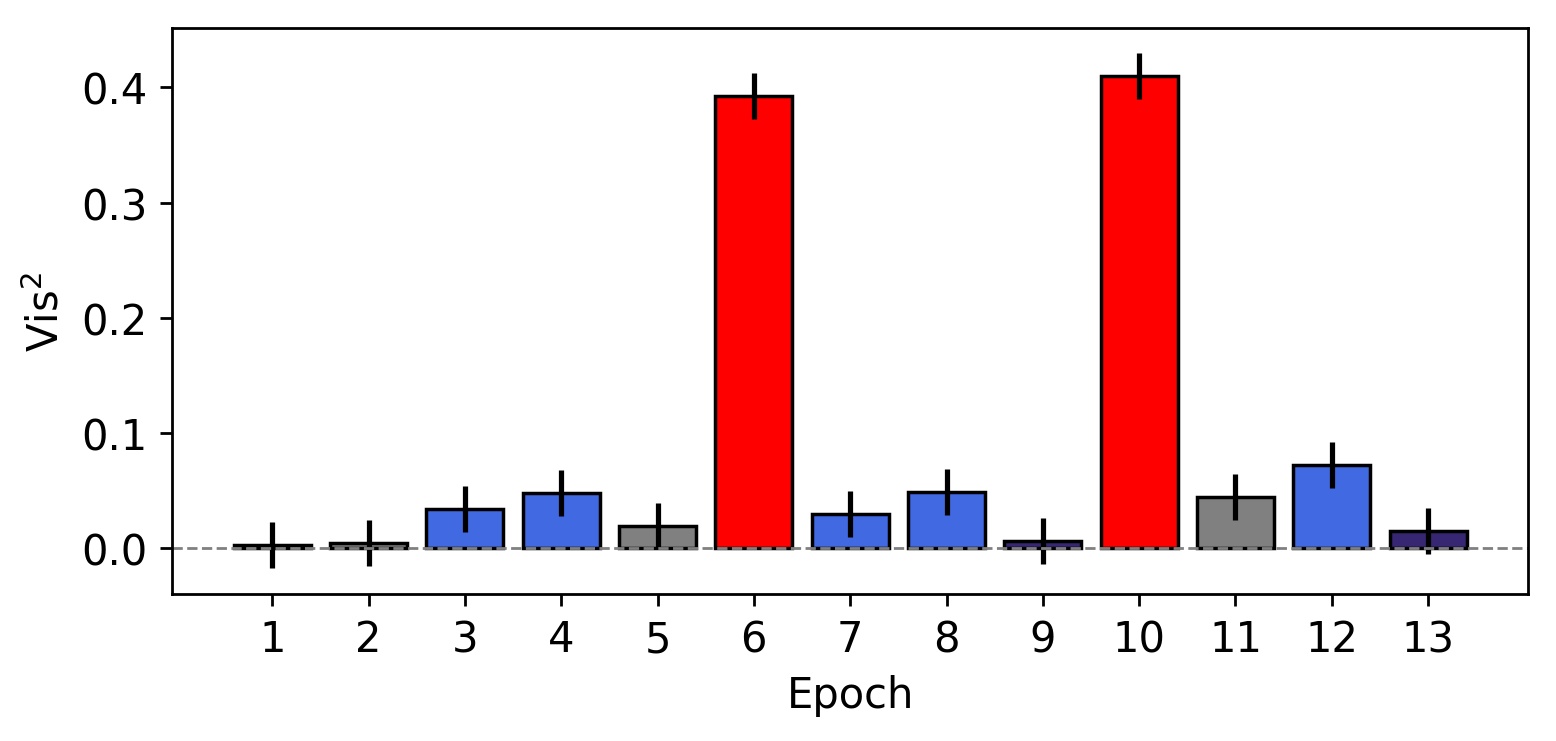}
    \caption{\object{HD\,98922} PIONIER and GRAVITY continuum data overview.
    Top row: 
    Overview diagram summarizing Fig.~\ref{fig:PIONIER-Data+Mod} and Fig.~\ref{fig:GRAVITY-FT-Data} and showing array configurations with comparable \textit{(u,$\varv$)} coverage. 
    Each line shows the spatial frequency and PA probed by the longest baseline of a given four-telescope configuration. 
    Two epochs have the same color and line style if they show a similar \textit{(u,$\varv$)} coverage. 
    The gray lines  refer to epochs for which the \textit{(u,$\varv$)} coverage cannot be reasonably compared to any other. Numbers identify epochs following Table~\ref{tab:ObsLogPIONIER} and~\ref{tab:ObsLogGRAVITY}; 
    Middle row: 
    Closure phase signal per epoch and instrument. The spatial frequency chosen is the one for which a strong variation in the CP signal between epochs with a comparable \textit{(u,$\varv$)} plane coverage is observed. Bottom row: Squared visibilities signal for each epoch and instrument, for the same spatial frequency as in the middle row. 
    }
    \label{fig:Baselines_PA}
\end{figure*}

Our data set on \object{HD\,98922} spans a 11 year time period for the continuum emission, which gives us a unique opportunity to monitor intrinsic variability effects at a scale of 1au over an extended period of time. 
The time intervals between successive epochs range from 10 to 20 months for the PIONIER data set, and from 5 to 15 months for the GRAVITY one (see Table~\ref{tab:ObsLogPIONIER} and Table~\ref{tab:ObsLogGRAVITY}). 
Signatures of temporal variability in interferometric data are ideally investigated with configurations having comparable \textit{(u,$\varv$)} plane coverage. For our data set, this is illustrated in Fig.~\ref{fig:Baselines_PA}, which provides a visual estimate of comparable $(u,\varv)$ coverage planes; although a more accurate assessment requires the comparison of the full configurations in Fig.~\ref{fig:PIONIER-Data+Mod} and Fig.~\ref{fig:GRAVITY-FT-Data}.
\\
\\
In the top row of Fig.~\ref{fig:Baselines_PA}, we report the length and PA of the longest baseline for a given configuration  for PIONIER (P)
and GRAVITY (G). The nomenclature can be followed in Table~\ref{tab:ObsLogPIONIER} and \ref{tab:ObsLogGRAVITY}. 
Following this approach, we note qualitatively that a comparable $(u,\varv)$ plane coverage is found for the following groups: (P1,15), (P2,6), (P3,4), (P7,8,9,10,17), (P11,12), (P13,14), and (P18,19,20). Depending on the group, the difference in PA of the longest baseline varies between $\sim$5$^{\circ}$ and 20$^{\circ}$. For GRAVITY, the configurations with a similar $(u,\varv)$ plane coverage are (G3,4,7,8,12), (G6,10), and (G9,13). The configurations of the remaining epochs differ more strongly. The central and bottom rows of Fig.~\ref{fig:Baselines_PA} are constructed as follows: first, we visually identify for a group of similar configurations the common spatial frequency for which a strong variation in the closure phase (CP) signal is observed and report the CP value (Fig.~\ref{fig:Baselines_PA}, central panels). For this same spatial frequency, we report in the bottom panels the value of the squared visibility. \\
\\
For those configurations with a similar \textit{(u,$\varv$)} coverage, we are able to detect significant variations in the CP values across the epochs. For instance, a clear difference in the CP signal of $\Delta \phi_{\rm max} \approx 25^{\circ}$ is observed for G9 and G13, as well as in the medium configuration epochs G3, G4, G7, G8, G12 where the CP signals vary between $\phi_{\rm max} \approx -7^{\circ}$ and $\approx 3^{\circ}$. 
Even though the error bars are larger, the CP variability in the PIONIER data set is also clearly observed. The epochs P1 and P15 show a significant closure phase variation ($\Delta \phi_{\rm max} \approx 10^{\circ}$), as does P17 with respect to P7, P8, P9, and P10 ($\Delta \phi_{\rm max} \approx 13^{\circ}$). On the contrary, when comparing observations taken with similar \textit{(u,$\varv$)} plane coverage, we note that the variability of the corresponding squared visibilities is typically smaller than $V^2$\,$\sim$\,0.05.\\

\clearpage

\subsection{Fit dependence on the \textit{(u,v)} plane coverage}
\label{ssec:AsproTest}

\begin{table}[t]
\centering
\caption{Results of the fit of the \texttt{Aspro} synthetic data for the small, medium, and large configurations of GRAVITY. Uncertainties are reported as 1$\sigma$ errors from the 16$^{th}$ and 84$^{th}$ percentile of the MCMC marginal distributions. The column {\it All} refers to the simultaneous fit of the three data sets combined. F$_c$ is not a free parameter, but is obtained according to $F_c = 1-F_h -F_s$.}
\begin{tabular}{lllll} 
\hline
\hline
& All & Small & Medium & Large\\
\hline
Parameter & Value & Value  & Value & Value  \\
\hline
F$_s$ [\%]      & $ 22.8^{+0.1}_{-0.1}$   & $ 42.3^{+1.3}_{-2.4}$   & $ 29.9^{+0.9}_{-1.9}$   & $ 16.7^{+0.4}_{-0.4}$ \\[1ex]
F$_h$ [\%]     & $ 0.3^{+0.1}_{-0.1}$    & $ 0.9^{+0.2}_{-0.3}$    & $ 0.3^{+0.7}_{-0.2}$    & $ 33.6^{+3.7}_{-4.2}$ \\[1ex]
F$_c$ [\%]     & $ 76.9^{+0.2}_{-0.2}$   & $ 56.8^{+1.5}_{-2.7}$   & $ 69.8^{+1.6}_{-2.1}$   & $ 49.7^{+4.1}_{-4.6}$ \\[1ex]
a$_r$ [mas]    & $ 3.46^{+0.01}_{-0.01}$ & $ 4.52^{+0.20}_{-0.15}$ & $ 3.76^{+0.30}_{-0.29}$ & $ 3.52^{+0.06}_{-0.06}$ \\[1ex]
a$_k$ [mas]   & $ 1.75^{+0.01}_{-0.01}$ & $ 0.25^{+0.40}_{-0.08}$ & $ 1.05^{+0.23}_{-0.19}$ & $ 1.59^{+0.07}_{-0.07}$ \\[1ex]
\textit{i} [deg]   & $ 10.7^{+1.0}_{-1.1}$   & $ 11.8^{+10.8}_{-8.1}$  & $ 8.5^{+5.2}_{-4.1}$    & $ 15.1^{+3.2}_{-4.4}$ \\[1ex]
PA [deg]     & $ 0.9^{+1.6}_{-0.7}$    & $ 37.0^{+48.9}_{-33.8}$ & $ 1.8^{+7.6}_{-1.5}$    & $ 2.5^{+4.5}_{-1.8}$ \\[1ex]
$f$Lor     & $ 0.01^{+0.01}_{-0.01}$ & $ 0.11^{+0.24}_{-0.10}$ & $ 0.49^{+0.16}_{-0.33}$ & $ 0.08^{+0.11}_{-0.06}$ \\[1ex]
$c_1$        & $ 0.61^{+0.01}_{-0.01}$ & $ 0.50^{+0.17}_{-0.17}$ & $ 0.52^{+0.10}_{-0.12}$ & $ 0.52^{+0.04}_{-0.04}$ \\[1ex]
$s_1$      & $ 0.03^{+0.02}_{-0.01}$ & $ 0.08^{+0.19}_{-0.10}$ & $ 0.04^{+0.16}_{-0.06}$ & $ 0.01^{+0.05}_{-0.02}$ \\[1ex]
$\chi^2_r$ & $3.0 \pm 0.1$ & $2.3 \pm 0.1$ & $4.6 \pm 0.4$ & $2.6 \pm 0.2$ \\[1ex]
\hline
\end{tabular}
\label{tab:Aspro-models}
\end{table}

\begin{figure}[t]
\centering
\includegraphics[width=0.32\columnwidth]{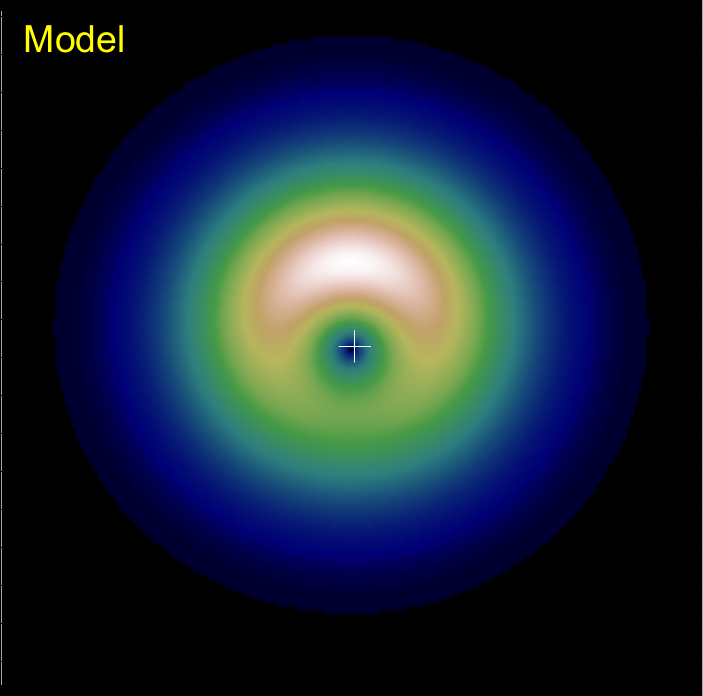}
\includegraphics[width=0.32\columnwidth]{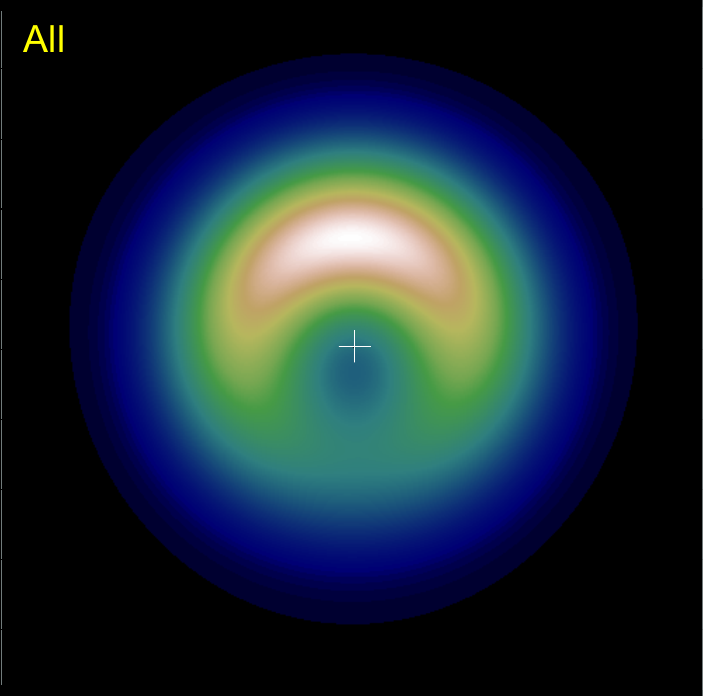}
\includegraphics[width=0.32\columnwidth]{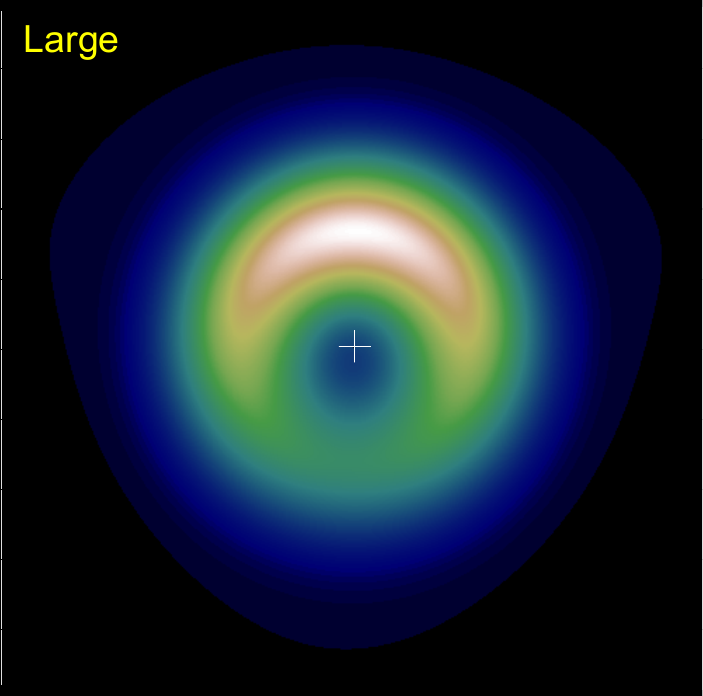}\\[0.1em]
\includegraphics[width=0.32\columnwidth]{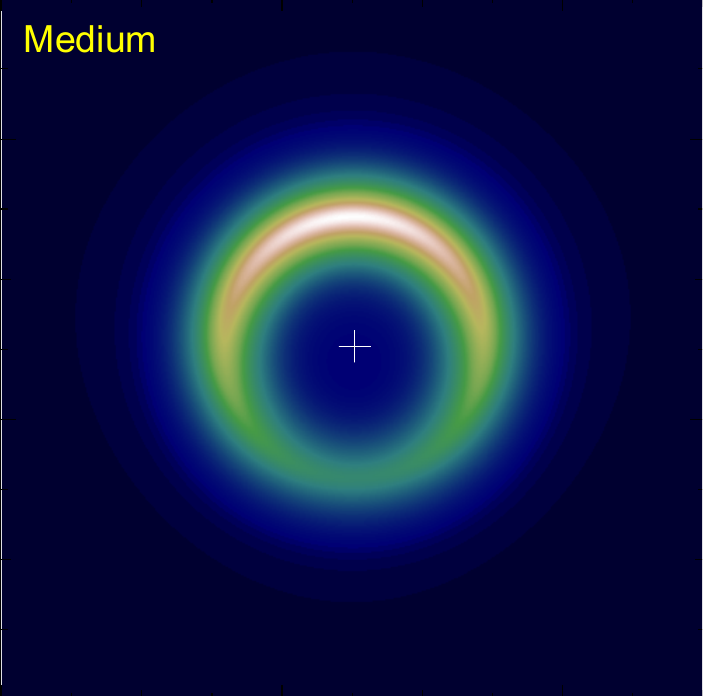}
\includegraphics[width=0.32\columnwidth]{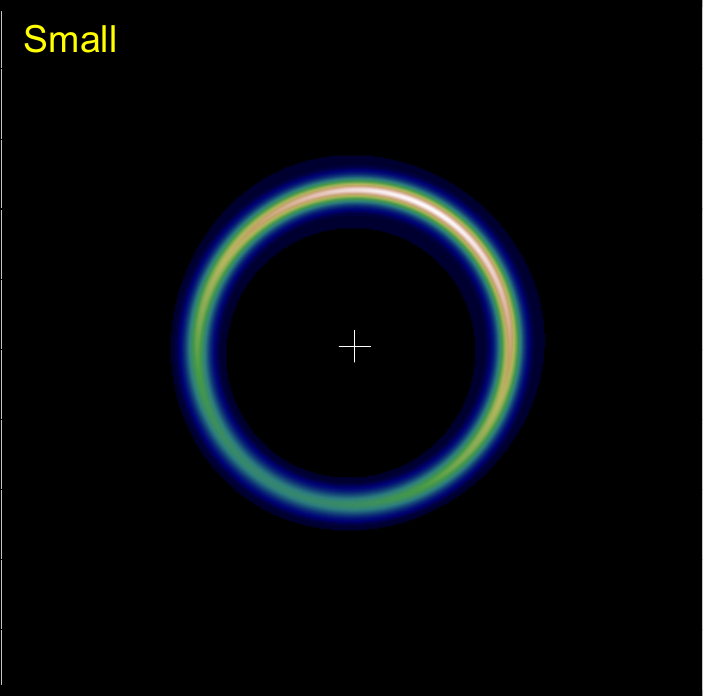}
\caption{\texttt{Aspro} synthetic data test images for the small, medium, and large configurations of GRAVITY. 
The top-left model was used to generate the synthetic visibilities and closure phases with \texttt{Aspro}. The resulting fitted models are shown for the respective configurations. 
The intensity scale is linear and ranges from zero to the peak pixel value of the asymmetry. 
The central star is not displayed and its position is marked with the white cross. 
The images have a size of 20$\times$20\,mas. It is visible from this figure and Table~\ref{tab:Aspro-models} that the small configuration properly constrains the absence of halo contribution in the system, but not the disk morphology, as opposed to the large configuration.
}
\label{fig:synthetic-imagets}
\end{figure}

\begin{figure*}[!ht]
    \centering
    \includegraphics[scale=0.38]{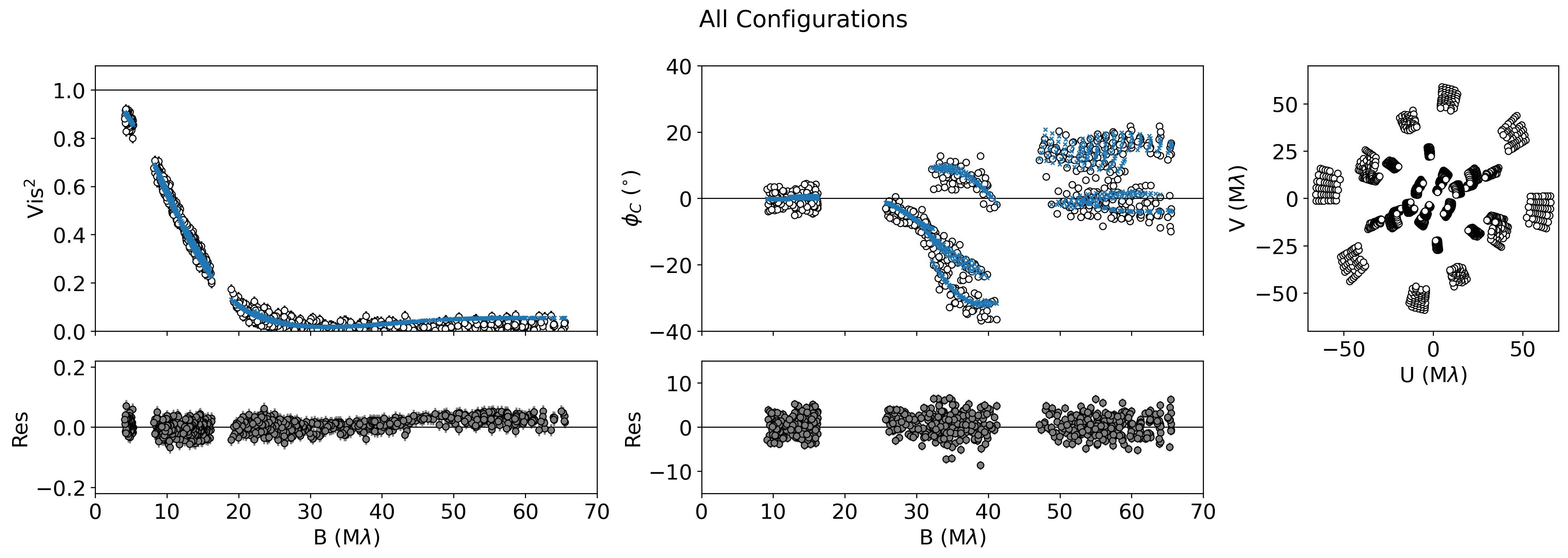}
    \includegraphics[scale=0.38]{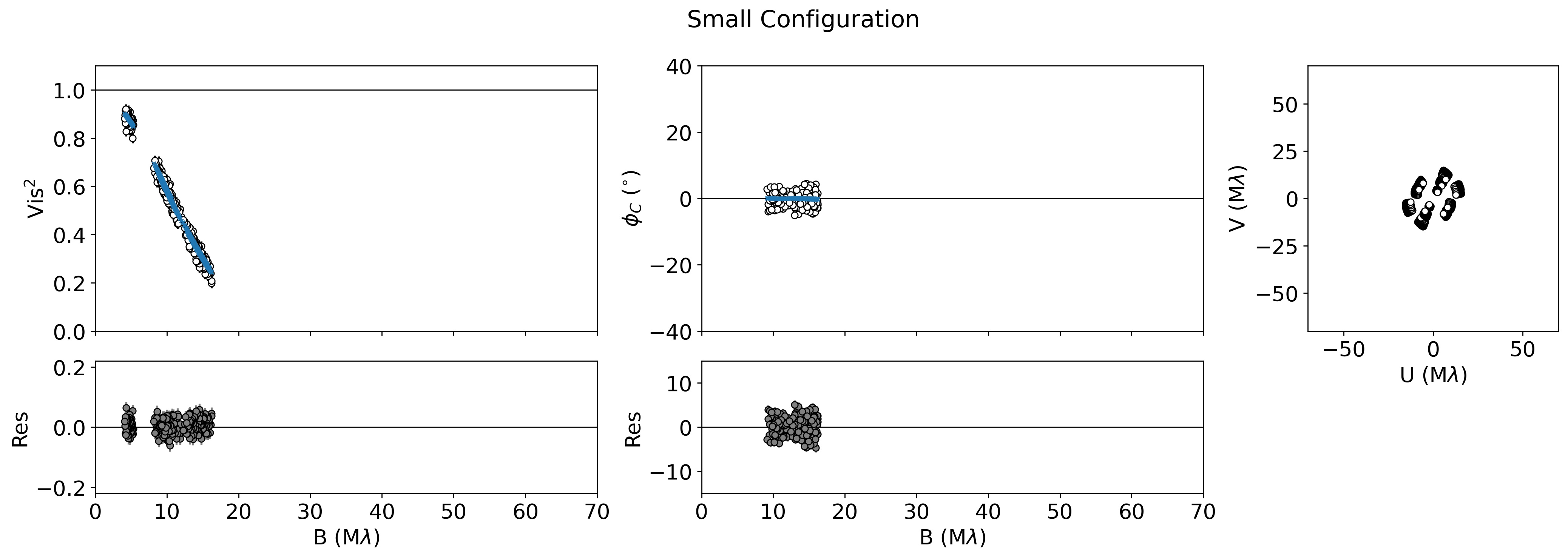}
    \includegraphics[scale=0.38]{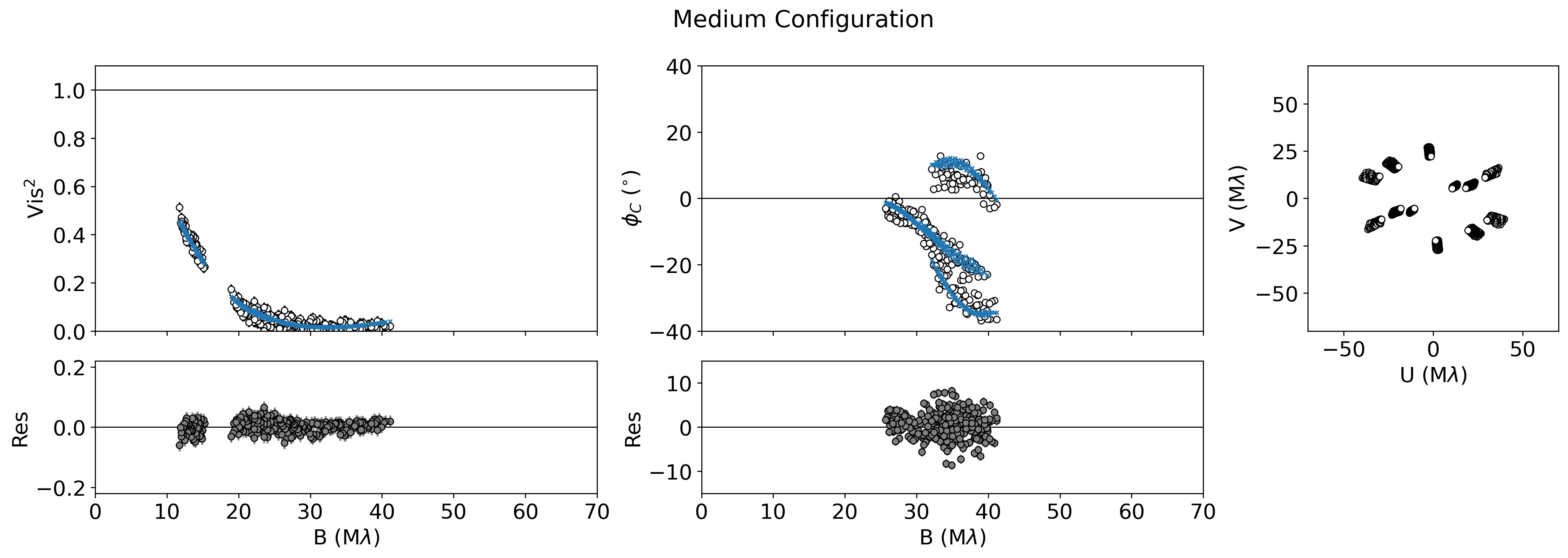}
    \includegraphics[scale=0.38]{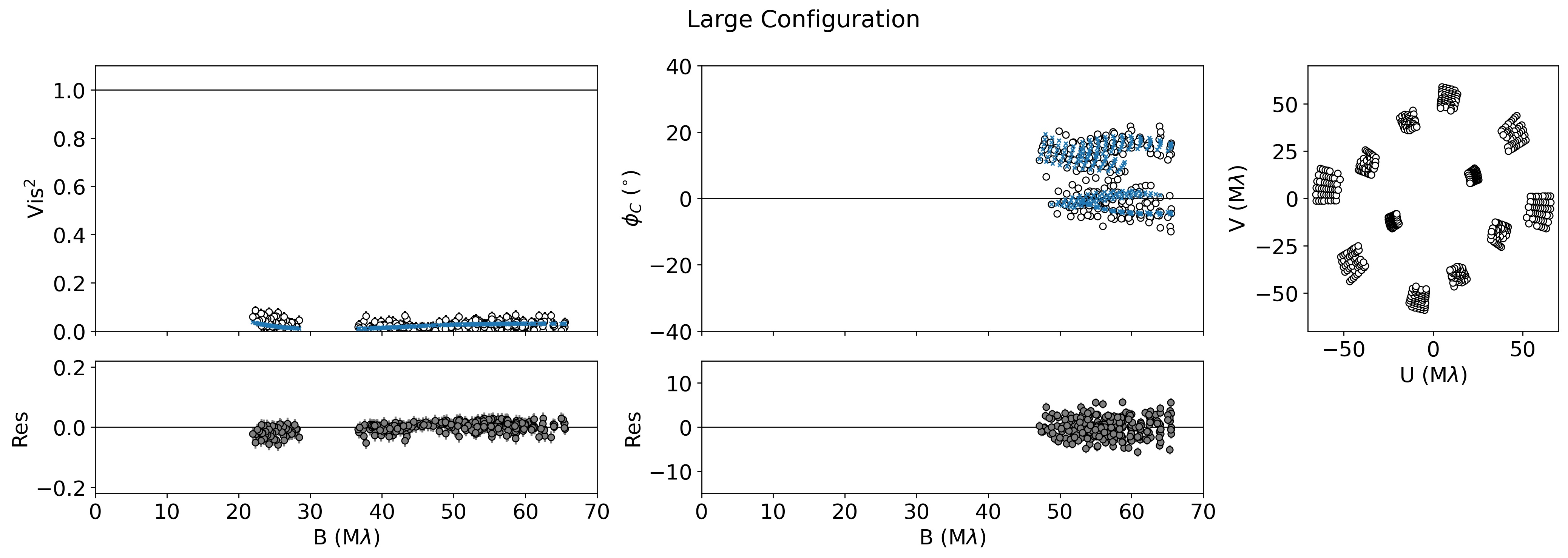}
    \caption{\texttt{Aspro} synthetic data (white markers), Vis$^2$, closure phases, and \textit{(u,$\varv$)} plane (left, center, and right panel, respectively) for the different baseline configurations. Blue markers represent the model described in Sect.~\ref{ssec:AsproTest}. Bottom panels show the residuals of the fitting process.}
    \label{fig:Aspro-Test}
\end{figure*}

In the context of the study of spatial variability, we first verified using synthetic data whether the fit of different data sets related to the same nonvariable system but obtained with different baseline configurations would lead to the same or to a different solution.
For this purpose, we generated with \texttt{Aspro}$\footnote{Available at \texttt{http://www.jmmc.fr/aspro}}$ three synthetic data sets from an input image formed by a star and an azimuthally modulated ring observed with the GRAVITY small, medium, and large configurations, respectively, on the  4$^{\rm }$ February 2020, 11$^{\rm }$ July 2019, and 14$^{\rm }$ February 2022. 
We then fitted the different synthetic data sets with the same parametric model and then compared the results obtained for each parameter.\\
The input model image, shown in Fig.~\ref{fig:synthetic-imagets}, corresponds to an azimuthally modulated torus around a central star that contributes 15\% of the  total flux of the system, and with no spatially resolved halo contribution. 
The synthetic interferometric visibilities and closure phases generated with \texttt{Aspro} are shown in Fig.~\ref{fig:Aspro-Test}. The same error bars as for the real data are adopted. 
The fit of the synthetic data implements the nine {\it free} parameters given in Table~\ref{tab:Aspro-models}. The spectral indices of the star and the disk were not implemented and were set to zero. 
Four cases are explored, corresponding to the individual small, medium, and large configurations plus a case where all the three configurations are combined. \\
From the results in Table~\ref{tab:Aspro-models}, we find that the flux parameters (F$_s$, F$_h$, F$_c$) are the most affected by some spurious variability due to the varying \textit{(u,$\varv$)} coverage, with differences of up to $\sim$30\,\%. In addition, the stellar contribution is slightly (large configuration) to significantly (small configuration) overestimated with respect to the input model. The geometrical parameters (a$_r$, a$_k$) describing the characteristic size are found to be more in agreement (though not necessarily within the error bars), except for the small configuration that only loosely constrains the solution. The inclination value is consistent within the error bars between the four cases, although modestly constrained by the small and medium configurations. Also, the results are found to be generally consistent with a low-inclination disk. 
The same conclusion applies to the PA, with the small configuration expectedly failing to constrain this parameter.
Finally, for the parameters $c_1$ and $s_1$ describing the azimuthal modulation, the fit converges towards very similar values for all four cases.\\
Simply accounting for the error bars on the fitted parameters, the modeling of the small-, medium-, and large-configuration data sets results in a consistent value for the ring inclination, the PA, $c_1,$ and $s_1$, but clearly these parameters are more stringently constrained when going from the small to the large configuration.
In Fig.~\ref{fig:synthetic-imagets}, we compare the input model image to the parametric models resulting from the fit of the different configurations. Besides the differences found in the relative flux contributions (see Table~\ref{tab:Aspro-models}), we can visually observe that the size properties are not in good agreement ---in particular for the small configuration--- when fitting the data of the different configurations  separately, although the general structure is retrieved. In contrast, the azimuthal position of the disk asymmetry appears to be well constrained. 
This simple test reminds us that the intrinsic sparsity of infrared interferometric data implies that different spatial frequencies are probed with different configurations, which may result in a spurious variability of the fitted parameters that may not reflect a physical temporal variability of the system. \\
Based on this observation, we adopted here a different strategy to more
robustly test the potential time-variability of the different parameters describing our model: we fitted the full PIONIER (respectively GRAVITY) continuum data set by forcing all the free parameters to be {constant} across the different epochs, except the parameter for which we wish to test the variability hypothesis. Throughout the paper we refer to this approach as the \textit{x global fit}, where \textit{x} is the selected time-variable parameter. This is motivated by the fact that a rich \textit{(u,$\varv$)} coverage remains desirable to constrain our model.

\clearpage

\section{Visualisation of the continuum geometrical models}\label{sec:visualisation}

\begin{figure*}
\centering
\includegraphics[width=\textwidth]{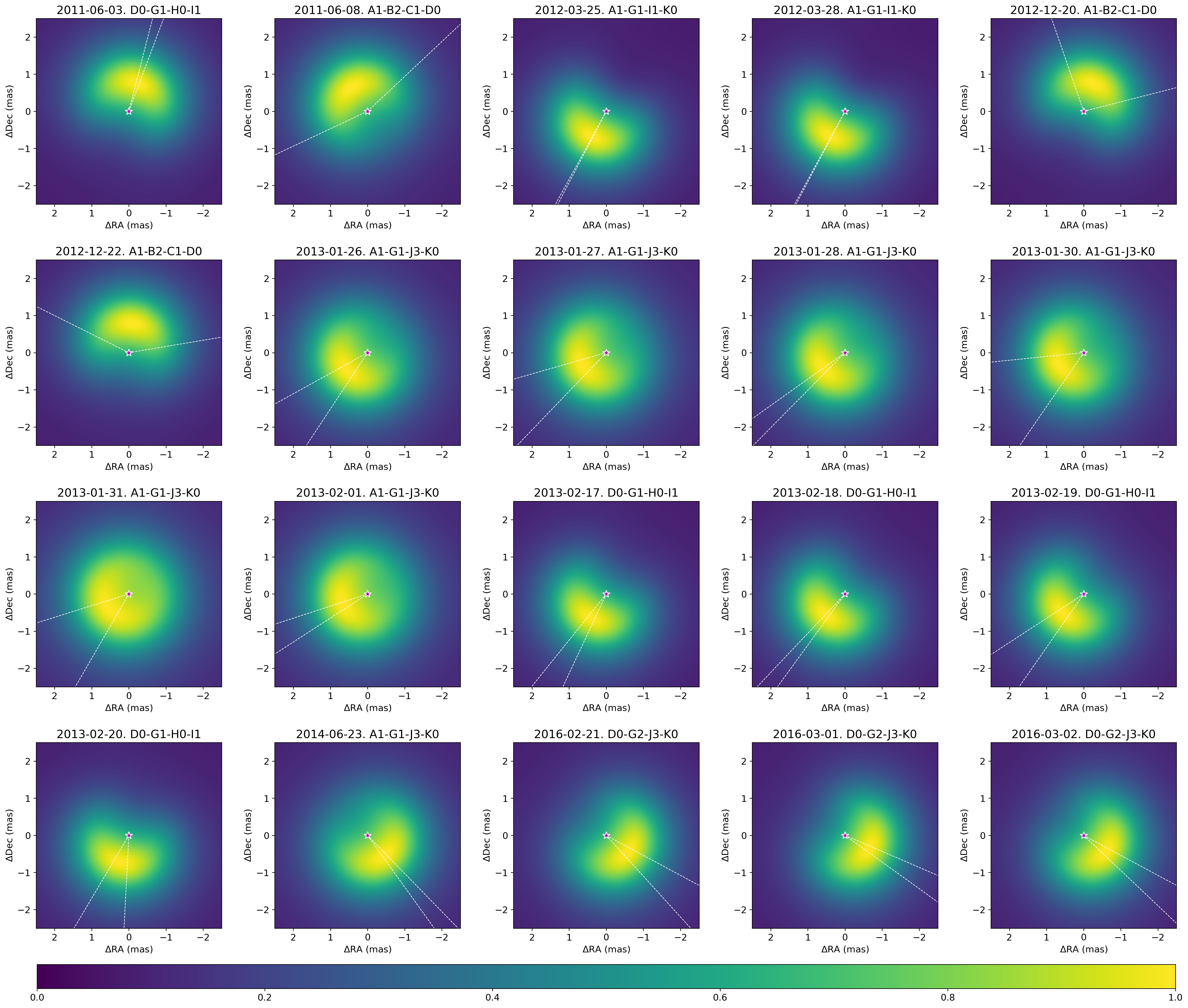}
\caption{Peak-normalized PIONIER continuum model images. The dashed white lines represent the $\pm$3$\sigma$ MCMC uncertainties on the PA of the azimuthal modulation. 
A description of the model is given in Sect.~\ref{sec:continuum_analysis}. The central object is not displayed in order to enhance the circumstellar emission.
The contrast reported in Sect.~\ref{subsubsec:disk-asymm} is calculated from the peak-normalized brightness distribution map as the ratio between the brightest pixel in the map and the corresponding centro-symmetric position in the disk. We report the standard deviation for the mean contrast value across the different PIONIER epochs.
}
\label{fig:PIONIER-Continuum-imgs_Full}
\end{figure*}

\begin{figure*}
\centering
\includegraphics[width=\textwidth]{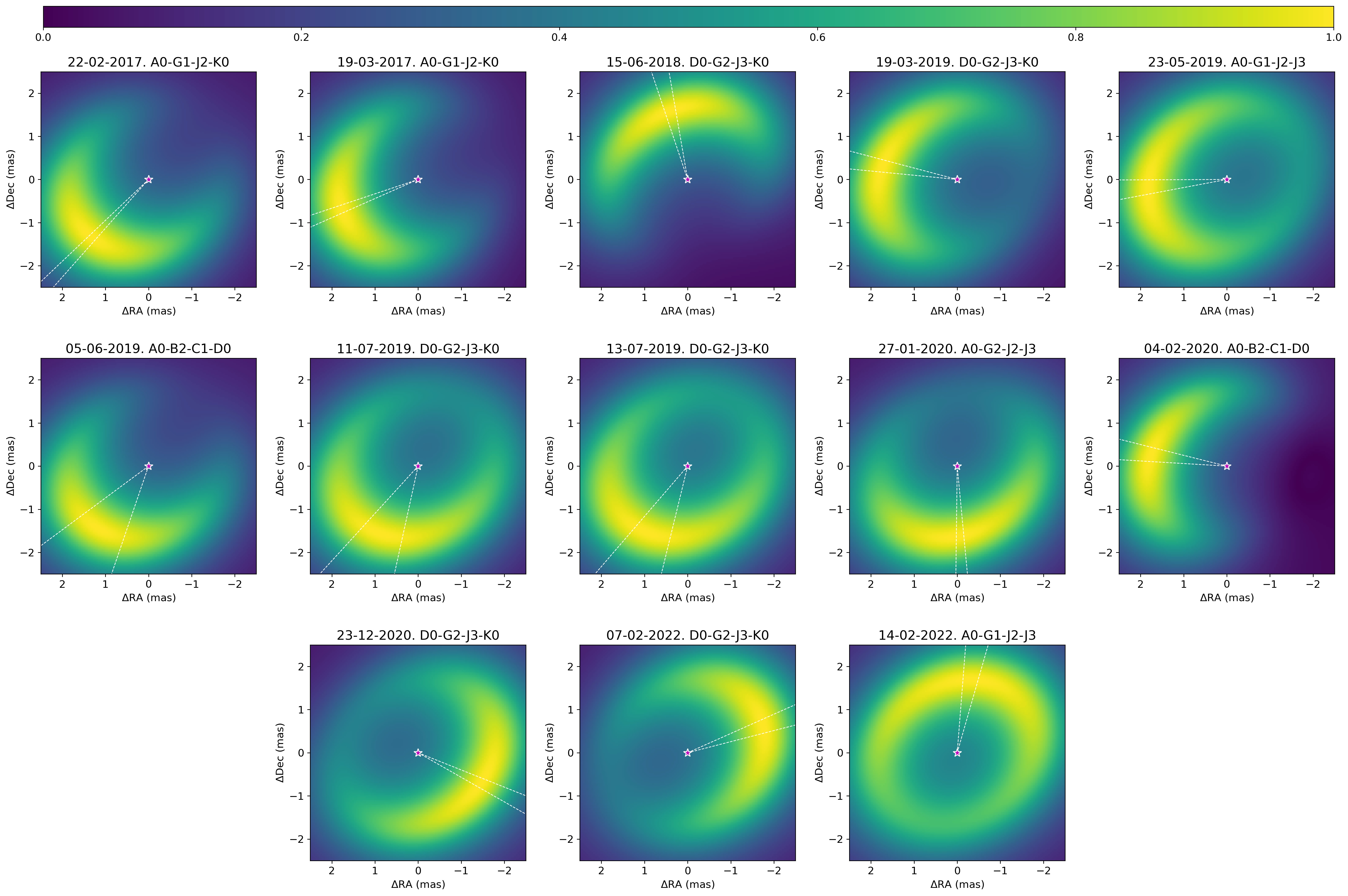}
\caption{Same as Fig.~\ref{fig:PIONIER-Continuum-imgs_Full} but for the GRAVITY data set.}
\label{fig:GRAVITY-Continuum+line-imgs_Full}
\end{figure*}


\clearpage

\section{Pure-line photocenter displacements}\label{apx:brg-visualisation}

\begin{figure*}
    \centering
    \includegraphics[width=0.32\textwidth]{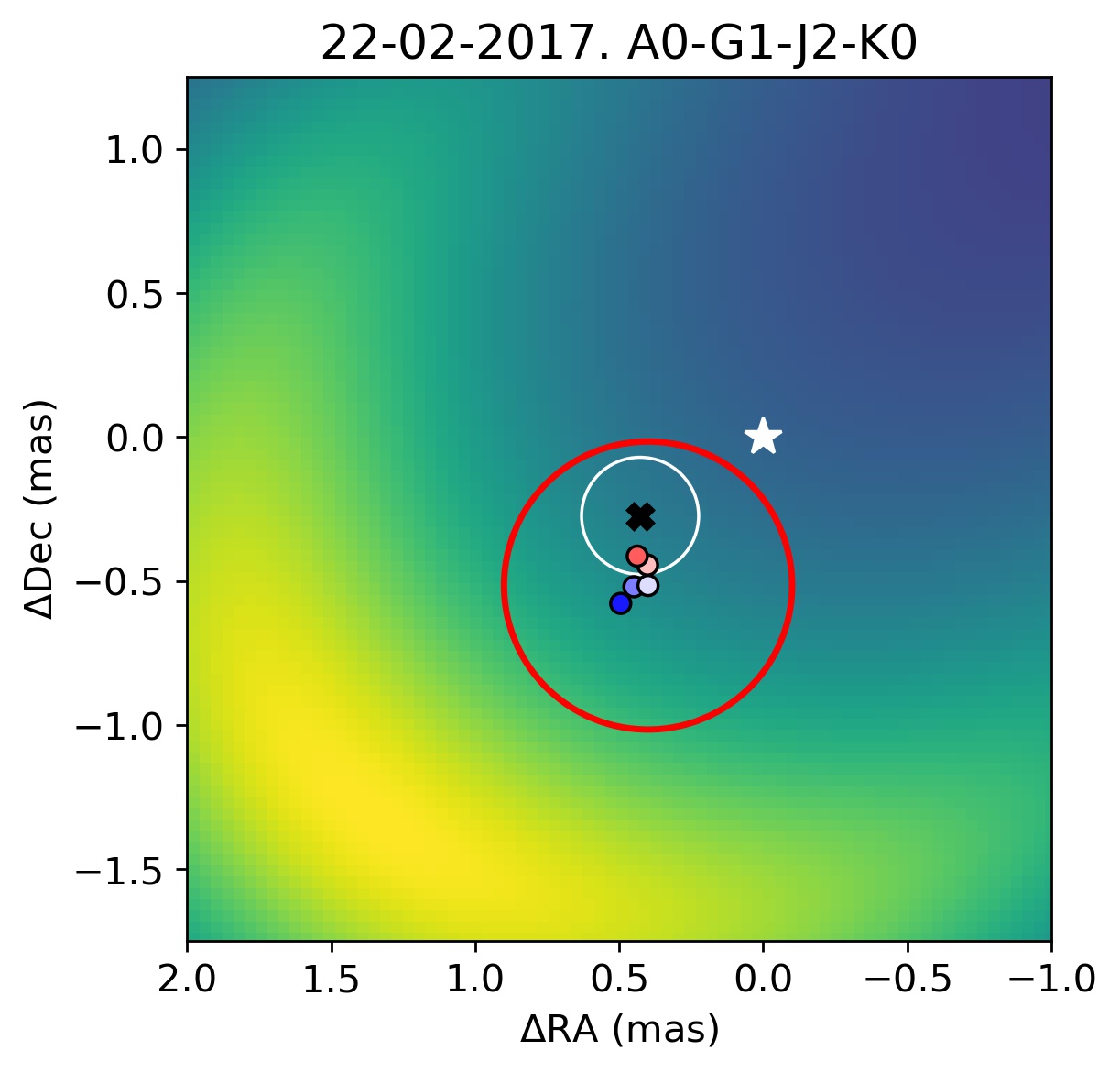}
    \includegraphics[width=0.32\textwidth]{Figures/Gas/PCenter_imgs/HD98922_GRAVITY_Gas_2_L.jpg} 
    \includegraphics[width=0.32\textwidth]{Figures/Gas/PCenter_imgs/HD98922_GRAVITY_Gas_3_L.jpg} \\
    \includegraphics[width=0.32\textwidth]{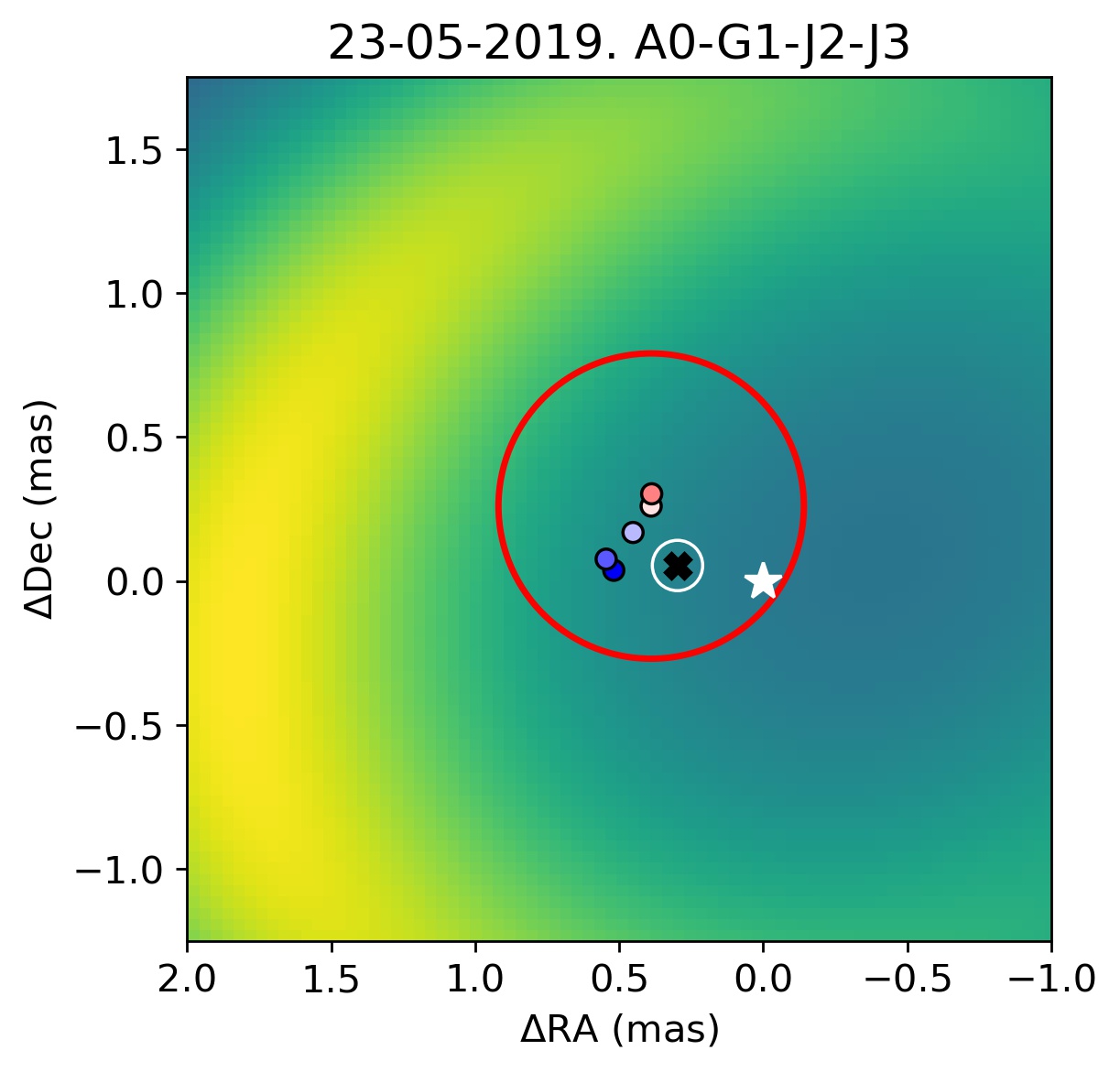} 
    \includegraphics[width=0.32\textwidth]{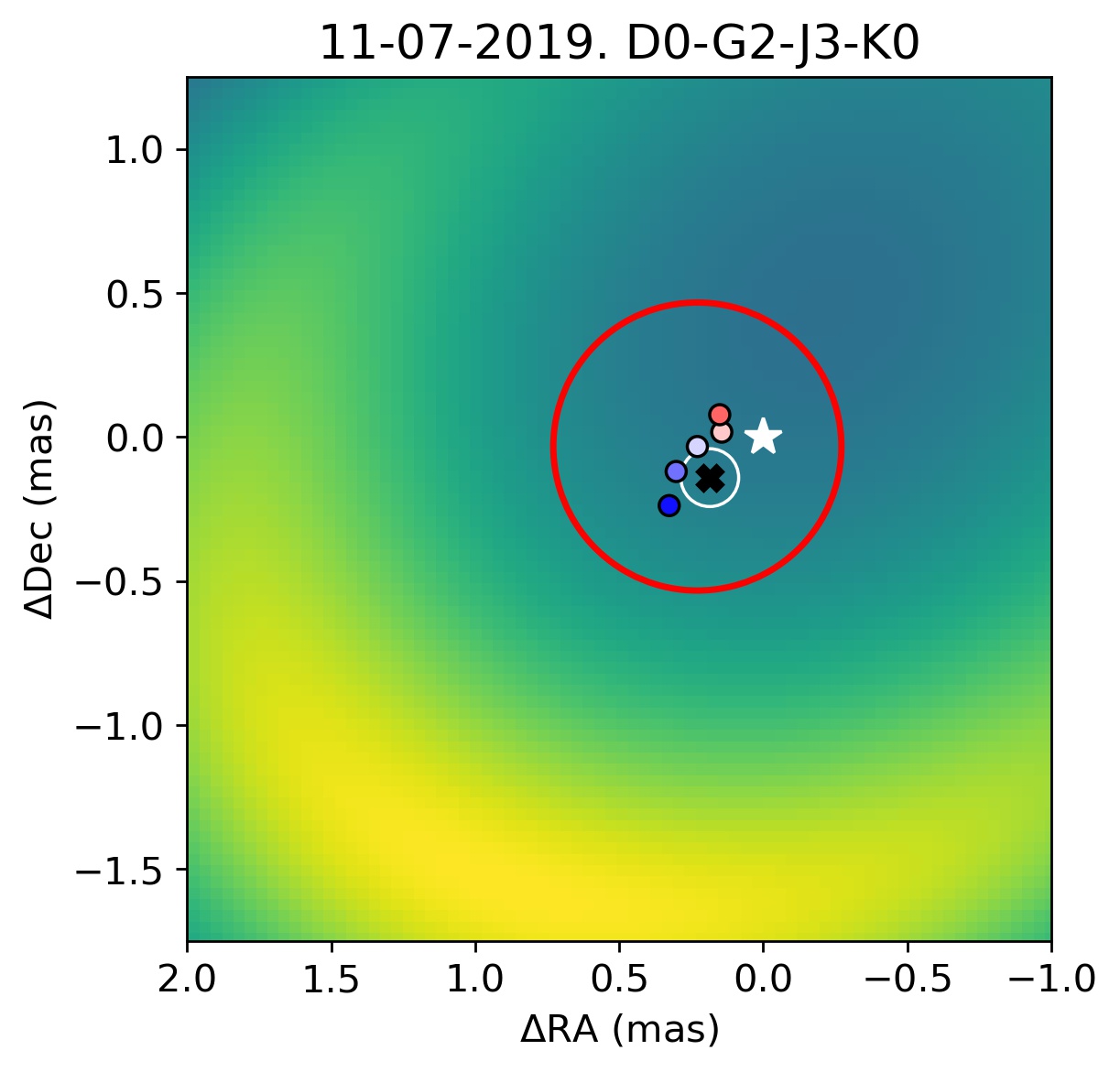}
    \includegraphics[width=0.32\textwidth]{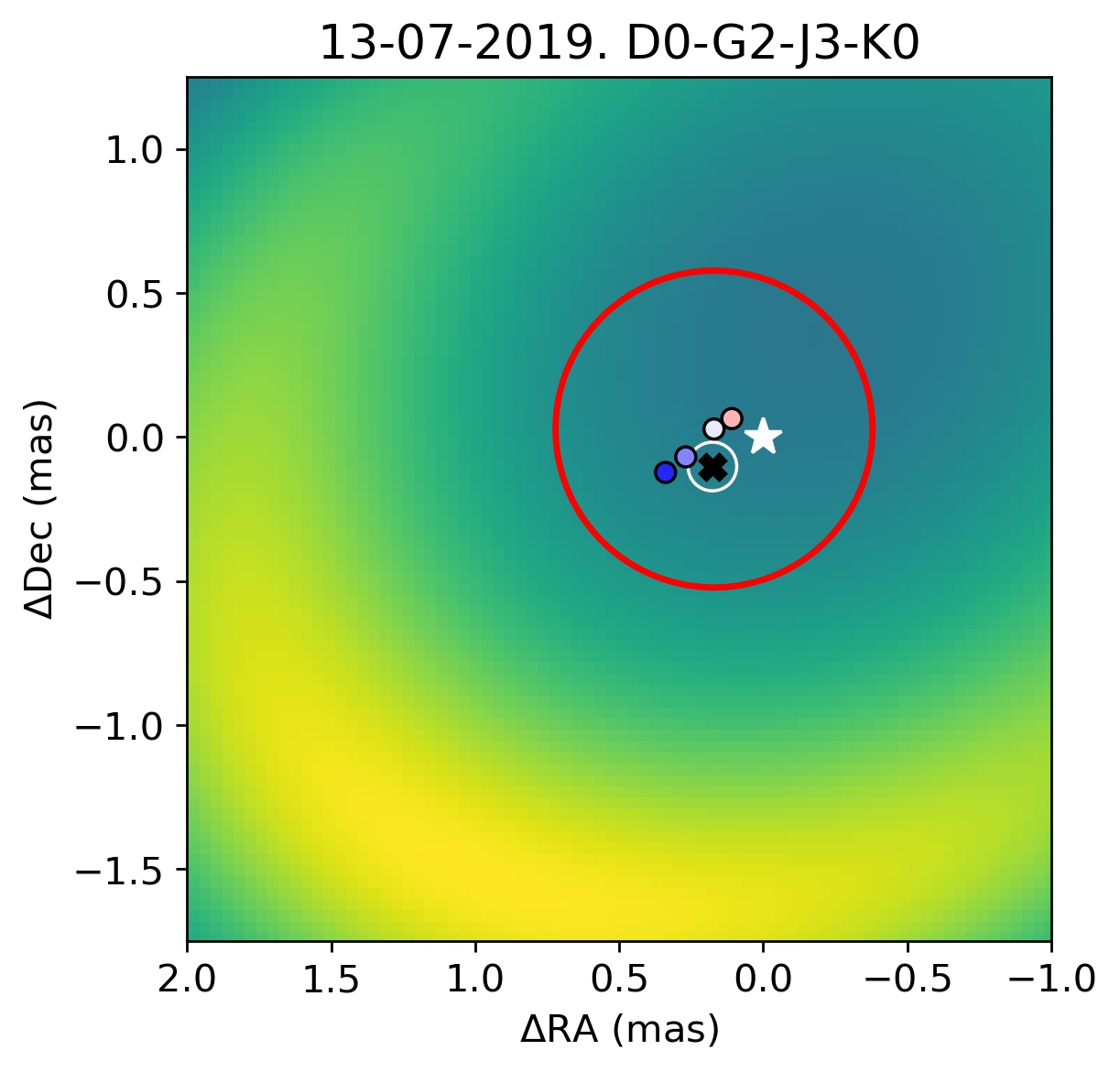} \\
    \includegraphics[width=0.32\textwidth]{Figures/Gas/PCenter_imgs/HD98922_GRAVITY_Gas_7_L.jpg}
    \includegraphics[width=0.32\textwidth]{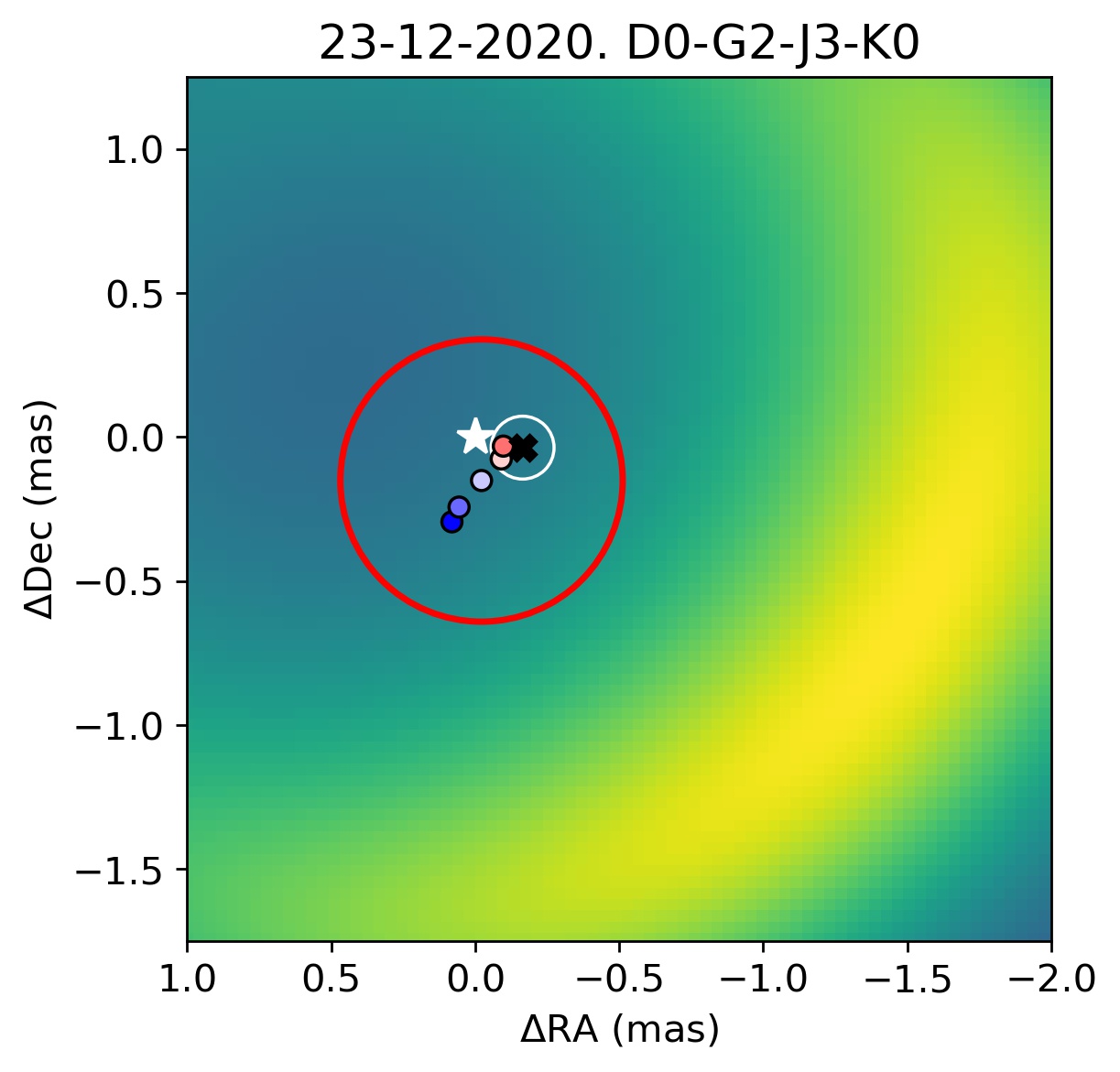} 
    \includegraphics[width=0.32\textwidth]{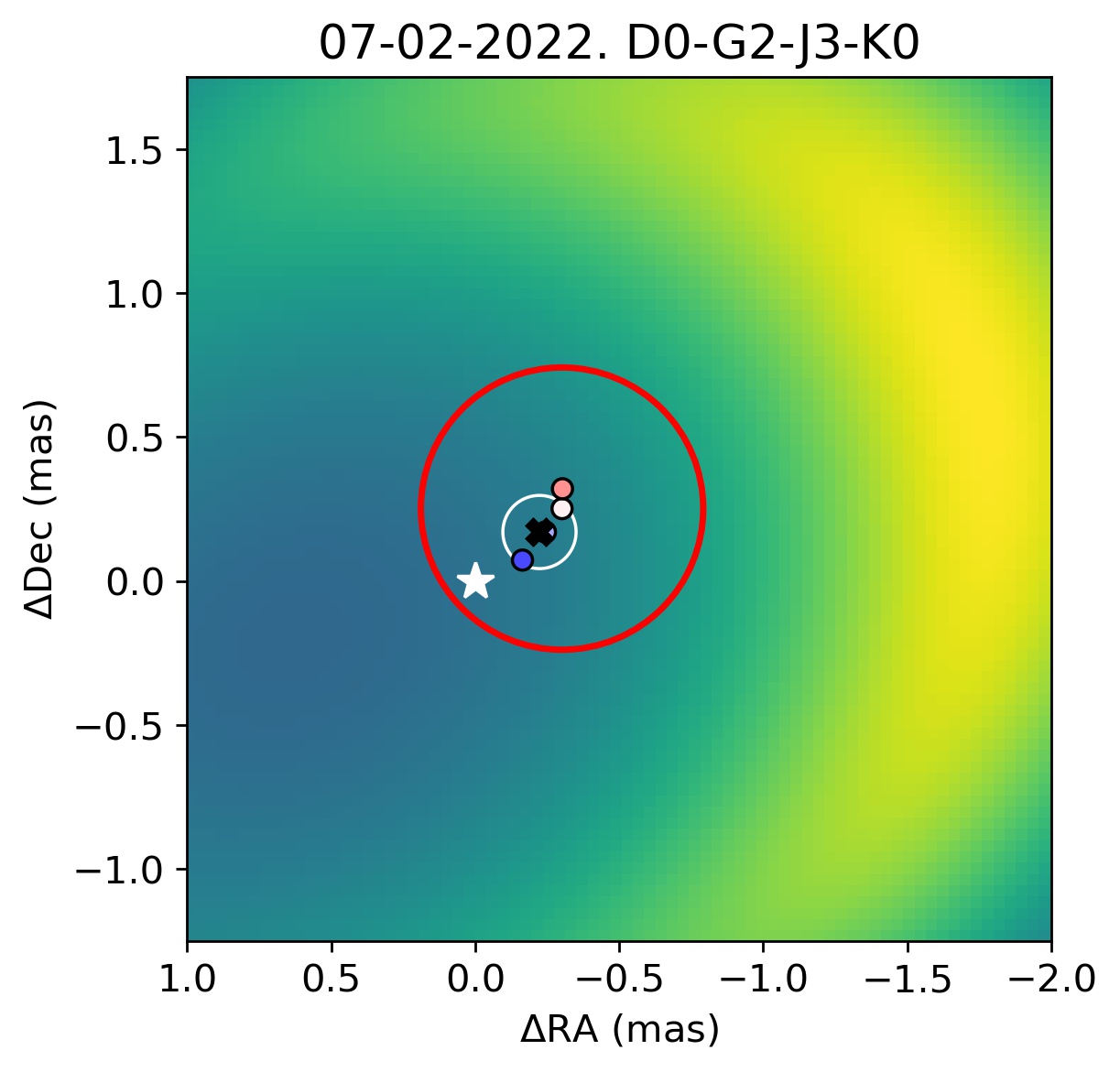} \\
    \includegraphics[width=0.32\textwidth]{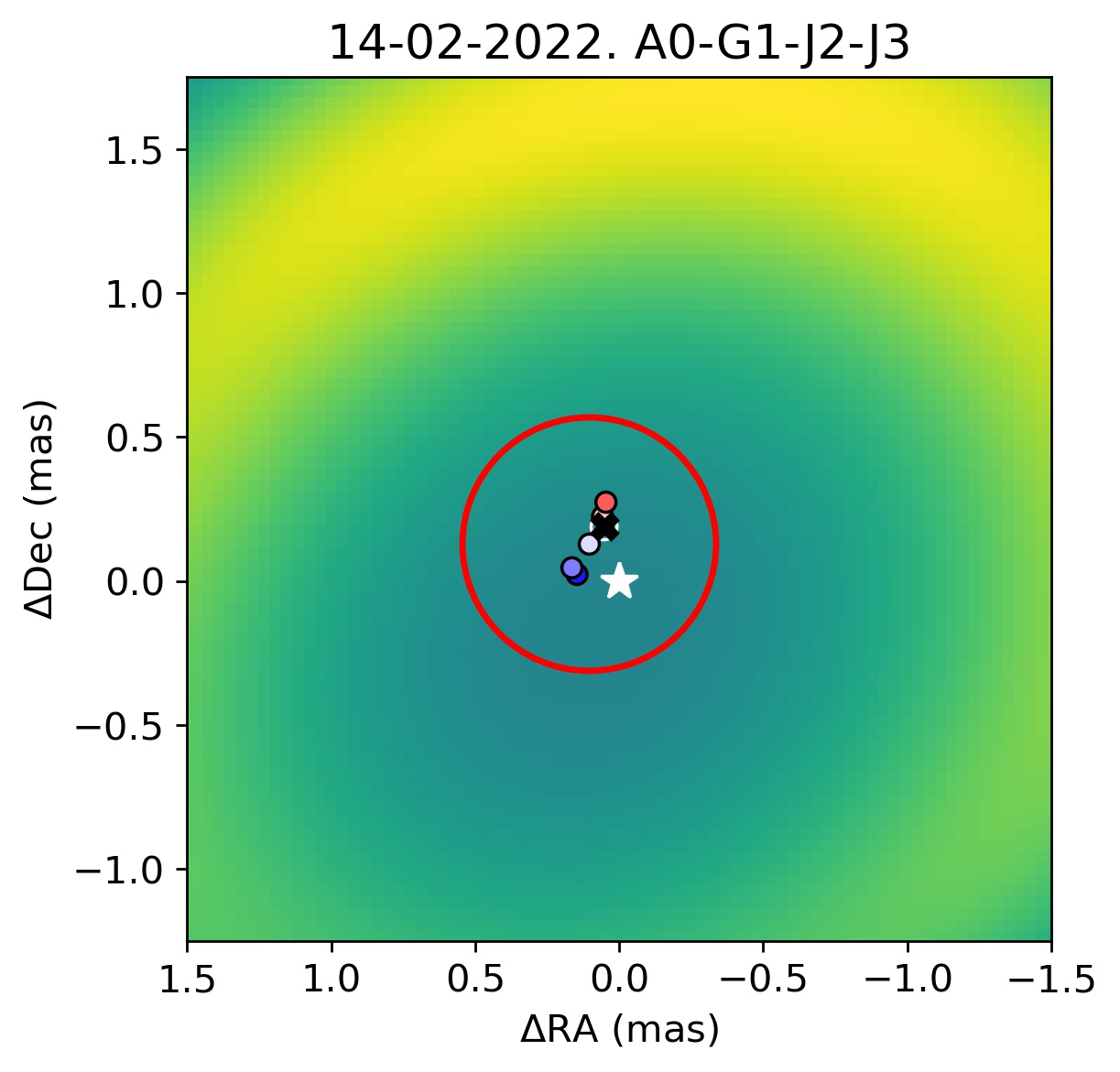}
    \caption{GRAVITY continuum model images and Br$\gamma$-line-emitting gas region. The black filled dot shows the continuum (star + halo + disk) photocenter. The white circle around that point gives the error on the continuum photocenter position estimated from the continuum modeling. The red circle represents the gas region size estimated at the peak emission and is centered on the $\sim$\,0\,km\,s$^{-1}$ point. The blue to red colored filled dots show the gas photocenters for different spectral channels, as in Fig.~\ref{fig:GRAVITY-SC-Data}.}
    \label{fig:BrG_PhCenters}
\end{figure*}

\end{document}